\documentclass{aastex62}
\usepackage{amsmath}
\usepackage{amssymb}
\usepackage{graphicx}
\usepackage{epstopdf}
\usepackage{bbold}
\usepackage{mathtools}
\usepackage[caption=false]{subfig}
\usepackage{xcolor}
\usepackage{hyperref}

\newcommand{\ikt}{{\it Kepler}}
\newcommand{\ik}{{\it Kepler~}}

\newcommand{\gaia}{{\it Gaia~}}
\newcommand{\gaiat}{{\it Gaia}}

\newcommand{\teff}{\ensuremath{T_{\rm eff}}}
\newcommand{\logg}{\ensuremath{\log{g}}}
\newcommand{\feh}{[Fe/H]}
\newcommand{\rstar}{\ensuremath{R_\star}}
\newcommand{\mstar}{\ensuremath{M_\star}}
\newcommand{\rsun}{R\ensuremath{_\odot}}
\newcommand{\msun}{M\ensuremath{_\odot}}

\newcommand{\rprs}{\ensuremath{R_{\rm p}/R_\star}}
\newcommand{\rhostar}{\ensuremath{\rho_\star}}
\newcommand{\rhoc}{\ensuremath{\rho_{\star {\rm c}}}}
\newcommand{\rpl}{\ensuremath{R_{\rm p}}}

\newcommand{\adrs}{\ensuremath{a/R_\star}}

\def\pnas{PNAS}

\received{2022 May 10}
\revised{2023 September 30}
\accepted{2023 October 2}
\submitjournal{\it The Planetary Science Journal}

\shorttitle{Orbital Periods and Other Properties of \ik Planet Candidates}
\shortauthors{Lissauer et al.}

\begin{document}

\title{Updated Catalog of \ik Planet Candidates: Focus on Accuracy and Orbital Periods }


\correspondingauthor{Jack J. Lissauer}
\email{ Jack.Lissauer@nasa.gov }

\author[0000-0001-6513-1659]{Jack J. Lissauer}
\affiliation{Space Science \& Astrobiology Division\\
MS 245-3\\
NASA Ames Research Center \\
 Moffett Field, CA 94035, USA}

\author[0000-0002-5904-1865]{Jason F. Rowe}
\affil{Department of Physics and Astronomy\\
Bishops University \\
2600 Rue College \\
Sherbrooke, QC J1M 1Z7,
Canada}

\author[0000-0002-6227-7510]{Daniel Jontof-Hutter}
\affiliation{Department of Physics\\
University of the Pacific\\ 
601 Pacific Avenue\\
Stockton, CA 95211, USA}

\author[0000-0003-3750-0183]{Daniel C. Fabrycky}
\affiliation{Department of Astronomy and Astrophysics\\
University of Chicago\\
5640 South Ellis Avenue\\
Chicago, IL 60637, USA}

\author[0000-0001-6545-639X]{Eric B. Ford}
\affiliation{Department of Astronomy and Astrophysics\\
525 Davey Laboratory\\
The Pennsylvania State University\\ 
University Park, PA 16802, USA}
\affiliation{Center for Exoplanets \& Habitable Worlds\\
525 Davey Laboratory\\
The Pennsylvania State University\\ 
University Park, PA 16802, USA}
\affiliation{Department for Astrostatistics\\
525 Davey Laboratory\\
The Pennsylvania State University\\ 
University Park, PA 16802, USA}
\affiliation{Institute for Computational \& Data Sciences\\
The Pennsylvania State University\\ 
University Park, PA 16802, USA}

\author[0000-0003-1080-9770]{Darin Ragozzine}
\affiliation{Department of Physics and Astronomy  \\
Brigham Young University\\ 
Provo, UT 84602, USA}

\author[0000-0003-2202-3847]{Jason H. Steffen}
\affiliation{Department of Physics and Astronomy\\
University of Nevada Las Vegas\\
4505 South Maryland Parkway\\
Las Vegas, NV 89154, USA}


\author[0000-0002-7217-446X]{Kadri M. Nizam}
\affiliation{Department of Physics\\
University of the Pacific\\ 
601 Pacific Avenue\\
Stockton, CA 95211, USA}
\affiliation{Department of Astronomy and Astrophysics\\
525 Davey Laboratory\\
The Pennsylvania State University\\ 
University Park, PA 16802, USA}

\begin{abstract}
We present a new catalog of \ik planet candidates that prioritizes accuracy of planetary dispositions and properties over uniformity. This catalog contains 4376 transiting planet candidates, including 1791 residing within 709 multi-planet systems, and provides the best parameters available for a large sample of \ik planet candidates. We also provide a second set of stellar and planetary properties for transiting candidates that are uniformly-derived for use in occurrence rates studies. Estimates of orbital periods have been improved, but as in  previous catalogs, our tabulated values for period uncertainties do not fully account for transit timing variations (TTVs). We show that many planets are likely to have TTVs with long periodicities caused by various processes, including orbital precession, and that such TTVs imply that ephemerides of \ik planets are not as accurate on multi-decadal timescales as predicted by the small formal errors (typically 1 part in 10$^6$ and rarely $>10^{-5}$) in the planets' measured mean orbital periods during the \ik epoch. 
Analysis of normalized transit durations implies that eccentricities of planets are anti-correlated with the number of companion transiting planets. Our primary catalog lists all known \ik planet candidates that orbit and transit only one star; for completeness, we also provide an abbreviated listing of the properties of the 
two dozen non-transiting planets that have been identified around stars that host transiting planets discovered by \ikt. 

\end{abstract}


\section{Introduction}\label{sec:intro}

NASA's \ik spacecraft monitored a single star field for four years during its prime mission, with a duty cycle of almost 90\%. The principal objective of the \ik Mission was to take a statistical census of planets having orbital periods of up to $\sim$~1 year. The \ik Project released eight catalogs of planet candidates found during the mission \citep{Borucki:2011a,Borucki:2011b,Batalha:2013,Burke:2014,Rowe:2015,Mullally:2015,Coughlin:2016,Thompson:2018}. Each new catalog used more sophisticated methods, and aside from the last one, each used more \ik data and listed more planet candidates than its predecessor. The Project's latter catalogs employed successively more automated procedures.

The primary goal of the \ik Project's final catalog of planetary candidates \citep{Thompson:2018}, often referred to as DR25 (which is an abbreviation of Data Release 25), was to produce a listing of planet candidates (PCs) found and vetted in a well-defined and reproducible manner for the exoplanet community to use as input for studies of planet occurrence rates (e.g., \citealt{Hsu:2019}).  As such, the data were processed in a highly automated manner, with uniformity and reproducibility prioritized over using all available information to identify and classify each individual potential planet signature.  {Previously found planet candidates that were not identified by the final search for transit-like patterns do not appear in the \cite{Thompson:2018} catalog. Furthermore, their vetting of individual candidates did not include the hands-on treatment (e.g., examination of lightcurves and centroid shifts by eye) that was a feature of the first six \ik planet catalogs.}

The neglect of hands-on vetting from previous catalogs ended up excluding dozens of \ik planets, (some of which are well-known and were verified with high confidence) from \cite{Thompson:2018}'s list of Kepler Objects of Interest (KOIs; the integer number refers to the target star, whereas the digits after the decimal point refer to the putative planet), and led to classifying others as false positives (FPs) rather than as viable planet candidates. One particularly relevant example is that the \cite{Thompson:2018} catalog is more biased against planets exhibiting significant transit timing variations (TTVs) than are previous catalogs.  This limitation leads to a bias against planet pairs with near-resonant orbits, {as foreseen by \cite{Garcia:2011}. An unpublished analysis of strategies for detecting planets with large TTVs by A.~Moorhead \& one of us (EBF) found that most such planets would naturally be detected by applying standard transit search algorithms to $\sim$one-year-long subsets of the data, since large near-resonant TTVs typically accumulate over multi-year timescales.  Therefore,} the use of candidates from multiple searches conducted with different amounts of data reduces the bias against planets with substantial TTVs.  {This improvement in sensitivity to planets with large TTVs (i.e., TTVs comparable to or exceeding the transit duration) is most significant for planets that would, in the absence of TTVs, likely be detected using only 13 or 16 months of data (the durations searched for the second and third \ik PC catalog releases).  }

Our new catalog is more analogous to the \ik Project's final cumulative catalog, DR25supp, which was not presented in a refereed publication\footnote{Documentation for DR25supp is available at: https://exoplanetarchive.ipac.caltech.edu/docs/PurposeOfKOITable.html\#q1-q17\_sup\_dr25.}, than to the DR25 catalog. Both the DR25supp catalog and our Table \ref{tab:planetcatalog}  differ from the DR25 catalog in two key respects: the use of manual vetting and including KOIs from multiple sources, including previous \ik project catalogs. For DR25supp, the cumulative DR24 catalog was combined with the DR25 catalog and the \ik False Positive Working Group re-dispositioned all KOIs whose dispositions were disputed  in the last six project catalogs (listed as a planet candidate in at least one catalog but listed as a false positive in at least one other); apart from dispositions, the most recent properties were listed. We began with the  DR25supp catalog, added other KOIs from various sources, and manually vetted select KOIs, as described in Section \ref{sec:selection}. 

The \ik mission defined a threshold crossing event (TCE) as a periodic signature with multiple event statistic (MES) at least 7.1, where the MES is effectively the signal-to-noise ratio of the putative planetary transits in the folded lightcurve, as measured by the \ik pipeline. The threshold of MES $\geq 7.1$ was chosen to keep the expected number of KOIs resulting from white noise small (see \S\ref{sec:selection} for details). The \ik pipeline looked for planet candidates in the lightcurve of an individual target star one by one, beginning with the signal possessing the largest MES and then performing the search again on a lightcurve from which data at the times of the transits of this KOI were removed. The process was repeated until no additional signal with MES $\geq 7.1$ was found. Because data were removed from the lightcurve prior to searching for additional candidates, the search for multis (systems with multiple transiting candidates) is less complete than that for singles (systems with only a single transiting candidate). This bias against multis is analyzed quantitatively by \cite{Zink:2019}. 

Various groups have published lists of additional \ik planet candidates; see \S\ref{sec:selection} for details.
Together with those appearing in the official \ik planet candidate tabulations, the total number of \ik planet candidates listed in one or more catalogs is $\sim$~5000, although several hundred of these have subsequently been re-classified as FPs. More than 2700 of these planet candidates have been verified (either confirmed using radial velocity data or via TTVs, or statistically validated as having a high likelihood of being true planets) and assigned official \ik planet designations such as Kepler-11 g \citep{Lissauer:2011a}. 
 
 We assembled our catalog using data from the final \ik project planet candidate catalog \citep{Thompson:2018}, previous planet candidate catalogs produced by the \ik project, and planet candidate lists from other groups (\S\ref{sec:selection}). We incorporated the improved stellar properties derived using distance measurements by ESA's {\it Gaia} spacecraft \citep{Gaia:2018,Berger:2020a} and, where available, spectroscopic measurements taken with the Keck I telescope \citep{Petigura:2017,Fulton:2018}.  Figure \ref{fig:perrad} shows the distribution of orbital period vs.~radius for our \ik planetary candidates based on the work reported herein.
 
Our goals in compiling a new catalog of \ik planet candidates are to provide a comprehensive listing of KOIs with significantly more accurate vetting and to give improved estimates of planet properties. As described in more detail in Section \ref{sec:catalog}, our new listing relies on more homogeneous and robust techniques to compute planetary parameters, removing previous biases such as the dependence on orbital period estimates of planets exhibiting TTVs with the amount of data analyzed when they were first announced. We also list more planet properties and use more robust techniques to compute the values and uncertainties of estimated planetary characteristics.

{The primary advantages of using our new catalog are: We present the most complete listing of \ik planet candidates to date, based on the \ik project's catalogs, community efforts, and our own analysis. We have provided initial dispositions for new KOIs in our sample. We have also revisited dispositions for those KOIs that were dispositioned as FP in DR25supp despite being listed as a PC in DR24 or DR25, or having a Kepler number according to NASA Exoplanet Science Institute (NExScI).   We provide additional disposition cuts based on S/N, mass measured from radial velocity variations, and derived planetary radius, $R_p$. 

The  parameter values listed in our catalog (Table \ref{tab:planetcatalog}) are more accurate than those in previous catalogs, with significant effort to systematically and uniformly improve transit models and calculations of posteriors for model parameters, including corrections for bias in impact parameter and the mean-stellar density computed from the photometric model, \rhoc~\citep{Gilbert:2022}.  Orbital periods have been revisited and in some cases recomputed to address the complications of transit timing variations by providing the best fit constant period to transit times observed by \ikt.  We investigated multiplanet systems with suspiciously close-period planet candidates and corrected period-aliasing.   Significant improvements in stellar parameters from ground based followup and the parallax survey from \gaia have been incorporated \citep{Fulton:2018,Berger:2020a}.  This has allowed for planetary parameters to be derived in a more uniform manner than in other cumulative catalogs.  We include a concise description of all planetary characteristics listed (see \S\ref{sec:unified}) to allow for the community to maximize the combined knowledge of exoplanets in the \ik field-of-view.
}




For ease of reading, we often refer to planet candidates simply as ``planets''. Planets that are the sole transiting candidate of their host star are referred to as ``singles'', whereas the term  ``multis'' is used for both systems with more than one transiting planets and individual planets in such systems.  

We present our catalog of \ik planet candidates in Section \ref{sec:catalog}. In Section \ref{sec:singles}, we {characterize the sample of planet candidates, }compare the ensemble of planet candidates in multis to those in singles, compare planets in two-planet systems with those in higher-multiplicity multis, and quantify the reliability of the sample of multis as representing true planetary systems.  
{We investigate the distribution orbital eccentricity for various subsets of \ik planet candidates (\S\ref{sec:ecc}), improving on previous studies thanks to 
the enhanced accuracy of the stellar densities and impact parameters in our catalog.   We find significant changes in the eccentricity distribution as a function of the inferred size of the planet candidates (\S\ref{sec:eccsize}) and the number of \ik planet candidates detected around a given host star (\S\ref{sec:eccmultiplicity}). We also find significant differences between planet candidates with orbital periods less than 6 days and those with longer orbital periods (\S\ref{sec:eccperiod}).}
We  {consider various factors that can lead to} orbital period variations of \ik planets   (Section \ref{sec:periodtheory}). {We show that planets on eccentric orbits have variations in the times between successive transits on time scales much longer than the 4-year duration of the \ik mission  (\S\ref{sec:precession}). Section \ref{sec:individual} analyses long-term variations in mean orbital periods of planets  within  several \ik systems  showing significant TTVs that have been solved for dynamically.}  We conclude the main text by summarizing our principal results in Section  \ref{sec:conclusions}.

We select which objects to include in our catalog of planet candidates and list their properties to maximize accuracy on an object-by-object basis.  Therefore our selection criteria and various planetary properties are not homogeneous, and \emph{our planet candidate list is not appropriate for use as input for planetary occurrence rate calculations}.  Nonetheless, some aspects of our derivation of planetary properties provide estimates that are more accurate and at least as uniform as those found in previous studies. Therefore, we also present a second set of planetary properties using a uniformly-derived set of stellar parameters in Table \ref{tab:planetcatalog} and outline a process for utilizing some of the information tabulated therein for studies of occurrence rates in Appendix \ref{sec:App_ORtables}. Our primary planet candidate catalog (Table \ref{tab:planetcatalog}) is restricted to transiting planets orbiting just one star.  An abbreviated catalog of non-transiting planets found (using TTVs and/or radial velocity measurements) around stars with transiting \ik planets 
is provided in Appendix \ref{sec:App_nontransiting}. 

\smallskip

\section{Planet Catalog} \label{sec:catalog}

In this section, we introduce our catalog of \ik planet candidates and describe the calculation of properties of the planets, as well as the sources used for characterizing their host stars.  Figure \ref{fig:perrad} displays the  radius vs.~period distribution of the planets in our \ik catalog and highlights the abundance of multiplanet systems discovered.  Multiplanet systems provide a special opportunity to study the potentially rich dynamical history of exoplanet formation and evolution.  Our construction and review of the \ik exoplanet catalog focuses on orbital periods and the prevalence of multi-planet discoveries.

We discuss stellar parameters in \S\ref{sec:star}, transit models in \S\ref{sec:transitmodel}, candidate selection and catalog unification in \S\ref{sec:selection} and the calculation and interpretation of orbital periods in \S\ref{sec:periods}. Our planet catalog is presented in \S\ref{sec:unified}.  Weaknesses of this catalog and its previous incarnations are discussed, including the impact of transit timing variations, relationships between the historical \ik catalogs, and the non-uniformity of candidate selection and biases of some derived properties. Section \ref{sec:2433} focuses in on the KOI-2433 system, which now has seven planet candidates. 

\subsection{Input Stellar Properties}\label{sec:star}

In preparing this catalog (Table \ref{tab:planetcatalog}) and throughout our study, we take stellar properties for the hosts of more than 99.7\% of the planet candidates from one of three sources.  When available, we select parameters from the latest catalog provided by California \ik Survey (CKS){\footnote{Stellar parameters for KOI-1792 were chosen from \cite{Berger:2020a} despite the availability of CKS parameters for that target for reasons specified in \S\ref{sec:selection}.}}, which lists properties of \ik planet hosts that have both spectral measurements from the Keck I telescope and well-determined \gaia properties, especially distances \citep{Fulton:2018}.  This list includes $\sim 60$\% of the hosts of multis as well as $\sim 60$\% of the single planet hosts brighter than \ik magnitude $Kp = 14.2$, but fewer than 6\% of the fainter hosts of singles. For stars with \gaia parallaxes/distances that were not included in the CKS sample, we use properties from  \cite{Berger:2020a}, which includes $\sim 95\%$ of the \ik targets; this list accounts for most of the remaining planet hosts.  For stars absent from both catalogs, we use the stellar parameters listed in \ik DR25 \citep{Thompson:2018}.  

Parameters for KOI-3206 were obtained from the \gaia online archive. We adopted custom parameters, as described in the following paragraph, for stellar hosts in two binary star systems.  No useful data were found for KOIs 2324, 4713, 5226, 5718, so we have adopted solar parameters with large uncertainties ($R_\star$ = $1\pm1$~\rsun, $M_\star = 1\pm1$~\msun, \logg = $4.5\pm4.5$) for these planet-hosting stars. 




Transit depths for both KOI-119 (Kepler-108) and KOI-284 (Kepler-132) suffer substantial dilution due to stellar companions.  For the case of KOI-119, we adopt the nominal dilution of 69.9\% as reported in \cite{Mills:2017a} for their (preferred) mutually inclined solution.  Observations indicate the KOI-284 system consists of 2 nearly identical stars with a total of 4 known transiting exoplanets.  From orbital stability considerations, the orbital periods of KOI-284.02 (6.41 days) and KOI-284.03 (6.17 days) are inconsistent with these planets orbiting the same star.  Thus, KOI-284 represents the special case of a {\it split multi}; see \S\ref{sec:falsemultis} for more details on this system and other \ik split multis.  For computing the planetary parameters presented in this paper, we adopted a dilution of 50\% for the KOI-284 transit models, which implies that half of the light in the photometric aperture is due to the companion star.
 
\subsection{Transit Models}\label{sec:transitmodel}

The calculation of transit models and the preparation of data products is worth reviewing in the context of potential biases and providing motivation for future improvements of the \ik catalog and its legacy value.  For all transit models, we use Presearch Data Conditioning (PDC) lightcurves \citep{Stumpe:2014} as reported in DR25.  {Observations with a quality flag set for any of bit 1,2,3,4,6,7,9,13,15,16,17 were rejected from our analysis for the reasons described in Table 2.3 of the {\it Kepler Archive Manual} \citep{Thompson:2016}.}  PDC lightcurves were prepared for transit analysis by detrending using a second-order polynomial filter with running window of 5~days.  The width of the running window was always truncated to avoid gaps larger than 10 long cadence (30-minute) observations of valid \ik photometry.  Thus, the filter does not cross large gaps in \ik photometry that arise from monthly data downloads or quarterly spacecraft rotations, and it avoids problems with significant jumps in the reported photometric flux from thermal settling of the spacecraft after attitude adjustments that are not fully captured by PDC \citep{vanCleve:2016}.  Observations within 1 transit duration of the mid-transit time for each observed transit were excluded from the polynomial fit.  Thus, the photometric baseline of in-transit data was interpolated based on out-of-transit observations only.  Outliers in the detrended data were identified and removed; outliers are defined herein as single long-cadence photometric measurements more than five standard deviations away from the mean after removal of a best fit transit model.


The DR25 transit models and updated models for this paper use the same software (\texttt{TRANSITFIT5} transit modeling software; \citealt{Rowe:2015,Rowe:2016}) and techniques for parameter estimation and the calculation of posteriors.  Additional details can be found in \cite{Rowe:2014, Rowe:2015, Thompson:2018}.  Briefly, a multi-planet transit model was calculated for each lightcurve and used to isolate the transits for each individual planet in the system by subtracting the model with the depth set to zero for the planet of interest.  This lightcurve was then used to fit each KOI separately with a photometric transit model using the analytic quadratic limb-darkening model from \cite{Mandel:2002}. Limb-darkening coefficients are based on the tables of \citet{Claret:2011} and were fixed to values used in the DR25 KOI catalog.  The photometric model parameterization uses the mean stellar density (\rhostar), fixed quadratic limb-darkening coefficients, photometric zero point, the mid-transit time ($T_0$), orbital period ($P$), impact parameter ($b$) and scaled planet radius (\rprs). Eccentricity was fixed at zero for these models; thus, \rhostar~is replaced by \rhoc.  The adoption of mean stellar density as a fitted parameter assumes the mass of the host star is much larger than the combined mass of the transiting planet and any other planet(s) orbiting closer to the star, whether transiting or not.  Errors from TTVs for specific systems (see additional discussion below) were corrected by adjusting the observation times based on a linear interpolation of measured center of transit times (TTs) to create an aligned ephemeris.  {We calculated the center of transit times for each observed transit by fitting two transit durations of \ik photometric data centered on the predicted or pre-calculated time of each observed transit seeded with the best fit transit model, and only allowing the mid-transit time to vary.}  Biases in TTs can be introduced from overlapping transits as the multi-planet models used for lightcurve preparation do not simultaneously fit transit parameters and center-of-transit times.

\begin{figure}[!hbt]
\includegraphics[scale=1.0,angle=0]{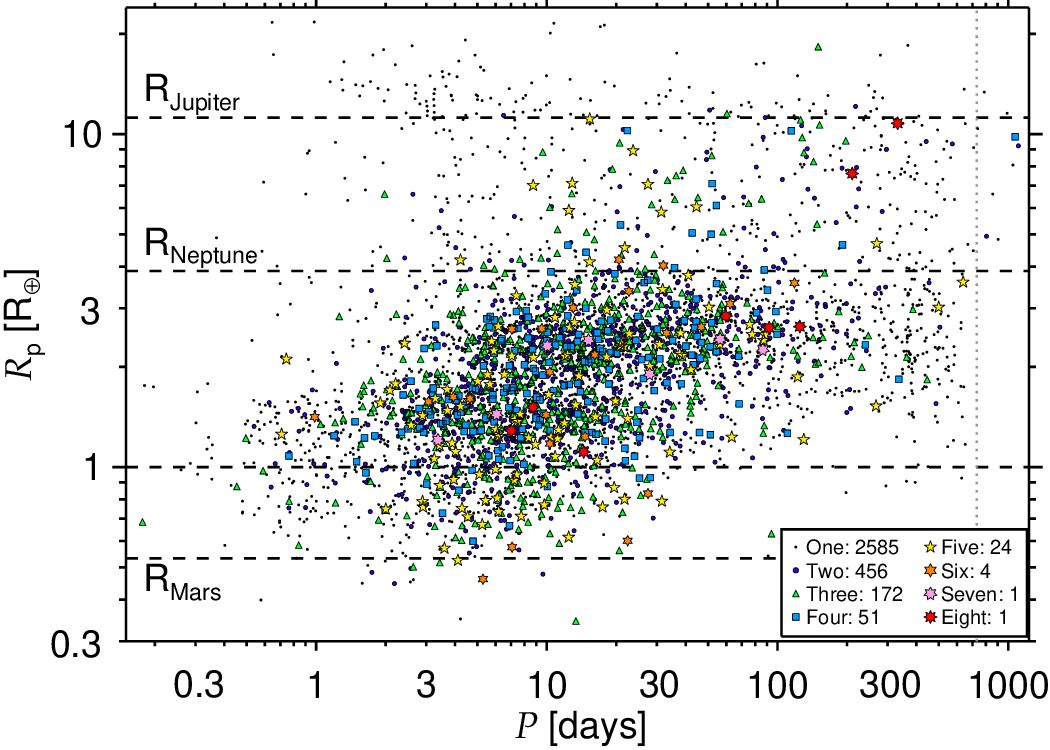}
\caption{Orbital period vs.~radius of \ikt's planetary candidates.  Those planets that are the only candidate for their given star are represented by black dots, those in two-planet systems as dark blue circles, members of three-planet systems as green triangles, those in four-planet systems as light blue squares, in systems of five PCs as yellow five-pointed stars, with six PCs as orange six-pointed stars, the seven PCs associated with KOI-2433 as pink seven-pointed stars, and the eight planets orbiting KOI-351 (Kepler-90) as red eight-pointed stars. The legend lists the number of stars hosting each multiplicity. The planetary candidates are listed in Table \ref{tab:planetcatalog} (First letter of disposition = ``P''). 
Non-transiting planets (listed in Table \ref{tab:nontransiting}) and circumbinary planets are not included. Planet candidates only observed to transit once (mono-transits) are not plotted because their orbital periods are highly uncertain (see item \#4 in the list presented in Section \ref{sec:unified}), but they are accounted for in the multiplicity designation of their companion planets and the total numbers of systems of each multiplicity given in the lower right of the figure. KOI-846.01, with a radius $R_p=30.043$ R$_\oplus$ (see \S\ref{sec:selection}), falls outside the plotting window. All planets to the right of the dotted gray vertical line at $P = 730$ days (as well as some of the planets with shorter periods) transited only twice during the \ik mission, and therefore were not detected by the standard \ik pipeline, which required a minimum of three transits for a detection.  It is immediately apparent that there is a paucity of giant planets in multi-planet systems, especially giants with short orbital periods $P<15$ days.  The upward slope in the lower envelope of the plotted points is caused by the low S/N of small transiting planets with long orbital periods, for which few transits occurred during the time intervals that \ik observed. Adapted from a previous figure generously provided by Rebekah Dawson. }
\label{fig:perrad}
\end{figure}

We adopted Markov Chains calculated for DR25, apart from KOI PCs with large impact parameters ($b > 1$) and a few other KOIs that required model updates.  The DR25 Markov Chain Monte Carlo (MCMC) sampler assumed a non-informative prior for the impact parameter, which works well for non-grazing transits.  However, when $b > 1$, the minimum value of scaled planetary radius, \rprs, required for a planetary transit grows linearly with $b$.   Uniform sampling would result in a bias towards very large impact parameters, which are unphysical for planetary transits of stars.  To sample correctly, a prior was introduced to disfavor large impact parameters by de-weighting the model likelihood by $b^2$ when $b > 1-$\rprs, i.e., multiplying the prior by $b^{-2}$ for such regions of parameter space and not allowing $b>10$.  We computed new MCMC models for all KOI PCs with $b > 1$  and used these models to calculate the values presented in Table \ref{tab:planetcatalog}.  Some KOIs in our cumulative catalog required updated best fit models, and new MCMC runs to allow for sufficient sampling of low $b$ parameter space.  New models are presented for the following KOIs: 1681.02, 1681.03, 1681.04, 2092.03, 2398.01, 2474.01, 2578.01, 2604.01, 2695.01, 2775.01, 2919.01, 2933.01, 3013.01, 3130.01, 3384.01, 3572.01, 3853.01, 4007.01, 4034.02, 4035.01, 4056.01, 4345.01, 4498.01, 4528.01, 4625.01, 4632.01, 4670.01, 4743.01, 4778.01, 4782.02, 4838.02, 4886.01, 4890.01, 5804.01, 5831.01, 6103.02, 6941.01, 7368.01 as well as for all of the new KOIs listed in Section \ref{sec:selection}.

The development of the \ik catalog introduced a few quirks that biased some of the reported best fit parameters. As the \ik mission progressed, transit modelling techniques were improved and the methodology of reporting parameters changed, such as the choice of reporting either maximum likelihood vs.~mode or median from Markov Chains without much consistency between iterations.  A more insidious consequence from model evolution were biases that resulted from how subsequent-generation models were initialized based on the ancestral adaptation of previous models as a reference starting point.  Prior to the \cite{Mullally:2015} catalog, each update introduced new models that incorporated new photometry, resulting in longer time-series observations that ideally should produce more accurate measurements of periods, transit-depths and overall better fidelity.  However, there are two identifiable deficiencies from this approach: a bias towards extreme values of impact parameter due to the nature of \ik observations and excessive dependence on the average period measured up to the time that the KOI was initially announced when appreciable TTVs occur.

In general, impact parameters are not well measured for \ik planets.  This is due to the low S/N of a majority of \ik discoveries and the long (30-minute) cadence, which is comparable to the ingress and egress durations of the typical observed transit.  Combined with a potentially simplistic limb-darkening model, it is common to see the posterior distribution of $b$ from the transit model skew towards 0 or 1.  From a probabilistic view, the data for most \ik transits are insufficient to confidently distinguish a model with $b=0$ from models with $b \sim 0.5$.  The evolution of the \ik models has meant that over time, a large number of models would always be initialized near the boundary of allowable values of $b$.  Using least-square methods such as Levenberg-Marquardt (e.g., \citealt{More:1980}), which explores the local gradient of the model parameter value, can result in many models remaining near $b=0$ when initialized there.  The solution to the problem is to recognize that distributions of model parameters are more robust when sampling from estimates of posteriors based on methods such as MCMC.   For this paper, we explicitly note how all model parameters are reported in \S\ref{sec:unified}. Additional improvements may result from improved modelling of limb-darkening in future studies.

\begin{figure}[!hbt]
\includegraphics[width=0.8\textwidth,angle=0]{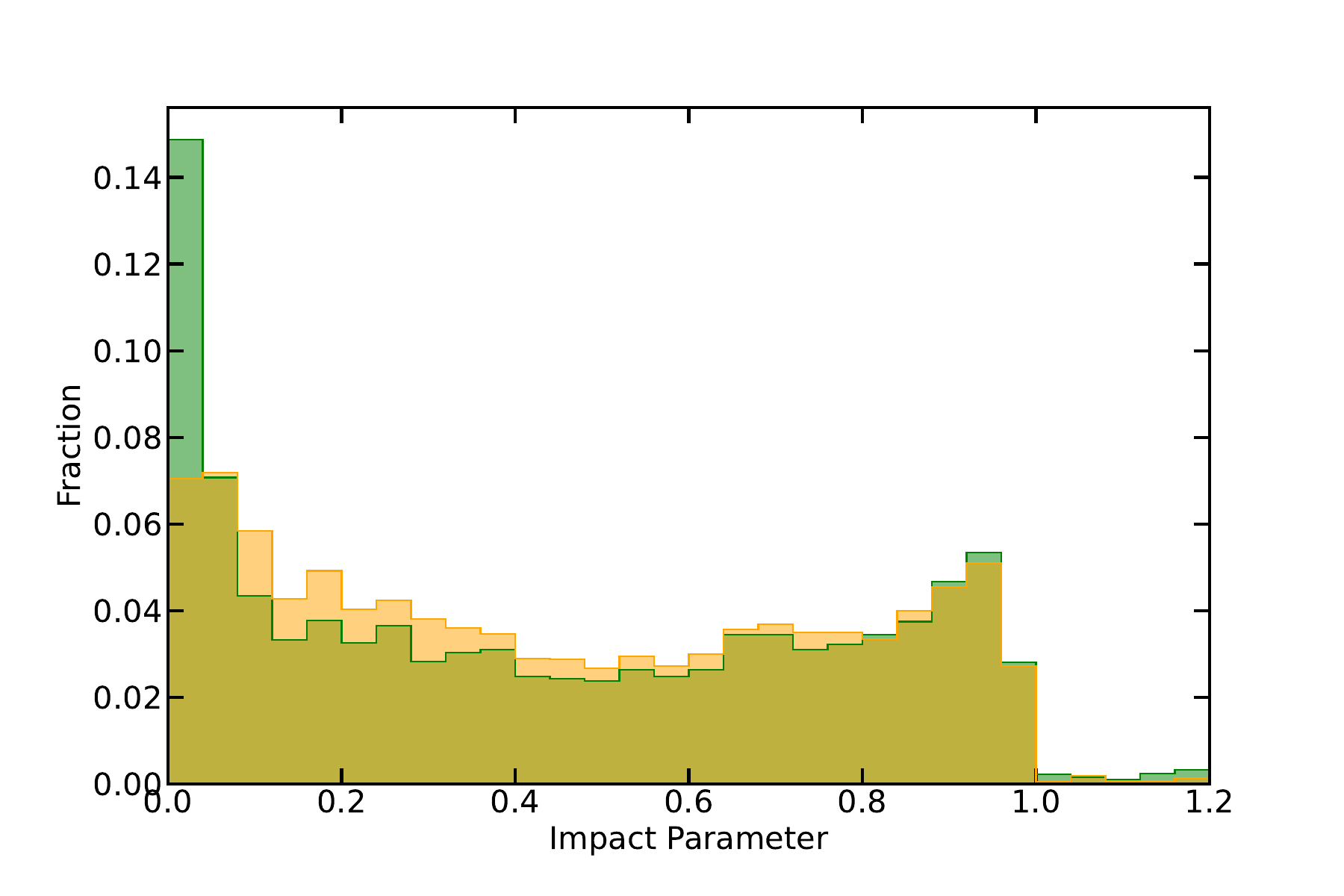}
\caption{{Comparison of the impact parameter ($b$) distributions for planet candidates from this work (orange) to the values listed for planet candidates in DR25 (green).  Only planetary candidates with $b<1.2$ are shown; PCs with $b>1.2$ represent $0.5\%$ of our sample and $1.4\%$ of the DR25 PCs. Restricting the samples to those KOIs that are classified PCs in both catalogs does not substantially alter either of the distributions. 
}
}\label{fig:b_hist}
\end{figure}

Figure \ref{fig:b_hist} compares the impact distribution of PCs presented in Table \ref{tab:planetcatalog} (orange) against that in DR25 (\citealt{Thompson:2018}; green). The DR25 catalog shows a pronounced excess of model fits with $b\sim0$ and $b>1$. Our efforts to reinitialize fits and include priors de-weighting large impact parameters produces a distribution of impact parameters more consistent with an isotropic inclination distribution \citep{Kipping:2016}, although there are still excess populations near $b = 0$ and for $b \gtrsim 1$.  

As recognized by \cite{Newton:1687}, the orbital period of a planet is, in general, not constant.  The orbital periods observed by \ik are the mean values that were observed during the 4-years of the primary mission, at least to a good approximation.  In the case of systems with large TTVs that were discovered early in the mission, the evolution of the models can result in the period reported in catalogs published by the \ik mission being only valid for the duration of data used for the initial discovery and characterization.  This occurred because the transit models assumed a non-interacting Keplerian orbit.  TTVs were handled by explicitly measuring the transit time of each individual event, then re-sampling the time-stamps with linear interpolation based on the TT of each event to be aligned.  Resampling used the observed-minus-calculated (O-C) values with the calculated transit time based on the reported mean period from the best fit model.  As the transit models evolved with each new data release, the transit times and O-C values were measured based on the transit model from the previous catalog. The new TTs were incorporated and the model updated.  Since the O-C values were fixed when the transit model parameters were updated, any significant change in the orbital period was captured in the O-C, not the reported transit model period.  Our solution to this problem is to calculate the mean orbital period, {\it as observed by \ikt}, directly from a straight-line fit to the measured times of each transit.  An O-C diagram (O-C vs.~time) should have no significant slope.  If a single planet demonstrated clear TTVs, then TTVs were typically calculated and included for all planets in the system.  See \S\ref{sec:unified} for an explicit description of how each model parameter is reported, \S\ref{sec:periods} for extended discussion of orbital periods during the \ik epoch, and \S\ref{sec:periodtheory} for an analysis of period variations over longer time scales.

Transit timing variations are commonly observed when two or more planets interact dynamically \citep{Agol:2005,Holman:2005}.  Other potential causes of TTVs include stellar binarity and astrophysical effects such as activity and star spots. (The latter two processes do not cause variations in acutal times of transit, but their observational signals can mimic TTVs.) The process by which TTVs have been accounted for in previous catalogs was inhomogeneous and overall {\it ad hoc}.  The primary criteria for selecting the solution with TTVs included was either the visual identification of TTVs from examination of O-C diagrams or for the inclusion with specific KOIs for the detailed study of individual systems.  For example, KOI-8298.01 is reported to use TTVs in the model but has a period less than 0.2 days.  The model with TTVs shows a significantly deeper transit, hence inclusion of TTVs.  However, this may indicate that either KOI-8298.01 is not a transiting planet (e.g., a manifestation of stellar variability) or that the apparent times of transit are significantly affected by spot crossings.  Thus, the TTV flag is not a definitive indication of whether or not TTVs are present and should not be used as evidence for the validity of a KOI being a true planet.  
The first digit of the TTV flag merely reports which KOIs have transit models that include TTVs and the second and third digits pull tabulations from the published TTV catalogs of \cite{Holczer:2016} and \cite{Kane:2019}, respectively. These external catalogs provide excellent assessments of TTVs for most KOIs.  

\begin{figure}[!hbt]
\includegraphics[width=0.49\textwidth,angle=0]{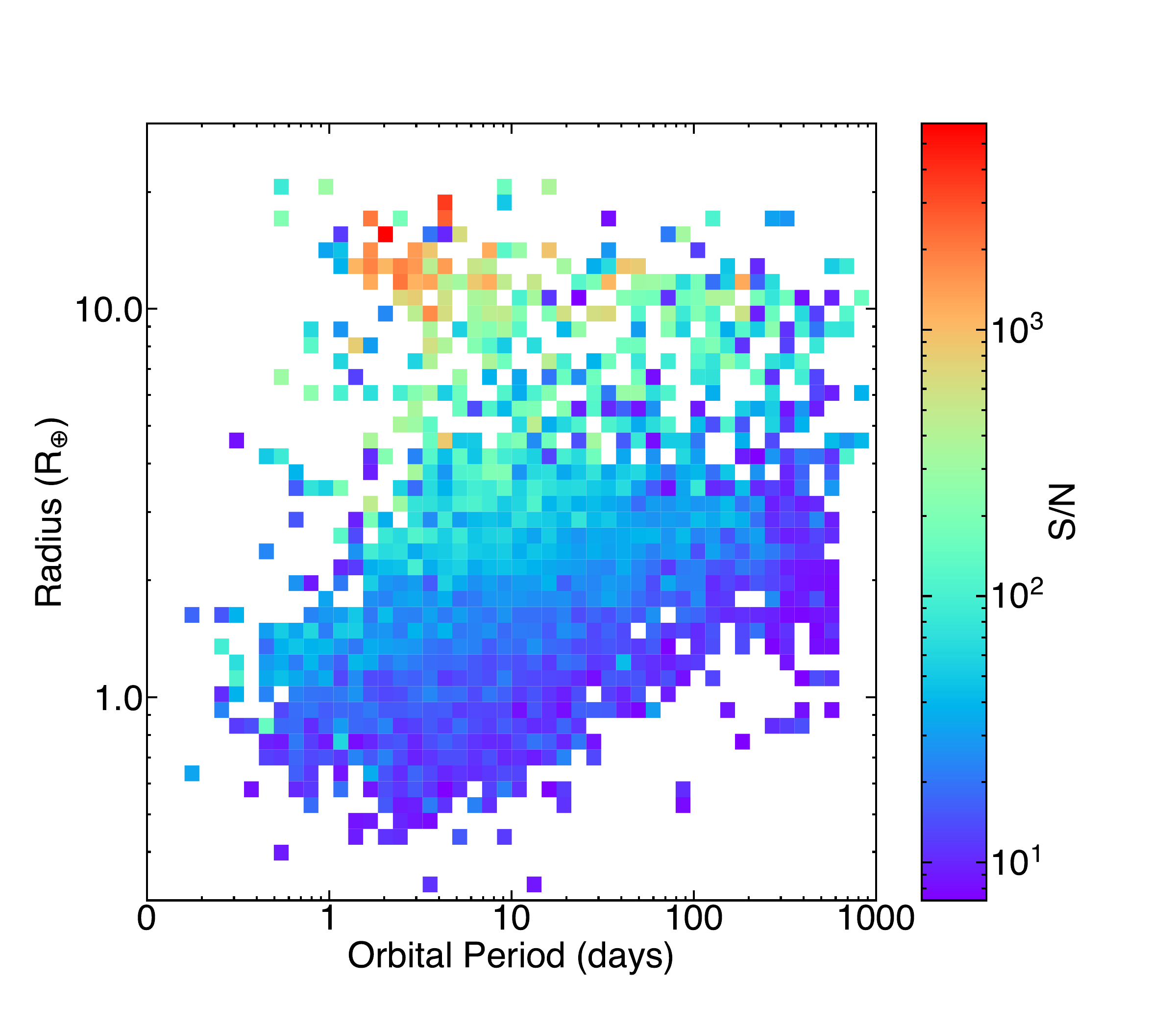}
\includegraphics[width=0.49\textwidth,angle=0]{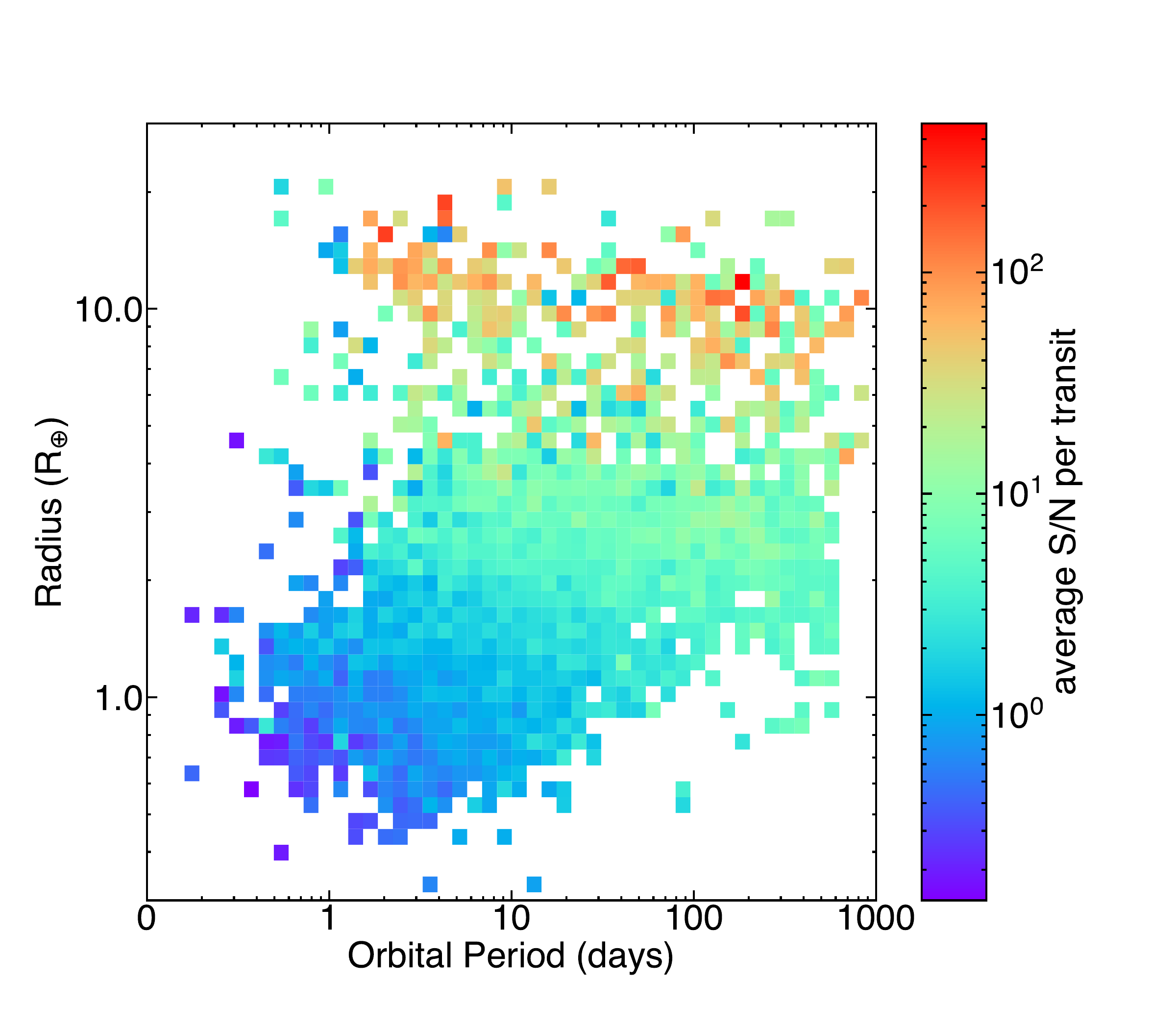}
\caption{Median signal to noise ratio for \ik planet candidates as a function of location in the period-radius plane. Each square represents a factor of $\approx 1.1$ in radius and $\approx 1.2$ in period. The left panel shows the total S/N, and the right panel shows the average (mean) S/N per transit.  The average S/N was calculated by dividing the total S/N by the square root of the number of observed transits for each planet candidate. Note the differences between the color scales of the two panels. The number of planets represented in each colored square ranges from 1 to 32. 
}\label{fig:perrad_snrmap}
\end{figure}

The S/N of the sum of all observed transits was calculated via
\begin{equation}\label{eq:S/N}
    {\rm S/N} = \sqrt{\sum^n_{i=1}\left ( \frac{m_i - 1}{\sigma}  \right )^2 },
\end{equation}
where $m_i$ is the calculated flux from the fitted transit model at each observation $i$ with a total of $n$ observations, and $\sigma$ is the standard deviation {based on out-of-transit observations using detrended PDC photometry with outliers removed}. The model is scaled to have the out of transit flux equal to unity.  Figure \ref{fig:perrad_snrmap} shows the complex dependence of the S/N of the population of exoplanet transit signatures on planetary radius and orbital period.  The trends in S/N are further discussed in \S\ref{sec:unified}.

\begin{figure}[!hbt]
\includegraphics[width=0.49\textwidth,angle=0]{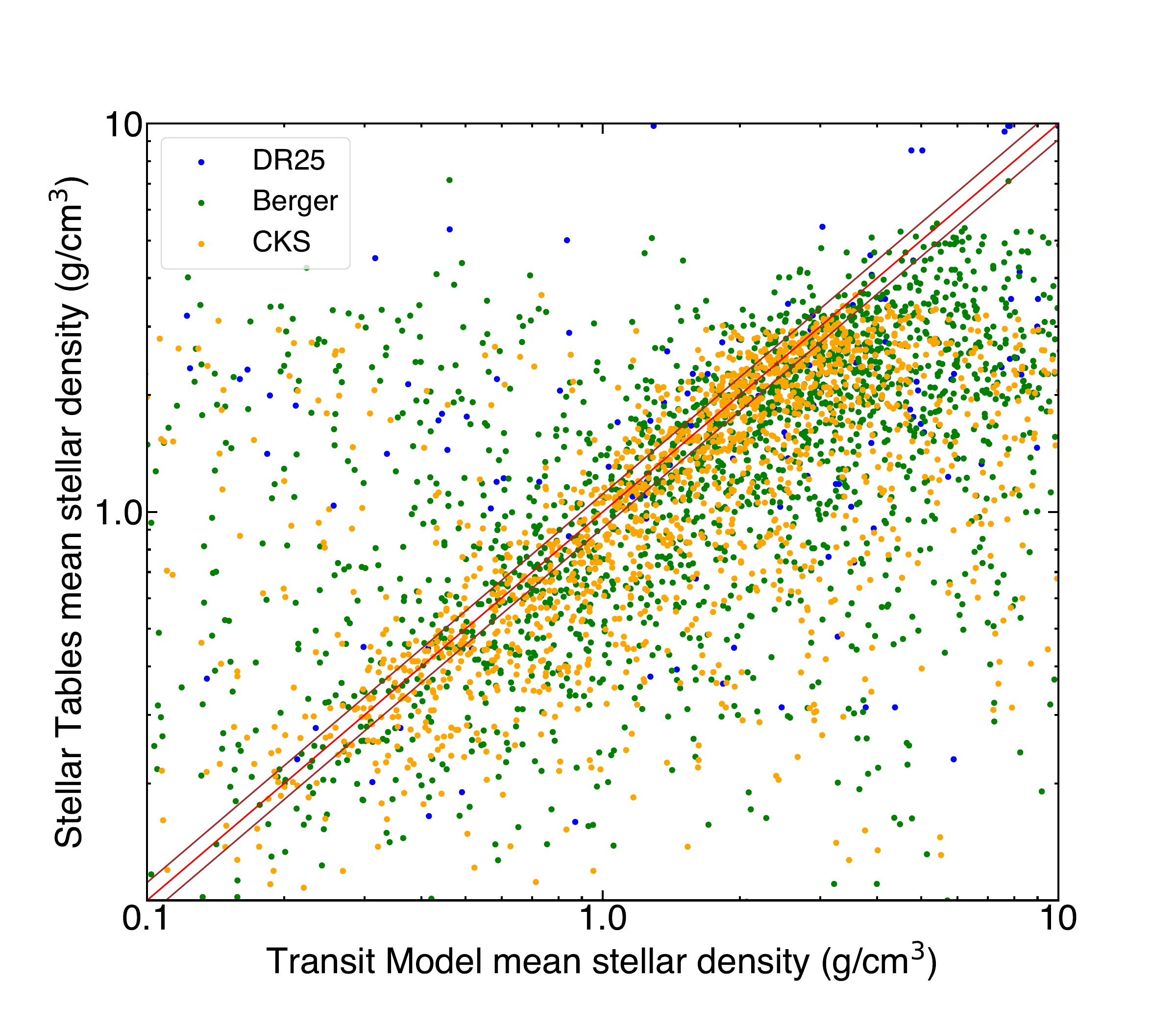}
\includegraphics[width=0.49\textwidth,angle=0]{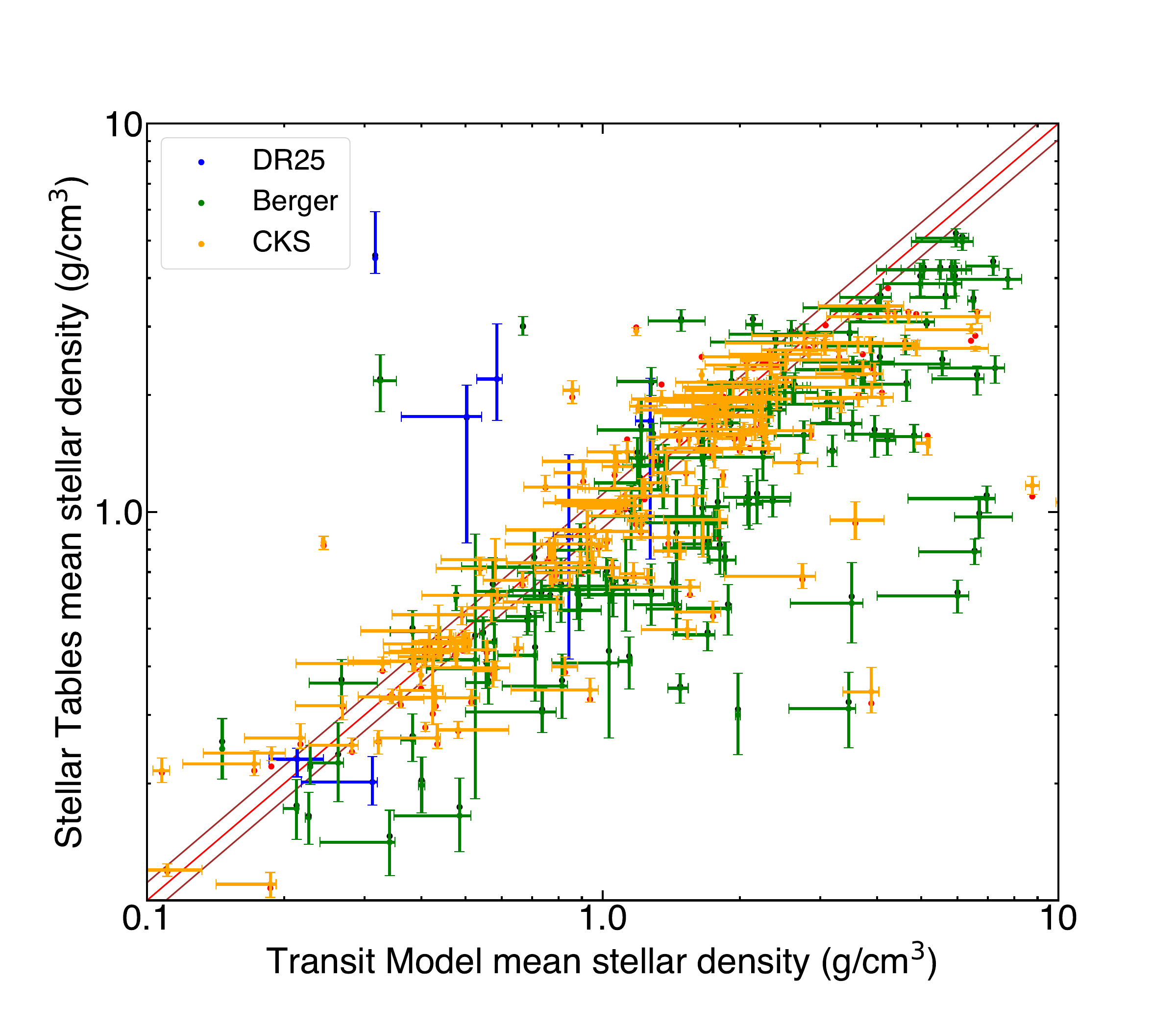}
\caption{Comparison of the mean stellar density from our circular orbit transit models (\rhoc) to the mean stellar density from our adopted stellar parameters (\rhostar). The red diagonal line is for \rhostar = \rhoc, and the two parallel darker red lines show a bias of 10\% in \rhostar.  The colored markers note the source of stellar parameter: blue = DR25 \citep{Thompson:2018}, green = \cite{Berger:2020a} and orange = CKS \citep{Fulton:2018}.  The left panel shows all planetary candidates from our sample with 0.1 $<$ \rhostar $<$ 10 g/cm$^3$.  The right panel is restricted to planetary candidates that have S/N $>$ 10, $b$ $<$ 0.9, and  uncertainty (average of $\sigma_+(\rhoc)$ and $\sigma_-(\rhoc)$) of less than 20\%.  Uncertainties are half-widths of the 68.27\% credible interval ($\pm 1 \sigma$). 
}\label{fig:mean_stellar_density}
\end{figure}

Figure \ref{fig:mean_stellar_density} compares the mean stellar density based on input stellar parameters to the mean stellar density calculated from our circular orbit transit models. Colors denote the source of the adopted stellar parameters. The left panel shows the complete sample.  The right panel shows only PCs with good S/N, non-grazing transits and well measured \rhoc.   The distribution is clearly skewed such that the mean stellar density estimated by fitting the lightcurve and assuming  circular orbits is generally larger than that from stellar parameters tables.  This trend is expected because detection bias favors planets that transit near the periastron of their orbits.  A simple application of Kepler's 2$^{\rm nd}$ law dictates that for a given impact parameter, transits observed near periastron are shorter than predicted from a circular orbit, and impact parameters should be roughly uniformly distributed for $b<1$. The distribution is also biased by impact parameter and dilution (see \S6.2 of \citealt{Rowe:2015b}).

A visual examination of the right panel in Fig.~\ref{fig:mean_stellar_density} appears to shows a bias in \rhostar~between the CKS \citep{Fulton:2018} and \cite{Berger:2020a} samples. It is important to note that our stellar parameters give preference to CKS, then \cite{Berger:2020a}, and use those of DR25 only for targets not appearing in either of the preferred catalogs. The CKS sample was skewed towards the inclusion of multiplanet systems.  As shown in \S\ref{sec:ecc}, planets in compact multiplanet systems tend to have more circular orbits (Figs.~\ref{fig:EccMulti} and \ref{fig:EccSpacing}). Thus, the observed bias is a selection effect.  Figure \ref{fig:EccStellarSource} shows that the measured transit duration distributions do not depend on the choice of stellar parameters catalog in any systematic manner.  {The vertical blue error bars tend to be large because of the high uncertainties in DR25 stellar parameters.  These stars were not characterized in \gaia DR2 or CKS, which suggests that they may have strong stellar blends or other observational challenges. This also explains the larger scatter of the blue points relative to the middle diagonal line.}

\subsection{Planet Candidate Selection} \label{sec:selection}

We pulled planet candidates from a variety of sources, including 9564 KOIs from the cumulative  DR25 supplement catalog\footnote{Retrieved from the NASA Exoplanet Archive on {2022/10/27}.}, ultra-short period (USP) planets from \cite{Sanchis-Ojeda:2014}, long period ($P \gtrsim 1$~year) and transit candidates {only observed to transit once, which we refer to as mono-transits}, from \cite{{Kawahara:2019}}, {the autoregressive planet search from \cite{Caceres:2019}}, PCs from the machine learning search by \cite{Shallue:2018} that sought additional planet candidates around targets already having 2 or more PCs {and low-S/N candidates found by revisiting marginal TCEs \citep{Bryson:2021}.}  The inclusion of new catalogs and discoveries yields new KOIs:  KOI-1843.03, 8298-8303 are ultra-short period planet candidates (including the 3 shortest-period PCs in our catalog) from \cite{Sanchis-Ojeda:2014}; KOI-8304~--~8335, 1108.04, 4307.02, 408.06, 2525.02, 847.02, 671.05, 3349.02, 693.03, 7194.02, 1870.02 are long-period planet candidates; KOI-500.06, 351.08, 691.03, 354.03, 191.05, 1165.03, 2248.05, 542.03, 1589.06, 2193.03, 1240.03, 1992.04, 1276.03, 416.05, 1889.03, 4772.04, 2433.08, 597.04 are from \cite{Shallue:2018}\footnote{We classified 2061.03 from the machine learning search as an FP because we recognized that it resulted from a poor masking of 2061.01, due to large TTVs. We did this by using Quasiperiodic Automated Transit Search (QATS, \citealt{Carter:2013}), determining an approximate ephemeris $T_n [BJD] = 2454949.27 + n \times 14.097 + 0.12 \times \cos( 2\pi/P_{\rm ttv} (t-2455154) )$, where $P_{\rm ttv}=1160$~days.}, and  {KOI-4246.03, 4302.02, 8336.01, 8337.01 and 8338.01 are from \cite{Bryson:2021}}. 

Finally, KOI-8339~--{~8394, the majority of which we have dispositioned as false positives/false alarms because their S/N $< 7.1$, are from \cite{Caceres:2019}. There are 30 potential candidates from Table 5 of \cite{Caceres:2019} not included as we were unable to compute a best fit model based on the their predicted ephemeris. Table 5 of \cite{Caceres:2019} also lists 11 identifications around targets with pre-existing KOIs, including 8302.01 from \cite{Sanchis-Ojeda:2014}, that are already in our table, some of which are discussed below because we corrected their periods to the values given in \cite{Caceres:2019}.}   

We provide new transit models and parameter posteriors for these new KOIs using the same models and methodology presented in \S\ref{sec:transitmodel}.  We corrected the period of KOI-1353.03 (Kepler-289d) to match the reported value in \cite{Schmitt:2014} rather than that in the DR25 catalog because the latter was the result of an aliasing problem. {One of our new long-period PCs, KOI-3349.02, was previously only observed to transit once (\citealt{Kawahara:2019} and references therein), but we located a second, nearly identical, transit in the SAP (Simple Aperature Photometry) lightcurve, showing it is a duo-transit (only two transits observed) planet with $P = 805$~days.}

%
%
%
%

{We revisited the dispositions of KOIs that have Kepler numbers but were classified as FPs in DR25supp.  Since there is no peer reviewed source for the reasoning leading to the dispositions, we do not know whether or not additional observations beyond \ik photometry and centroids were used.  Based on our assessment of TCERT data validation reports downloaded from the NASA Exoplanet Archive, we reverted the following KOIs from FP disposition to PC: KOI-125.01, 129.01 (but see below), 631.01, 3138.02 and 3184.02.  DR25supp listed these KOIs as showing evidence of stellar eclipses; however, we found no evidence of a secondary eclipse in the photometry lightcurve, either directly or through measurement of an odd-even effect.

We confirmed that the following KOIs with Kepler numbers are indeed false positives: KOI-3032.01, 126.01 (see \citealt{Carter:2011}), 1416.01 and 1450.01.  Their photometric lightcures show clear evidence of secondary eclipses that are indicative of stellar companions.  Despite an analysis of putative TTVs by \cite{Hadden:2014}, Kepler-37e (KOI-245.04) is a false alarm that was never validated \citep{Barclay:2013}. 

Table 7 of \cite{Carmichael:2019} lists five KOIs, all of which are listed as PCs in DR25supp and three of which have been given Kepler numbers, as having measured masses above the planet/brown dwarf dividing line  (see also \citealt{Carmichael:2023}). We have reclassified as false positives all five of these KOIs: KOI-423.01 = Kepler-39~b, KOI-189.01 = Kepler-486~b, KOI-205.01 = Kepler-492~b,  KOI-415.01 and KOI-607.01. Note that Kepler-39~b has a mass of $\sim 20$~M$_{\rm Jupiter}$, whereas the others have masses $\gtrsim 40$~M$_{\rm Jupiter}$.

\cite{Santerne:2016} analyze radial velocity measurements of more than 100 KOIs that were listed as giant planet candidates in one or more of the \ik project's catalogs. They present convincing evidence that the following five KOIs, which we would have otherwise classified as planet candidates, are produced by eclipsing binary stars: 129.01, 969.01, 1465.01, 1784.01 and 3787.01. Furthermore, the following three KOIs, which we would have been classified as planet candidates had they not failed our upper size limits, are also produced by eclipsing binary stars: 
3411.01, 3811.01 and 5745.01. Table \ref{tab:planetcatalog} dispositions the thirteen KOIs listed in this and the previous paragraph with the letter `M' to distinguish them from other false positives.}

{We undertook a photometric analysis of new candidates presented in Table 5 of \cite{Caceres:2019} to assign dispositions based solely on \ik photometry.  Photometry was processed in a manner similar to DR25 \citep{Thompson:2018}, by detrending the data with a second-order Savitzky-Golay filter.  Data in the transit window as predicted from the reported transit duration, period and center-of-transit time were excluded from polynomial fits.  This means that we did not complete an exhaustive test against false alarms due to in-phase periodicity (e.g., {\it depth-test} as described in \citealt{Coughlin:2016} for uniqueness).  We attempted to compute best fit transit models through $\chi^2$ minimization.  For 33 of the 86 proposed new candidates, the model either failed to converge or returned a fit consistent with a flat line.  In these cases, we then ran a BLS ({box} least squares; see \citealt{Kovacs:2002}) search restricted to $\pm 0.1$ days around the reported period{, which allowed us to recover 3 of them. As we could not find evidence of the remaining 30 proposed PCs, we do not include them in our KOI table.}  The ephemerides reported in \cite{Caceres:2019} are only accurate to one \ik long-cadence ($\sim$~30 minutes), which combined with uncertainties in the reported period and barycentric drift may result in poor recovery with our methods.  If the BLS search found a candidate event with $P$ within 0.01 days of the reported event, we adopted the period and center-of-transit times from our localized search.  Thus, there is risk that we have not recovered all events as previously reported.  With best fit models we ran our MCMC algorithm to compute posteriors.  All new candidates from this activity that had a S/N as determined by our transit models to be less than 7.1 have been flagged as false positives in  Table \ref{tab:planetcatalog}.  The S/N reported in Table 5 of \cite{Caceres:2019} is a detection statistic that appears to not be strongly related to the folded transit S/N that we report (see Equation \ref{eq:S/N} above).  We did not assess the photometric centroids for this sample; however, we noted that there is a substantial mismatch when comparing \rhoc~from our transit models to \rhostar~from stellar properties catalogs for KOI-8345.01 and KOI-8366.01, and for this reason we have flagged these two KOIs as false positives.} 

{We investigated several cases in which the \cite{Caceres:2019} table listed signals around existing KOIs with periods that are a small integer multiple or fraction of those listed in DR25supp. This investigation led us to revise the orbital periods of two PCs downward by a factor of two, KOI-6262.01 from $P = 0.673$ days to 0.3365 days and KOI-4777.01 (which was independently identified with the correct period by \citealt{Canas:2022}) from $P = 0.824$ days to  0.412 days. Additionally, we revised the periods of FPs (both of which were previously identified as having centroid offsets) KOI-4305.01 (from 0.935 to 0.234 days) and KOI-4872.01 (from 1.035 to 0.207 day). Reported detections for KOI-6749, KOI-6984, KOI-2431, KOI-4788, KOI-2642 are the secondaries; these targets are already FPs.} 

KOIs have been vetted into new categories using several criteria.  Nonetheless, the vast majority of KOIs are dispositioned as one of the two categories: Planetary Candidates (PC) and False Positives (FP, which here, as in previous catalogs, includes false alarms).  False positives are defined as transit-like astrophysical events that are not produced by a planet.  Eclipsing binaries are the primary source of false positives.  False alarms are spurious detections caused by features in the target star’s lightcurve that are not transit-like.  False alarms can be caused by stellar variability and/or instrumental systematics.  We provide additional vetting criteria based on S/N and planetary radius. A description of all vetting flags is presented in Section \ref{sec:unified}. 

Initial FP and PC classification is adopted from the DR25 and DR25supp catalogs. DR25supp is represented by the cumulative catalog from the NASA Exoplanet Archive retrieved on {2022/10/27}. If vetting classification is not available in either DR25 or DR25supp, we use DR24 \citep{Coughlin:2016}, which includes KOIs from previous searches that were not seen in the TCE search conducted for DR24.

Consistent with previous \ik catalogs, generally we require the total transit S/N~$\geq 7.1$ for a KOI to be considered a PC; KOIs below this threshold are considered false alarms. {We allowed four exceptions to this rule, KOI-2022.02 = Kepler-349~c, KOI-2034.02 = Kepler-1065~c, KOI-4024.01 = Kepler-1541~b and KOI-7368.01 = Kepler-1974~b, all of which have been both validated as planets and passed our visual inspection---which revealed marginal evidence of a visible transit event in their lightcurves.

The  S/N~$\geq 7.1$ } criterion was based on limiting the number of false alarms considered to be PCs to 1 in the 90 day -- 2 year period range per $10^5$ stars searched the \ik sample in the presence of white noise on 6-hour timescales.  However, much of the noise in \ik lightcurves is correlated, so multiple false alarms become problematic for S/N in the range of $10 \gtrsim$~S/N~$\geq 7.1$, as has been verified through injection tests and reliability studies \citep{Thompson:2018,Hsu:2019}.  Moreover, low S/N KOIs do not have sufficient signals to allow for precise centroiding and other vetting procedures used to distinguish astrophysical signals such as eclipsing binaries (EBs) from transiting planets.  Therefore, higher S/N cuts are needed to obtain the purer (higher confidence) samples of PCs that are required for some studies, including most of our analyses of the characteristics of the population of \ikt's planet candidates. 

Some KOIs have all transit model parameter uncertainties listed as zero.  This indicates that the MCMC computations did not converge.  This can happen when the S/N of transit event is low or the event is non-transit-like in shape.  While we did not consider MCMC convergence when assigning PC or FP (P or F) dispositions, there is a high probability that these KOIs are indeed false alarms.  {There are 220 KOIs flagged as `P' or `S' without MCMC computed posteriors, 24 are members of candidate multiplanet systems and a total of 26 are PCs, 14 of which have a S/N $>$ 10.}

There is increasing degeneracy when distinguishing between ultra-cool stars, brown dwarfs and planets for transiting objects with radii approaching the size of Jupiter that lack mass information.  We have flagged KOIs with an estimated planetary radii $R_p > 21.947$~R$_\oplus \approx 2$ R$_{\rm Jupiter}$, as well as KOIs with period $P>20$ days and a $1 \sigma$ lower limit for the planetary radii $R_p$ that exceeds 13.17~R$_\oplus \approx 1.2$~R$_{\rm Jupiter}$, as likely FPs, but as these boundaries are not precisely defined, KOIs that pass all criteria for being classified as PC apart from size are given the disposition ``R''. {
The following KOIs have Kepler numbers but radii that exceeded our limit on planetary size using nominal stellar parameters: KOI-846.01, 855.01 and 1792.01.  There is no strong evidence from \ik photometry or in the literature that contradicts their status as a confirmed planet.  Thus, no change in their disposition as PC is warranted based on our study. The radius of the star KOI-1792 estimated by \cite{Fulton:2018} is almost three times as large as that estimated by \cite{Berger:2020a}; the former leads to a radius estimate for KOI-1792.01 of $R_p = 31.1$~R$_\oplus$, so we used the \cite{Berger:2020a} parameters for this star. The other two cases both appear to be evolved stars (KOI-846 being slightly evolved, whereas KOI-854 is a giant), whose radii are difficult to estimate, so we suspect that the stellar radii have been overestimated, but in these cases, we only have one \gaiat-constrained radius estimate (from \citealt{Berger:2020a}), so we retained the probably-overestimated nominal values of $R_\star$ and $R_p$.}

{We give new KOIs dispositions based on an assessment of the photometric transit.  All new KOIs are assigned a PC status unless they failed to meet our S/N, radius cuts or visual inspection.  Visual inspection includes determination whether observed transit duration is in very rough agreement with the stellar parameters and the phased lightcurve has a transit-like shape.  We did not attempt to measure photometric centroids, thus it is possible that many of the new short-period KOIs from the \cite{Caceres:2019} sample (8339~--~8394) may be background binary blends.  As most of the orbital periods from this sample are relatively short, there is reasonable expectation that the modelled mean stellar density should match stellar tables because the orbits should be nearly circular.  For example, KOI-8386.01 is likely a background binary due to the significant order-of-magnitude mismatch in the transit-duration and stellar classification, but we nonetheless dispositioned it as a PC as we have not conducted an analysis of photometric crowding or photometric centroid shift during the transit event. None of the KOIs from the \cite{Caceres:2019} sample failed {the radius cuts, so all were dispositioned either ``S'' (36 cases), ``P'' (18 cases) or ``F'' (2 cases that were discarded based upon visual inspection of the lightcurve); only 5 of the PCs have S/N $> 12$, and only 1 has S/N $> 14$.}} 

We visually re-examined KOIs that were classified as PCs in either DR24 or DR25 but subsequently classified as FPs in DR25supp. 
Based on our analysis, we reverted the following KOIs back to PCs: 82.06, 198.01, 1693.01, 1796.01, 1902.01, 2306.01, 2307.01. These KOIs do not exhibit measurable centroid shifts from examination of DR25 or DR25supp vetting reports and show visually identifiable transits that can be modeled with MCMC that meet convergence criteria.  

\subsection{Orbital Periods During the \ik Epoch}\label{sec:periods}

The orbital periods listed in Table \ref{tab:planetcatalog} are estimated by fitting the \ik lightcurve assuming a Keplerian, non-interacting orbit and using a limb-darkened transit model. As such, we are estimating something akin to the planet's mean period over the interval of \ik observations. 

{Transit times are the measure of when mid-transit occurs.  We define the mid-transit as when the project distance between the center of the star and planet is minimized.  For a circular orbit, this is equivalent to when the transit model is deepest. We measure transit times (TT) and then report transit timing variations relative to the orbital period and reference time.  The latter has typically been the first or mid-observation transit. The transit model is used as a template to measure the center-of-transit time for each transit, and the transit model orbital period from the initial fit is used to calculate the difference between the observed transit and the calculated transit (O-C).}   If TTVs are present, then the transit model is updated by {\it deTTVing} the lightcurve, whereby the time stamps from \ik observations are adjusted to have all transits aligned.  Time stamp adjustments use a linear interpolation based on the O-C transit times such that the new time stamps have an effective O-C of zero.  A new transit model is then fit to the data with updated time stamps. 

If TTVs are not properly accounted for, the ingress and egress of sequential transits become misaligned, leading to errors in the reported characteristics of the transit, especially the transit's duration, depth and the impact parameter.  Since the scaled planetary radius ($R_p/R_\star$) and transit model mean stellar density (\rhoc) are correlated with the impact parameter ($b$) due to geometry and stellar limb-darkening, those parameters also have increased errors.  The models based on deTTVed lightcurves better estimate the transit shape, {often leading to substantially improved estimates of the above mentioned parameters, especially} for $b$ and $R_p/R_\star$. However, this process {previously led to errors in the reported period that were not accounted for in estimates of period uncertainties.}

TTVs are calculated independent of the fitted transit model, and when possible initial TTVs were adopted from previous catalogs.  Examples where the period reported in previous catalogs are significantly different from the observed mean period (averaged over the time interval during which transits were observed by \ikt) include KOI-142.01 (Kepler-88~b) and KOI-377.01 and .02 (Kepler-9~b and c).  Early KOI catalogs used only a few Quarters of observations, and the mean period over that time frame differed significantly from estimates using the entire \ik observational baseline.  The model parameters for each previous catalog were used as a seed for the updated model.  Thus, the period from a previous catalog was used as the seed for the best fit in the subsequent catalog in which it appears.  If additional TTVs (O-C) were measured, the mean-period from a previous catalog was held fixed before the new model fit was made.  In some cases, the previously reported period did not represent the true mean period in the presence of strong TTVs. 

For the catalog presented herein, the mean observed period from \ik observations is of interest as it is the best approximation to the long-term mean period that can be straightforwardly estimated for all KOIs apart from those that only transited once (mono-transits). The value of $P$ is calculated by applying a correction to the best fit period, $P_{bf}$, from the transit models. For each KOI with four or more measured transit times, a straight line model is fit to the Observed-Calculated (O-C) vs.~Observed (O) values using standard least-squares minimization weighted by measurement uncertainties in O-C values.  The slope, $m$, from the fit gives the correction, with
\begin{equation}
P = P_{bf} (1 + m).
\end{equation}
We report $P$ and propagate the uncertainty in $m$ to compute the orbital period uncertainty thereof for all KOIs in Table \ref{tab:planetcatalog}.  The measured O-C values are based on $P_{bf}$, thus any non-zero slope indicates that $P_{bf}$ does not represent the mean period.  This correction specifically addresses the issue of incorrect periods for planets in systems such as Kepler-9 and Kepler-88.  For non-TTV planets, there is good agreement between $P_{bf}$ and $P$.  {If a planet has just 2 or 3 observed transits, we adopt $P$ from $P_{bf}$.}  To avoid additional inhomogeneities in our catalog, only $P$ is reported.

The incorporation of TTVs in orbital period calculations, as described above, yields {more robust estimates of orbital periods of planets with observed TTVs than provided by previous catalogs.  For most KOIs, no  TTVs  are detected, and the revised period is statistically similar to the reported period from DR25.  However, there are extreme examples, such as Kepler-9~b and c and Kepler-88~b, where the period changes significantly ($\sim 1$ hour) relative to the DR25 catalog.  Overall, the distribution of fractional changes in period ratio is non-Gaussian, with substantially larger numbers in the tails of the distribution. }  

However, we account for TTVs only to the extent that they average out over the time in which \ik observed the planets to transit.  Typically, this cancellation is incomplete for planets with observed TTVs. The periods and uncertainties quoted in Table \ref{tab:planetcatalog} thus do not account for TTVs that have periodicities that are long compared to the \ik observations, and do not fully account for TTVs with periodicities comparable to the amount of time between \ikt's first and last observations of transits of a particular planet. Some types of TTVs are of very small amplitude over four years (and thus unlikely to have been detected), but grow to become substantially larger on time scales of decades to centuries, affecting the long-term mean orbital period.  These differences in orbital period are important for understanding three-body resonances and producing ephemerides that are accurate far into the future.  We discuss these issues in more detail in Section \ref{sec:periodtheory}.

We searched the lightcurves of the 10 duo-transit PCs with $P > 730$~days to determine whether or not there are \ik data that can exclude the possibility that additional transit(s) were missed because they occurred during a data gap(s) and the actual period of the planet is either half or one-third of the reported value. Data points with SAP\_QUALITY equal to 16 or greater were excluded from the examined lightcurves. The period of KOI-375.01 (= Kepler-1704~b) could, indeed, be half of the reported value of 988.9 days, but \ik data exclude the possibility of any of the other 9 PCs having periods equal to  half or one-third of the values listed in Table \ref{tab:planetcatalog}.

\subsection{Unified Planet Candidate Catalog} \label{sec:unified}

Our catalog of \ik planet candidates is presented in Table \ref{tab:planetcatalog}.  We list, from left to right, catalog numbers of the target star and planet (columns 1~--~3), fundamental transit model parameters (4~--~19), properties derived from the transit model and \ik photometry (columns 20~--~38), parameters that depend on the transit model and stellar parameters (39~--~44), stellar parameters (45~--~62)\footnote{Columns 14~--~16 list stellar parameters derived from the transit model. Data in column 45 are taken from DR25.  Data in columns 46~--~61 are taken from the source specified in column 62.}, and vetting dispositions (63).  Data in columns 64~--~85 list the same properties as provided in columns 39~--~44 and 46~--~61 with stellar parameters taken from \cite{Berger:2020a}; these columns are left blank if properties of the target star are not given by \cite{Berger:2020a}.

Table \ref{tab:astro_constants} presents our adopted values for astrophysical constants.  The mean radius of the Sun (R$_{\odot}$) and Earth (R$_{\oplus}$) are used to calculate the absolute radius of planets (\rpl) and incident flux, $S$, normalized to that of the Earth.  The gravitational constant (G) is used to calculate transit durations, the scaled semimajor axis (\adrs) from the mean stellar density (\rhoc) and orbital inclination ($i$).  The astronomical unit (AU) and solar effective temperature \teff$_{,\odot}$ are used in the calculation of the flux incident upon Earth, S$_\oplus$.  

For many properties, we report the location of the maximum posterior density and give additional information about the uncertainty.  For some properties, these uncertainties are typically well-described by a symmetric distribution, so we report the half-width of the 68.27\% credible interval, assuming a symmetric distribution. For other properties, the posterior distribution is often significantly asymmetric, so we report separate uncertainties in the positive and negative directions based on the 68.27\% credible interval that minimizes the width of the marginal distribution for that parameter. The units for all uncertainties are the same as for the quantities themselves.  

Overall we have updated transit models for more than 1000 KOI systems. Included among these are 121 KOIs that previously had unreasonably large impact parameters.  We have updated the dispositions of 11 KOIs from FP to PC and 6 KOIs from PC to FP.  We flagged 361 cases with very weak S/N transit signals as likely false-alarms, flagged 100 KOIs with large planet radii that may be due to stellar binaries and flagged 13 KOIs that previously had a PC flag that now have a mass measurement inconsistent with the exoplanet hypothesis.

\begin{longrotatetable}
\begin{table}[!hbt]
    \scriptsize
    \centering
    \begin{tabular}{l c c c c c c c c}
\hline
KOI          & KIC         &   Period [d]           & T0 [MJD]          & \rprs       & $b$            & \rhoc [g/cm$^3$]      & u$_1$    & u$_2$ \\
 Disp        & Kepler-Name &   \rpl [R$_\oplus$]       & $d_{\rm transit}$ [ppm]      & $T_{\rm dur}$ [h]     & $T_{1.5}$ [h]      & \#TT$_{\rm obs}$                 & \#TT      & TTVflag \\
             & \adrs      &   $i$ [deg]              & $S$ [S$_\oplus$]        & S/N          & MES          & $\Delta$S/N$_{ttv}$          & S/N$_{wTTV}$  & S/N$_{woTTV}$ \\
             & kepmag      &   \rhostar [g/cm$^3$] & \teff [K]        & \rstar [\rsun] & \mstar [\msun] & \logg                 & \feh   & sparflag \\
\hline
1.01 & 11446443 & 2.47061338 $\pm$ 0.00000002 & 787.064865 $\pm$ 0.000009 & 0.123865 $\pm$ $^{ 0.000066 } _{ 0.000062 }$ & 0.8179 $\pm$ $^{ 0.0006 } _{ 0.0006 }$ & 1.8365 $\pm$ $^{ 0.0062 } _{ 0.0085 }$ & 0.3860 & 0.2724 \\
PPPP & Kepler-1 b & 14.1 $\pm$ $^{ 0.4 } _{ 0.3}$ & 14230.5 $\pm$ 4.6 & 1.7430 $\pm$ 0.0012 & 1.2264 $\pm$ 0.0010 & 431 & 431 & 000 \\
& 8.4001 $\pm$ $^{ 0.0095 } _{ 0.0130 }$ & 84.4130 $\pm$ $^{ 0.0096 } _{ 0.0140 }$ & 874 $\pm$ $^{ 56 } _{ 40 }$ & 4360.4 & 6468.0 & -0.19 & 7070.7 & 7323.0 \\
& 11.338 & 1.219 $\pm$ $^{ 0.062 } _{ 0.061 }$ & 5815 $\pm$ 66 & 1.04 $\pm$ $^{ 0.02 } _{ 0.02 }$ & 0.99 $\pm$ $^{ 0.03 } _{ 0.03 }$ & 4.390 $\pm$ $^{ 0.078 } _{ 0.078 }$ & 0.013 $\pm$ 0.041 & 2 \\
[.2cm]
2.01 & 10666592 & 2.20473545 $\pm$ 0.00000005 & 788.535435 $\pm$ 0.000017 & 0.075182 $\pm$ $^{ 0.000010 } _{ 0.000008 }$ & 0.22 $\pm$ $^{ 0.16 } _{ 0.22 }$ & 0.40810 $\pm$ $^{ 0.00024 } _{ 0.00030 }$ & 0.3043 & 0.3129 \\
PPPP & Kepler-2 b & 16.18 $\pm$ $^{ 0.39 } _{ 0.36 }$ & 6668.3 $\pm$ 1.4 & 3.87410 $\pm$ 0.00052 & 3.59923 $\pm$ 0.00047 & 602 & 602 & 006 \\
& 4.7156 $\pm$ $^{ 0.0009 } _{ 0.0012 }$ & 89.9596 $\pm$ $^{ 0.0323 } _{ 0.2344 }$ & 4146 $\pm$ $^{ 183 } _{ 183 }$ & 5952.0 & 3862.2 & 13.12 & 20695.8 & 20905.0 \\
& 10.463 & 0.2788 $\pm$ $^{ 0.0088 } _{ 0.0069 }$ & 6447 $\pm$ 64 & 1.97 $\pm$ $^{ 0.04 } _{ 0.04 }$ & 1.51 $\pm$ $^{ 0.03 } _{ 0.02 }$ & 4.15 $\pm$ $^{ 0.11 } _{ 0.11 }$ & 0.192 $\pm$ 0.042 & 2 \\
[.2cm]
3.01 & 10748390 & 4.88780321 $\pm$ 0.00000024 & 786.095757 $\pm$ 0.000081 & 0.05786 $\pm$ $^{ 0.00045 } _{ 0.00012 }$ & 0.054 $\pm$ $^{ 0.186} _{ 0.041}$ & 3.693 $\pm$ $^{ 0.039 } _{ 0.286 }$ & 0.64 & 0.10 \\
PPPP & Kepler-3 b & 4.914 $\pm$ $^{ 0.080 } _{ 0.088 }$ & 4315.6 $\pm$ 6.5 & 2.3639 $\pm$ 0.0054 & 2.2327 $\pm$ 0.0030 & 184 & 184 & 007 \\
& 16.699 $\pm$ $^{ 0.081 } _{ 0.405 }$ & 89.79 $\pm$ $^{ 0.18 } _{ 0.63 }$ & 85.3 $\pm$ $^{ 4.8 } _{ 4.4 }$ & 862.1 & 2034.6 & 0.02 & 4445.0 & 4449.0 \\
& 9.174 & 2.31 $\pm$ $^{ 0.15 } _{ 0.10 }$ & 4541 $\pm$ 54 & 0.77 $\pm$ $^{ 0.01 } _{ 0.01 }$ & 0.77 $\pm$ $^{ 0.03 } _{ 0.02 }$ & 4.542 $\pm$ $^{ 0.023 } _{ 0.014 }$ & 0.410 $\pm$ 0.084 & 1 \\
[.2cm]
4.01 & 3861595 & 3.8493714 $\pm$ 0.0000012 & 787.26293 $\pm$ 0.00053 & 0.03960 $\pm$ $^{ 0.00072 } _{ 0.00129 }$ & 0.914 $\pm$ $^{ 0.016 } _{ 0.035 }$ & 0.22168 $\pm$ $^{ 0.12742 } _{ 0.06371 }$ & 0.3150 & 0.3029 \\
PPPP & Kepler-1658 b & 13.073 $\pm$ $^{ 0.563 } _{ 0.563 }$ & 1298 $\pm$ 14 & 2.639 $\pm$ 0.091 & 2.108 $\pm$ 0.041 & 279 & 279 & 001 \\
& 5.56 $\pm$ $^{ 0.89 } _{ 0.37 }$ & 80.8 $\pm$ $^{ 1.4 } _{ 0.9 }$ & 5158 $\pm$ $^{ 588 } _{ 452 }$ & 132.7 & 235.6 & -0.27 & 6624.3 & 6706.0 \\
& 11.432 & 0.0868 $\pm$ $^{ 0.0080 } _{ 0.0081 }$ & 6776 $\pm$ 139 & 3.05 $\pm$ $^{ 0.09 } _{ 0.08 }$ & 1.78 $\pm$ $^{ 0.06 } _{ 0.14 }$ & 3.716 $\pm$ $^{ 0.026 } _{ 0.037 }$ & -0.076 $\pm$ 0.082 & 1 \\
[.2cm]
5.01 & 8554498 & 4.78032745 $\pm$ 0.00000087 & 787.80350 $\pm$ 0.00016 & 0.03654 $\pm$ $^{ 0.00022 } _{ 0.00019 }$ & 0.9517 $\pm$ $^{ 0.0017 } _{ 0.0020 }$ & 0.346 $\pm$ $^{ 0.018 } _{ 0.015 }$ & 0.3746 & 0.28 \\
PPPP & ----- & 8.70 $\pm$ $^{ 0.50 } _{ 0.42 }$ & 959.3 $\pm$ 4.2 & 2.020 $\pm$ 0.015 & 1.383 $\pm$ 0.019 & 282 & 282 & 002 \\
& 7.48 $\pm$ $^{ 0.12 } _{ 0.11 }$ & 82.69 $\pm$ $^{ 0.13 } _{ 0.13 }$ & 1387 $\pm$ $^{ 142 } _{ 129 }$ & 380.8 & 360.2 & -0.50 & 5139.5 & 5446.0 \\
& 11.665 & 0.182 $\pm$ $^{ 0.030 } _{ 0.026 }$ & 5936 $\pm$ 109 & 2.19 $\pm$ $^{ 0.11 } _{ 0.10 }$ & 1.37 $\pm$ $^{ 0.13 } _{ 0.10 }$ & 3.894 $\pm$ $^{ 0.049 } _{ 0.054 }$ & 0.235 $\pm$ 0.158 & 1 \\
[.2cm]
5.02 & 8554498 & 7.05201 $\pm$ 0.00010 & 785.646 $\pm$ 0.031 & 0.00316 $\pm$ $^{ 0.00038 } _{ 0.00051 }$ & 0.7272 $\pm$ $^{ 0.0001 } _{ 0.6674 }$ & 0.35 $\pm$ $^{ 1.72 } _{ 0.34 }$ & 0.3746 & 0.28 \\
FFNN & ----- & 0.74 $\pm$ $^{ 0.13 } _{ 0.11 }$ & 10.8 $\pm$ 2.6 & 4.3 $\pm$ 1.1 & 4.3 $\pm$ 1.1 & 191 & 187 & 000 \\
& 11.9 $\pm$ $^{ 4.3 } _{ 4.7 }$ & 89.45 $\pm$ $^{ 0.55 } _{ 3.57 }$ & 827 $\pm$ $^{ 81 } _{ 81 }$ & 6.2 & 0.0 & -9.27 & 6127.5 & 6329.0 \\
& 11.665  & 0.182 $\pm$ $^{ 0.030 } _{ 0.026 }$ & 5936 $\pm$ 109 & 2.19 $\pm$ $^{ 0.11 } _{ 0.10 }$ & 1.37 $\pm$ $^{ 0.13 } _{ 0.10 }$ & 3.894 $\pm$ $^{ 0.049 } _{ 0.054 }$ & 0.235 $\pm$ 0.158 & 1 \\
[.2cm]
6.01 & 3248033 & 1.3341070 $\pm$ 0.0000032 & 788.4519 $\pm$ 0.0011 & 0.0101 $\pm$ $^{ 0.0003 } _{ 0.0002 }$ & 0.15 $\pm$ $^{ 0.35 } _{ 0.14 }$ & 1.23 $\pm$ $^{ 0.11 } _{ 0.42 }$ & 0.3306 & 0.3007 \\
FFFF & ----- & 1.443 $\pm$ $^{ 0.066 } _{ 0.051 }$ & 120.0 $\pm$ 3.7 & 2.106 $\pm$ 0.035 & 2.081 $\pm$ 0.033 & 804 & 803 & 100 \\
& 4.88 $\pm$ $^{ 0.16 } _{ 0.62 }$ & 89.03 $\pm$ $^{ 0.97 } _{ 5.84 }$ & 3812 $\pm$ $^{ 192 } _{ 177 }$ & 43.0 & 21.5 & -8.89 & 16014.0 & 16685.3 \\
& 12.161 & 0.800 $\pm$ $^{ 0.045 } _{ 0.049 }$ & 6344 $\pm$ 66 & 1.30 $\pm$ $^{ 0.03 } _{ 0.03 }$ & 1.20 $\pm$ $^{ 0.04 } _{ 0.02 }$ & 4.323 $\pm$ $^{ 0.060 } _{ 0.06 }$ & 0.041 $\pm$ 0.042 & 2 \\
[.2cm]
7.01 & 11853905 & 3.2136689 $\pm$ 0.0000010 & 786.11467 $\pm$ 0.00030 & 0.024505 $\pm$ $^{ 0.000138 } _{ 0.000078 }$ & 0.021 $\pm$ $^{ 0.216 } _{ 0.016 }$ & 0.4637 $\pm$ $^{ 0.0080 } _{ 0.0350 }$ & 0.4025 & 0.2645 \\
PPPP & Kepler-4 b & 4.05 $\pm$ $^{ 0.12 } _{ 0.11 }$ & 727.6 $\pm$ 2.7 & 3.986 $\pm$ 0.010 & 3.8879 $\pm$ 0.0083 & 338 & 338 & 005 \\
& 6.326 $\pm$ $^{ 0.040 } _{ 0.161 }$ & 89.40 $\pm$ $^{ 0.55 } _{ 1.64 }$ & 1191 $\pm$ $^{ 62 } _{ 73 }$ & 346.8 & 294.5 & -0.44 & 10568.0 & 10881.0 \\
& 12.211 & 0.475 $\pm$ $^{ 0.023 } _{ 0.032 }$ & 5833 $\pm$ 64 & 1.51 $\pm$ $^{ 0.04 } _{ 0.04 }$ & 1.15 $\pm$ $^{ 0.05 } _{ 0.05 }$ & 4.12 $\pm$ $^{ 0.11 } _{ 0.11 }$ & 0.171 $\pm$ 0.042 & 2 \\
[.2cm]
8.01 & 5903312 & 1.1601530 $\pm$ 0.0000016 & 787.92102 $\pm$ 0.00072 & 0.01201 $\pm$ $^{ 0.00035 } _{ 0.00028 }$ & 0.7483 $\pm$ $^{ 0.0001 } _{ 0.5935 }$ & 4.28 $\pm$ $^{ 0.32 } _{ 1.47 }$ & 0.3792 & 0.2764 \\
FFFF & ----- & 1.230 $\pm$ $^{ 0.052 } _{ 0.044 }$ & 169.6 $\pm$ 4.5 & 1.324 $\pm$ 0.020 & 1.306 $\pm$ 0.019 & 1153 & 1152 & 100 \\
& 6.73 $\pm$ $^{ 0.18 } _{ 0.89 }$ & 89.32 $\pm$ $^{ 0.68 } _{ 4.34 }$ & 2025 $\pm$ $^{ 93 } _{ 110 }$ & 48.4 & 36.6 & -8.99 & 15328.0 & 16106.1 \\
& 12.450 & 1.731 $\pm$ $^{ 0.042 } _{ 0.071 }$ & 5883 $\pm$ 66 & 0.94 $\pm$ $^{ 0.02 } _{ 0.02 }$ & 1.00 $\pm$ $^{ 0.02 } _{ 0.03 }$ & 4.460 $\pm$ $^{ 0.044 } _{ 0.044 }$ & -0.063 $\pm$ 0.042 & 2 \\
    \end{tabular}
    \caption{Abbreviated Catalog of planet candidates. The positions of the KOI Disposition and Kepler-Name columns have been moved relative to the electronic table to facilitate viewing the examples presented in the typeset manuscript. The tabulated information is described in the numbered list in Section \ref{sec:unified}, where the numbers refer to the column numbers in the electronic table. (Alternative values of stellar and planetary parameters listed in Columns 63-84 are not presented in this abbreviated sample.)}
    \label{tab:planetcatalog}
\end{table}
\end{longrotatetable}

\begin{table}[!hbt]
    \centering
    \begin{tabular}{c c c}
    \hline
        Name & Value & units \\
    \hline
    R$_{\odot}$     & 6.957$\times$10$^8$     & m \\
    R$_{\oplus}$    & 6.371$\times$10$^6$     & m \\
    G               & 6.674$\times$10$^{-11}$ & m$^3$ kg$^{-1}$ s$^{-2}$ \\
    au              & 1.4956$\times$10$^{11}$ & m \\
    \teff$_{,\odot}$ & 5778                    & K \\
    \hline
    \end{tabular}
    \caption{Adopted astrophysical constants for derived planetary parameters.  The mean radius of the Sun (R$_{\odot}$), Earth (R$_{\oplus}$), the gravitational constant (G), astronomical unit (au), and solar effective temperature (\teff$_{,\odot}$) are used to calculate absolute planetary radii, orbital inclinations and incident flux.}
    \label{tab:astro_constants}
\end{table}

We determine the following transit parameters and their uncertainties by fitting transit models of \citet{Mandel:2002} to \ik lightcurves assuming circular orbits (and adjusted for TTVs if TTVflag = 1): mean stellar density, \rhoc; orbital period, $P$; impact parameter, $b$; scaled radius, \rprs.  Limb-darkening coefficients are held fixed. The transit model is used calculate the depth of transit, $d_{\rm transit}$, and the signal to noise ratio, S/N.  The fitted transit model parameters, together with equations (3) and (2) from \citet{Seager:2003}, are used to calculate transit duration measured from $1^{\rm st}$ to $4^{\rm th}$ contact, $T_{\rm dur}$, and an alternate measure of transit duration (see items 24 -- 27 in the numbered list below for details).
Planetary inclinations are calculated using the formula: 
\begin{equation}
i = \arccos{\left(b\frac{R_\star}{a}\right)}.
\end{equation}

\begin{enumerate}
    \item \textbf{KIC}: The \ik input catalog number of the target star.
    \item \textbf{KOI}: The \ik object of interest number, with the integer portion referring to the target star (source of light) and the decimal portion referring to the particular signal, i.e., the putative planet's transits. 
    \item \textbf{Kepler ID}: The number and letter assigned to verified planets. Blank if the KOI has not been assigned a \ik planet number/letter.  Kepler IDs are do not automatically qualify a KOI to be dispositioned a PC (see Section \ref{sec:selection}) and are supplied for cross-identification purposes only. 
    \item \textbf{Period}, $P$ [days]: Mean orbital period (the period that gives the best fit to observed TTs). For candidates with only one transit observed, the period is estimated based on transit duration and impact parameter, assuming a circular orbit, and given as a negative value to distinguish it from periods computed for multiple transit objects. Note that mono-transit planet candidates (those observed to transit just once) are not included in any of our analyses, figures or tabulations that require knowledge of the orbital period, but are accounted for in assessing multiplicity of planetary systems in all cases.
    \item \textbf{$\sigma(P)$}: Uncertainty (half-width of 68.27\% confidence interval for the average period during the epoch of \ik observations) of $P$.  See discussion in Section \ref{sec:periods}. Based on the uncertainty in fitted slope of the measured TTs. As discussed in the second and third paragraphs of Section \ref{sec:periodtheory}, the actual mean orbital periods of \ik planets over time scales much longer than the four years of \ik observations can differ from the values given in Table \ref{tab:planetcatalog} by many times as much as the listed uncertainties. The reported periods and uncertainties for mono-transit candidates assume circular orbits and are considered unreliable.  
    \item \textbf{Epoch}, T0 [BJD~--~2454900]:  Time, calculated using the constant period reported in Column \#4, at which the center of the planetary disk is closest to the center of the stellar disk for the last transit that occurred prior to halfway between the start of the first Quarter and end of the final Quarter that \ik observed the target star in question, whether or not said transit was actually observed by \ikt. This is the mode value from MCMC, calculated using the SciPy kernel density estimator \texttt{stats.gaussian\_kde} with default settings \citep{Virtanen:2020}.  
    BJD $\equiv$ Julian Date viewed from the barycenter of the Solar System. 
    \item \textbf{$\sigma({\rm T0})$}: Uncertainty in the epoch.
    \item \textbf{Planet/star radius ratio}, $R_p/R_\star$: This is the mode value from MCMC of the ratio of the planet's radius to the stellar radius.
    \item \textbf{$\sigma_+(R_p/R_\star)$}: Upwards (to higher values) uncertainty of the ratio of planet's radius to stellar radius.
    \item\textbf{$\sigma_-(R_p/R_\star)$}: Downwards uncertainty of the ratio of planet's radius to stellar radius.
   \item \textbf{Impact parameter}, $b$:  We report the best fit value. See Section \ref{sec:transitmodel}.
   \item \textbf{$\sigma_+(b)$}: Upwards uncertainty of $b$.
   \item \textbf{$\sigma_-(b)$}: Downwards uncertainty of $b$.
   \item \textbf{Stellar density from lightcurve}, \rhoc~[g/cm$^3$]: The mean-stellar density computed from the photometric model.  A \emph{circular orbit} model has been assumed, and a separate fit is done for each planet in a multi. This is the mode value from MCMC.
   \item \textbf{$\sigma_+(\rhoc)$}: Upwards uncertainty of \rhoc.
   \item \textbf{$\sigma_-(\rhoc)$}: Downwards uncertainty of \rhoc.
   \item \textbf{$u_1$}: First quadratic limb-darkening parameter.
   \item \textbf{$u_2$}: Second quadratic limb-darkening parameter.
   \item \textbf{TTV flag}: A three digit number preceded by `T' represents the results of three separate searches for TTVs. The first digit is 1 if TTVs have been included in our transit model, 0 otherwise. The second digit refers to results listed in \citet{Holczer:2016}'s catalog, with 2 signifying sinusoidal TTVs, 1 polynomial TTVs, 0 means no TTVs found, and - means not investigated. The third digit is  the overall rating from \citet{Kane:2019} catalog, with 9 signifying the strongest TTVs, 8 strong TTVs, 7 weak and/or noisy TTVs, and 6 and below no TTVs of interest; we use `-' for KOIs not rated by \citet{Kane:2019}.
   \item \textbf{\# transits}: Total number of transits observed by \ikt.
   \item \textbf{\# TTs}: Number of transits for which the transit time has been measured.  This number is always $\leq$ the number in the previous column.
    \item \textbf{$d_{\rm transit}$} [ppm]: Depth of transit. Specifically, the depth of the transit model evaluated at the mid-transit time assuming a circular orbit. We report the mode of the MCMC distribution.  
   \item \textbf{$\sigma(d_{\rm transit})$}: Uncertainty of $d_{\rm transit}$.
   \item \textbf{Transit duration}, $T_{\rm dur}$ [hr]:  Transit duration, measured from first contact to fourth contact ($T_{\rm dur}\equiv$ T$_{1,4}$), which is the standard measurement for exoplanet transit duration. Based on transit model parameters ($\rhoc$, $P$, $R_p/R_\star$, $b$), using equation (3) of \citet{Seager:2003}.  The mode of the MCMC distribution is reported.  
   \item \textbf{$\sigma(T_{\rm dur})$}: Uncertainty of $T_{\rm dur}$.
   \item \textbf{Alternate measure of transit duration}, $T_{1.5,3.5}$ [hr]: Mode of (T$_{1,4}$ + T$_{2,3})/2$. T$_{2,3}$ uses equation (2) from \citet{Seager:2003}.  If $b> 1-(R_p/R_\star)$ then T$_{2,3}$ is set to zero. Note $T_{\rm dur} > T_{1.5,3.5} \ge 0.5~T_{\rm dur}$.
   \item \textbf{$\sigma(T_{1.5,3.5})$}: Uncertainty of $T_{1.5,3.5}$.
    \item \textbf{Signal to noise ratio}, S/N: The ratio of the signal, S, which is the integral of the transit model over all transits, to the noise, N, estimated as the standard deviation of the photometric lightcurve out of transit.  Calculated assuming a constant period if  TTVflag=0, and incorporating TTVs if the first digit of the TTVflag~=~1. S/N~$\geq$ 7.1 is a generally necessary (but not sufficient) for a KOI to be dispositioned as a PC in our catalog. {Four exceptions to this general requirement are made for KOIs with Kepler numbers whose lightcurves also passed our visual inspection (see Section \ref{sec:selection}).}
   \item \textbf{MES}: Multiple event statistic, as reported in DR25.  KOIs that were not listed in DR25 have `0.0' listed in this column. 
   \item \textbf{Improvement in S/N when transit model allows for TTVs}: S/NwTTV~--~S/NwoTTV: Difference in S/N calculated without TTVs and with TTVs (e.g., \citealt{Ofir:2018}). Positive for most KOIs; typically large when the preferred fit uses TTVs (since less out-of-transit data obfuscates the signal) and small when it does not. 
   \item \textbf{$\chi^2_{\rm wttv}$}: Chi-squared calculated with TTVs based on the best fit transit model.  If the first digit of TTVflag = 0, then photometric uncertainties have been scaled to have $\chi^2_{\rm wottv}$ = DOF (degrees of freedom).  If the first digit of TTVflag = 1, then photometric uncertainties have been scaled to have $\chi^2_{\rm wttv}$ = DOF. 
   \item \textbf{$\chi^2_{\rm wottv}$}: Chi-squared calculated without TTVs.  The values are normalized as for $\chi^2_{\rm wttv}$. 
   \item \textbf{Scaled semimajor axis, $a/R_\star$}: Scaled semimajor axis.
   \item \textbf{$\sigma_+(a/R_\star)$}: Upward uncertainty of the scaled semimajor axis.
   \item \textbf{$\sigma_-(a/R_\star)$}: Downward uncertainty of the scaled semimajor axis.
   \item \textbf{Inclination, $i [^\circ]$}: Inclination of the planet's orbit relative to the plane of the sky; an edge-on orbit has $i = 90^\circ$. 
   \item \textbf{$\sigma_+(i)$}: Upwards uncertainty of the inclination. Note that $i+\sigma_+(i) \le 90^\circ$.
   \item \textbf{$\sigma_-(i)$}: Downwards uncertainty of the inclination.
   \item \textbf{Planet radius, $R_p$} [R$_\oplus$]:  Radius of the planet.
   \item \textbf{$\sigma_+(R_p)$}: Upwards uncertainty of $R_p$.
   \item \textbf{$\sigma_-(R_p)$}: Downwards uncertainty of $R_p$.
   \item \textbf{(Bolometric) Incident flux, $S$ [S$_\oplus]$}: Amount of flux intercepted by the planet relative to that intercepted by the Earth.
   \item \textbf{$\sigma_+(S)$}: Upward uncertainty of incident flux.
   \item \textbf{$\sigma_-(S)$}: Downward uncertainty of incident flux.
     \item \textbf{Kepmag}: Target star magnitude in the \ik passband.
   \item \textbf{Stellar density}, $\rhostar$ [g/cm$^3$]: The mean stellar density from the stellar parameter tables. (Not the estimate derived from the transit model, which is given in Column 14).  
   \item \textbf{$\sigma_+(\rhostar)$}: Upwards uncertainty of \rhostar.
   \item \textbf{$\sigma_-(\rhostar)$}: Downwards uncertainty of \rhostar.
   \item \textbf{Stellar temperature}, $\teff$ [K]: The target star's effective temperature, taken from the stellar properties catalog.
   \item \textbf{$\sigma(\teff)$}: Uncertainty of \teff.
   \item \textbf{Stellar radius}, $\rstar$ [R$_\odot$]: Radius of the star as given in the stellar properties catalog.
   \item \textbf{$\sigma_+(\rstar)$}:  Upwards uncertainty of \rstar.
   \item \textbf{$\sigma_-(\rstar)$}:  Downwards uncertainty of \rstar.
   \item \textbf{Stellar mass}, $\mstar$ [M$_\odot$]: Mass of the star. 
   \item \textbf{$\sigma_+(\mstar)$}:  Upwards uncertainty of \mstar.
   \item \textbf{$\sigma_-(\mstar)$}:  Downwards uncertainty of \mstar.
     \item \textbf{Stellar surface gravity}, $\logg$ [cgs]: Surface gravity of the star as given in the stellar properties catalog. For stellar parameters taken from DR25 and \citet{Berger:2020a}, \logg\ is from isochrone models.  For stellar parameters taken from \citet{Fulton:2018}, \logg\ is based on spectroscopy. 
   \item \textbf{$\sigma_+(\logg)$}:  Upwards uncertainty of \logg.
   \item \textbf{$\sigma_-(\logg)$}:  Downwards uncertainty of \logg.
   \item \textbf{Metallicity}, [dex]:  Metallicity of the target star.
   \item \textbf{$\sigma$}(Metallicity):  Uncertainty of the metallicity of the target star. 
   \item \textbf{Stellar model source}:  0 = solar parameters, 1 = DR25, 2 = \citet{Berger:2020a}, 3 = \citet{Fulton:2018}, 4 = special (see Section \ref{sec:star}). {Aside from KOI-1792 (for reasons discussed in Section \ref{sec:selection}) and two binary star systems for which special parameters are used, }\citet{Fulton:2018} values are used where available, and the \citet{Berger:2020a} values are always preferred over those listed in DR25.
 \item \textbf{Dispositions}: This four letter code gives the dispensations (status flags) from this work, DR25supp, DR25, and DR24 in reverse chronological order, with F = False positive (or False alarm), P = Planet/Candidate, N = not included in specified catalog,  S = rejected by us because S/N $<7.1^($\footnote{Note, the rounding done to produce the S/N numbers displayed in Column 31 of this table is done \emph{after} assignment of `S' in the flags. }$^)$, M = Jupiter-sized objects for which mass measured via RV clearly exceeds the 13 M$_{\rm Jupiter}$ limit for classification as a planet but otherwise would have been dispositioned `P' or `R' and R = Radius too large (see Section \ref{sec:selection}), but meets all other criteria. `N' is never applicable to the first column (our catalog provides dispositions for all KOIs listed) and `S', `M' and `R' are used exclusively for our dispositions. {The seven validated planets (with Kepler numbers listed on Nexsci) that failed either the radius cut or the S/N cut but passed visual inspection of the lightcurve were dispositioned `P' (see Section \ref{sec:selection} for details).}
\end{enumerate}

64~--~85. Values of the same properties reported in columns 39~--~44 and 46~--~61 with stellar parameters taken from \cite{Berger:2020a}. All zeros if host star not characterized by \cite{Berger:2020a}.\\

\bigskip
We treat KOIs with dispositions beginning with `S' and `M' as false positives throughout our study; KOIs whose dispositions beginning with `R' are treated as false positives for most purposes, but are included together with planet candidates in our study of the distribution of planetary radii (\S\ref{sec:sizes}).  The total number of planet candidates is 4376, including 35 mono-transits; additionally, there are 100 KOIs, including 2 mono-transits, that we vetted as `R'. There are 709 multiple transiting planet systems, which account for 1791 candidates, including {7 mono-transits}.

\begin{figure}[!hbt]
\begin{center}
\includegraphics[width=0.7\textwidth,angle=0]{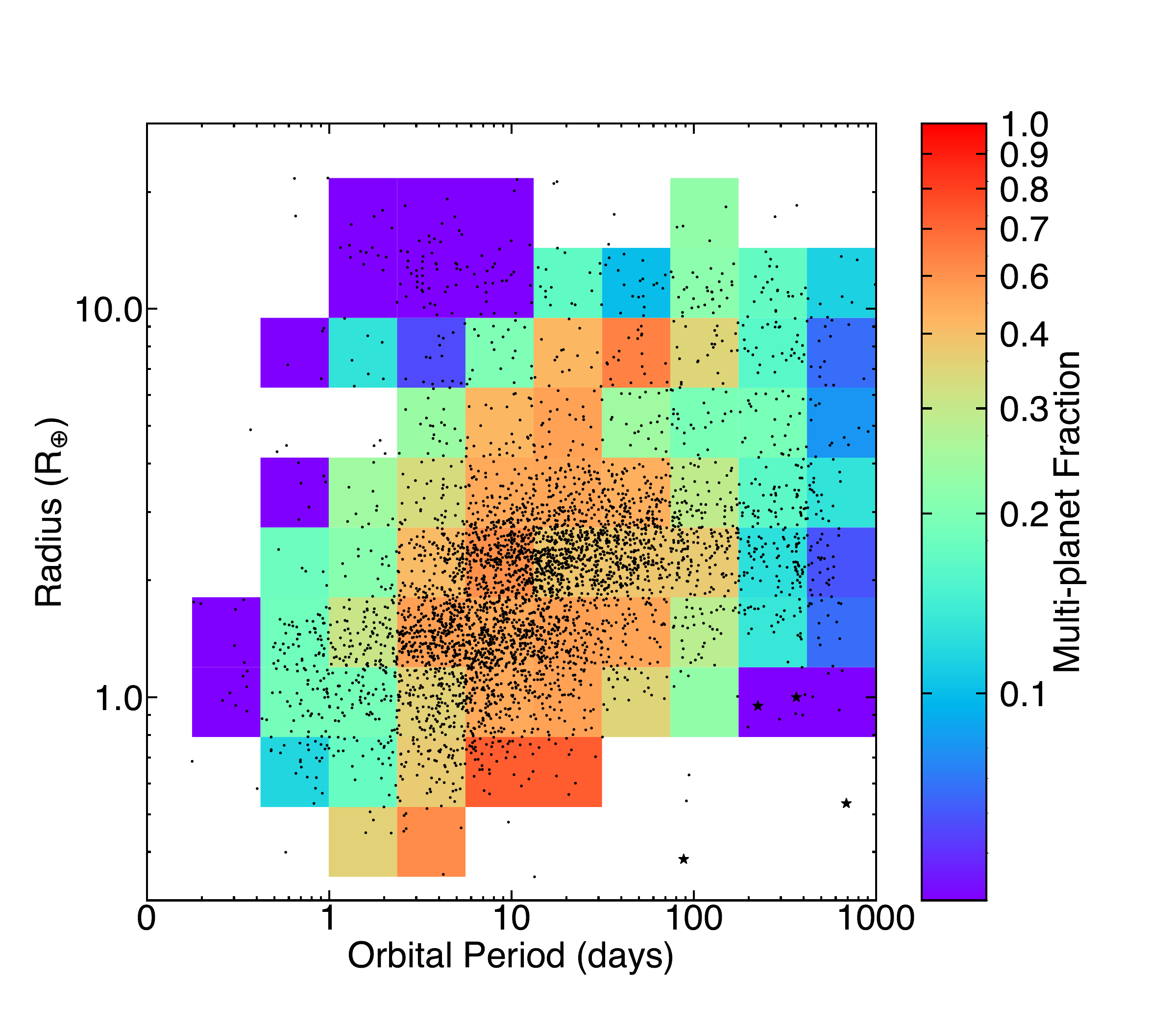}
\caption{The observed multiplicity of transiting planets from the \ik sample as a function of period and radius, with individual squares representing a factor of $10^{3/8} \approx 2.371$ in $P$ and $12.5^{1/6} \approx 1.523$ in $R_p$.  Black dots represent individual \ik planets and black stars represent the terrestrial planets in our Solar System.  The colored squares give the multiplicity fraction for each area that contains at least 4 planets.  A planet is considered to be part of a multiplanet system if more than one transiting planet candidate is seen in the photometric lightcurve.  If a square were to contain four planets, three of which were from multi-planet systems,  then the multi-planet fraction would be equal to 0.75.  The multiplicity fraction shows a paucity in multiplanets observed for hot jupiters, ultra-short periods ($P<1$ day) and long period planets ($P > 100$ days), with an observed increase in multiplicity as the planet radius decreases in the well-populated population with periods from 1~--~40 days.  
}\label{fig:perrad_multimap}
\end{center}
\end{figure}

\begin{figure}[!hbt]
\includegraphics[width=0.49\textwidth,angle=0]{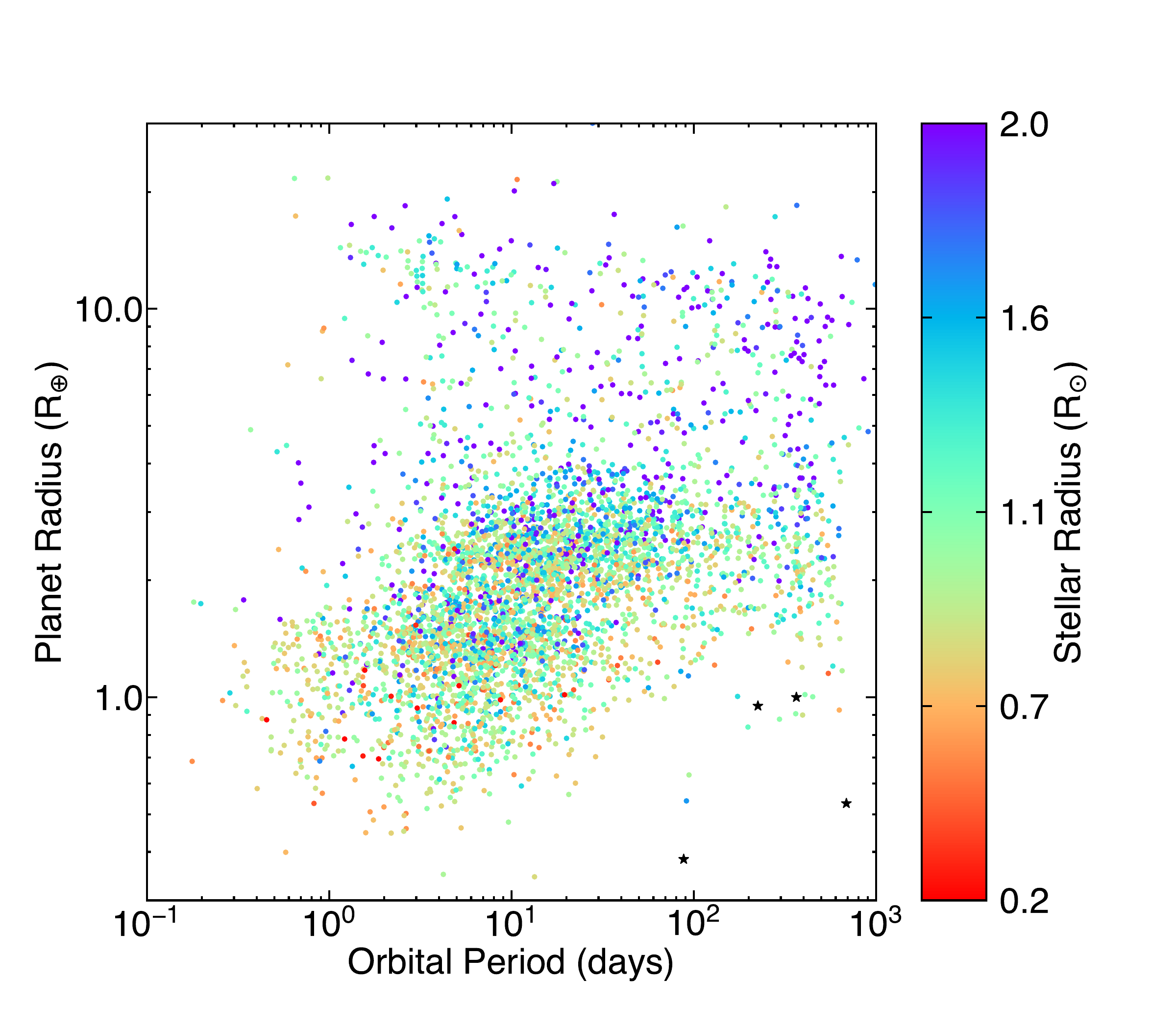}
\includegraphics[width=0.49\textwidth,angle=0]{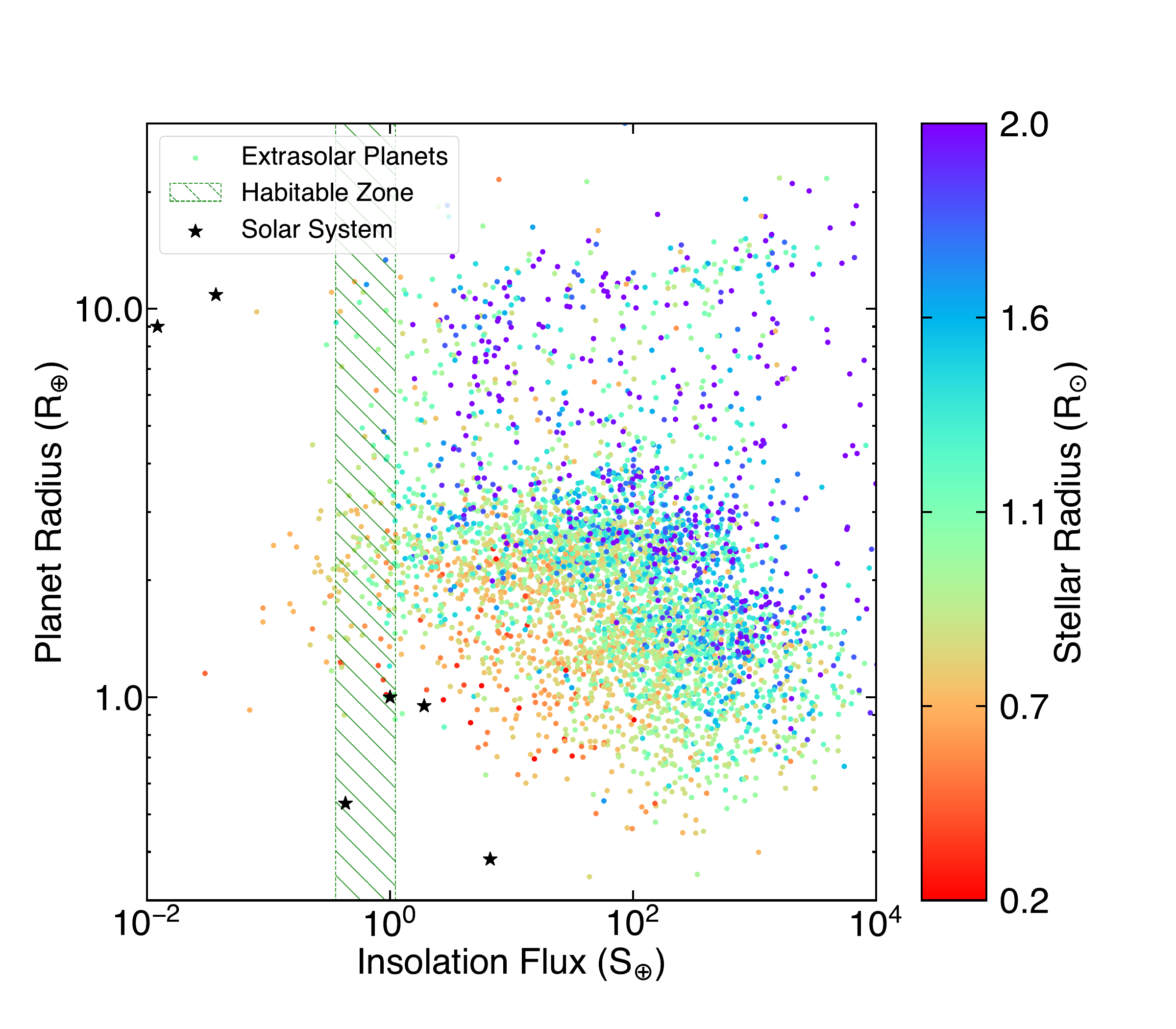}
\caption{The \ik planet population showing radius ($R_p$) vs.~period ($P$) on the left and radius vs.~incident flux (S$_{\oplus}$) on the right.  The points have been colored by the radius of the host star ($R_\star$).  Solar System planets are marked with black stars, and the conservative habitable zone for an Earth-like planet around a Sun-like star \citep{Kopparapu:2014} is shown by the hashed green lines. 
}\label{fig:rpvsperwrstar}
\end{figure}

Figure \ref{fig:perrad} displays the sizes and radii of almost all of the planet candidates in our catalog, and indicates the multiplicity of the planetary systems in which they are observed.  The bulk of multi-planet candidates have radii $R_p \lesssim 4$~R$_\oplus$ and orbital periods $2 \lesssim P \lesssim 100$ days.  Figure \ref{fig:perrad_multimap} confirms these general observations and demonstrates the paucity of additional transiting planets in systems hosting transiting hot jupiters (e.g., \citealt{Steffen:2012}) and planets with periods longer than $\sim 100$ days. The period-radius valley separating super-Earths from sub-Neptunes slopes downward from $\sim 2$~R$_\oplus$ at $P=2$~days to $\sim 1.5$~R$_\oplus$ at $P=40$~days (left panel of Figure \ref{fig:rpvsperwrstar}). 

Figure \ref{fig:perrad_snrmap} shows the distribution of transit S/N for planet candidates.  The left panel shows the total S/N calculated using Equation \ref{eq:S/N}, and the right panel shows the average S/N per transit.  The total S/N has a floor of 7.1 as discussed in \S\ref{sec:selection} and marks the boundary where the smallest planets can be found for a specified range in orbital period before the population is dominated by false alarms.  The rate of false alarms drastically increases for periods close to 1-year because of the {\it rolling band noise}, which is video cross-talk between detectors that produces a slowly drifting band of noise {and static {\it star-like} artifacts}. Rolling-band is most prevalent on detectors 22, 26, 44 and 58 and creates a mismatch in the sky estimate leading to a semi-periodic instrumental systematic that can resemble a photometric signature of a transit-like signal (see Table 13 from \citealt{vanCleve:2016}).  The 372 day heliocentric orbit of \ik and quarterly change of the spacecraft roll results in rolling band only being present once or twice per year for any given target star.  Significant effort was invested to identify and eliminate false-alarms with periods centered on 372 days without being overly aggressive against low S/N Earth-like transits \citep{Thompson:2018}. Nonetheless, visual examination of the left panel of our Figure \ref{fig:rpvsperwrstar} and Figure {7} of \citet{Thompson:2018} shows an overall increase in planet candidates centered on 372 days.  As noted by \cite{Burke:2019}, extra care and analysis is needed when assessing the statistical properties of planets in this regime, and this analysis would strongly benefit from a directed study of noise properties and follow-up observations from facilities such as {\it HST} and {\it Plato}. {Moreover, low-S/N long period PCs can also be produced by a few systematic dips in a lightcurve that line up to produce a signal that looks transit-like. Such chance alignments are common for TCEs that appear to transit just 3 or 4 times, but their frequency declines with 5 or more transits \citep{Mullally:2018}.}

 {We computed minimum periods of each of the 7 mono-transit PCs in multis by examining their target's lightcurve to determine the shortest possible period for which no additional transits would have occurred at times when \ik obtained good data (i.e., was observing and photometric noise was not excessive). These lower bounds, as well as the (crude) estimates of the orbital periods using the duration and shape of the lightcurve together with the stellar properties and an assumed circular orbit (the absolute value of the negative  $P$ listed for these planets in Table~\ref{tab:planetcatalog}) and the periods of longest-period observed companion planet, are given in Table~\ref{tab:minp}. 
 None of these mono-transiting planets in multis have minimum periods significantly shorter than the period estimates from lightcurve analysis, which would be the case if their transverse orbital velocity at the time of transit was larger than that of a planet on a circular orbit with the minimum calculated period. Note that all of the multi-transiting PCs in these seven systems, apart from KOI-2525.01 and 4307.01, have been verified as planets and given \ik numbers.}

\begin{table}[]
    \centering
    \begin{tabular}{|c|c|c|c|c|}
        \multicolumn{5}{c}{}\\
        \hline
        KOI & $P$ [minimum] & $P$ [estimate] & system & $P$ [neighbor]\\
         & (days) & (days) & multiplicity & (days)\\
        \hline
        $435.02$  & $528.5$ & $934$ & 6 & $62.30$\\
        $671.05$  & $691.5$ & $4865$ & 5 & $16.26$\\
        $693.03$  & $588.0$ & $719$ & 3 & $28.78$\\
        $1108.04$ & $507.0$ & $1289$ & 4 & $18.93$\\
        $1870.02$ & $490.0$ & $550$ & 2 & $7.96$\\
        $2525.02$ & $418.5$ & $562$ & 2 & $57.29$\\
        $4307.02$ & $483.0$ & $993$ & 2 & $160.85$ \\
        \hline
        \end{tabular} 
 \caption{ {The minimum orbital period of each of the mono-transit planet candidates in multi-planet systems based on \ik lightcurve coverage (the smallest value, stepping by 0.5 days, in which the data could not rule out a second transit) is listed together with the orbital period estimated from the transit duration and shape and the stellar properties, system multiplicity and the period of the transiting planet orbiting immediately interior to it.  The uncertainties of the estimated periods are poorly-quantified and likely to be substantial. }}
    \label{tab:minp}
\end{table}

\begin{figure}[!hbt]
\includegraphics[scale=1.2]{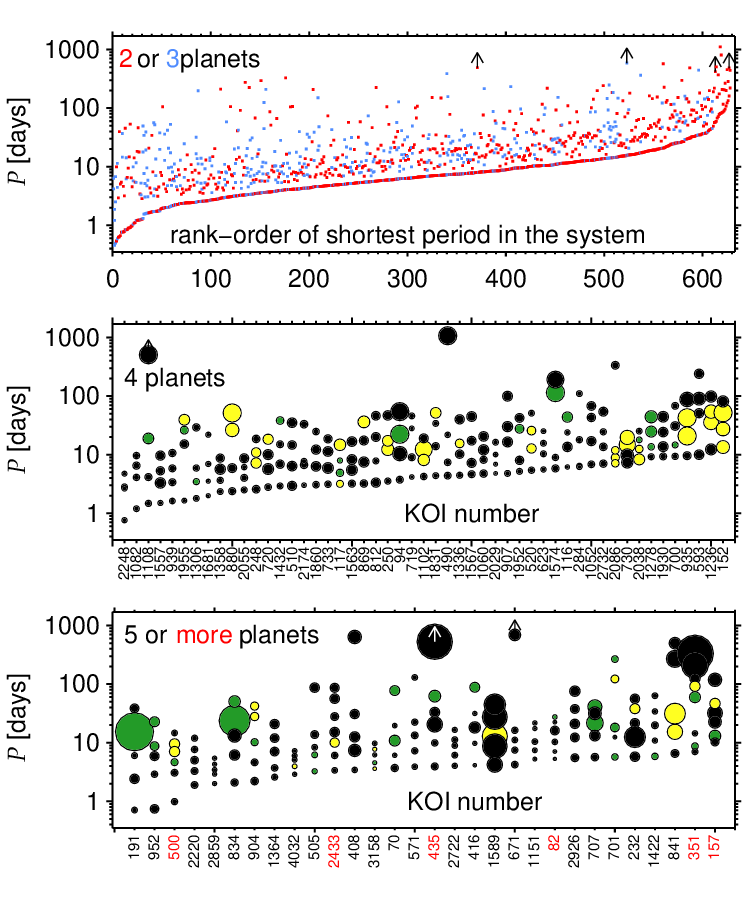}
\caption{ These plots show the orbital periods of each of the planets in every \ik multi-planet system.  The three panels show systems of differing multiplicity. Within each panel, all symbols along a given vertical line represent planets belonging to the same star, and the systems are ordered horizontally according to the  orbital period of the innermost planet.  The top panel presents systems with two (red) or three (blue) detected transiting planets
, the middle panel systems with four transiting planets, and the bottom panel systems with at least five transiting planets. The top panel does not plot KOI-1843.03 due to its short orbital period, $0.177$~day, which is $<40\%$ as long as that of any other \ik PC in a multi; however, its two companion planets are represented by blue points touching the vertical axis. In the middle and bottom panels, the symbol size is proportional to planetary radius, although in the middle panel planets larger than 5 R$_\oplus$ are shown as if $R_p = 5$ R$_\oplus$, and the systems are labeled by KOI number at the bottom, with red numbers being used for systems with 
6 -- 8 transiting planets. Colors of planets in the lower two panels denote the TTV flag given by \cite{Kane:2019} -- light yellow is 8 or 9 (moderate or strong TTVs), dark green is 7 (probable TTVs), and black is no TTVs (note, however, that recently-identified PCs were not examined by \citealt{Kane:2019}). The periods of mono-transit objects are constrained to be greater than the plotted values by not showing a second transit in the recorded data; see Table~\ref{tab:minp}. 
}
\label{fig:gallery} 
\end{figure}

Figure \ref{fig:gallery} provides a compact sketch of the architectures of all 709 multi-planet systems discovered by \ikt.  Systems are grouped into panels by number of planets detected and sorted within each panel by the orbital period of the innermost planet. Planetary radii and the presence of detected TTVs are also indicated for planets in systems with 4 or more PCs.  Note that all 81 systems with 4 or more planets have at least one planet with $P<13.5$ days, and only two such systems lack planets with $P<10.4$~days. The smallest period ratio of planet candidates plausibly orbiting the same star is 1.1167, between the two small $\sim 0.75$~R$_\oplus$ PCs KOI-3444.04 ($P = 14.15$~days) and  KOI-3444.01 ($P = 12.67$~days); no other pair of \ik PCs orbiting the same star has a period ratio smaller than 1.15.

\subsection{KOI-2433: A Candidate Seven-Planet System}\label{sec:2433}

Figures \ref{fig:perrad} and \ref{fig:gallery} show that there are now two \ik systems with more than six transiting planet candidates. One of these is the familiar 8-planet KOI-351 (Kepler-90) system (e.g., \citealt{Lissauer:2014, Shallue:2018}). With the addition of KOI-2433.08 from \cite{Shallue:2018}'s list, a second \ik target now has eight KOIs, which we elaborate upon in the following paragraphs. 
According to Table \ref{tab:planetcatalog}, the 10 and 15 day planets, which were validated by \cite{Rowe:2014} as Kepler-385~b \& c, both have S/N $\sim$ 28. The 56 day planet has an S/N = 16.4 and was validated as Kepler-385~d by \cite{Armstrong:2021}. The S/N of the other five KOIs range from 11.1~--~14.3, above the S/N~$> 10$ required as one of many tests for planet validation of PCs in multis by \cite{Rowe:2014}. 
 
The 0.6 day KOI 2433.05 has disposition FFFF\footnote{See the item \#63 in the numbered list in \S\ref{sec:unified} for an explanation of the dispositions given in the last column of Table \ref{tab:planetcatalog}.}. Its lightcurve clearly exhibits secondary events indicative of an eclipsing binary. Furthermore, pixel-level data show the periodic dimming to be spatially offset from the target star. Thus, we do not consider this ultra-short period KOI further.

The {three validated planets plus the 28 day} KOI have dispositions PPPP. The disposition of the 6 day KOI is PPNP; its S/N~=~12.4. The 86 day KOI has PPFN; its S/N~=~11.2. The 3.4 day KOI-2433.08 has PNNN; its S/N~=~11.1.

The outer three planet candidates have neighboring pairs with period ratios nominally placing them just wide of first-order MMRs ({as does the pair of validated planets orbiting interior to this threesome}), increasing the likelihood of them being real planets \citep{Lissauer:2014}. Indeed, it is possible that the validated pair of planets are locked within a two-body resonance and that the three outer planets librate within a three-body resonance. Period ratios between the planets all exceed 1.5, which is sufficient for system stability provided the planets have masses typical for their sizes and small eccentricities, which are the norm for \ik planets in systems of high multiplicity (Fig.~\ref{fig:EccMulti}).  In sum, we consider KOI-2433 as having seven strong planet candidates. Nonetheless, validation of all of these candidates to well above 99\% probability of representing true planets is beyond the scope of this paper.

\smallskip

\section{Characteristics of the Planet Population: Multis vs.~Singles} \label{sec:singles}

\ik found far more multiple planet candidate systems (multis) than would be the case if candidates were randomly 
distributed among target stars \citep{Lissauer:2011b, Latham:2011}. \cite{Lissauer:2012} presented a statistical analysis that combined the large numbers of multis observed by \ik that were listed in \citealt{Borucki:2011b} (as modified by \citealt{Lissauer:2011b}) together with the assumption that false positives are nearly randomly distributed among \ik targets to demonstrate that the fidelity of \ik multiple planet candidates is far higher than that for singles.  \cite{Lissauer:2014} expanded upon the statistical analysis of \cite{Lissauer:2012} and developed techniques that \cite{Rowe:2014} used to validate more than 700 of \ikt's multiple planet candidates as true planets. 

\begin{figure}
    \centering
    \includegraphics[width=0.7\textwidth]{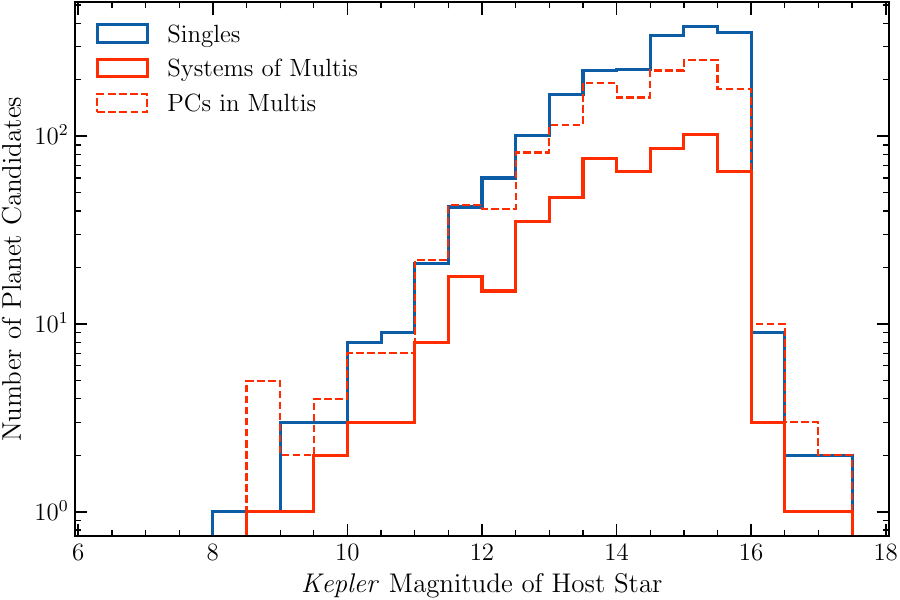}
    \caption{Histogram of \ik planet hosting star magnitudes, with hosts of a single PC in blue and stellar hosts of multis in red.  All PCs that were observed by \ik are represented in this plot.}
    \label{fig:magnitudes}
\end{figure}

The distribution of stellar host magnitudes for singles and multis are shown in Figure \ref{fig:magnitudes}. We compare the S/N distributions of \ik singles and multis in \S\ref{sec:snr} and estimate the fraction of apparent \ik multis in Table \ref{tab:planetcatalog} that do not represent planets orbiting the same star in \S\ref{sec:falsemultis}.

Many groups, dating back to the aforementioned \citet{Lissauer:2011b} and \citet{Latham:2011}, have compared characteristics of the distributions of single planet candidates in the \ik sample with those of individual planets in multis. We compare size distributions (in \S\ref{sec:sizes})   and orbital period distributions (in \S\ref{sec:periodDist}) for ensembles of planets in singles to those within multis and between planets in two-planet systems to those in systems of higher multiplicity. The distributions of normalized transit durations and eccentricities of subsets of planets segregated by multiplicity, orbital period, size and orbital spacing are presented and analyzed in \S\ref{sec:ecc}. 

Up through Subsection \ref{sec:periodDist}, all PCs are accounted for in the determination of the multiplicity of a planetary system. For our analysis of planetary size distributions (\S\ref{sec:sizes}), we treat KOIs dispositioned as ``R'' as if they were PCs, but do not count PCs with high impact parameter or  S/N~$< 12$. In \S\ref{sec:periodDist}, we omit PCs with S/N~$< 12$ as well as those with only one observed transit. In \S\ref{sec:ecc}, we employ somewhat different restrictions described in the introduction to that subsection.

\subsection{Signal to Noise Distributions {\& Reliability of the Sample}}\label{sec:snr}

\begin{figure}
    \centering
    \includegraphics[width=0.7\textwidth]{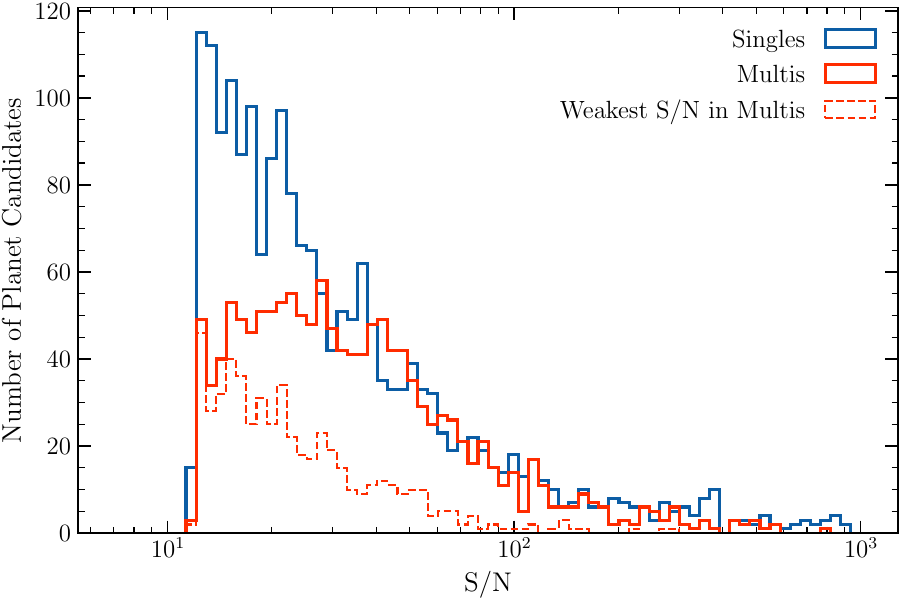}
    \caption{Number of singles (blue) and multis (red) as functions of S/N. The dotted red curve is restricted to the lowest S/N planet in each multi-planet system. All PCs that were observed by \ik are represented. See Fig.~\ref{fig:perrad_snrmap} for a plot of how typical S/N varies with planetary radius and orbital period.  }
    \label{fig:snr_comparison}
\end{figure}

Figure \ref{fig:snr_comparison} compares the histograms of singles vs.~multis as functions of S/N. Note that there are similar numbers of multis and singles over a wide range in S/N, but singles dominate for both the smallest values of S/N and the largest ones. Examining the distributions more quantitatively, \ik found almost 50\% more planet candidates in singles than in multis, but Figure \ref{fig:snr_comparison} shows that the numbers of singles and multis are nearly equal across the S/N range 25~--~180.  Below S/N~=~12 and above S/N~=~300, well over twice as many singles as multis have been identified; intermediate ratios are found in transition regions (see Figure \ref{fig:snr_comparison}).  The predominance of singles at high S/N is primarily accounted for by the paucity of large planets, especially hot Jupiters, in multis \citep{Latham:2011, Steffen:2012}. 

The excess of singles at low S/N is probably caused by a combination of the following effects, {the first two of which are related to \ik multis being a highly reliable subsample of the \ik planet candidate population \citep{Latham:2011, Lissauer:2011b, Lissauer:2012, Lissauer:2014, Rowe:2014}}: (1) A larger fraction of singles being false alarms (this wasn't the situation in 2014 because of the more aggressive search for PCs in multis \citep{Rowe:2014}, but probably is the case now because of the more automated procedure used to find and vet KOIs in recent catalogs); (2) Planet candidates with low S/N cannot be tested as rigorously as can PCs with high S/N, so a larger fraction of EBs and other FPs are included in the PC list, since they are more common among targets with a single transit-like pattern than around targets with more than one such pattern \citep{Latham:2011, Lissauer:2011b, Lissauer:2012, Lissauer:2014, Rowe:2014}; (3) When multiple planets transit, the detectability of planets other than the highest S/N candidate by the \ik pipeline is reduced significantly \citep{Zink:2019}, especially if the highest S/N planet has low S/N, so it is less likely that other transiting planets in the system have been detected; (4) TTVs, which are detected in more than twice as large a fraction of multis compared to singles (Table \ref{tab:planetcatalog} and \S\ref{sec:periodDist}), reduce MES, but if they are accounted for in our model fits, they don't reduce S/N; and, perhaps, (5) Real differences between the populations, such as longer period planets having lower S/N, all other factors being equal, combined with neighboring companion planets typically being within a factor of $\sim 2 - 3$ in period and planets less likely to transit at long periods unless typical relative inclinations of planetary orbits decrease significantly as periods increase.  Determining what fraction of the difference is caused by (3) is important both to a comparison between multis and singles and to understanding the fidelity of the sample of single planet candidates, but it is beyond the scope of this work. 

\smallskip
\subsection{Split Multis \& Orbital Period Aliases}\label{sec:falsemultis}

Searching through photometric time series for transiting planets may yield false positives, a term that conventionally means a real astrophysical signal within the same detector pixels, but caused by something other than a planetary transit.   For the purpose of computing occurrence rates, a subclass of false positives is signals resulting from transits of real planets  that are \emph{not} hosted by the intended target star, such as the planets observed to transit KOI-119 (\S\ref{sec:star}). Such blending causes the planetary radius to be significantly under-estimated (as the target star is invariably the one providing the most photons absorbed by the pixels), and the stellar host type and other parameters can be incorrect, to the detriment of statistical occurrence efforts. These real planets, which we classify as PCs, should therefore be termed false positives for the purposes of computing planetary occurrence rates. {(The number of \ik planet candidates identified as likely to orbit stars other than the \ik target star is quite small, and all prime suspects that we know of are discussed within the first half of this subsection.)}

In the case of candidate multi-planet systems, it could be that each periodic signal is due to a real planet, but these planets are not orbiting the same star. Recognizing the planets' reality, we do not use the word ``false'' here, but instead call these systems ``split multis". Although they are real planets, split multis can be a source of contamination for dynamical studies. Given that the orbital periods of planet candidates span a very large range, random planets around different stars will not necessarily appear strange when (mis)interpreted as orbiting the same star. 
In some fraction of cases, however, we would notice their periods are too close together to be stable if interpreted as being around the same star. 
Given the steep dependence of occurrence rates on planet size, split multis are expected to be most common for binary stars of similar luminosities.  
Therefore, when studying multiple planet systems, it can be advantageous to restrict one's attention to a subset of \ik targets that have been filtered to minimize contamination from binary stars with similar luminosities \citep[e.g.,][]{Hsu:2019,He:2020}.

Here we discuss a few potential split multis.  
KOI-284, first listed in the catalog of \cite{Borucki:2011b} and introduced here in \S\ref{sec:star}, includes the planet candidates KOI-284.02 (with orbital period is $P = 6.415$~days) and KOI-284.03 (with $P = 6.178$~days), both of which appear to be larger than Earth.  \cite{Lissauer:2011b} noted that if both of these planet candidates represented planets orbiting the same stellar host, then for any reasonable densities their proximal orbits would lead to dynamical instability on a short time scale. Thus, they do not represent planets in the same planetary system; this system is the prototypical \ik example of a split multi. Further investigation revealed that the target ``star'' is actually a binary system with nearly identical components \citep{Lissauer:2012}, and both of these PCs were subsequently validated as true planets, one orbiting each star, by \cite{Lissauer:2014}.  These two planetary systems are now known collectively as Kepler-132. There are two additional validated planets, with orbital periods of 18 and 110 days.  However, it is not yet known which member of the stellar binary any of the planets orbit, apart from the two 6 day period planets needing to orbit different stars.

Two planet candidates apparent in the lightcurve of KOI-2248, denoted KOI-2248.01 ($P = 2.818$~days) and KOI-2248.04 ($P = 2.646$~days), were first listed in the \cite{Batalha:2013} catalog, and they were highlighted as a split multi by \cite{Fabrycky:2014}. Neither of these KOIs has ever been dispositioned as a false positive in any KOI catalog, including our own (Table \ref{tab:planetcatalog}). Two other planet candidates in the system were listed in DR25supp, KOI-2248.02 ($P = 9.49$~days) and KOI-2248.03 ($P = 0.762$~days). \cite{Shallue:2018} listed a ``new'' candidate associated with the same \ik target and $P = 4.745$~days that we have included in our catalog and dubbed KOI-2248.05 (\S\ref{sec:selection}).  None of the KOIs in the system has yet been verified as a {\it bona fide} planet. One member of the pair of planet candidates with similar periods, KOI-2248.04, has S/N = 8.4 in our catalog, which is quite peculiar given that it was identified so early in the \ik mission, calling its veracity into question.  The other member of that nominally-unstable pair, KOI-2248.01, has fairly low DR25 disposition score of 0.895, but we find that it has a respectable S/N = 16.5.  The 9.5-day signal, KOI-2248.02, has an unacceptably low S/N = 4.5, so we classify it as a false alarm, probably an alias of KOI-2248.05, which has a period almost exactly half as long.

Thus, KOI-2248 hosts four planet candidates, two of which (both somewhat suspect for other reasons) could not represent planets orbiting the same star, as a system containing both of these putative planets would be dynamically unstable. High-resolution imaging of the target star KOI-2248 has been carried out by  \cite{Furlan:2017}, whose Table 8 (accessed by Vizier, 16 July 2018)  lists KOI-2248 with two stellar near neighbors on the sky plane detected by WFC3 (Hubble Space Telescope): one (denoted B) that is 0.169 magnitude fainter in the F555W filter band (a nearly-equal brightness companion) with a separation of just 0.148 arcsec, so it could well host one of the $P \sim 2.7$~day PCs.  The other stellar sky-plane neighbor (denoted C) is 3.872 arcsec away and 5.318 magnitudes fainter, so it is not likely to be the host of either planet. The only survey using WFC3 listed in Table 1 of \cite{Furlan:2017} is that by \cite{Gilliland:2015} and \cite{Cartier:2015}, but neither of these latter papers addresses KOI-2248, so we suspect that the results were posted to the KFOP (\ik Follow-up Observing Program) website, but not published in the refereed literature.

Although Table \ref{tab:planetcatalog} does not list any proximate-period split multis other than those in KOI-284 and KOI-2248 as discussed above, an early draft version of this table listed the planet candidate KOI-521.02 with an orbital period $P = 10.82$~days, which would place it too close to the Neptune-size PC KOI-521.01 ($P = 10.16$~days) for dynamical stability. We therefore re-examined the \ik lightcurve of this target and found that alternate transits of KOI-521.02 had been missed, so the actual period of this PC is only half of what is listed in previous catalogs, $P = 5.41$~days. The Q1-Q16 KOI catalog was the first to list  KOI-521.02, and dispositioned it as an FP \citep{Mullally:2015}.  In the subsequent catalogs DR24, DR25 and DR25supp, it was listed as a PC, albeit one with weak S/N (initially 8.7 and later 7.3; with the corrected period it has a more respectable S/N = 9.7).  The estimated orbital periods of both KOI-521.01 and 521.02 changed by less than 1 part in $10^4$ between previous catalogs, and were similarly close to these estimates in early drafts of our catalog.  The TCE searches for DR24 and DR25 both identified KOI-521.02 with the correct 5.41 day period. However, because this TCE had a period that differed by a factor of two from a previously-cataloged KOI on the same target and the orbital phases matched, it was assigned the previous period in both of those PC catalogs. 

A systematic search for period aliasing was done in \S5.4 of the \cite{Batalha:2013} catalog paper. However, it is unlikely that the error in the initial period estimate for KOI-521.02, which was not identified as a KOI until a later catalog search, would have been recognized had we not given it special scrutiny because it appeared to be part of a split multi. This example suggests that some nontrivial fraction of low S/N PCs in our catalog could have listed periods that err by a factor of two. Several other PCs have had their estimated periods changed by a factor of roughly two subsequent to their first appearance in an official \ik KOI catalog, most notably KOI-730.03, which was initially listed with a period of half its true value \citep{Borucki:2011b} that would have placed it within a 1:1 (co-orbital) resonance with KOI-730.02, but the additional scrutiny that this putative pair of Trojan planets received quickly led to its estimated period being corrected by \cite{Lissauer:2011b}.  By the end of the mission, the DR24 TCE table, KOI-730.03 was found with its correct period, with a large MES of 17. So in the calculation below, we do not count it as an alias corrected by dynamical considerations. 

Note that PCs with estimated periods (nearly) identical to or a factor of two different from the period of another KOI of the same target may suffer from the discarding of data surrounding the transit of the first candidate \citep{Schmitt:2017}, a type of aliasing not an issue when periods are not commensurate.  This effect contributed in the case of KOI-2248.02, but not for KOI-730.03, whose transit phase differed substantially from that of its commensurate sibling.  Both KOI-730.03 and KOI-521.02 had their periods adjusted only after receiving additional attention due to apparent co-orbital planets or unstable systems. We do not know of other PCs that were corrected for dynamical reasons.

Motivated by these examples, we next estimate the number of period aliases that likely remain in the sample of multis. The number of PCs {with measured periods in multis that have at least one other PC with measured period is 1781. We took each of the pairs of planets in the same target, 1610 in total, multiplied the lower orbital period by 2, and counted how often that change would make the pair unstable. Following \cite{Lissauer:2011b} and \cite{Fabrycky:2014}, planetary masses were taken as  $(R_p/$R$_\oplus)^\alpha$M$_\oplus$, where $\alpha=3$ for $R_p<$~R$_\oplus$ (sub-Earths), $\alpha=2.06$ for R$_\oplus<R_p<$~R$_{\rm Saturn}$, and finally, M$_{\rm Saturn}$ for $R_p>$~R$_{\rm Saturn}$. Pairs of planets were deemed unstable if their difference in semimajor axis was less than $2\sqrt{3}$ their mutual Hill separation \citep{Gladman:1993}. No higher-order accounting of stability (for triples, for instance) was performed for this calculation. Stability requirements for systems of $\geq 3$ planets are more restrictive, but cannot be expressed in such simple terms (see, e.g., \citealt{Petit:2020} and \citealt{Lissauer:2021}).


Of these 1610 mock pairs, 281 ($17.5$\%) were unstable. Since only 1 out of 1610 real pairs of planets were found (via instability) to be an alias, then the estimated rate for a planet candidate to have an aliased period is (1/281) $\approx 0.36$\%.  
This rate is small enough that we can safely neglect the rate of systems becoming unstable due to multiple PCs being affected by  aliases in the same system.   
%
Among planets with multiple transits in multi-transiting systems, we expect roughly $0.0036\times1781 \approx 6$ aliased planets, $5$ of which are yet-undiscovered, though since this is based on only $1$ such detected system, it should be considered an order of magnitude statement. If the systems with just a single transiting planet with a quoted period also have this aliasing problem at the same rate, then $0.0036\times2550 \approx 9$} are also aliased.  Aliasing may occur at a different rate among singles, however, as the signal to noise distribution differs (Fig.~\ref{fig:snr_comparison}), and the dataset when searching for additional transits needs to be censored in certain areas. 


Using a similar approach, we update the estimated rate of split multis from \cite{Fabrycky:2014}. 
We choose pairs of ($P$, $M_p/M_\star$) values of {all of the planet candidates, and determine that 453,281/9,419,770~$\approx 4.8$\% are unstable\footnote{{\cite{Fabrycky:2014} restricted their choice to pairs of planet candidates in multis. Applying that prescription to our data set results in $88,758/1,590,436 \approx 5.6$\% being unstable, a slightly larger percentage since the period distribution of PCs overall is broader than that of PCs in multis (Fig.~\ref{fig:period_multiplicities_split}). This yields an estimate of $\sim 36$ split multis. We prefer using all PCs because split multis can include planets that are single and/or those that are in multis.}}. Having two detected unstable pairs (via the split multi channel) out of 1610 PC pairs, we estimate that $2/0.048\sim 42$ of our pairs may actually be split multis. Thus, split multis are likely a larger contaminant for the study of planetary system architectures than are period aliases, even though only $\sim2.6$\% of sampled pairs in multis are expected to be planets orbiting different stars.} 

The above estimates suggest that, among planets in multis, there are $\sim 7$ times as many split multis as period aliases. A factor of 2 comes from twice as many split multis having been identified.  A slightly larger factor is because period aliases are more likely to result in instability since it's fairly common for a pair of planets in the same system to have a period ratio near two (Fig.~\ref{fig:gallery}; see also \citealt{Fabrycky:2014}), but the overall distribution of periods is quite broad (Fig.~\ref{fig:period_multiplicities_split}), so few pairs of randomly-selected planets have a period ratio near unity.

The estimates of the number of period aliases calculated in this section do not account for the aliases found for planets with periods $\lesssim 1 $~day analogous to those noted in Section \ref{sec:selection}, which are caused by \ik automated pipeline searches for TCEs being limited to signals with periods $> 0.5$ day. Aliases among ultra-short period planets are very unlikely to be found via the techniques discussed in this section because only one planetary system (Kepler-42) is known to have more than a single planet with $P < 2.25$~days (e.g., \citealt{Steffen:2013b,Lissauer:2023}).

\subsection{Size Distribution} \label{sec:sizes}

A total of 100 KOIs, none of which are in multis and 73 of which have $P > 10$~days, are dispositioned as ``R'' in Table \ref{tab:planetcatalog},
indicating that they were rejected as planet candidates solely due to their estimated size.  The placement of our upper bound on the estimated radius of a body that we classify as a planet candidate (\S\ref{sec:selection}) is somewhat arbitrary, so we consider KOIs that are rejected based on size alone together with planet candidates (KOIs vetted as ``P'') for all studies presented in this subsection.  

\begin{figure}
    \centering
    \includegraphics[width=0.7\textwidth]{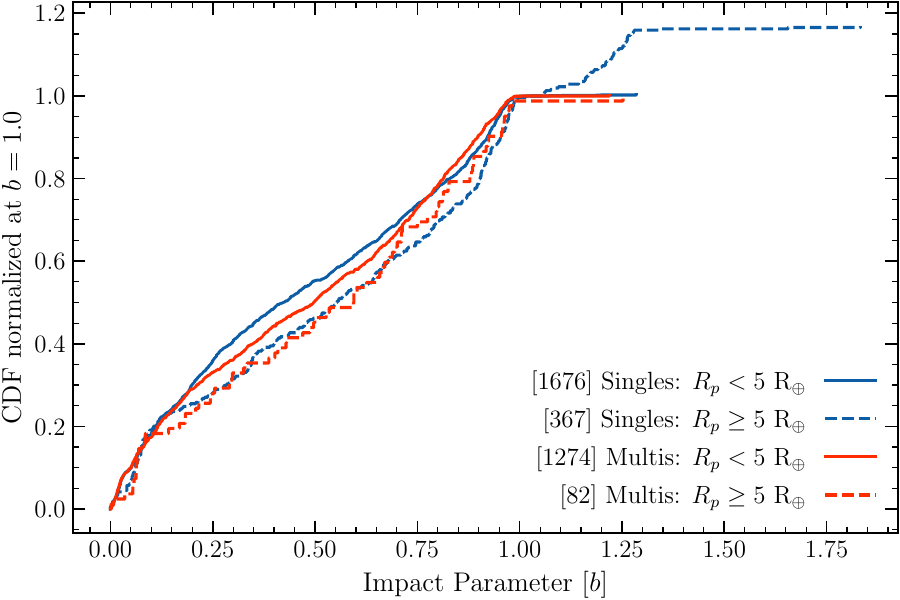}
    \caption{The cumulative impact parameter distributions of various subsets of planet candidates, as well as KOIs rejected solely because they are too large (dispositioned as R in Table \ref{tab:planetcatalog}), that were observed by \ikt. Red curves represent planet candidates in multis, blue curves single planet candidates; small planets are shown by solid lines and large planets by dashed  lines. There are 6 small ($R_p < 5$~R$_\oplus$) and 54 large singles that have $b > 1$. In multis, there are 2 small and 1 large planets  that have $b > 1$.  PCs with S/N~$< 12$ are not counted for the CDFs, but they are considered when determining the multiplicity of the systems in which the counted planets reside.}
    \label{fig:cdfimpactparambgt1}
\end{figure}

Figure \ref{fig:cdfimpactparambgt1} contrasts the fractional cumulative distributions of impact parameters of small and large planets in singles and multis. Small planets are clearly underrepresented for $b$ very close to unity because near-grazing transits of small planets are difficult to detect, and they are quite rare for larger impact parameters because $R_p/R_\star \gtrsim b-1$ is a requirement for a transit to be observable. Therefore, when comparing radius distributions, we only consider planets with estimated impact parameters $b + \sigma_+(b) <$ 0.95. (Because we report planet sizes using the modes of the distributions, we do not need to adopt the stricter $b < 0.8$ cutoff used by \citealt{Petigura:2020}.) We also exclude from our analysis in this subsection those candidates with S/N $< 12$ because the population of low S/N candidates has more FPs and typically higher fractional uncertainties on estimates of $R_p$; this cut removes the 4 PCs around targets for which solar parameters were assumed, for which planetary radii are especially poorly estimated. Planets that are excluded from the counts by these cuts are nevertheless included when determining the multiplicity of the system in which companion planets that meet these criteria reside.

\begin{figure}
    \centering
    \includegraphics[width=0.48\textwidth]{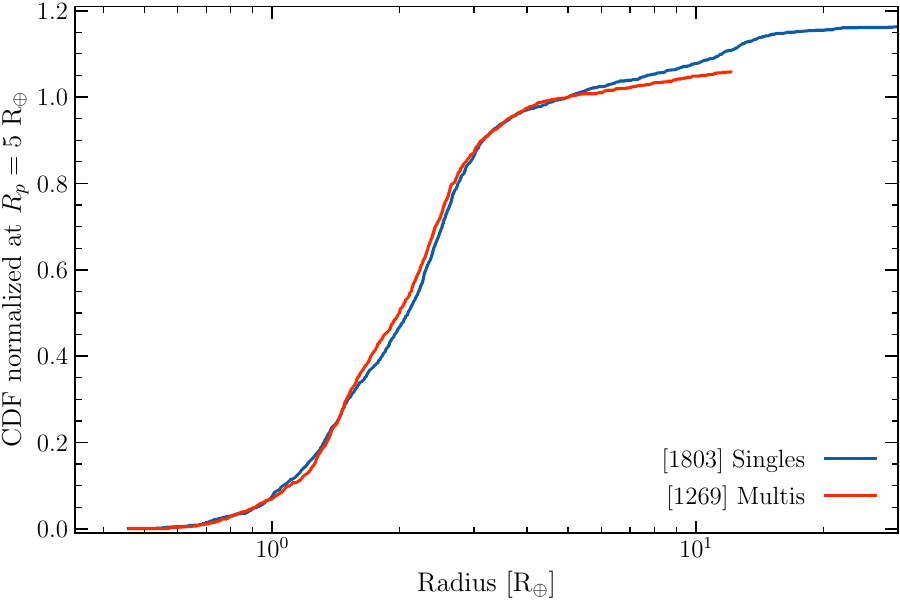}
    \includegraphics[width=0.48\textwidth]{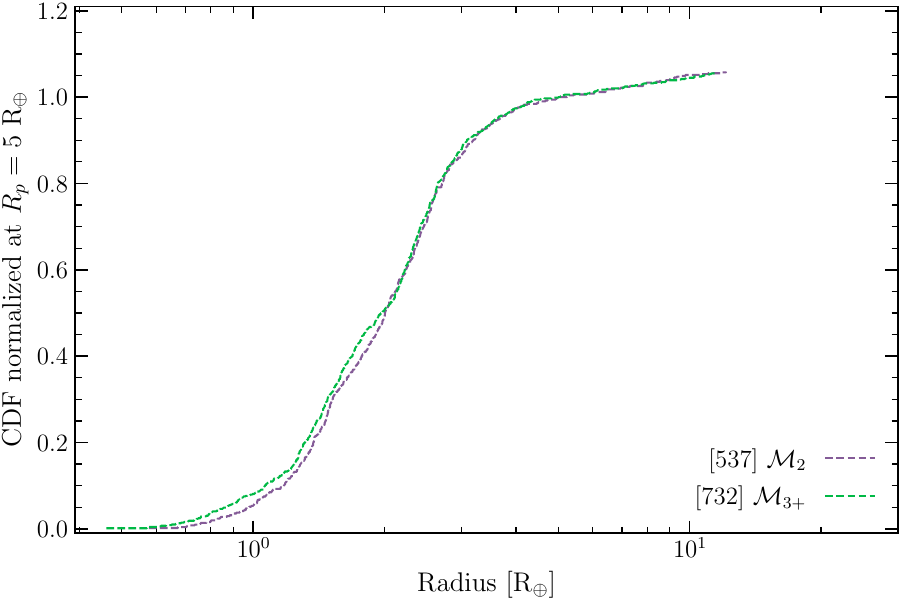}
    \caption{Normalized cumulative distribution function (CDF) of planetary radii of planets for specified system multiplicity. {The panel on the left compares singles with planets in multis, whereas the panel on the right compares planets in two-planet systems with those in systems of higher multiplicity.} We normalized the CDFs to unity at $R_p = 5$~R$_{\oplus}$. Candidates with S/N $< 12$, or $b + \sigma_+(b) \geq$ 0.95 are excluded from this radius distribution, although planets are considered to be in multis even if all of their companions fail to meet one or both of these cuts. These plots include KOIs rejected because they are too large (dispositioned as R in Table \ref{tab:planetcatalog}). Only the portion of the distributions with  $R_p < 30$~R$_{\oplus}$ are shown, although larger bodies that satisfy our criteria are included in the computation of the numbers given within brackets. No candidates in multis satisfying our upper limit on impact parameter have $R_p > 30$~R$_{\oplus}$, whereas 1 singles with status P and 12 with status R are larger than this value and not shown. (The green triangle with $P\approx150$ days in Fig.~\ref{fig:perrad} represents KOI-1426.03, which has V-shaped transits and $b>1$.)
    \label{fig:radius_multis_singles}}
\end{figure}

\begin{figure}
    \centering
    \includegraphics[width=0.48\textwidth]{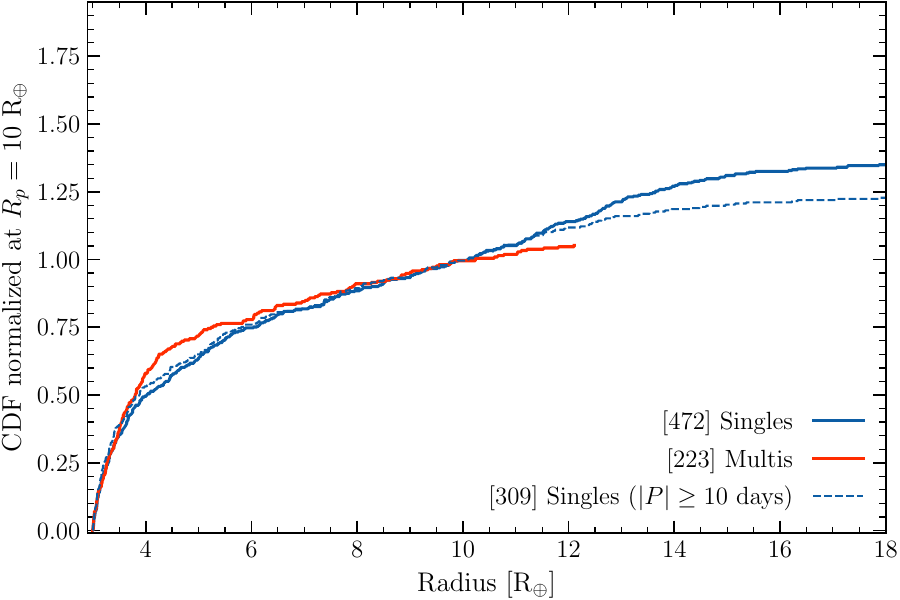}
    \includegraphics[width=0.48\textwidth]{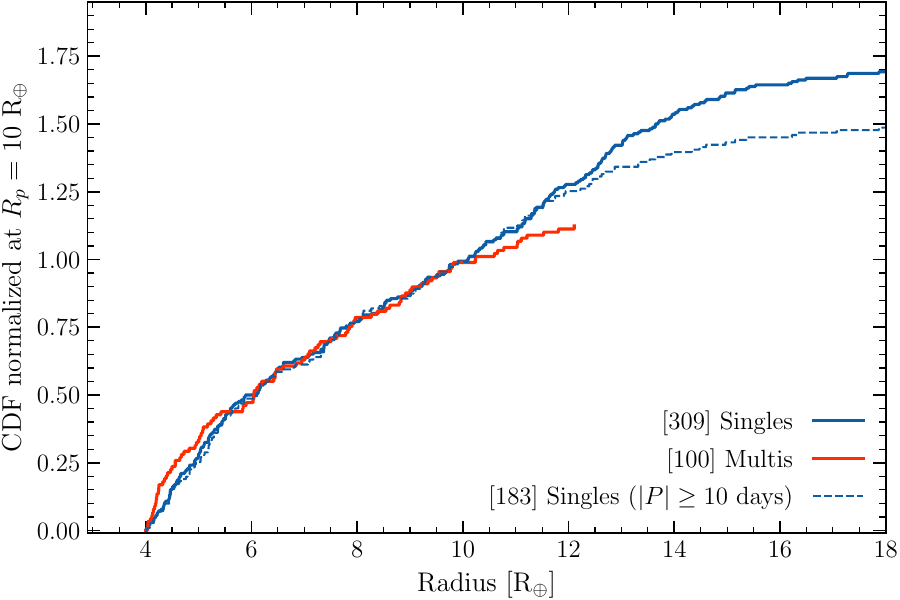}\\
    \includegraphics[width=0.48\textwidth]{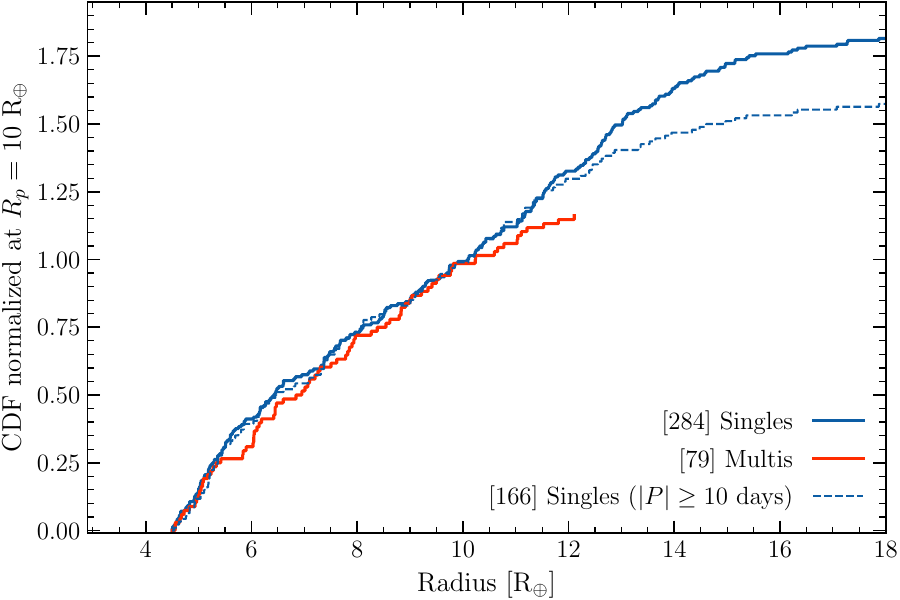}
    \includegraphics[width=0.48\textwidth]{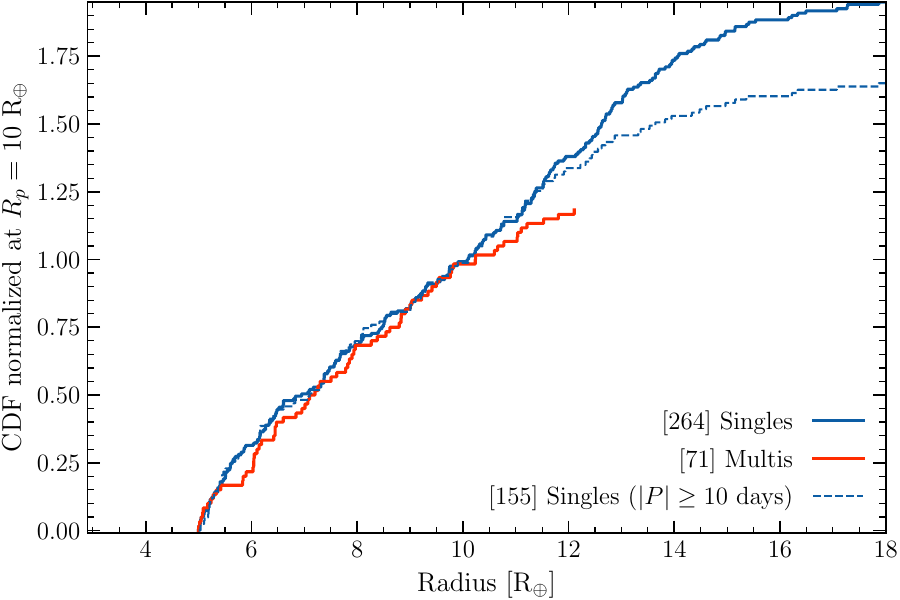}
    \caption{Cumulative distribution function of planetary radii comparing ``large'' planets in multiplanet systems with similarly-sized planets in singles, as well as to planets in singles with the additional constraint $P >$ 10 days; all curves are normalized to unity at 10 R$_{\oplus}$. The four panels, upper left, upper right, lower left and lower right, consider planets larger than 3, 4, 4.5 and 5 R$_\oplus$, respectively. The numbers in square brackets represent planets (plus KOIs rejected solely because they are too large and therefore dispositioned as ``R''  in Table \ref{tab:planetcatalog}) larger than the minimum value represented in the particular panel (no upper size cutoff). This plot uses the same criteria for inclusion as used in Fig.~\ref{fig:radius_multis_singles}.}
    \label{fig:large_radius_multiplicities_split}
\end{figure}

Figures \ref{fig:radius_multis_singles} and \ref{fig:large_radius_multiplicities_split} compare the size distributions of ensembles of planets in systems of different multiplicity.  In all cases, we lump together systems with three or more planets to have adequate numbers of planet candidates for statistically robust results. 

Giant planets are more common among \ik singles than among \ik multis  \citep{Latham:2011}. Nonetheless, when the cumulative size distributions for singles and planets in multis are normalized to unity at $R_p = 5$~R$_\oplus$ (Figure \ref{fig:radius_multis_singles}), the curves for planets up to this size are very similar{. Over the the size range $R_p < 5$~R$_\oplus$,   differences between the distributions are of  marginal statistical significance, and for $R_p < 2.5$~R$_\oplus$ the distributions of singles and planets in multis are consistent with being drawn from the same population.}

There are, however, various biases in the discovery of planets in multis as opposed to singles that could allow the actual distributions of transiting planets with $R_p < 2.5~$R$_\oplus$ to have some size dependence. Those tilting the size distribution of small planets towards singles include: detecting  a transiting planet in a \ik lightcurve reduced the amount of data used by the \ik pipeline to search for additional planets and thereby lowered the efficiency of detecting any other transiting planets associated with the same target star, reducing the probability of finding more planets, especially small ones \citep{Zink:2019}; geometric factors imply that a larger fraction of planets of a given orbital period transit large stars than small stars, and small planets are difficult to detect around large stars. By contrast, less photometrically noisy, brighter and/or smaller stars make all (but especially small) planets easier to detect, yielding a bias towards detecting multiple small planets around the best target stars.  Searching for additional planet candidates has at times been more aggressive for targets with at least one candidate already identified, and some searches for transiting planet signatures such as that of \citet{Shallue:2018} have focused exclusively on lightcurves of targets already known to possess planet candidates.


Plotting a similar comparison between the size distributions of planets in systems with two planet candidates vs.~those in higher-multiplicity systems (the panel on the right in Figure \ref{fig:radius_multis_singles}) shows a slight excess of super-Earth size planets relative to \hbox{(sub-)Neptunes} in systems with three or more PCs, with the distribution of two-planet systems lying between those for single planets and those for high multiplicity systems. While the differences between singles and multis over the entire range in radii are highly significant, both the Kolmogorov-Smirnov test and the Anderson-Darling test {show marginally significant differences between 2's and 3+'s. These marginal differences persist when restricting to the range $R_p < 5$~R$_\oplus$, but such differences are no longer evident if restricting to $R_p < 2.5$~R$_\oplus$, even though the majority of planets in multis are smaller than 2.5~R$_\oplus$.}


The divergence between the curves in the left panel of Fig.~\ref{fig:radius_multis_singles} comes in gradually as $R_p$ increases. There is no sign of divergence below  4~R$_\oplus$, but it is plainly there above 5~R$_\oplus$.  Nonetheless, it is clear that multis are deficient relative to singles for planets that are larger than Neptune. Note that because \ik has detected far more small planets than large ones, errors in estimates of $R_p$ could well mean that a small fraction of $\sim 3$ R$_\oplus$ planets with overestimated sizes contribute a substantial fraction of the population of apparently Neptune-size objects, and the actual transition between the multis-rich population of ``small'' planets that are $\lesssim 1$\% H/He by mass and the multis-poor population of gas-rich planets may occur closer to 4~R$_\oplus$. Indeed, although this transition {\it appears to} occur at a somewhat larger radius than that of the radius cliff, which is the sharp reduction in the overall occurrence rate of \ik planets observed near 3~R$_\oplus$ \citep{Kite:2019, Hsu:2019} that manifests as a reduction in slope in all of the curves in Fig.~\ref{fig:radius_multis_singles}, the break in the ratio of number of multis to that of singles could be coincident with the radius cliff.

Figure \ref{fig:large_radius_multiplicities_split}
compares cumulative size distributions of large planets (and KOIs that we dispositioned as R because they failed our upper radius cutoff)  in multis vs.~singles and in two-planet systems vs.~higher multiplicity systems. The size distributions of singles and multis are essentially indistinguishable  over the range 4.5~--~10~R$_\oplus$, but the number of planets in multis  in this range 
is only a little more than half the number in singles, whereas similar numbers are present for smaller planets.  Very few planets in multis have $R_p > 12~$R$_\oplus \approx$~1.1~R$_{\rm Jupiter}$, but plenty of singles have radii larger than 12~R$_\oplus$ (e.g., \citealt{Santerne:2016}).  The divergence of the curve for singles with $P > 10$ days from that for all singles shows that almost half of the members of the plotted population with  $R_p > 12$~R$_\oplus$ orbit within the period range of inflated hot Jupiters.  Most of the excess at longer periods (as well as some at short periods) is probably caused by false positives, which could be nearby (on the sky plane) eclipsing binaries or transits of the target star by ultracool dwarfs that are too faint to show an occultation (sometimes referred to as a secondary eclipse) deeper than can be explained by heating of the dayside by radiation from the primary star, or which travel on sufficiently eccentric and inclined orbits that no such occultation occurs. This was our motivation for classifying KOIs that otherwise would have been considered PCs that have $P>20$~days and $R_p > 1.2$~R$_{\rm Jupiter} \approx 13$~R$_\oplus$, i.e., significantly, albeit not substantially, larger than this boundary, with the disposition ``R''. Further investigation of KOIs vetted ``R'' is a topic worthy of investigation by observers interested in small stars within eclipsing binary systems, but is beyond the scope of this work.

Overall, there appears to be an abundant population of planets with sizes less than 3~R$_\oplus$ (in agreement with \citealt{Kite:2019,Hsu:2019}), 44\% of which are in multis.  \ik found a much less abundant population of giant planets; 29\% within the size range 5~R$_\oplus < R_p < 10$~R$_\oplus$ reside in multis, whereas the population of larger objects is dominated by singles that includes HJs and FPs.  Boundaries are smoothed over by a combination of radius errors and true fuzziness.  Note that the size range from 5~--~10~R$_\oplus$ includes a huge diversity of planets from super-puffs (currently only known in systems with multiple transiting planets because no cool very low-mass/low-density \ik transiting planets have RV mass measurements) to exo-jupiters that are more enriched in heavy elements than the prototype present in our Solar System.  However, all planets in this size range are clearly gas-rich, with H/He abundances by mass of the same order as or larger than that of astrophysical metals.

\subsection{Period Distribution} \label{sec:periodDist}

We now compare the distributions of the orbital periods of \ik single planet systems, that of planets in two-planet systems, and planets in systems of higher multiplicity.  As in  \S\ref{sec:sizes}, we consider only planets with S/N $> 12$, but nonetheless count objects that are classified as planet candidates yet are rejected from these samples due to S/N below this threshold in determining the multiplicity of a planetary system.  However, here we do not include mono-transit PCs and KOIs with dispositions of R in our sample,  and apart from the portion of our analysis wherein we compare the period distributions of different ranges of planetary sizes, we do not impose any restriction on estimated impact parameters.

\begin{figure}
\centering
\includegraphics[width=0.48\textwidth]{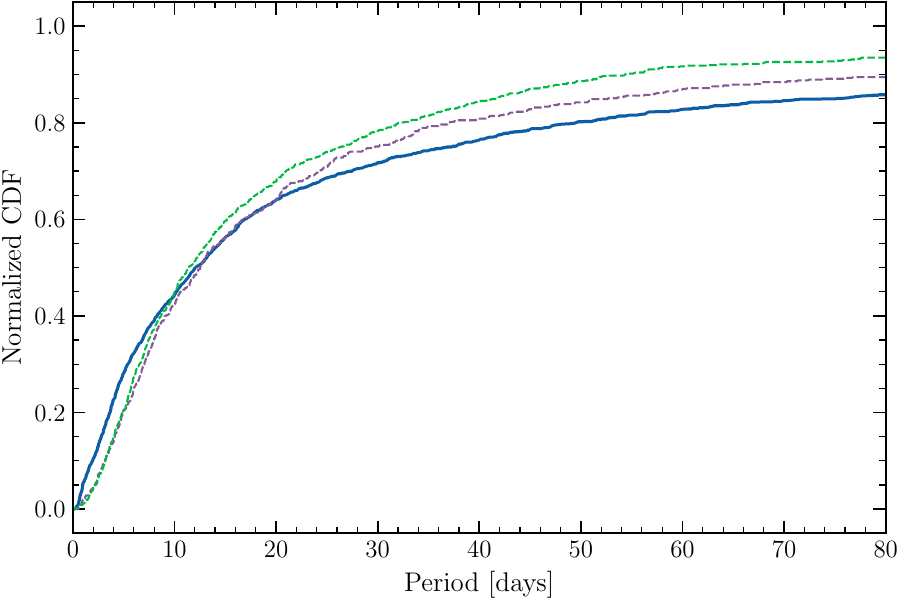}
    \includegraphics[width=0.48\textwidth]{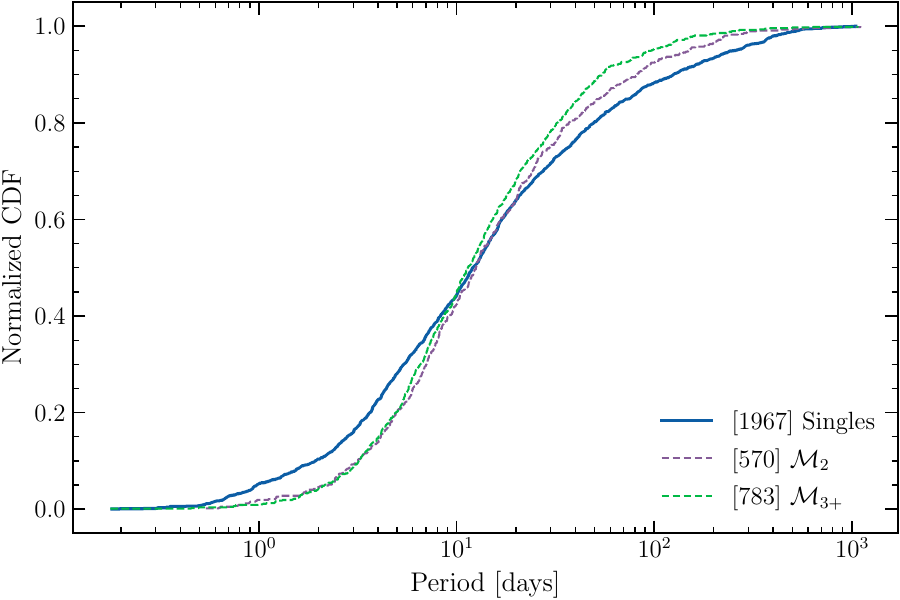}
\caption{The value of the vertical coordinate gives the fraction of planets within systems of specified multiplicity  having orbital periods less than the value of the horizontal coordinate. 
Although multis are deficient in planets relative to singles at both the shortest and longest orbital periods, the ensembles of planets in multis and those in singles both have median periods of $\sim$10 days. This plot uses the same criteria for inclusion as stated in the first paragraph of \S\ref{sec:sizes}. 
}
\label{fig:period_multiplicities_split}
\end{figure}

Figure \ref{fig:period_multiplicities_split} displays the orbital period distributions of single \ik PCs, that of planets in two-planet systems, and PCs in systems with three or more transiting planets. The vast majority of planets within multis, $\sim 90\%$, have orbital periods between 2 and 100 days, whereas 78\% of singles are in that same period range ($10\%$ of singles have periods less than 2 days and $\sim 12\%$ have periods longer than 100 days).  The difference is even larger restricting to the shortest periods ($P<1.6$ days), at which only $\sim 20\%$ of the \ik PCs have known transiting siblings. A KS test shows highly-significant differences between the distributions of singles and multis ($p$-value $\approx 10^{-6}$) and one comparing two-planet systems with higher multiplicity shows significance with $p$-value $\approx 0.27$. 
 These results reinforce the findings of \cite{Lissauer:2014} and  \cite{Rowe:2014}. Geometric factors reduce the probability of longer-period planetary siblings of a transiting planet also to be transiting when viewed from the Solar System. The tendency of hot Jupiters to lack nearby companions (Fig.~\ref{fig:perrad_multimap}) contributes to the smallness of the fraction of short-period planets residing in multis; larger period ratios for neighboring planets with short periods (perhaps due to tidal decay of the orbits of very short-period planets), and tides driving very short-period planets to the host star's equatorial plane also likely contribute. See Appendix B of  \cite{Lissauer:2014} for a more comprehensive discussion of possible additional causes of the paucity of very short period planets in multis. 



The normalized distribution of orbital periods of planets with exactly one companion PC is very similar to that of planets in higher multiplicity systems for periods up to $\sim 20$~days (Fig.~\ref{fig:period_multiplicities_split}). The  fraction of planets with two or more companions having periods in the range 20~days $< P < 80$~days is larger than that of planets possessing one transiting sibling, with a compensating deficit of PCs with multiple companions for $P > 80$~days, but despite this visual divergence, the overall differences between the two period distributions are not statistically significant (neither the KS test nor the AD test allows us to reject the hypothesis that the curves were drawn from the same distribution at the 95\% confidence level). Nonetheless, since the 2-planet systems are intermediate between single planets and high-multiplicity systems, there may well be real differences that are obscured by the small number statistics. \cite{He:2021} compute conditional occurrence rates of an additional putative planet as a function of both the period and radius of the detected and putative planets.

\begin{figure}[!hbt]
    \centering
    \includegraphics[width=0.75\textwidth]{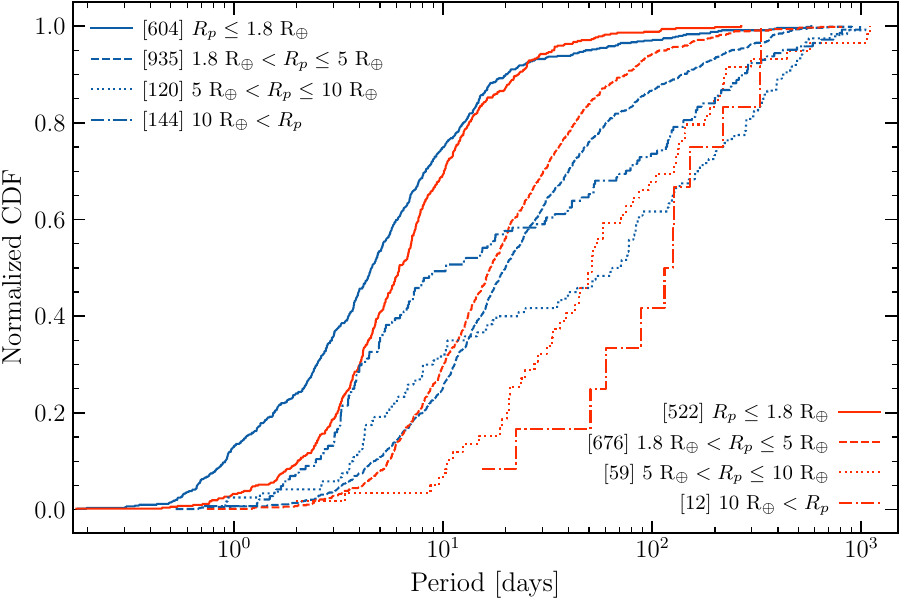}
    \caption{Cumulative period distribution of \ik planet candidates, considering the singles (blue curves) and multis (red curves) independently and separating the planets by radius into 4 bins ($R_p < 1.8$~R$_\oplus, \, 1.8$~R$_\oplus \leq R_p < 5$~R$_\oplus,\,5$~R$_\oplus \leq R_p < 10$~R$_\oplus$,  10~R$_\oplus \leq R_p$). Only planets with $b + \sigma_+(b) < 0.95$, S/N~$> 12$ and more than one transit observed are included, as radii of planets with grazing transits are not well-constrained and the false positive rate is relatively high among PCs with low S/N.}
    \label{fig:cdf_log3binradiisplit}
\end{figure}

Figure \ref{fig:cdf_log3binradiisplit} shows the period distributions of various subsets of \ik PCs that have been grouped by planet size and system multiplicity.  Of the eight curves in Fig.~\ref{fig:cdf_log3binradiisplit}, seven conform to the following trends: For a given size range, singles have a broader period distribution (more planets at both very short and very long periods) than do multis, with the cumulative fractions crossing ``near'' 10 days.  For all size ranges considered in multis and for non-giant singles, larger planets tend to have longer periods.  The exceptional curve is for giant planets in singles, a larger fraction of which are detected at short periods than is the case for mid-sized planets in singles.  The tendency to observe smaller transiting planets at shorter periods is consistent with observational biases.  Mid-sized planets with $P \lesssim 2$~days are quite uncommon (the hot Neptune desert), consistent with H/He envelopes being stripped from close-in planets with $M_p \lesssim 20$~M$_\oplus$. 

\begin{figure}
\centering
\includegraphics[width=0.48\textwidth]{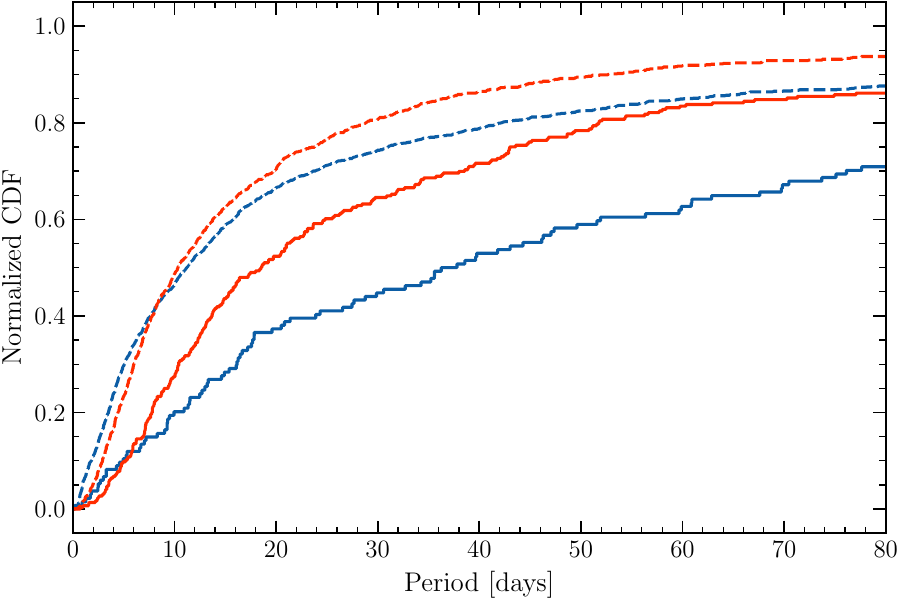}
\includegraphics[width=0.48\textwidth]{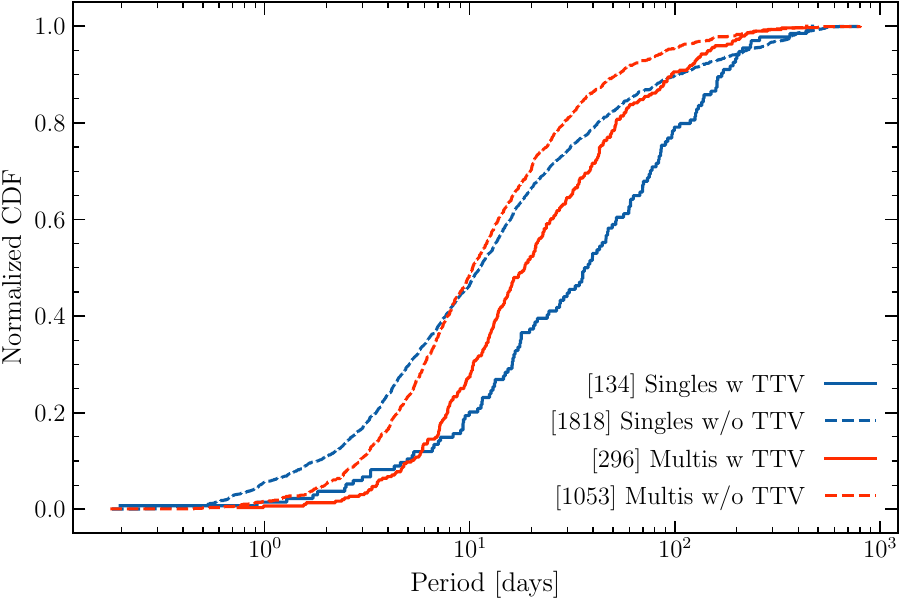}
\caption{The normalized cumulative distributions of orbital periods of single planets with TTVs, singles without TTVs, multis with TTVs and multis lacking TTVs.  Here we consider planets to have TTVs if the first digit in their TTV flag in Table \ref{tab:planetcatalog} is 1 (which somewhat overestimates the number of planets in multis with TTVs because TTVs are used for fitting all planets in a system if any planet in the system is found to exhibit TTVs), the second digit is 1 or 2, and/or the third digit is 8 or 9.  Only PCs with at least three TTs measured and S/N~$> 12$ are counted for these distributions. The plot on the left uses a linear scale in period and is truncated at 80 days, whereas the one on the right has a logarithmic period scale and extends to the longest-period \ik planets displaying at least 3 transits. }
\label{fig:period_TTVs}
\end{figure}

Figure \ref{fig:period_TTVs} shows the distributions of orbital periods of planets with and without Transit Timing Variations, with the planets also being categorized as being a single or a member of a multi.  Here, we consider planets to have TTVs if our lightcurve fitting prefers solutions with TTVs, and/or they are listed as having TTVs in the \cite{Holczer:2016} catalog, and/or they are classified as having strong or moderate TTVs by \cite{Kane:2019} (8 or 9 overall rating)\footnote{The criteria for being counted as having TTVs in Fig.~\ref{fig:period_TTVs} differ from those used for coloring planets to denote TTVs in the bottom two panels of Fig.~\ref{fig:gallery}. Here, we do not count PCs based on a \emph{tentative} TTV signature identified by \cite{Kane:2019}, but we include PCs with TTVs identified by \cite{Holczer:2016} as well as TTVs being used in our fits (which is the case for all PCs in a multi if we detect TTVs in any of the PCs associated with that target star). The latter criterion includes PCs in multis that don't show TTVs themselves, but we prefer including a small number of planets in multis without TTVs in the distribution of planets with TTVs to the alternative of not testing for TTVs in PCs added to the KOI table in recent years and small number statistics.}.  Note that only 7\% of single PCs with S/N~$> 12$ have TTVs that meet our criteria, whereas 15--20\% of such planets in multis have TTVs\footnote{ Our decision to use of TTVs in fitting is made on a system by system basis. The lower limit quoted here for multis only counts planets with Holczer or Kane TTVs plus one for each \emph{system} for which we used TTVs in the fits that does not have any planets with Holczer or Kane TTVs, whereas the upper limit includes all planets in systems for which our fits used TTVs.}. This result is consistent with \cite{2014Xie}, who found TTV rates that grow with transit multiplicity. For both singles and multis with $P < 200$~days
, we find that TTVs are more likely to be observed in planets having longer periods, with the period differences being larger in singles than in multis.  For self-similar (scale-invariant) planetary system architectures, both TTV amplitudes and near-resonant superperiods increase linearly with orbital period, so the dependence of the fraction of planets with TTVs on orbital period may be explained in whole or in part by observational selection effects (small TTV amplitudes for the shortest period planets and observational baseline small compared to TTV superperiods for long period planets).

\begin{figure}[!hbt]
\centering
\includegraphics[width=0.75\textwidth]{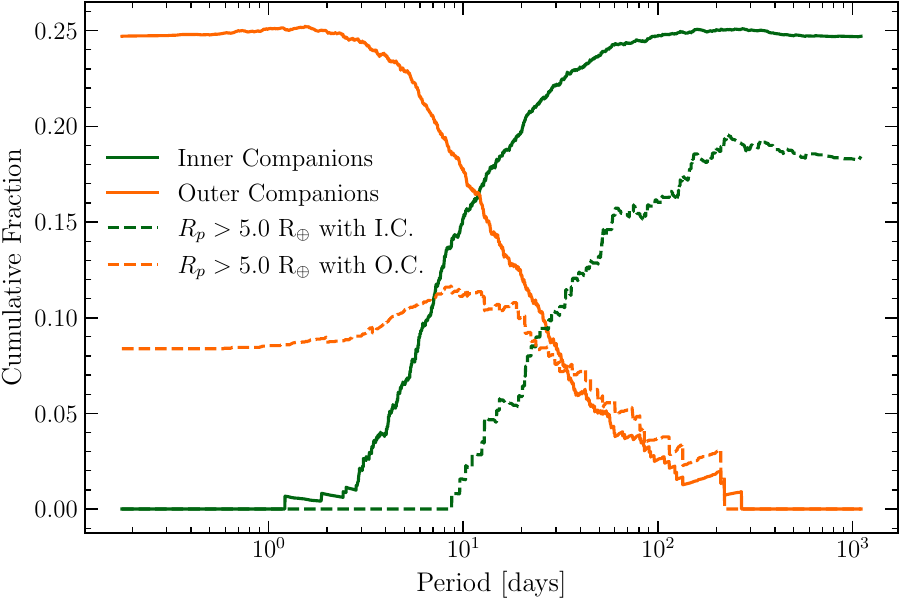}
\caption{Cumulative fraction of candidates with inner (outer) companions as function of orbital period, as specified by Equation \ref{eqn:innercompanionfraction} (Equation \ref{eqn:outercompanionfraction}). The solid green curve shows the cumulative fraction of PCs up to the plotted orbital period that have inner planetary companions (shorter-period PCs associated with the same target star). The solid orange curve marks the cumulative fraction of planets with $P$ \emph{larger than} the specified period that have even longer-period transiting companions. The dashed curves show the analogous curves for large planets (PCs with $R_p > 5$~R$_\oplus$) having inner or outer planetary companions, with PCs of all sizes except mono-transits counting as companion planets. Again only PCs with S/N~$> 12$ are included for these distributions, although this requirement isn't enforced for companions.}
\label{fig:neighbors}
\end{figure}

We next examine the ``cumulative'' fraction of planets with \emph{inner} companions as a function of planetary orbital period.  The green curves in Figure \ref{fig:neighbors} show ${\cal F}_{\rm inner}(\textrm{Period})$, the fraction of transiting planets with periods less than the value specified on the horizontal axis that have at least one transiting sibling on an interior orbit. This fraction can be calculated from the formula:

\begin{equation}
{\cal F}_{\rm inner}(\textrm{Period}) \equiv \dfrac{N_{\rm inner}^{P<\textrm{Period}}}{N^{P<\textrm{Period}}} ~~,
\label{eqn:innercompanionfraction}
\end{equation}

\noindent where $N^{P<\textrm{Period}}$ gives the total number of PCs with $P < \textrm{Period}$ and $N_{\rm inner}^{P<\textrm{Period}}$ represents the number of such planets having one or more transiting companions with period shorter than its own period.  The resulting distribution (in green) increases at any orbital period where a planet has an inner sibling, and (once its value exceeds zero) decreases where a planet lacking inner companions is added to the count, since the denominator accounts for all PCs, including singles and planets within multis with no inner neighbors. 

Similarly, we examine the fractions of planets with \emph{outer} companions. The orange curves in Fig.~\ref{fig:neighbors} show ${\cal F}_{\rm outer}(\textrm{Period})$, the fraction of transiting planets with at least one transiting sibling with $P>\textrm{Period}$, which is given by: 


\begin{equation}
{\cal F}_{\rm outer}(\textrm{Period}) \equiv \dfrac{N_{\rm outer}^{P>\textrm{Period}}}{N^{P>\textrm{Period}}} ~~. 
\label{eqn:outercompanionfraction}
\end{equation}


\noindent The terms on the right hand side of Equation (\ref{eqn:outercompanionfraction}) are defined analogously to those in Equation (\ref{eqn:innercompanionfraction}), but note that in this case, planets are added to the distribution starting from the longest period and moving to the shortest.  To zeroth order, the solid orange and green curves appear to be mirror images of one another, albeit a bit stretched out towards longer periods. The symmetry and ``reflection'' near 10 days and somewhat broader shape at long periods all mimic trends visible in the right panel of Fig.~\ref{fig:period_multiplicities_split}.

The solid curves in Fig.~\ref{fig:neighbors} can be compared to subsamples (dashed  curves) that show the fraction of large planets ($R_p > 5$~R$_\oplus$) that have transiting siblings, where siblings of any size are included.  It is well-known that giant transiting planets, especially hot jupiters, rarely have companions that also transit \citep{Latham:2011, Steffen:2012}. The dashed curves in Fig.~\ref{fig:neighbors} quantify the differences, showing (among other things) that large planets are only one-third as likely to have outer companions and two-thirds as likely to have inner companions as are \ik planets overall (note that no effort has been made to attempt to correct these numbers for possible detection biases).  Although large planets overall are substantially less likely to have outer companions, large and small planets with $P \gtrsim 30$ days are equally likely to have one or more outer companions.

\subsection{Eccentricity \& Transit Duration Distributions}\label{sec:ecc}

{The improved estimates of stellar, planetary and transit parameters in our new catalog (Table \ref{tab:planetcatalog}) enable us to} characterize the eccentricity distribution of various subsets of the \ik planet candidates using the distribution of period-normalized transit durations.
{As accurate period estimates are required for this analysis, mono-transits (planets candidates with only a single detected transit) were excluded from the analysis in this section.  Our analysis presented below shows that there is not evidence for changes in the eccentricity distribution of \ik planet candidates as a function of the host star effective temperatures over the range from 4000~K to 6200~K.  Similarly, we see no dependence on period for the eccentricity distribution for planets with orbital periods greater than 6 days.  In contrast, we show that there are significant differences in the eccentricity distribution as function of planet size and the number of planets detected by \ik in a given system.  We find marginal evidence suggestive of a trend for planets with a known companion within a factor of 2.04 in orbital period to have a smaller eccentricity than more widely spaced planets.  Below, we describe the specific comparisons and statistical tests performed to support these findings.}

In order to compare the eccentricity distributions of various subsets of \ik planet candidates, we begin by measuring the transit duration for each PC. 
We calculate the posterior mode of the average of the full transit duration, {$T_{1,4}$}, and the duration of the ``flat bottom'' portion of the transit, {$T_{2,3}$},  normalized by the analogous predicted transit duration for the measured orbital period and host star properties, assuming a circular orbit and central transit ($b=0$). Using the notation introduced in Section \ref{sec:catalog}, the normalized transit duration, $\tau$, is given by:

\begin{equation}
    \tau \equiv \frac{T_{1.5}}{T_{1.5}(e = b = 0,P,\rho_\star)} = \frac{T_{1,4} + T_{2,3}}{{ T}_{1,4}(e = b = 0,P,\rho_\star) + {T}_{2,3}(e = b = 0,P,\rho_\star)}  .
    \label{eqn:tdur}
\end{equation}  

\noindent For planet candidates with {$b \ge 1-R_p/R_\star$, $T_{2,3}=0$}, and we exclude these cases from our analysis due to difficulty in precisely constraining physical parameters for grazing and near-grazing transits. The particular definition of the normalized transit duration given in Equation (\ref{eqn:tdur}) was chosen for robustness in measurement as well as its independence (to first-order) of the value of $R_p/R_\star$.  {By defining $\tau$ relative to the duration for $b=0$ (rather than the best estimate of $b$), we avoid a dependence on $b$, which is advantageous because estimates of the value of impact parameters can have significant measurement uncertainties (\S\ref{sec:transitmodel} and Table \ref{tab:planetcatalog}).}

Planets with $\tau > 1$  (when accounting for measurement uncertainties) must transit while the planet-star separation exceeds the semimajor axis, and there can be an interesting constraint on the pericenter direction $\omega$ (e.g., \citealt{DawsonJohnson:2012}).  
In most cases, $\tau$ is comparable to or less than unity and there is only a minimal constraint on the marginal distribution of $\omega$, since the transit duration could be shortened due to either the planet being near pericenter or the impact parameter $b \neq 0$ (or both).  

Star-planet orbital planes are isotropically distributed and thus randomly oriented relative to \ikt's line of sight, hence the \emph{intrinsic} distributions of the impact parameters (within the range $b \lesssim 1$) and the pericenter angles (of all planets, not just transiting ones) are nearly uniform.  
Therefore, the distribution of $\tau$ provides a useful probe of the eccentricity distribution of a population of transiting planets \citep{Ford:2008}.  
In practice, the distribution of observed impact parameters is affected by detection biases, since shorter transits that occur for larger $b$ provide less time in transit to accumulate signal.  
Fortunately, this is a relatively modest effect for most planets, since the changes in transit duration, and hence transit signal, are typically small, and most detected planets have a S/N much greater than required for detection.

Multiple studies have begun to characterize the eccentricity distribution based on the observed period-normalized transit duration ($\tau$) distribution.  \citet{Moorhead:2011}, \cite{Kane:2012}, \cite{Plavchan:2014} and \cite{Xie:2016} focused on using the $\tau$ distribution to characterize the eccentricity distribution (as opposed to modeling the several observed properties at once,  as in, e.g., \cite{Mulders:2018,He:2019,Sandford:2019,He:2020,Zhu:2018,MacDonald:2020,Yang:2020}).  
The strength of conclusions from these early studies was limited due to the uncertainty in the host star density.  While random measurement errors can be easily incorporated into the analysis \citep{Ford:2008}, the potential for systematic error is more concerning.  For example, \citet{Moorhead:2011} showed a trend for the $\tau$ distribution to broaden with increasing host star temperature, which could be attributed to either the eccentricity distribution changing with host star temperature or the errors in host star densities increasing for stars that have had time to evolve far from the zero age main sequence.  Such concerns helped to motivate follow-up campaigns to characterize host star properties using high-resolution spectroscopy \citep{Fulton:2018} and more detailed stellar modeling \citep{Berger:2020a}, both of which  incorporate improved distance measurements from \gaiat.

To assess the effects of potential systematic errors in the host star density on the eccentricity distribution, we focus our analysis on planets whose host star properties are available from either \citet{Fulton:2018} or \citet{Berger:2020a}, both of which represent dramatic improvements of stellar properties from those derived using photometric data (e.g., the KIC).  (Asteroseimic densities are expected to be even more accurate, but are available for only a substantially smaller subset of planet candidates \citep{vanEylen:2015,vanEylen:2019}.)  
While the stellar properties estimated using high-resolution spectroscopy \citep{Fulton:2018} are likely more accurate than those with lower resolution spectroscopy, they are not available for a substantial fraction of the stars in the full sample (e.g., most faint stars with a single known transiting planets, stars whose planets were identified after the CKS survey).  However, the \cite{Fulton:2018} properties are available for a substantial majority of the hosts of systems with multiple transiting planets.

\begin{figure}
    \centering
    \includegraphics[scale=.3]{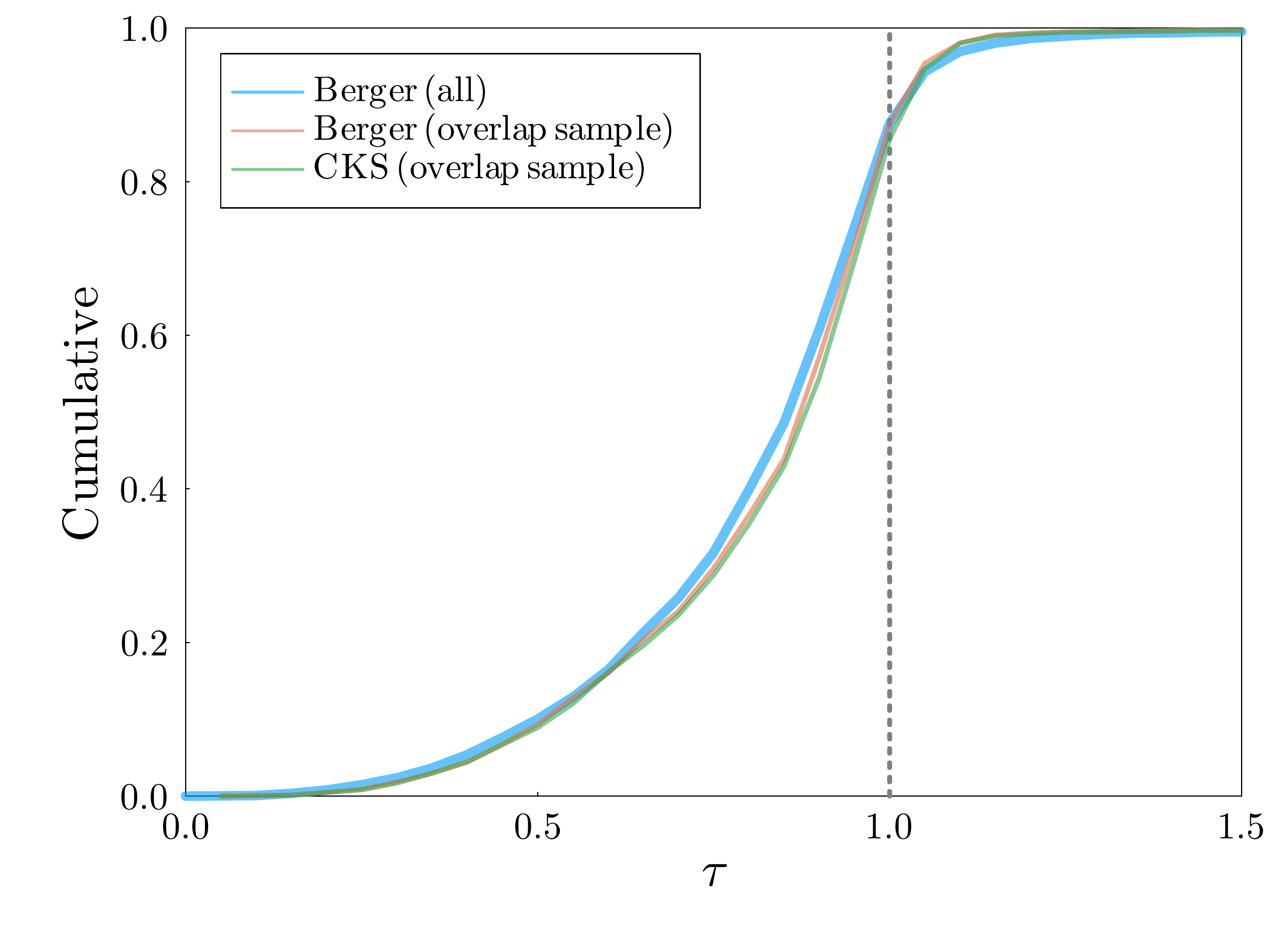}
    \caption{Cumulative distribution of $\tau$, the normalized transit duration (Equation \ref{eqn:tdur}), based on stellar properties from either \citet{Berger:2020a} or \citet[][CKS]{Fulton:2018}.  
    The thick blue curve is the sample used for our primary analysis of transit durations.  
    The orange and green curves are for a smaller sample for which both Berger and CKS stellar parameters are available.  
    The period-normalized transit duration distributions based on the different stellar properties do not differ significantly if we restrict the comparison to the same sample. The difference between the full and overlap samples is primarily due to the CKS sample favoring stellar hosts of multi-planet systems (see Fig.~\ref{fig:EccMulti}) and bright target stars, which implicitly affects both the stellar properties and \ikt's sensitivity to planets around those stars.}
    \label{fig:EccStellarSource}
\end{figure}

Figure \ref{fig:EccStellarSource} compares the cumulative distribution of normalized transit durations ($\tau$ values) based on using the  host star properties from \citet{Fulton:2018} (CKS) and those from \citet{Berger:2020a}.  The orange and green curves are for an overlap sample for which both sets of stellar parameters are available.  The wide blue curve shows the results using the \citet{Berger:2020a} properties for the full sample.  (The results for all transiting planet candidates for which CKS parameters are available is not plotted because it is visually indistinguishable from the overlap sample using stellar properties from CKS.)  ~The distributions of $\tau$ values for planets using host star properties from \citet{Fulton:2018} show small, but statistically significant, differences from those using the stellar properties of \citet{Berger:2020a}.  Most of this difference arises due to the fact that the CKS sample is significantly less complete than the \citet{Berger:2020a} sample, particularly for targets with a single transiting planet candidate, and the distributions using the overlap sample are statistically indistinguishable.  

In order to minimize the risk of systematic biases, we perform nearly all of our subsequent analyses of the $\tau$ distribution based on stellar properties from \citet{Berger:2020a}, even when parameters from \citet{Fulton:2018} are available. The single exception is for comparing the distributions between multiple planet systems of differing multiplicity, where we make comparisons using each set of stellar parameters (since the CKS sample has much better completeness for hosts of multi-planet systems than for single planet hosts).   We also applied cuts so that our subsequent analysis only includes PCs with a measured orbital period, transit S/N $\geq 12$ (so other transit parameters are well measured), an impact parameter $b < 1- R_p/R_\star$, host star temperature between 4000~K and 6600~K, planet radii estimated to be smaller than 12.5 R$_\oplus$, and reported uncertainty in the stellar density less than 25\%.   

Having defined a sample of 2762 planets for which systematic biases should be minimal, we performed several checks. We verified that there were not significant differences in the $\tau$ distribution if we increased the thresholds for the minimum transit S/N to be included in our primary analysis.  Similarly, we confirmed that using a fixed maximum impact parameter of 0.95 or greater did not affect our conclusions.  

\begin{figure}
    \centering
    \includegraphics[scale=.3]{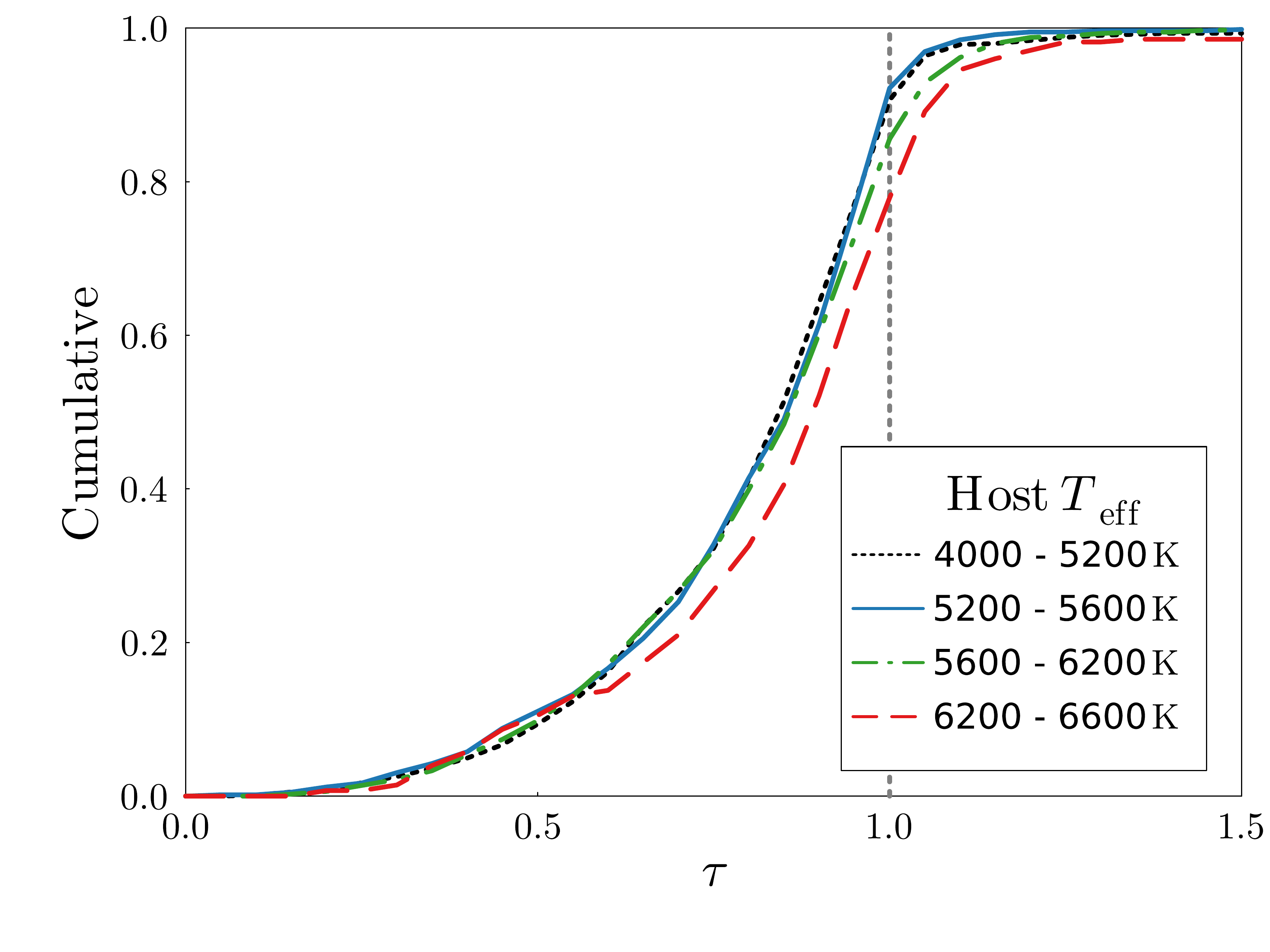}
    \caption{Cumulative distribution of $\tau$, the normalized transit duration, for four subsets of planets based on the effective temperature of their host star reported in \citet{Berger:2020a}.  The subsets are: 4000~--~5200~K (black, dotted), 5200~--~5600~K (blue, solid), 5600~--~6200~K (green, dash-dot), and 6200~--~6600~K (red, dashes).  A 4-sample AD test strongly rejects the null hypothesis that the $\tau$ distributions are drawn from a single population.  Excluding the 6200~--~6600~K bin,  a 3-sample AD test still nominally rejects the null hypothesis that the remaining bins are drawn from a common distribution, but with weak significance.}
    \label{fig:EccTeff}
\end{figure}

Motivated by \citet{Moorhead:2011}, we divided the sample into four bins based on host star temperature (Figure \ref{fig:EccTeff}) and performed 4-sample Anderson-Darling (AD) test of the null hypothesis that each subsample was drawn from the same distribution.  
(The AD test is usually more powerful than the more common Kolmogorov-Smirnov (KS) test.  The KS test is most sensitive to a shift of the distribution, but significantly less sensitive to differences in the shape of the wings.)  The AD $p$-value was $<3\times10^{-5}$.   The highest temperature bin is clearly the most discrepant from the other three.  If we exclude this bin,  the $p$-value from a 3-sample AD test is 0.016, much less extreme, but still low enough to reject, at the level comparable to $2.4 \sigma$, the null hypothesis that the $\tau$ distribution is the same for the three remaining subsets of host stars with $4000 \le T_{\rm eff} \le 6200$~K.  

\citet{Moorhead:2011} also saw significant differences between planets orbiting stars with temperature above and below the Kraft break ($\approx$~6200~K).
{(Main sequence stars with effective temperature greater than the Kraft break have negligible convective envelopes, resulting in dramatically reduced tidal dissipation in the star compared to cooler main sequence stars.)}
However, uncertainty in stellar parameters led them to focus on stars with $T_{\rm eff} \le 5100$~K, rather than risk confusing changes in the $\tau$ distribution due to the eccentricity distribution with those caused by measurement uncertainties and selection effects. 
Thanks to improved stellar parameters made possible by \gaia distance measurements, we no longer find differences in the $\tau$ distribution of stars cooler than the Kraft break and see a more significant difference between the $\tau$ distribution of planets with stars on either side of the Kraft break.  

These differences are unlikely to be due to uncertainties in stellar properties for stars with $6200 \le T_{\rm eff} \le 6600$~K, since imposing strict cuts on uncertainty in stellar densities has a minimal effect on this subset.
The observed differences could be partially due to the reduced efficiency of tidal dissipation in hotter stars with no significant convective envelope.  
However, we caution that the differences do not go away, even if we focus only on planets with periods greater than 10 days, for which tidal effects during the main sequence would be small.


Therefore, subsequent analyses of the $\tau$ distribution are focused on planets with host stars with $4000 \le T_{\rm eff} \le 6200$~K,
so as to minimize the risk of systematic bias (e.g., due to small planets being more readily detectable around bright stars). These restrictions leave us with 2485 planets. For the analyses below, we only consider those PCs that passed all of these cuts in determining system multiplicity. However, we do consider PCs that were rejected by these cuts in determining the smallest period ratio to a neighboring planet in \S\ref{sec:eccspacing}.

\subsubsection{Fitting the Eccentricity Distribution}
\begin{figure}
    \centering
    \includegraphics[scale=.15]{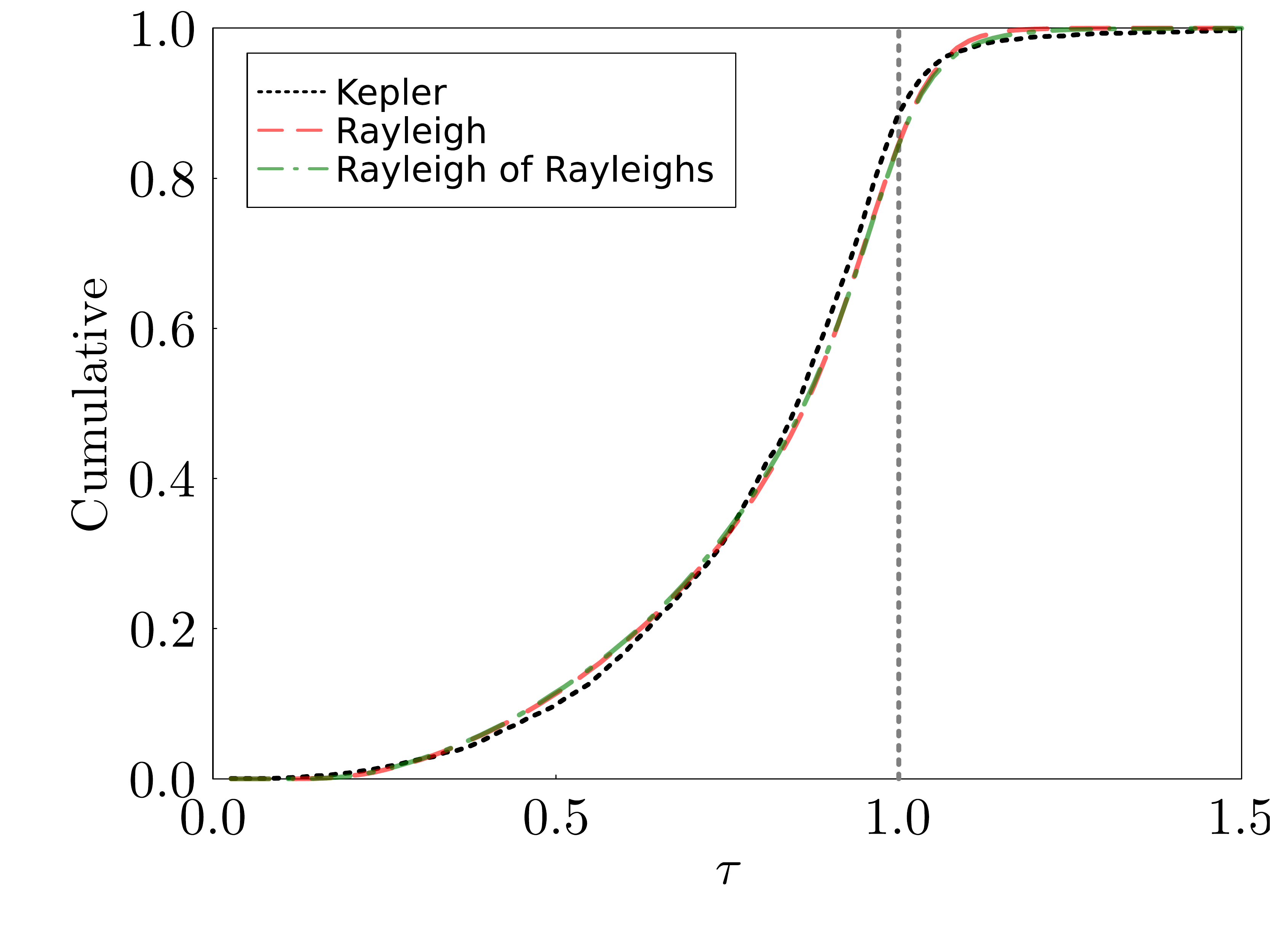}
    \includegraphics[scale=.15]{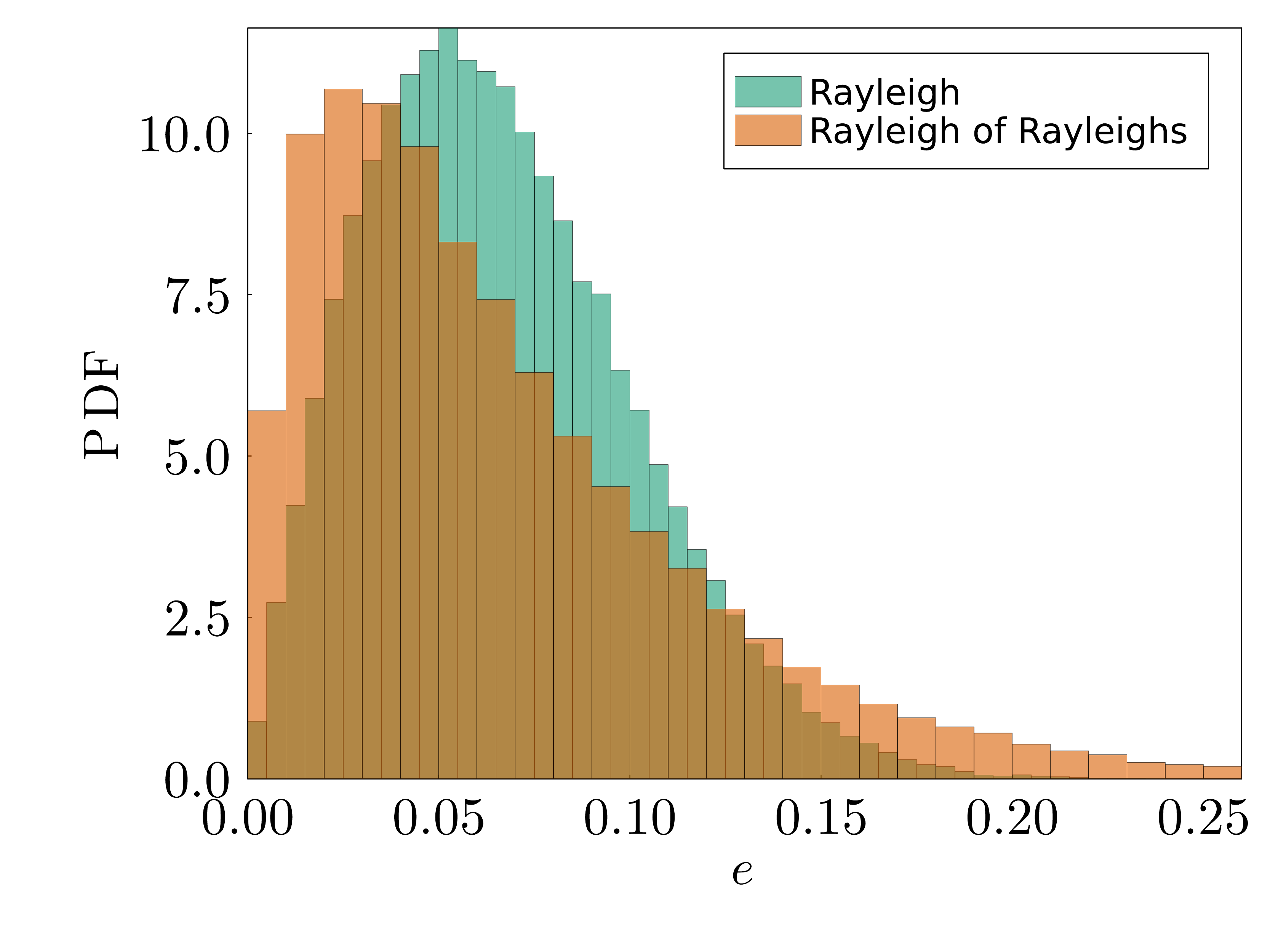}
    \caption{Left:  Cumulative distributions of $\tau$, the normalized transit duration, for our sample of \ik planets (dotted black curve) and two simulated populations: The eccentricity is drawn from a single Rayleigh distribution with Rayleigh parameter $\sigma = 0.053$ (red dashed curve), and the eccentricity distribution is drawn from a continuous mixture of Rayleigh distributions with weights given by a Rayleigh distribution with $\sigma=0.043$ (green dot-dashed curve).  Neither model reproduces both the rapid rise of the $\tau$ distribution due to low eccentricity planets and the tail of planets with $\tau$ larger than unity seen in the \ik sample.  Right:  The eccentricity distributions implied by either the best fit Rayleigh distribution (green) or Rayleigh of Rayleigh distributions (orange) described above.  While the two models make very similar predictions for the distribution of normalized transit durations, they have substantially different implications for the tail of the eccentricity distribution.}
    \label{fig:EccModel}
\end{figure}

We attempted to fit multiple simple analytic models to the observed $\tau$ distribution.  
For each analytic model, we generate simulated populations of planets assuming a uniform distribution of $\omega\sim U\left[0,2\pi\right)$ and $b\sim U\left[0, b_{\max}\right)$, {where $b_{\max} =  1 - R_p/R_{\star}$ is based on the value reported in Table \ref{tab:planetcatalog} for each planet.}  
We sample the planet-star radius ratio, orbital periods and stellar densities from those of the observed sample and add measurement noise based on the reported uncertainties (parameterized as mixture of two half-Gaussians).  
We weight each simulated planet by its geometric transit probability, which accounts for the dependence on the $e$ and $\omega$ drawn for each planet.  {\citep[However, this does not account for how differences in transit duration affect the detection probability conditional on the planet transiting, as done in][]{He:2019,He:2020}.} 
We find that a Rayleigh distribution of eccentricities that is  truncated to be less than one, which is best fit with Rayleigh parameter of 0.053, is not sufficient to reproduce the observed $\tau$ distribution (see Fig.~\ref{fig:EccModel}, red dashed curve).  
Using a small Rayleigh parameter under-predicts the number of extreme $\tau$ values, while using a large Rayleigh parameter results in too broad a distribution.  
This led us to consider a continuous mixture of Rayleigh distributions, where the weights for each Rayleigh parameter are proportional to a Rayleigh distribution (with Rayleigh parameter 0.043), i.e., a Rayleigh of Rayleighs, as in \S6.1.2 of \cite{Lissauer:2011b}.  
While this results in a slight improvement in the fit for long-duration transits, it did not significantly improve the fit for $\tau$ in the range of 0.8~--~1.0 (see the left panel Fig.~\ref{fig:EccModel}, green dot-dashed curve). %

More detailed modeling of the joint {probability density function (PDF)} of multiplicity, inclination and eccentricity distributions suggests that the joint {PDF} is not simply the product of the distribution for each quantity individually \citep{He:2019,He:2020,Yang:2020,Millholland:2017}.
Therefore, we do not attempt to perform a detailed characterization of the uncertainties in such a model.

{While the transit duration distribution can provide a powerful constraint on parameters given an assumed eccentricity distribution, it has much less statistical power for comparing different functional forms for the eccentricity distribution.
This is illustrated by the substantial differences between the two histograms in the right panel of Fig.~\ref{fig:EccModel}, despite the very similar predictions of the two models for the transit durations (long dashed and dot-dashed curves in left panel).}

\subsubsection{Transit Duration vs.~Multiplicity}\label{sec:eccmultiplicity}
\begin{figure}
    \centering
    \includegraphics[scale=.3]{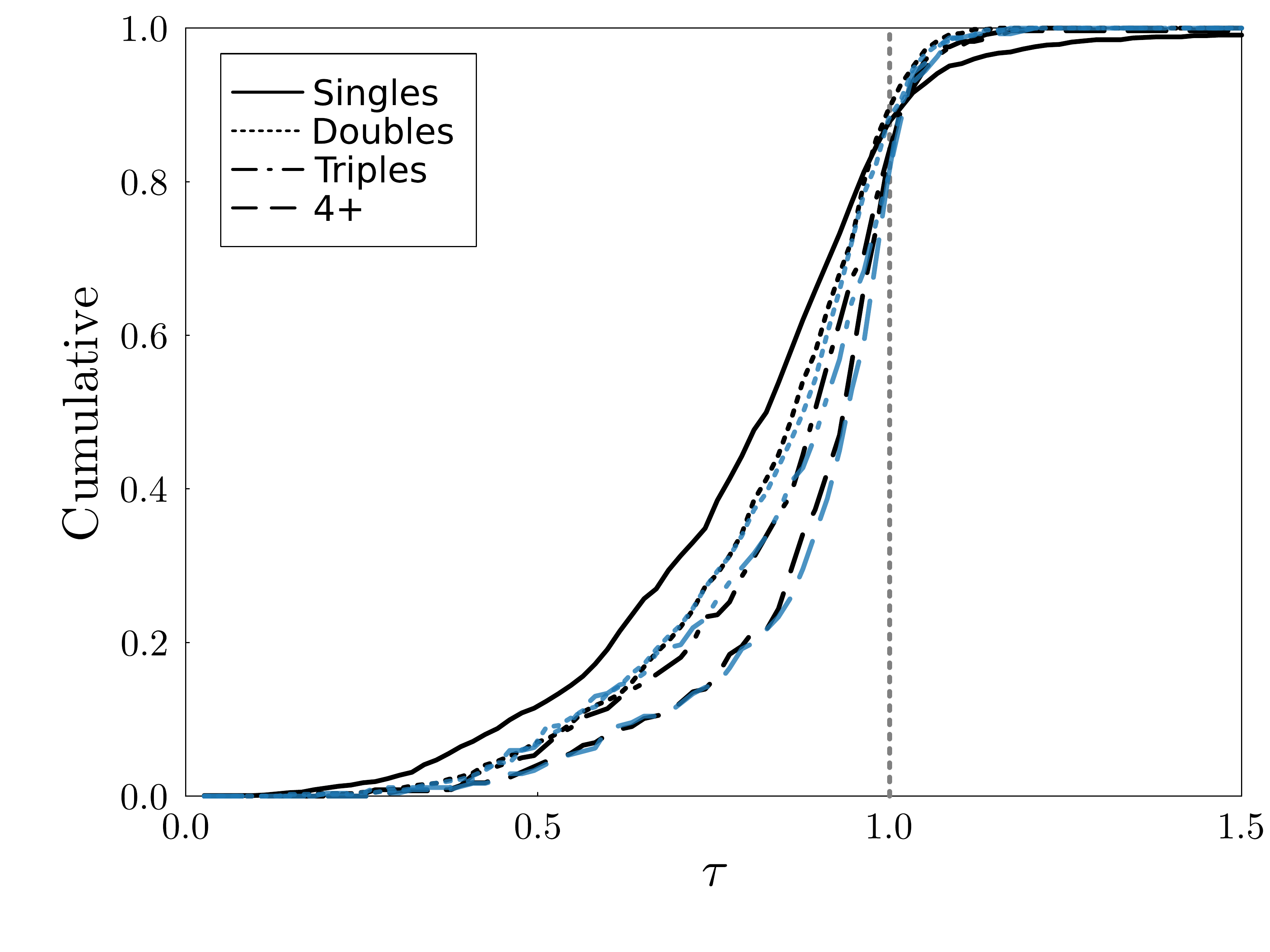}
    \caption{Comparison of the $\tau$ distribution of planets with $R_p\le 5$~R$_{\oplus}$ in systems with a single known transiting planet (``singles''; solid curve), two known transiting planets (``doubles''; dotted curves) and three planet systems (``triples''; dot-dashed curves) or systems with four or more known transiting planets (``4+'' systems; dashed curves). Light blue curves are based on stellar properties from CKS \citep{Fulton:2018} instead of the \cite{Berger:2020a} catalog (shown in black).  Since CKS parameters are not available for all stars, the planet samples used to compute the blue curves  differ somewhat from those used for the the black curves. No blue curve is shown for singles because CKS only analyzed a small fraction of the fainter stars hosting just one planet candidate. In contrast, the CKS survey devoted extra effort to survey hosts of multi-planet systems \citep{Petigura:2017}, so the differences in the samples are smaller and more random.  Both the KS and AD tests strongly reject the null hypothesis that the distributions for planets in singles and doubles are drawn from the same population.  Similarly,  we reject the null hypothesis that any pair of the $\tau$ distributions for doubles, triples and higher multiplicity systems are drawn from the same distribution.  
    }
    \label{fig:EccMulti}
\end{figure}

Long-term orbital stability and planet formation models suggest that the eccentricity and mutual inclination distributions of planets depend on the multiplicity of their host planetary system \citep{Pu:2015,Gratia:2021,Bartram:2021}.
Indeed, previous studies of the observed transit duration ratio distribution find evidence that the mutual inclination distribution decreases with multiplicity of the inner planetary system \citep{He:2020,Yang:2020}. 
Therefore, in Figure \ref{fig:EccMulti} we compare the $\tau$ distributions for: ({\it i}) systems with a single known transiting planet (``singles''), 
({\it ii}) systems with two known transiting planets (``doubles''),  ({\it iii}) systems with three known transiting planets (``triples'') and ({\it iv}) systems with four or more known transiting planets (``high multiplicity'').  


Both the KS or AD tests strongly reject the null hypotheses that the $\tau$ distributions of planets smaller than $5 R_{\oplus}$ are the same when comparing 
singles and doubles ($p_{KS}\approx 3.3\times10^{-4}$, $p_{AD}\approx 1.5 \times 10^{-5}$).  Comparing all multiples to singles, both the  KS and AD tests very strongly reject the null hypothesis that the $\tau$ distributions are the same ($p<10^{-14}$), confirming the result that \cite{Xie:2016} obtained for a smaller sample of \ik planets whose host stars had been characterized using spectra obtained by LAMOST. 
When comparing the $\tau$ distributions of planets in doubles to those of planets in higher multiplicity systems, the $p$-values for the KS and AD tests are $\approx 1.3\times10^{-5}$ and $\approx 1.4\times10^{-6}$, respectively.  (If using the CKS stellar properties, then the $p$-value for the KS test decreases to $\approx 2.4\times10^{-6}$, but the $p$-value for the AD test does not change significantly.) The significance is strengthened by the clear pattern of the $\tau$ distribution becoming more highly concentrated near unity as one moves from singles to doubles to higher multiplicity systems. 
When comparing the $\tau$ distributions of planets in triples to those of planets in systems with $\ge 4$ detected planets (using CKS stellar parameters), the $p$-values for the KS and AD tests are $\approx 0.01$ and $\approx 0.009$, respectively.  

Note that the observed transit duration distributions for multiple planet systems are subject to complex observational biases, due to a combination of transit signal-to-noise and geometric transit probabilities that can be correlated within a planetary system.  Properly accounting for these biases requires a full forward model that accounts for the joint distribution of planet sizes, orbital periods, eccentricities and inclinations. For example, \cite{He:2019,He:2020} 
find that the mutual inclination distribution is narrower for systems with more detected transiting planets.  However, a narrow distribution of mutual inclinations does not imply a narrow distribution of impact parameters.  Even small mutual inclinations can cause significant $\Delta b$ between planets (by construction, since \ik detects planets with $b\lesssim1$).  Further, the $\Delta b$ can have either a positive or negative sign.  Indeed, we have verified that the distribution of maximum likelihood estimate of $b$'s is not more narrowly peaked for widely-spaced multiple planet systems than for closely spaced multi-planet systems (nor that of systems with only a single detected planet).  Therefore, we conclude that the difference in normalized transit duration distribution shown in Fig. 20 is primarily due to differences in the eccentricity distributions between systems with one, two or more detected transiting planets.  The strength of these differences is consistent with the change in eccentricity distribution as a function of number of detected transiting planets resulting primarily from the constraints imposed by long-term orbital stability \citep{He:2020}. 

\subsubsection{Transit Duration vs.~Orbital Period}\label{sec:eccperiod}
\begin{figure}
    \centering
   \includegraphics[scale=.3]{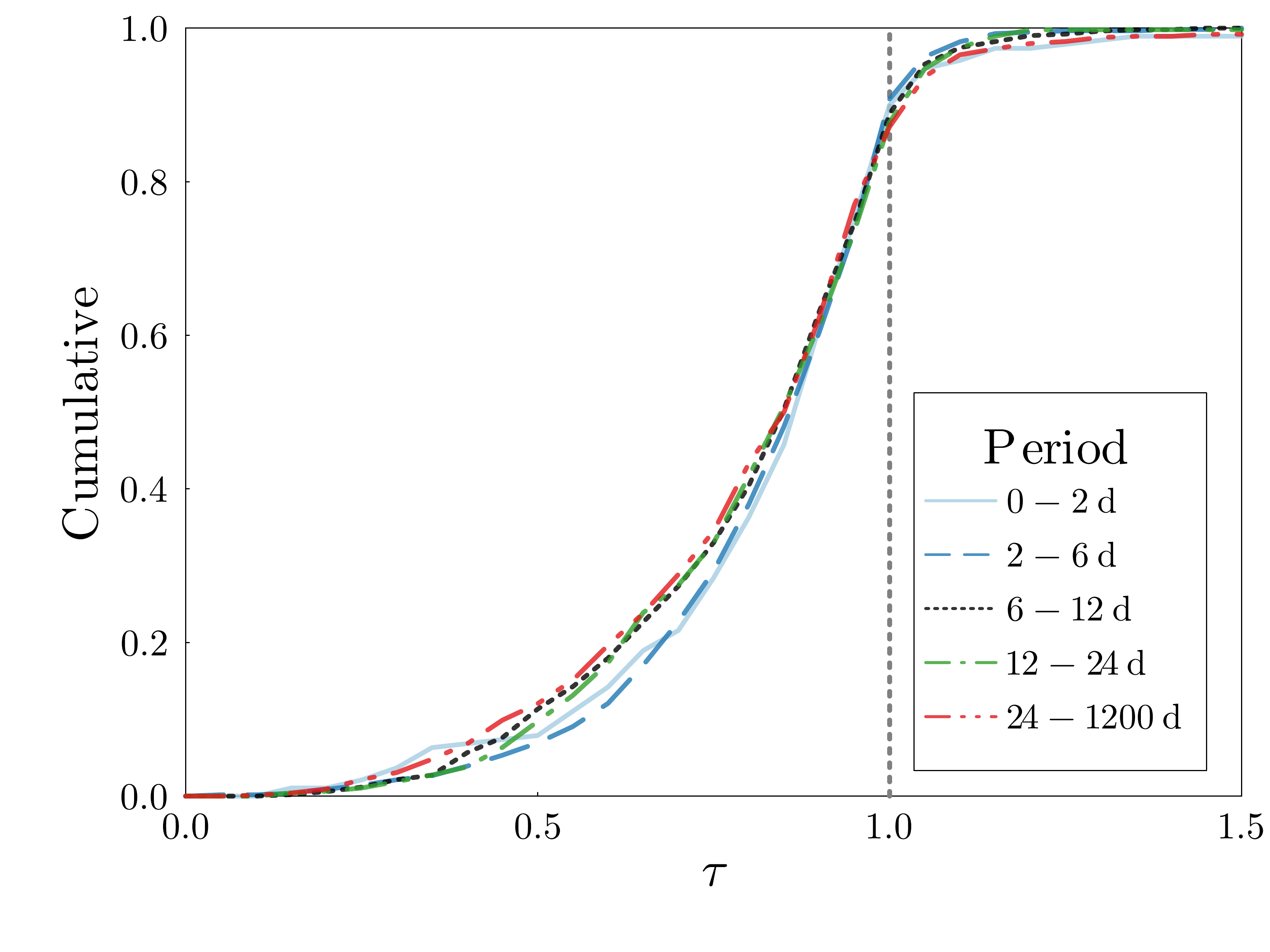}
    \caption{Comparison of the cumulative $\tau$ distribution for planets divided into five subsets based on their orbital period.  The curves correspond to  $P < 2$~days (solid light blue), $2 < P < 6$~days (dashed royal blue), $6 < P < 12$~days (dotted black), $12 < P < 24$~days (single dot-dashed green curve)), and $24 < P < 1200$~days (double dot-dashed red).  A 5-sample AD test fails to reject the null hypothesis that the $\tau$ distribution for all subsets is drawn from a common distribution.  However, if we compare planets with $P<6$~days to planets with $24 < P < 1200$~days, then a 2-sample AD test rejects the null hypothesis that the $\tau$ distribution for all subsets is drawn from a common distribution with a $p$-value of 0.01.  
    This is consistent with expectations if tidal effects are effective at circularizing a significant fraction of planets with $P<6$~days.  }
    \label{fig:EccPeriod}
\end{figure}

We next compare the $\tau$ distribution for planets as a function of orbital period. Dividing the sample at 11.8~days (near the median orbital period of our sample), the $p$-values for the KS and AD tests are $\approx~0.092$ and 0.045, respectively.  
Figure \ref{fig:EccPeriod} partitions the sample into five subsets with boundaries at 2, 6, 12 and 24 days, for which the 5-sample AD test $p$-value is 0.15.  
The biggest difference across subsets is that two shortest period subsets have the sharpest $\tau$ distributions (implying lower eccentricities).
If we combine planets with period in 0~--~2~days and 2~--~6~days, then the $p$-value for a 4-sample AD test is 0.08.  
If we perform a 2-sample AD test for planets with period in 0~--~6~days and planets with period in 6~--~1200, then the $p$-value for an AD test is 0.012. 
It is tempting to attribute the differences in transit duration differences as primarily due to the increased efficiency of orbital circularization for small orbital periods.
However, we caution that the size distribution of the planets with $P < 6$ days is weighted towards significantly smaller values than the size distribution of planets with larger periods, since  \ik has greater sensitivity for detecting planets at shorter orbital periods.  


\subsubsection{Transit Duration vs.~Planet Size}\label{sec:eccsize}
Splitting the cumulative $\tau$ distributions for planets by planet radius  at 2.16~R$_\oplus$ (near the median planet size of our sample), the $p$-value for the KS test is 0.15, and the $p$-value for the AD test is 0.07. Thus, neither of these tests finds statistically-significant differences between the two samples.

\begin{figure}
\centering
\includegraphics[width=0.48\textwidth]{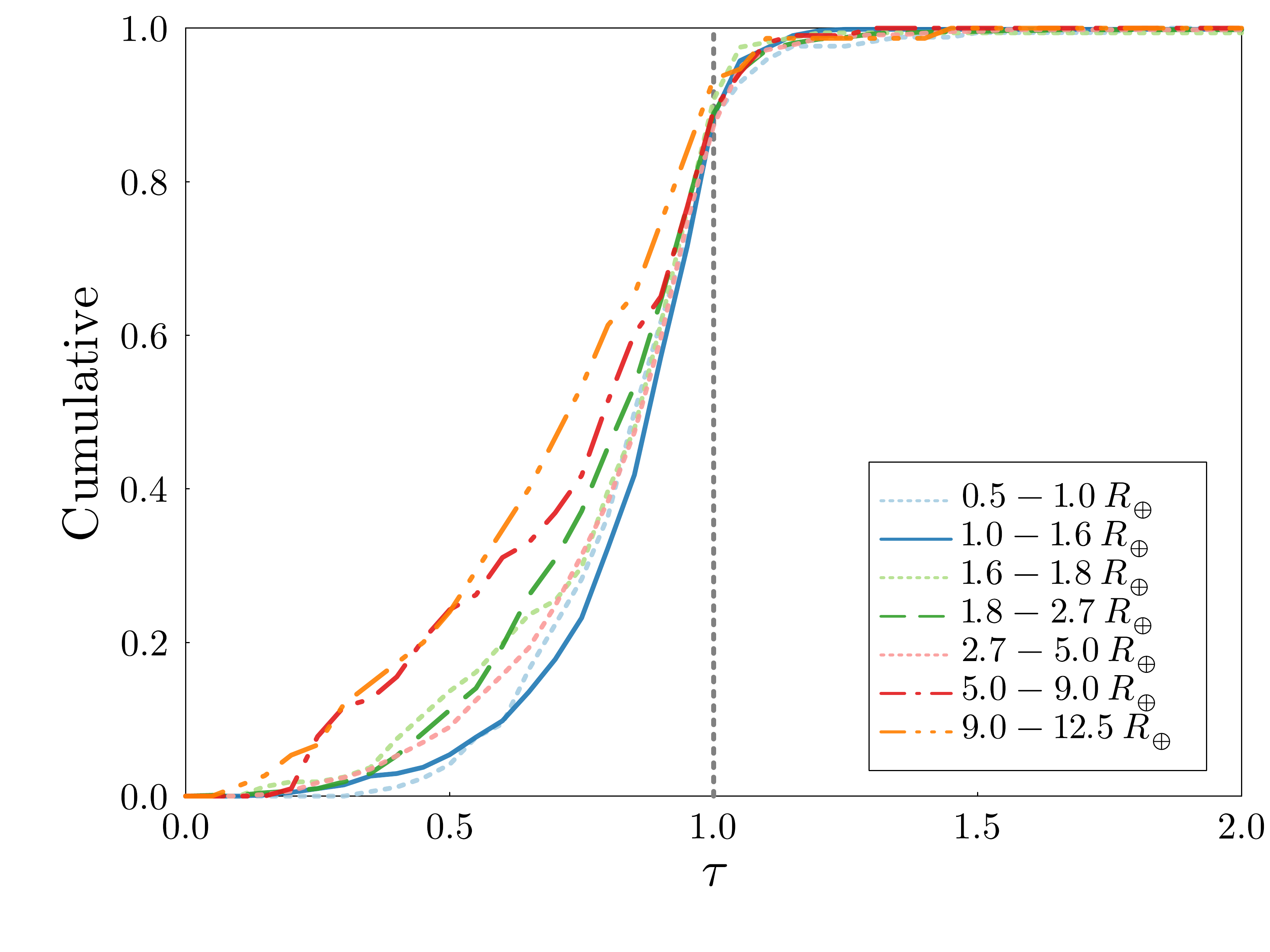}
    \includegraphics[width=0.48\textwidth] {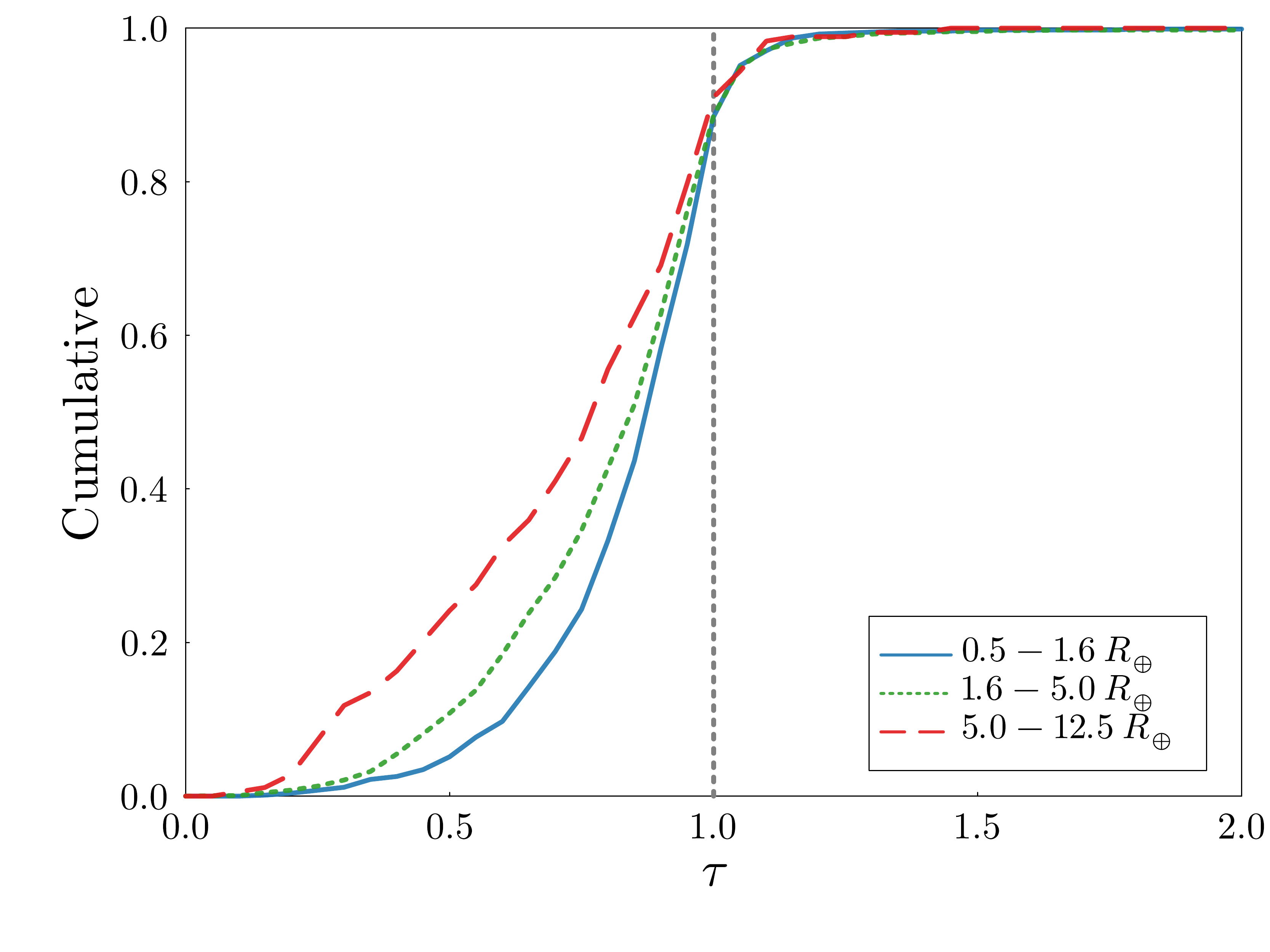}
    \caption{Comparison of the normalized cumulative $\tau$ distributions for planets grouped by size. Left panel: Subsets are:  0.5~R$_\oplus < R_p \le 1.0$~R$_\oplus$ (dotted light blue),
    1.0~R$_\oplus < R_p \le 1.6$~R$_\oplus$ (solid dark blue), 
    1.6~R$_\oplus < R_p \le 1.8$~R$_\oplus$ (dotted light green), 
    1.8~R$_\oplus < R_p \le 2.7$~R$_\oplus$ (dashed dark green), 
    2.7~R$_\oplus < R_p \le 5.0$~R$_\oplus$ (dotted light red), 5~R$_\oplus < R_p \le 9$~R$_\oplus$ (dash-dotted dark red), 9~R$_\oplus < R_p \le 12.5$~R$_\oplus$ (dash-double-dotted orange).  A 7-sample AD test strongly rejects the null hypothesis that each subset is drawn from a common distribution ($p$-value $\approx~4\times 10^{-11}$).  The two smallest radius curves are among the three subsets with the most highly concentrated $\tau$ distribution, whereas the two largest radius curves are the subsets with the largest tails.  Both of these would be expected if larger (and thus more massive) planets have typically experienced more significant dynamical excitation (e.g., planet-planet scattering followed by secular evolution).   Right panel: Subsets are: 0.5~R$_\oplus < R_p \le 1.6$~R$_\oplus$ (solid blue), 1.6~R$_\oplus < R_p \le 5$~R$_\oplus$ (dotted green), and 5~R$_\oplus < R_p \le 12.5$~R$_\oplus$ (dashed red).  A 3-sample AD test strongly rejects the null hypothesis that each subset is drawn from a common distribution ($p$-value $\sim 10^{-11}$). }
    \label{fig:EccRadius}
\end{figure}

For the left panel of Figure \ref{fig:EccRadius}, we  divide the population into subsets based on theoretically-motivated size bins with boundaries 0.5, 1.0, 1.6, 1.8, 2.7, 5.0, 9.0 and 12.5~R$_\oplus$. Visual inspection shows similarity between the distributions of the two smallest size bins; the {three} middle size bins also appear to have similar size distributions to one another, as do the distributions for two largest size bins. Furthermore, statistical tests do not find any significant differences among the distributions within any of these three subsets of planet size bins.

Thus, we combine planets into three size bins,  0.5~--~1.6~R$_\oplus$, 1.6~--~5~R$_\oplus$, and 5~--~12.5~R$_\oplus$, in the panel on the right of Fig.~\ref{fig:EccRadius}. The hypothesis that all three of these bins have the same underlying distributions can be strongly rejected, with $p$-value $\sim10^{-11}$ (using 3-sampled AD test), and
comparing any pair of these three results in a $p$-value of $<10^{-3}$ (KS tests) or $<10^{-4}$ (AD tests), showing highly significant evidence for differences in the eccentricity distributions between these three broad size bins. 
Further supporting this interpretation, planets with $R_p\le 1.6$~R$_\oplus$ have a particularly concentrated $\tau$ distribution (i.e., small eccentricities), while that for planets with $R_p > 5$~R$_\oplus$ has a relatively large tail (i.e., larger eccentricities), with planets 1.6~R$_\oplus < R_p \le 5$~R$_\oplus$ having an intermediate distribution of  $\tau$ values.

These differences are consistent with the observation that planet sizes (and likely masses) are correlated within a planetary system \citep{Ciardi:2013,Weiss:2018,Millholland:2017,He:2020}, together with the theoretical idea that more massive planets excite larger eccentricities in neighboring planets via planet-planet scattering and/or secular perturbations \citep{FordRasio:2008,Johansen:2012,Pu:2015,Laskar:2000}.  However, they run counter to (although do not necessarily conflict with) the expectation that within a given planetary system, equipartition of angular momentum deficit would typically lead to more massive planets having smaller eccentricities \citep{Lissauer:1995}. To contrast the eccentricities of large and small planets within the same system, we looked at all systems with two or more planets with $6 < P < 1200$ days (the lower limit being chosen to minimize the effects of tidal damping, see Fig.~\ref{fig:EccPeriod}) that meet our criteria for analysis in this section and found that in 136 out of 273 cases (50\%) the largest such planet has a larger value of $\tau$ than the smallest planet.  This implies that there is not a strong preference for planets  within a given inner planetary system to have different eccentricity distributions (after one removes planets potentially affected by tidal circularization).
These conclusions are not affected by restricting the sample to planets with orbital periods greater than 8 days (rather than 6 days).


\subsubsection{Transit Duration vs.~Spacing Between Orbits}\label{sec:eccspacing}

\begin{figure}
    \centering
    \includegraphics[scale=.3]{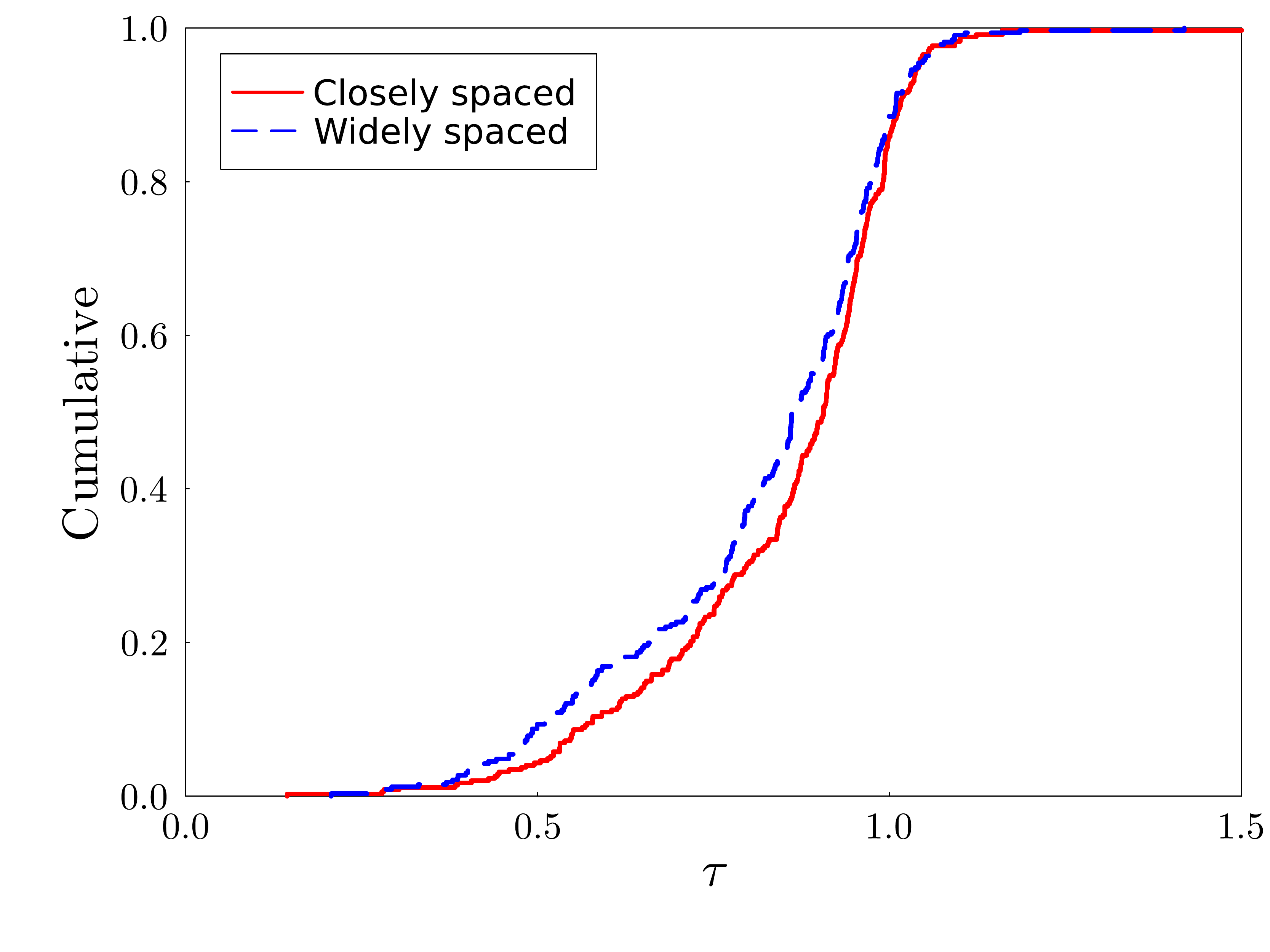}
    \caption{Comparison of cumulative $\tau$ distributions for planets in multi-planet systems with orbital periods greater than 8 days.  The two curves show results for subsets chosen based on the orbital period ratio.  Here, the closely-spaced subset of planets is defined as those having a detected companion planet (which could have $P<8$ days) with period within a factor of 2.06, a value selected so that the each curve represents nearly the same number of planets.   The widely-spaced subset is the complement of the closely-spaced planets.  A 2-sample AD test rejects the null hypothesis that the $\tau$ distributions for the closely and widely spaced planets are drawn from a common distribution ($p$-value $=0.037$).    }
    \label{fig:EccSpacing}
\end{figure}

We divided planets in multiple planet systems (excluding the split multis KOI-284 and KOI-2248; see \S\ref{sec:falsemultis}) into two or four nearly equal-size subsets based upon their period ratios with their nearest detected neighbor in $\log P$.  
The period ratio boundaries are the first three quartiles of the period ratio distribution, 1.65, 2.06 and 2.94. 
Very closely spaced planet pairs typically need to have small eccentricities to avoid close encounters (with potential for exceptions related to resonances).  
As expected, the period-normalized transit duration distribution is more sharply peaked for planets with a nearby neighbor.  However, there was not a statistically significant difference in the $\tau$ distribution between these subsets of planets based on performing a 2 or 4 sample AD test using our nominal sample of planets.  
However, if, as shown in Figure \ref{fig:EccSpacing}, we exclude planets with orbital periods less than 8 days (intentionally larger than 6 days to be confident that we exclude all planets which are likely to be significantly affected by tidal circularization), then {the 2-sample} AD tests rejects the null hypothesis that the observed normalized transit duration distributions for the different subsets of planets (based on period ratio to nearest detected neighbor) are drawn from the same underlying distribution with a $p$-value of 0.037.  

We also compared the duration distributions of planets in multiple planet systems near first-order MMRs with those of other planets in multiple planet systems and did not find statistically-significant differences. 
However, we note that the number of planets near resonances is small (214), so nontrivial differences in the distributions may become evident when larger samples become available for study (e.g., combining \ikt, {\it K2}, {\it TESS} and {\it PLATO} data).  




\newpage
\section{Long-term Average Planetary Orbital Periods}\label{sec:periodtheory}

The fractional uncertainties quoted for the periods of the vast majority of planets listed in Table \ref{tab:planetcatalog} are $<10^{-5}$, with $\sim 10^{-6}$ being typical (corresponding to 2 minutes per 4 years). These values represent the formal uncertainties in the best fit, constant-period ephemeris computed using the measured midpoints of transits and adjusted for the motion of the spacecraft relative to the rest frame of the barycenter of our Solar System (\S\ref{sec:periods}).  However, as discussed below, the actual mean orbital periods of the planets can differ from the values given in Table \ref{tab:planetcatalog} by many times the listed uncertainties. For studies of the architectures and dynamics of planetary systems, the mean orbital period, $\Bar{P}$, is generally far more important than the mean time interval between transits measured directly from \ik data, $P$. 

Typical radial velocities of \ik target stars relative to the barycenter of the Solar System are of order $10^{-4}$ times the speed of light, so the actual periods of the planets in the rest frame of their planetary system differ from the measured orbital periods by that fractional amount due time dilation.  This small error in tabulated orbital periods is not important to understanding the dynamics of an exoplanetary system because relativistic effects within these systems are small and the periods of all planets in a given system are altered by the same factor, so period ratios remain unchanged.  Moreover, the radial velocities of these planetary systems do not vary substantially, so ephemerides are also not significantly affected\footnote{If the planet's host star is a member of a binary star system, then its radial velocity relative to the Solar System can vary by a nontrivial amount. Consider a binary of two 1 M$_\odot$ stars on a circular orbit with semimajor axis of 100 AU that is viewed edge-on. The orbital period is 707 years, and when the stars are near one of the eclipses, their relative radial velocity observed from our Solar System is changing at a rate of $\sim~18$ m/s/yr $\approx 6.1~\times 10^{-8} c/$yr, where $c$ represents the speed of light in vacuo. Thus, ignoring other factors, the ratio of periods of planets around one of these stars to those around the other should be changing by $\sim 6\times 10^{-8}$/year, which for a precision of 1 part in $10^6$ would be detectable from two sets of observations taken a few decades apart.}.

In contrast to time dilation caused by uniform stellar motion relative to our Solar System,  TTVs may produce errors in estimated planetary orbital periods that must be accounted for in some dynamical investigations and also when making ephemeris predictions. 
Periodic TTVs with timescales that are short compared to the interval of \ik observations largely average out and do not produce significant errors in our estimates of orbital periods.  However, when the TTV timescale is long compared to the \ik observations, the period estimated from a hypothetical future set of transit time observations of similar quality and duration to the \ik data could differ significantly.  For example, using the integrations of planets in the Kepler-80 (KOI-500) system performed by \citet{MacDonald:2016}, we see that the observed orbital period changes by as much as a few~$\times 10^{-4}$ days -- several times larger than the reported uncertainties -- when averaged over eight years of data instead of four. The cause of this discrepancy is that the \ik observations cover less than half of a TTV cycle (Fig.~\ref{fig:Kep80}), and therefore the observed orbital period is not the same as the long-term mean period. Note that in this case, some of the planets' periods were  observed near the turning points in their evolution, far from the mean, as should often be the case.  This is compounded by other subtle issues like uncertain apsidal/nodal precession and differences between the measured anomalistic period and the true orbital period. 

\begin{figure}[!hbt]
 \includegraphics [height = 2.6 in]{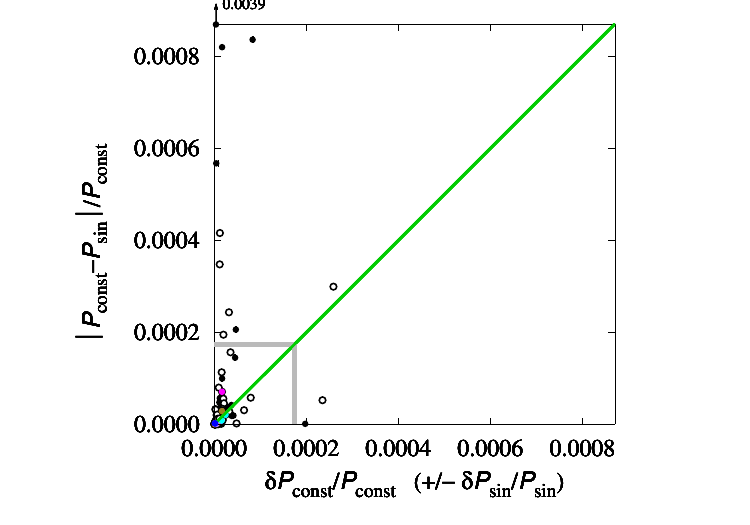}
 \hspace{-0.6 in}
 \includegraphics [height = 2.6 in]{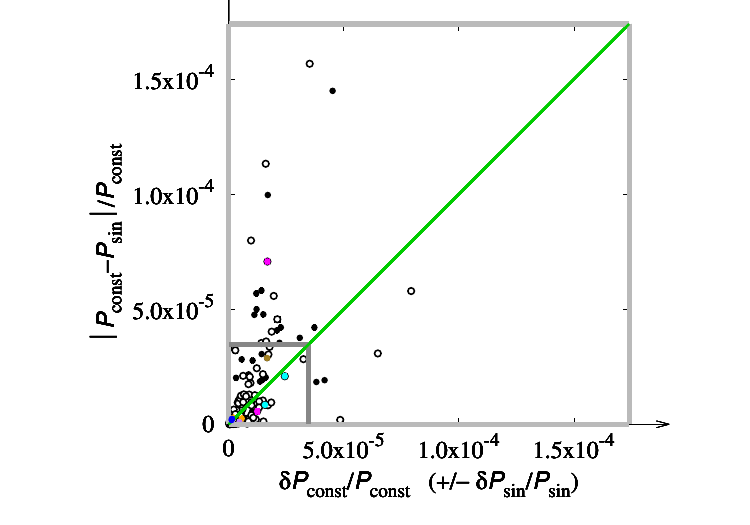}
  \hspace{-0.6 in}
 \includegraphics [height = 2.6 in]{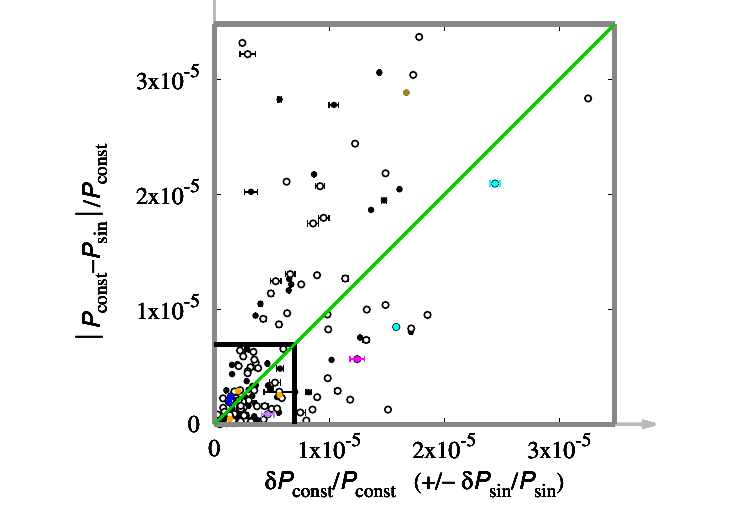}
  \hspace{-0.6 in}
 \includegraphics [height = 2.6 in]{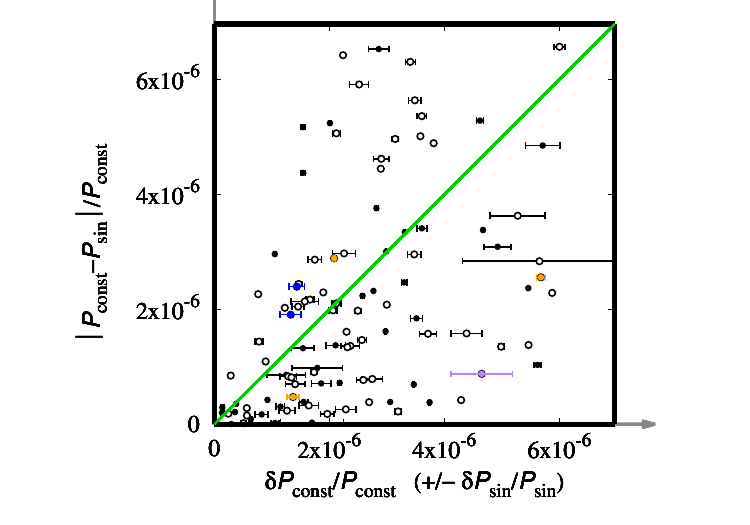}
 \caption{Comparison between the orbital periods listed in (our) Table \ref{tab:planetcatalog} ($P$, denoted $P_{\rm const}$ here) and those presented in Table  6 of \citet{Holczer:2016}, $P_{\rm sin}$.  The horizontal coordinate represents the fractional uncertainty given in Table \ref{tab:planetcatalog}, the length of the bars shows \citet{Holczer:2016}'s fractional uncertainty, and the vertical coordinate represents the fractional differences between our periods and those of \citet{Holczer:2016}. Open circles represent planets in multis; black filled circles represent single planets. The plot on the upper left shows the largest scale, and successive plots (upper right, lower left and lower right) zoom in by a factor of five relative to the previous plot. The green diagonal lines divide planets whose period estimates differ between the two tables by more than their uncertainties in our table (above and to the left) from those whose period differences are less than our uncertainties (below and to the right). One planet, the single KOI-1209.01, which has an orbital period of about nine months, lies outside the range of the plots, with fractional period difference of $3.9 \times 10^{-3}$, far larger than either the fractional uncertainty in Table \ref{tab:planetcatalog} ($3.6 \times 10^{-6}$) or that in  \citet{Holczer:2016}'s table ($ 5.2\times 10^{-6}$); its location is pointed to by the arrow at the upper left of the largest-scale plot. Colored points indicate planets in the dynamically ``solved'' systems of interest explored in this section:
 Kepler-11 (orange), Kepler-36 (cyan), Kepler-60 (purple), Kepler-80 (blue), Kepler-223 (magenta) and Kepler-419 (olive). 
 \label{fig:HolczerPeriods}} 
 \end{figure}

\cite{Holczer:2016} produced a catalog of \ik planets displaying periodic TTVs with timescales that are comparable to or shorter than the interval of \ik observations.  Their Table 6 lists estimated long-term mean orbital periods for 199 planet candidates based upon fitting a period plus a sinusoidal modulation to observed transit times.  
Figure \ref{fig:HolczerPeriods} compares our estimates of orbital periods to those in their Table 6. Inspection of these plots shows that the differences between \citet{Holczer:2016}'s sinusoidal-fit estimates of planetary periods and the ones that we list in Table \ref{tab:planetcatalog} are positively correlated with the uncertainties listed in our table.  In most cases the difference between the estimates given in the two papers is less than our tabulated uncertainties, although the difference greatly exceeds our error estimates for some planets. The uncertainties given in \citet{Holczer:2016}'s Table 6 are typically much smaller than ours; however, since the period uncertainties that they obtained using constant-period fits (their Table 2) are even smaller for most planets, we caution the reader against overinterpreting their quoted uncertainties.  

 \subsection{Apse Precession and TTVs}\label{sec:precession}

Precession causes the time interval between successive periapse passages to differ from the time interval required to travel 360$^\circ$ in azimuth (which is the reason that anomalistic periods differ from orbital periods).  This non-Keplerian behavior leads to eccentric planets spending a little less time at some radial distances between successive transits, resulting in variations in time intervals between successive transits that average out only over timescales much longer than that of \ik observations \citep{Agol:2005}.

The eccentricity of a planet near resonance can be viewed as a superposition of its free and forced eccentricity vectors in the ($e \sin \omega$, $e \cos \omega$) plane.  The precession of the forced eccentricity is relatively rapid, and accounts for some of the TTVs that are found among \ik planets; its effects are partially accounted for in period estimates.  The precession of the free eccentricity is much slower.  It is another source of TTVs, which we refer to as secular TTVs.  However, if $ \Delta \varpi T_{\rm obs}/P \ll 1$, i.e., precession is $\ll 2\pi$ during the \ikt's prime mission, then the secular TTVs won't be recognized  and thus can't be removed from/accounted for in calculations of the planet's period. 
Secular precession of the planets' free eccentricities is unrelated to resonances and generally has a period much longer than the baseline of the \ik observations, so it is not accounted for in our estimates of mean periods and uncertainties thereof (nor was it in previous \ik planet catalogs). 
Consider a planet with eccentricity $e \ll 1$ whose longitude of periapse precesses by $\Delta \varpi \ll 1$ radians per orbit and which transits near periapse.  During the interval of time between successive periapses passages, the planet moves through $2\pi+\Delta \varpi$ radians.  As the planet completes one radial oscillation during this interval, its average angular velocity is equal to the long-term average value.  However, from Kepler's second law, we know that the planet is sweeping out angle at a rate $1+e$ times as fast near periapse.  Thus, the (measured) time interval between successive transits, $P$, is related to the mean orbital period $\Bar{P}$ and the difference between the longitude of periapse and that of the transit mid-point by:


\begin{equation}\label{eqn:periodsII}
\frac{P}{\Bar{P}} = 1 - \frac{e\Delta \varpi}{2\pi}\cos{\varpi}.
\end{equation}

\noindent As the cosine term in Equation (\ref{eqn:periodsII}) integrates to zero, this effect averages out over one precession period of the eccentricity
. \emph{Nonetheless, the  long-term mean orbital period can differ significantly from the mean period between successive transits during the \ik era even for planets having TTVs that are too small to be observable during the epoch of \ik observations.}  

As planets in \ik multis typically have eccentricities of one to several percent \citep{Xie:2016, Jontof:2016}, $\Bar{P} - P \ll P$, but can nonetheless be much larger than the formal uncertainties in measured orbital periods. The differences between these periods is so small that it can be ignored for our comparisons of the period distributions of various subsets of the \ik sample (\S\ref{sec:periodDist}) and for most studies of period ratios of two planets. However, the (ill-quantified) errors in our estimates of $\Bar{P}$ are important for studies of the distribution of three-body resonances (Lissauer et al., in preparation).

\subsection{Case Studies of Select Dynamically-Solved Planetary Systems}\label{sec:individual}


Transit timing variations have been used for detailed dynamical analyses of a small fraction of the \ik multi-planet systems.  Most of the publications presenting these studies list osculating orbital periods at an epoch near the midpoint of the \ik data.  For several landmark systems, we integrated numerous (in most cases 101) samples of system parameters from the MCMC chains deduced via photodynamical or TTV analyses of the \ik data to compute estimated transit times during and after the \ik epoch.  Table \ref{tab:periods} compares orbital periods (in most cases at an epoch near the midpoint of \ik observations) from the dynamical solutions to long-term average periods of these planets that we computed by integrating these dynamically-solved systems, our standard estimates of the orbital periods obtained via a best constant-period fit to observed transit times (Table \ref{tab:planetcatalog}), and  (where available) to the average  orbital periods estimated in the sinusoidal fits of \citet{Holczer:2016}. The values of ``$P$ at epoch'' in Table \ref{tab:periods} for planets of Kepler-29 and Kepler-60 represent osculating orbital periods at the epochs listed in the subsections below, following our dynamical fits to the transit times reported by \citet{Rowe:2015b}.

\begin{table}[]
    \centering
    \begin{tabular}{|c|c||c|c|c|c|}
        \multicolumn{6}{c}{\textbf{COMPARISON OF ESTIMATES OF ORBITAL PERIODS OF SELECT PLANETS}}\\
        \hline
        KOI & Kepler- & $P$ [Table \ref{tab:planetcatalog}] & $P$ [\citealt{Holczer:2016}] & $P$ at epoch & $\Bar{P}$\\
        \hline
        157.06 & 11b & 10.304031~$\pm$ 0.000026 & -- & $10.30260\pm0.00027$ & $10.30391\pm0.00004$\\
        157.01 & 11c & 13.024917~$\pm$~0.000018 & 13.0249115 $\pm$ 0.0000014 & $13.02555\pm0.00018$ & $13.02507\pm0.00004$\\
         157.02 & 11d & 22.687141~$\pm$~0.000037 & -- & $22.68546\pm0.00037$&$22.68708\pm0.00003$ \\
         157.03 & 11e & 31.995517 $\pm$ 0.000067 & 31.9954254 $\pm$ 0.0000009 &  $31.99834\pm0.00052$ & $31.99555\pm0.00004$  \\
        157.04 & 11f & 46.68587 $\pm$ 0.00027 & 46.6857474 $\pm$ 0.0000036 &  $46.6933\pm0.0018$ & $46.6855\pm0.0005$ \\
         157.05 & 11g & 118.37857 $\pm$ 0.00025 & -- &  $118.38089\pm0.00057 $ & $118.3782\pm0.0005$  \\
         277.02 & 36b & 13.84899 $\pm$ 0.00034 & 13.848692 $\pm$ 0.000006 & $13.849194\pm0.00004$ & 
         $13.848063\pm0.0002$
         \\
         277.01 & 36c & 16.231949 $\pm$ 0.00026 & 16.232080 $\pm$ 0.000001 & $16.231774\pm0.00002$ &
         $16.232628\pm0.0002$
         \\
         500.05 & 80f & 0.9867860 $\pm$ 0.0000013 & -- & $ 0.9867873 \pm 0.0000044 $  & $ 0.9867862  \pm  0.0000012 $ \\
         500.03 & 80d & 3.0721523 $\pm$ 0.0000058 & -- &  $ 3.07253 \pm 0.00029 $ & $ 3.0721293  \pm  0.0000086 $ \\
         500.04 & 80e & 4.6453934 $\pm$ 0.0000084 & -- &  $ 4.64474 \pm 0.00022 $ & $ 4.645410  \pm  0.000014 $  \\
         500.01 & 80b & 7.0535287 $\pm$ 0.0000094 & 7.0535152 $\pm$ 0.0000013&  $ 7.05357 \pm 0.00036 $ & $ 7.053570 \pm 0.000025 $   \\
         500.02 & 80c & 9.521646 $\pm$ 0.000014 & 9.5216221 $\pm$ 0.0000013 &  $ 9.52330 \pm 0.00030 $ & $ 9.521525  \pm  0.000047 $  \\
         500.06 & 80g & 14.64538 $\pm$ 0.00011 & -- &  $ 14.6503 \pm 0.0018 $ & $ 14.6457  \pm  0.0013 $  \\
         
        730.04 & 223b &  7.384456 $\pm$ 0.000072 & -- &  7.38449 $\pm$ 0.00022& $ 7.3845\pm0.0002$ \\
        730.02 & 223c & 9.848211 $\pm$ 0.000082 & -- &  $9.84564\pm0.00052$ & $9.84934\pm0.00014$  \\
        730.01 & 223d  & 14.78701 $\pm$ 0.00018 & 14.7869296 $\pm$ 0.0000095 & $14.78869\pm0.00029$  & $14.7841\pm0.0002$ \\
        730.03 & 223e & 19.72434 $\pm$ 0.00033 & 19.725722 $\pm$ 0.000018 &  $19.72567\pm0.00055$ & $19.7256\pm0.0007$  \\
        738.01 & 29b  & 10.339236 $\pm$ 0.000056 & -- & 10.33842 $\pm$ 0.00029  & $10.336927\pm 0.000025$  \\
        738.02 & 29c & 13.28712 $\pm$ 0.00011 & -- & 13.28841 $\pm$ 0.00053 & $13.290961\pm 0.000037$ \\
        1474.01 & 419b & 69.7262 $\pm$ 0.0012 & 69.7281819 $\pm$ 0.0000004 &  $69.7960\pm$ 0.0042 &    $69.7869 \pm 0.0454 $  \\ 
        2086.01 & 60b & 7.132950 $\pm$ 0.000041 & -- & 7.13335 $\pm$ 0.00013 & 7.1325157 $\pm$ 0.000025 \\ 
        2086.02 & 60c & 8.918977 $\pm$ 0.000041 & 8.91867 $\pm$ 0.00020 &  8.91866 $\pm$ 0.00018 & 8.919029 $\pm$ 0.000004 \\
        2086.03 & 60d & 11.89825 $\pm$ 0.00010 & -- &  11.89810 $\pm$ 0.00020 & 11.899566 $\pm$ 0.000070 \\
        \hline
    \end{tabular} 
 \caption{Orbital periods, in days, of selected well-studied planets estimated using various different methods. From left to right, the columns give the values presented in Table \ref{tab:planetcatalog} of this work, Table 6 of \cite{Holczer:2016}, period at epoch (typically near the mid-time of \ik observations) from dynamical fits to TTs, and long-term (averaged over the same intervals used for ordering the samples to select representative systems shown in Figures \ref{fig:Kep11}~--~\ref{fig:Kep60Periods};  $10^4$ years for Kepler-419~b, $10^6$ days~$\approx 2738$ years for Kepler-11's planets, 100 years for Kepler-80, 1000 years for planets in other systems) average periods computed by integrating the systems using samples of the initial conditions at epoch obtained from dynamical studies.}
    \label{tab:periods}
\end{table}

Period variations for one or more of the planets in each of these dynamically active multi-planet systems are shown in Figs.~\ref{fig:Kep11}~--~\ref{fig:Kep60Periods}.  Each panel within these figures shows three samples, which we selected by ordering the solutions by the  long-term average period of the first KOI found in the system (KOI number ending in .01), then selecting  the median member of the list (usually the sample with the 51$^{\rm st}$ longest value, in black), one with that planet's period roughly one standard deviation shorter than the mean (17$^{\rm th}$ sample, in red), and one with this planet's mean period one standard deviation longer than the mean (85$^{\rm th}$ sample, in blue).

All of the systems that we analyze in this subsection have strong interactions among all or most of the known planets, leading to substantial TTVs.  Three of these systems, Kepler-11, Kepler-36 and Kepler-419, don't have any planets known to be librating within orbital resonances.  One system, Kepler-29, has two planets that are locked in a 9:7 (second-order) MMR.  The other three systems considered below, Kepler-60, Kepler-80 and Kepler-223, each have three or more planets locked into a resonant chain that includes (zeroth-order) three-body resonances and probably first-order two-body mean-motion resonances (in some cases, it isn't known whether or not the two-body resonance variables are librating).

To estimate long-term periods of all known transiting planets in the abovementioned seven \ik planetary systems, we ran simulations with an Embedded Runge-Kutta Prince-Dormand integrator (\citealt{PD:1981}; gsl\_odeiv2\_step\_rk8pd within the GNU Scientific Library, \citealt{Gough:2009}). From the transit times (more precisely, the times near inferior conjunction when the distance between the center of the planet and that of the star projected onto the plane of the sky reaches a minimum, as we typically do not have sufficient information on the impact parameter to know whether or not a transit actually occurs) simulated over the specified interval (typically 1000 years) beginning with the start of \ik observations, we determine the best fitting linear ephemeris. The transit periods thus derived are given in the final column of Table~\ref{tab:periods}.   Due to numerical dissipation, our non-symplectic code causes the orbital periods to decrease slightly; in a run for 1 million days with only the 10-day planet Kepler-11~b, $P_b$ decreased $-1.16\times10^{-6}$~day.  This effect on simulations with all 6 planets of Kepler-11, $P_b$ averages a loss of $-5.8\times10^{-7}$~day, which is a systematic bias, but is dwarfed by the statistical uncertainty of $\sim4\times10^{-5}$~day for that system. (We would advocate using a symplectic algorithm for integrations longer than those reported here.)

We present results of our modeling in the rightmost columns of Table \ref{tab:periods} and in Figs.~\ref{fig:Kep11}~--~\ref{fig:Kep60Periods}. Note the differences in mean period and period at epoch of the various solutions that fit the observed data, as well as differences of periods averaged over four years when taken starting at differing times for three representative solutions for each of the planets.

\subsubsection{Kepler-11 = KOI-157}

Six transiting planets are known to orbit Kepler-11, all larger and more massive than the Earth but less massive than Uranus. Each of the five inner planets is located near but not in a first-order mean-motion resonance with one or two of its neighbors: The inner pair, b and c, are slightly wide of the 5:4 MMR; the third and fifth planets, d and f, are just wide of the 2:1 resonance, whereas the fourth and fifth planets, e and f, are orbit a bit closer to one another than the 3:2 resonance. The outer planet, g, orbits well exterior to the inner five. Values of the osculating orbital periods for Kepler-11's planets, taken from \cite{Bedell:2017}, are at epoch $T_{\rm BJD} = 2455700$.

\begin{figure}[!hbt]
 \includegraphics [width =  2.2in, angle =0]{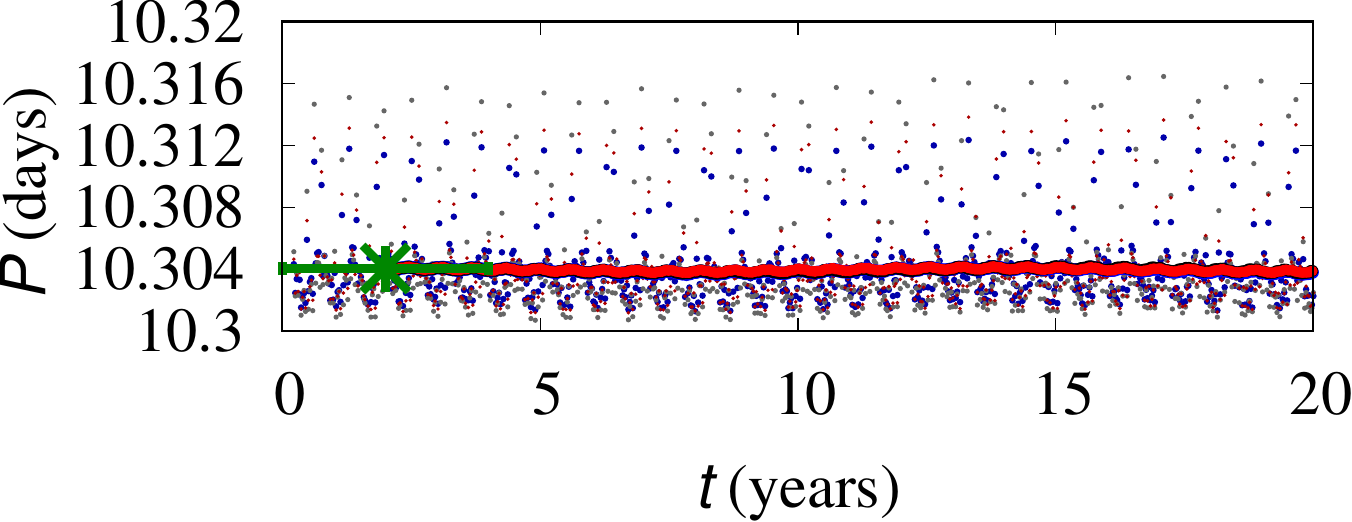}
  \includegraphics [width =  2.2in, angle =0]{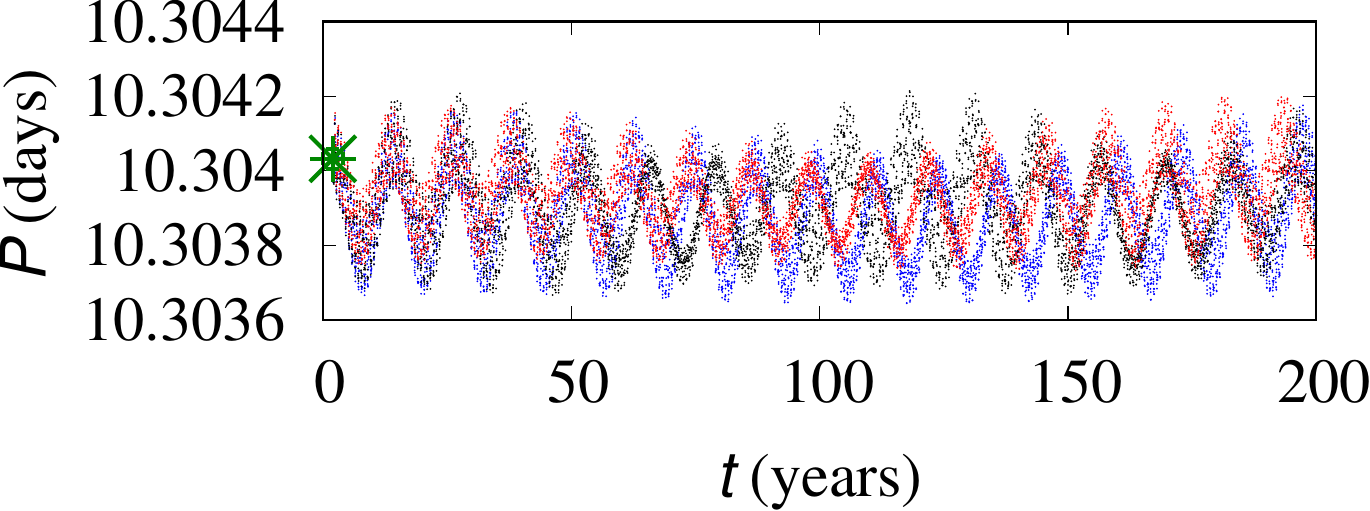}
   \includegraphics [width =  2.2in, angle =0]{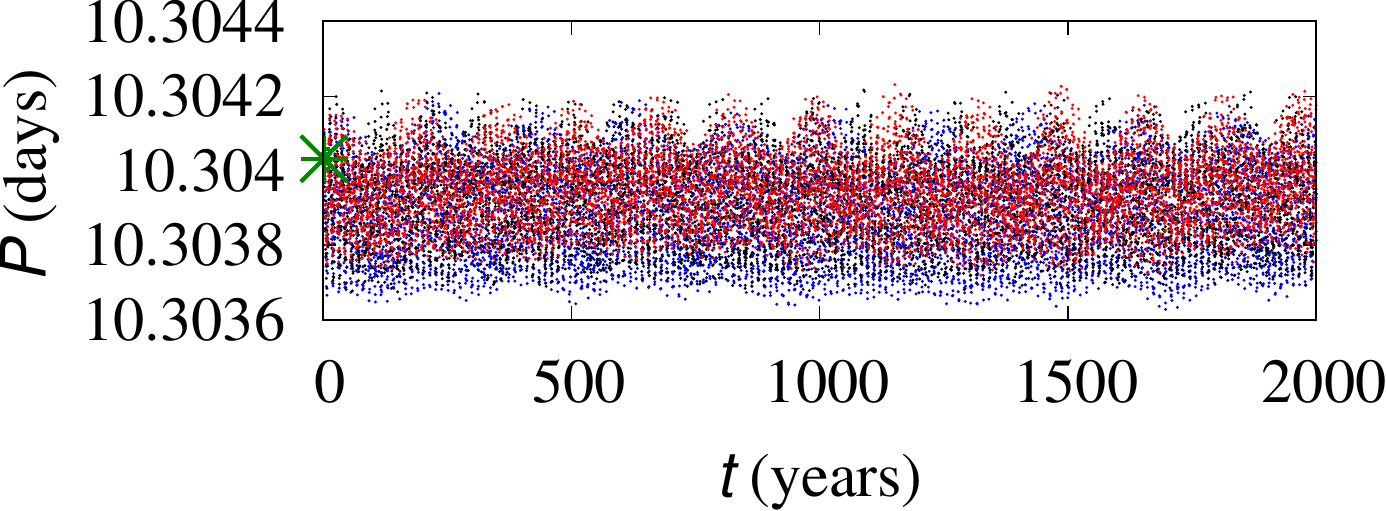}
\newline
 \includegraphics  [width =  2.2in, angle =0]{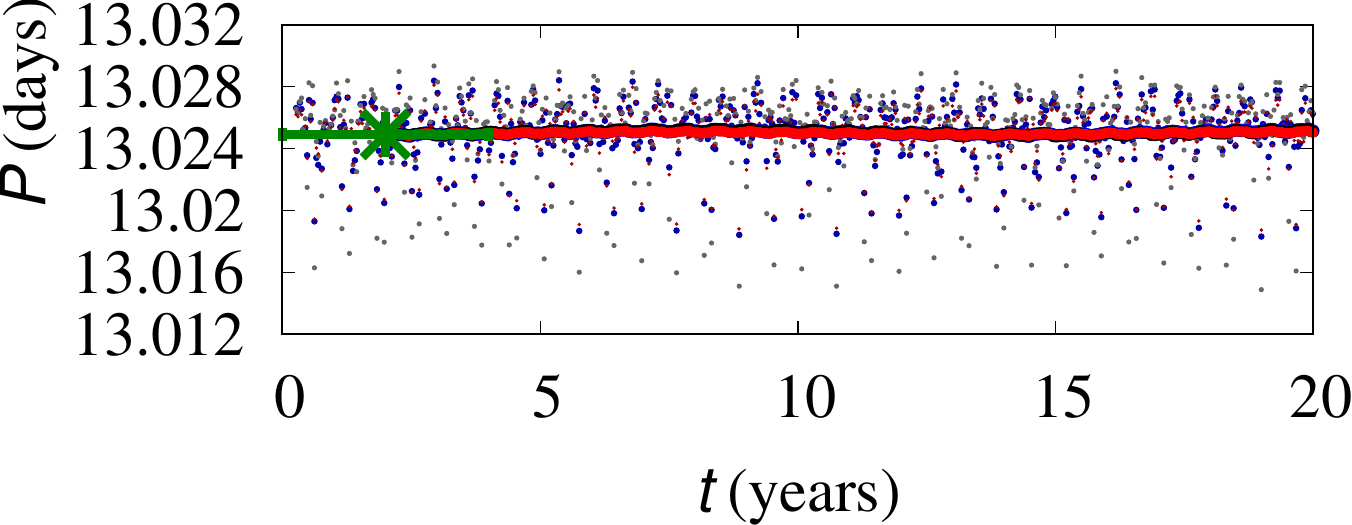}
  \includegraphics   [width =  2.2in, angle =0]{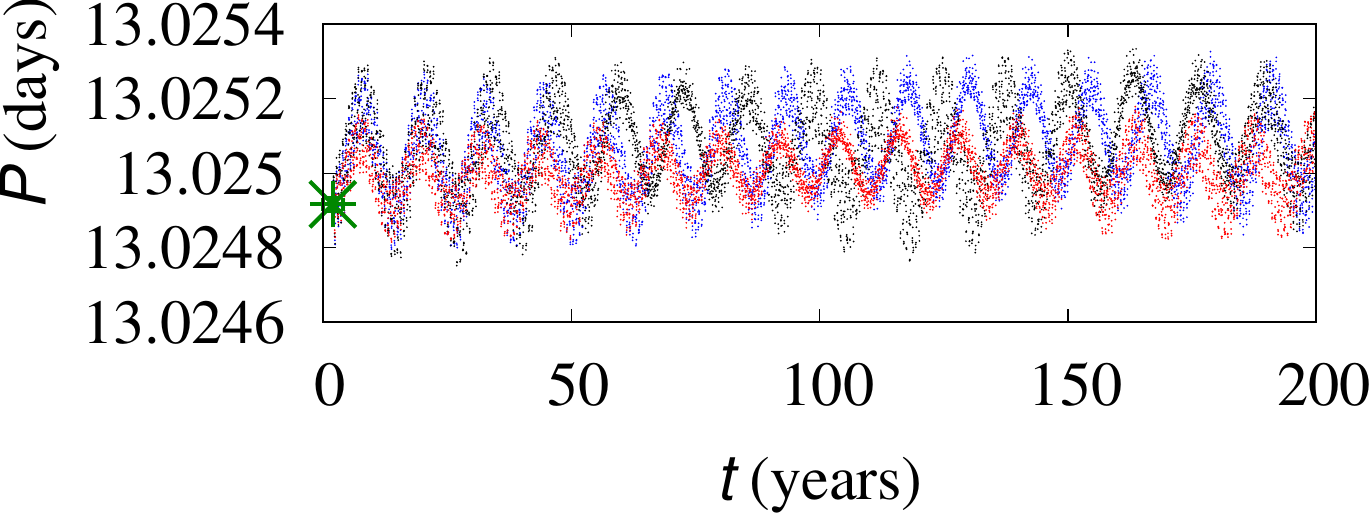}
    \includegraphics [width =  2.2in, angle =0]{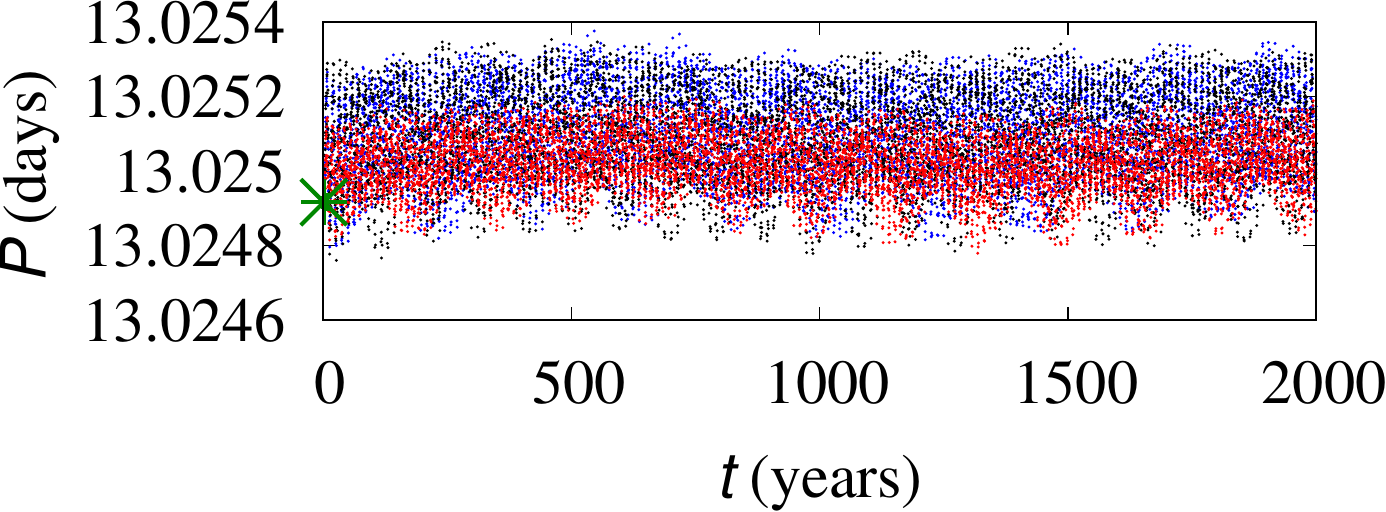}
     \newline
  \includegraphics [width =  2.2in, angle =0]{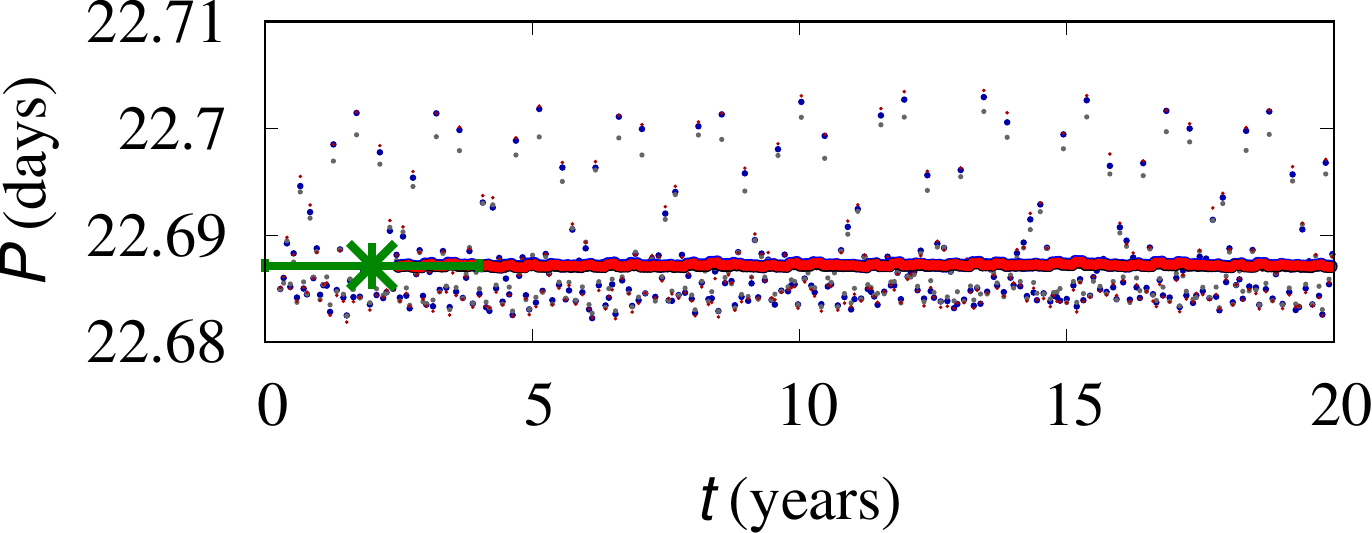}
    \includegraphics [width =  2.2in, angle =0]{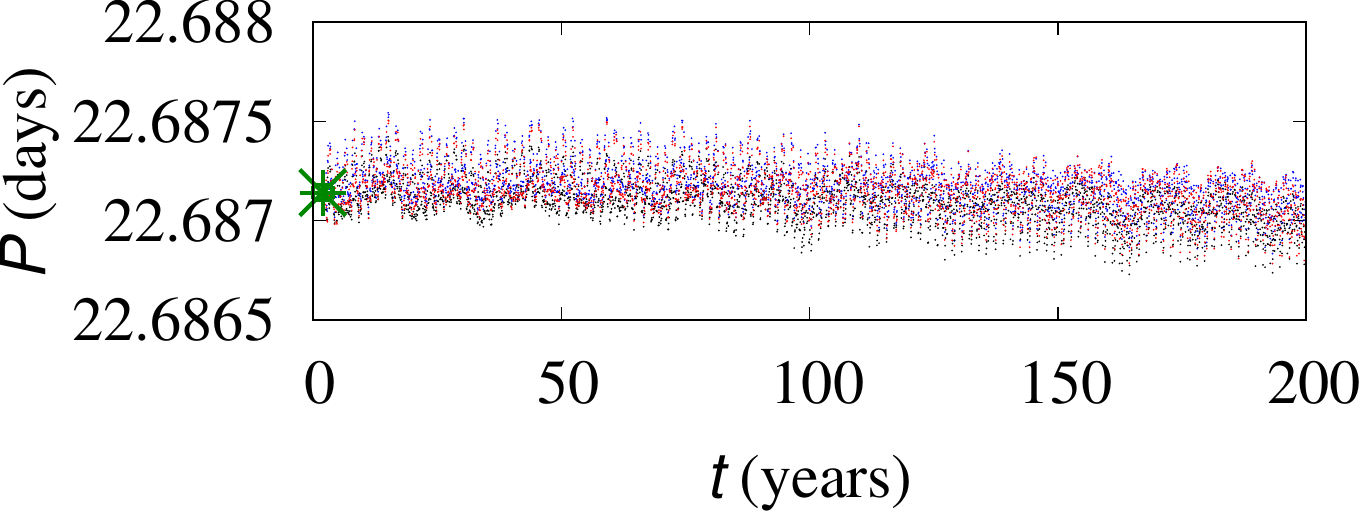}
 \includegraphics [width =  2.2in, angle =0]{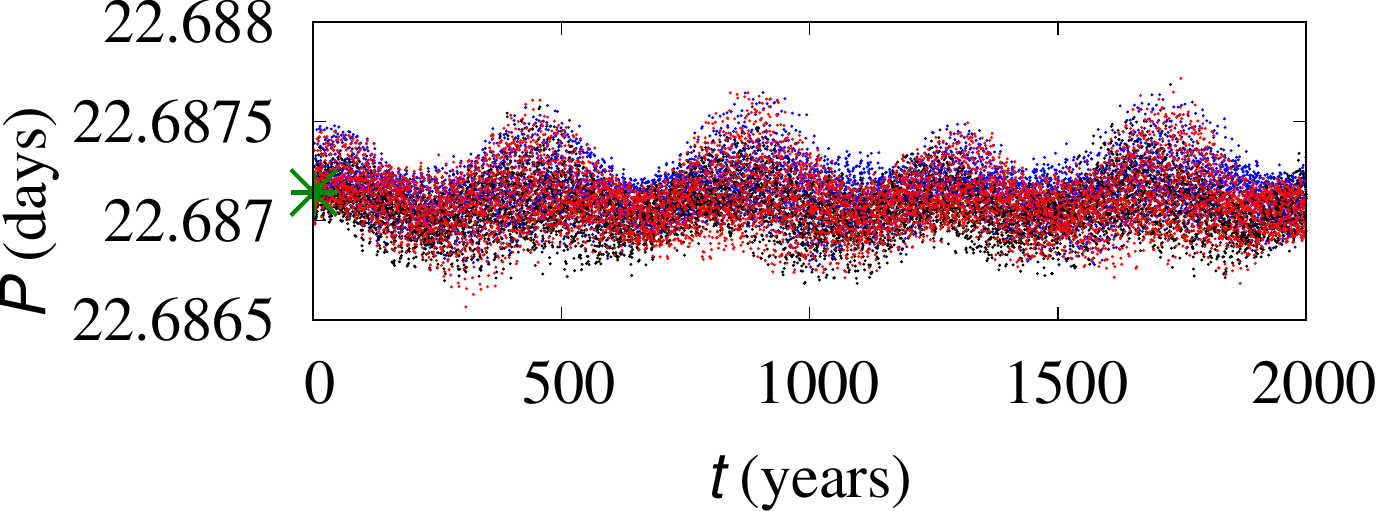}
      \newline
 \includegraphics [width =  2.2in, angle =0]{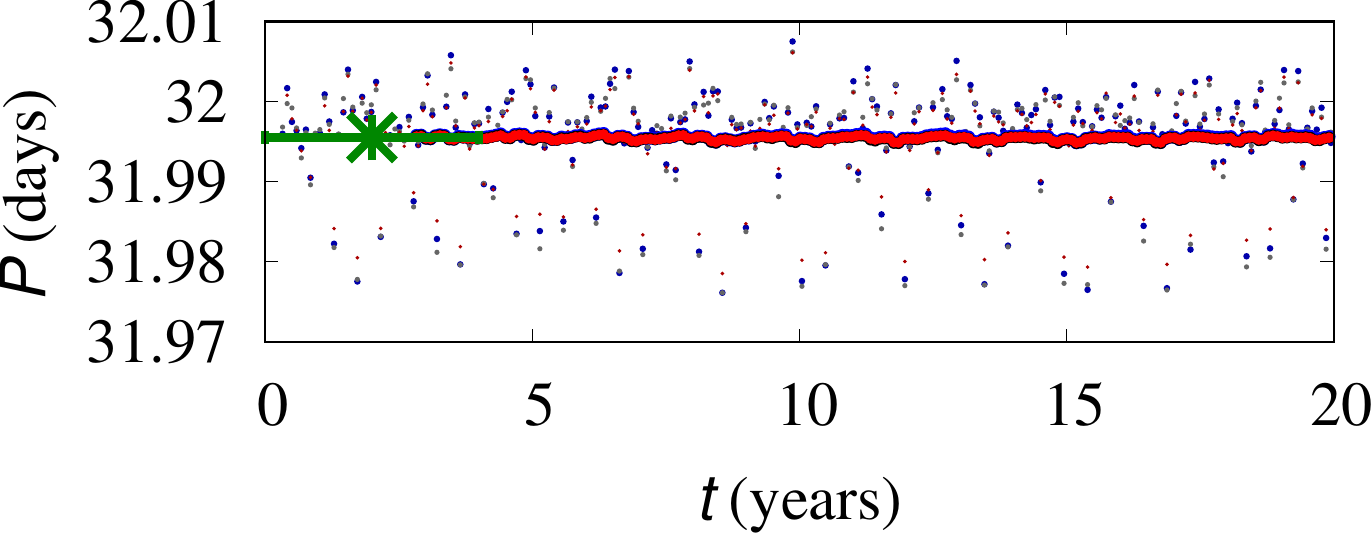}
   \includegraphics [width =  2.2in, angle =0]{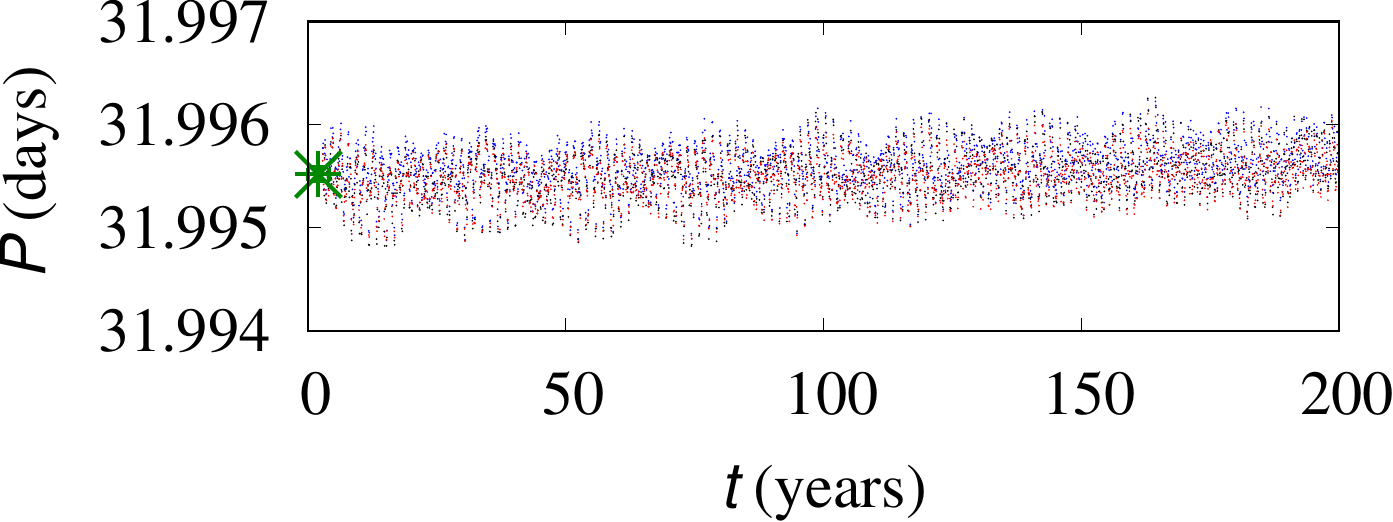}
      \includegraphics [width =  2.2in, angle =0]{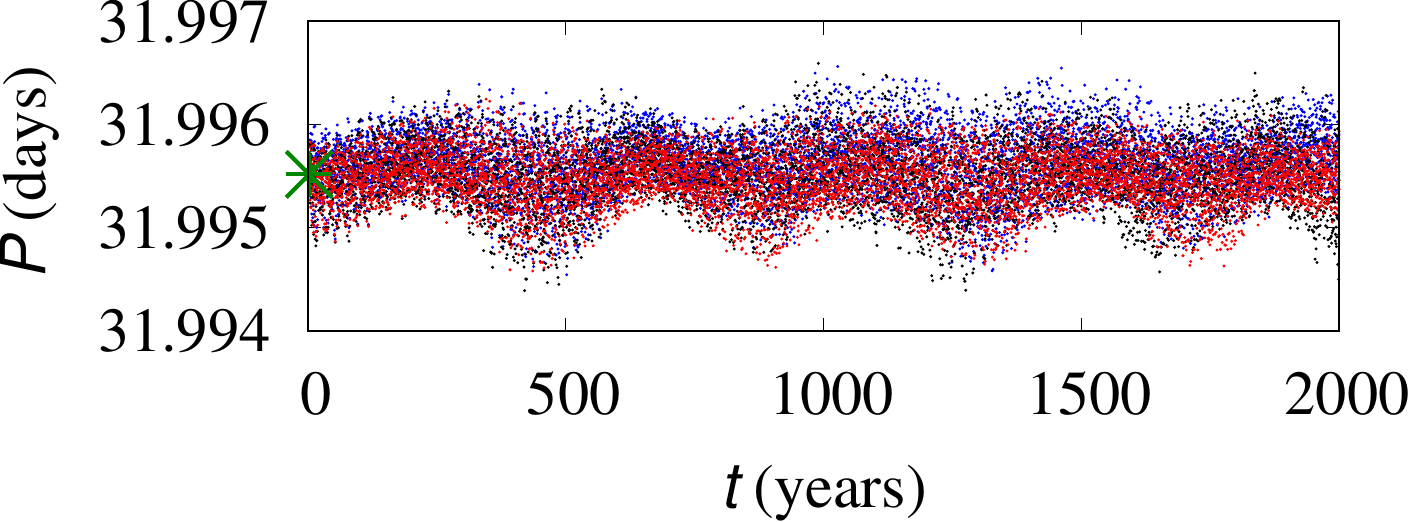}
\newline
  \includegraphics [width = 2.2in, angle =0]{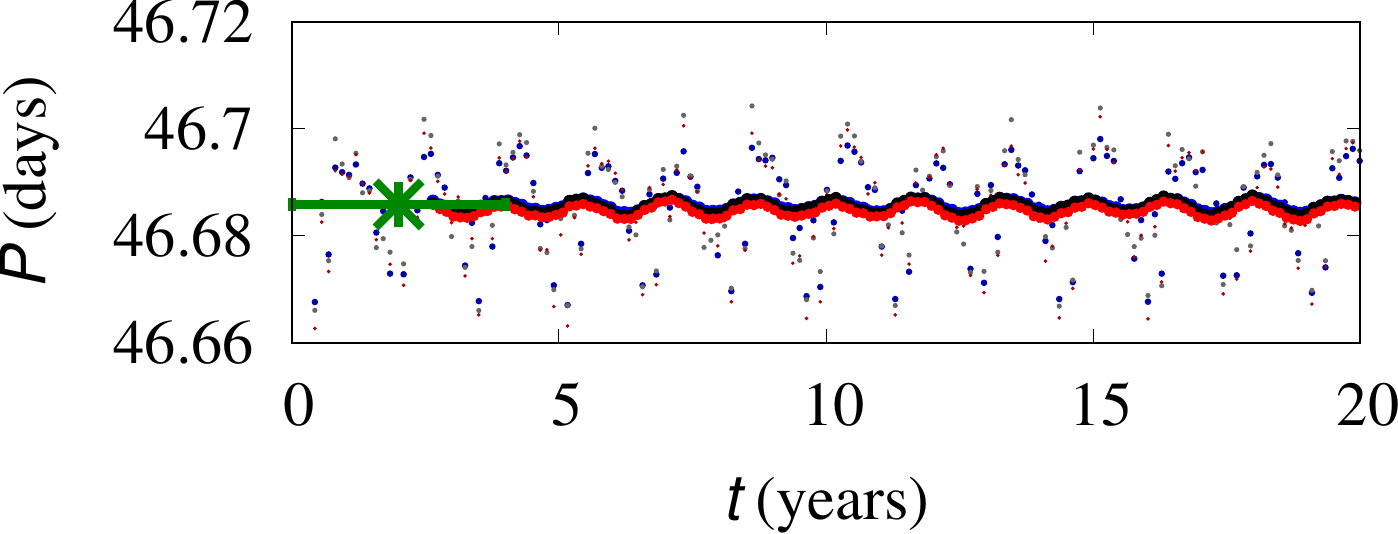}
    \includegraphics [width = 2.2in, angle =0]{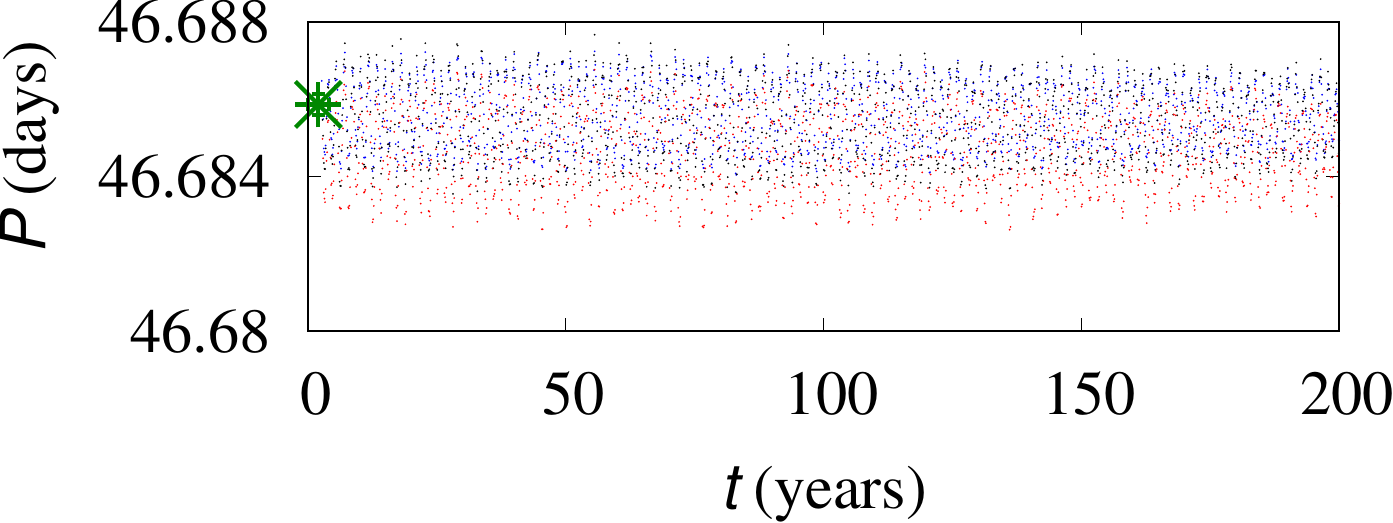}
        \includegraphics [width = 2.2in, angle =0]{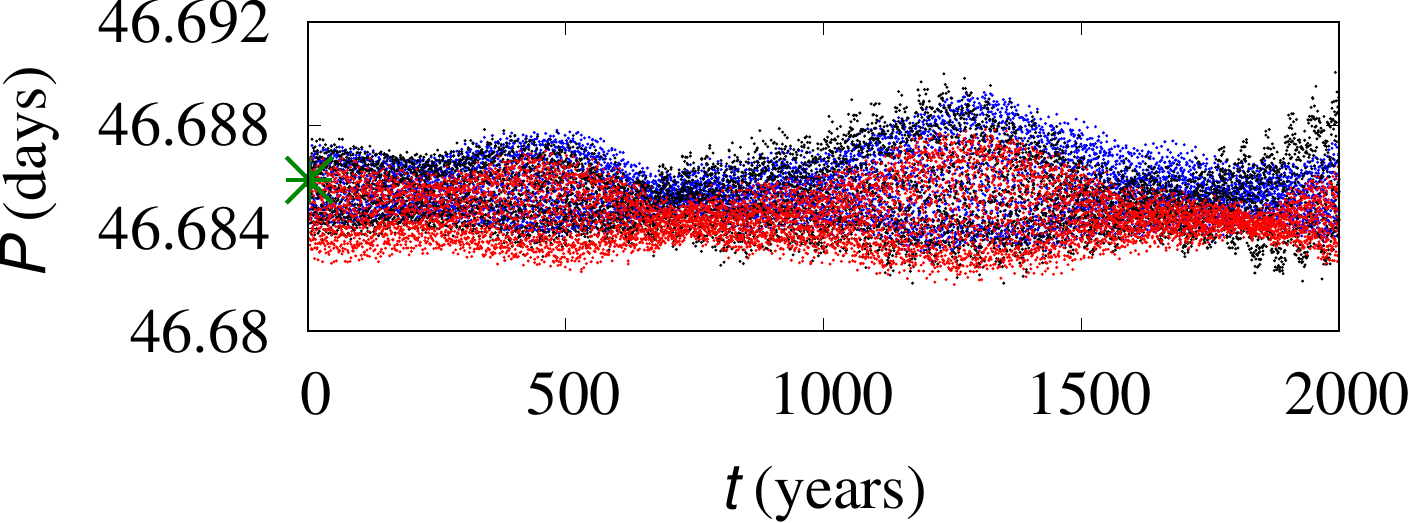}
      \newline
 \includegraphics [width = 2.2in, angle =0]{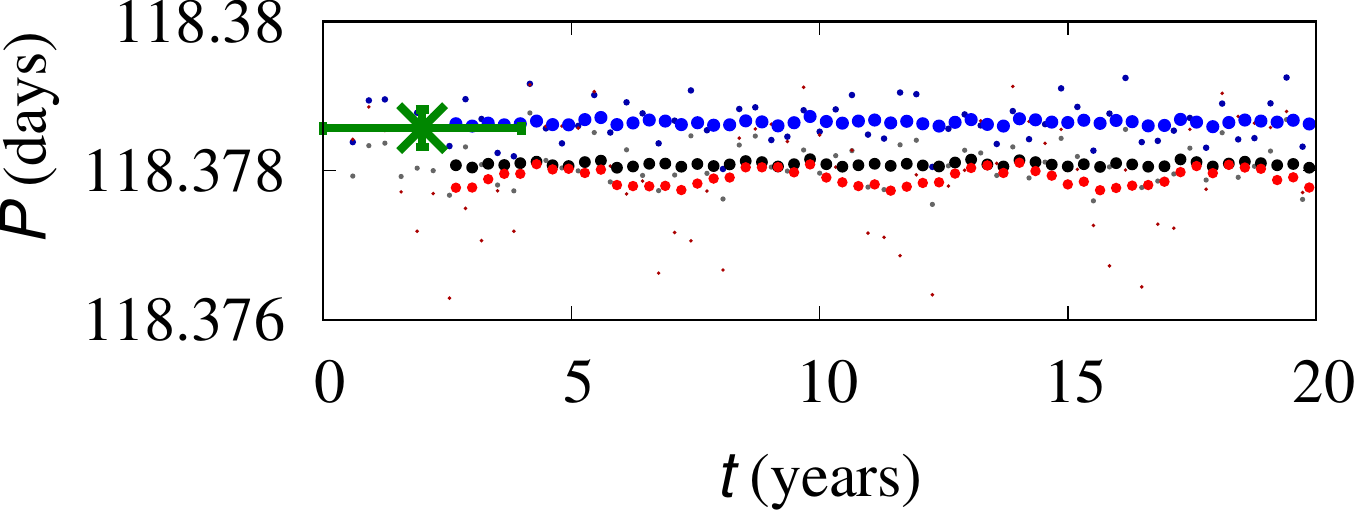}
   \includegraphics [width = 2.2in, angle =0]{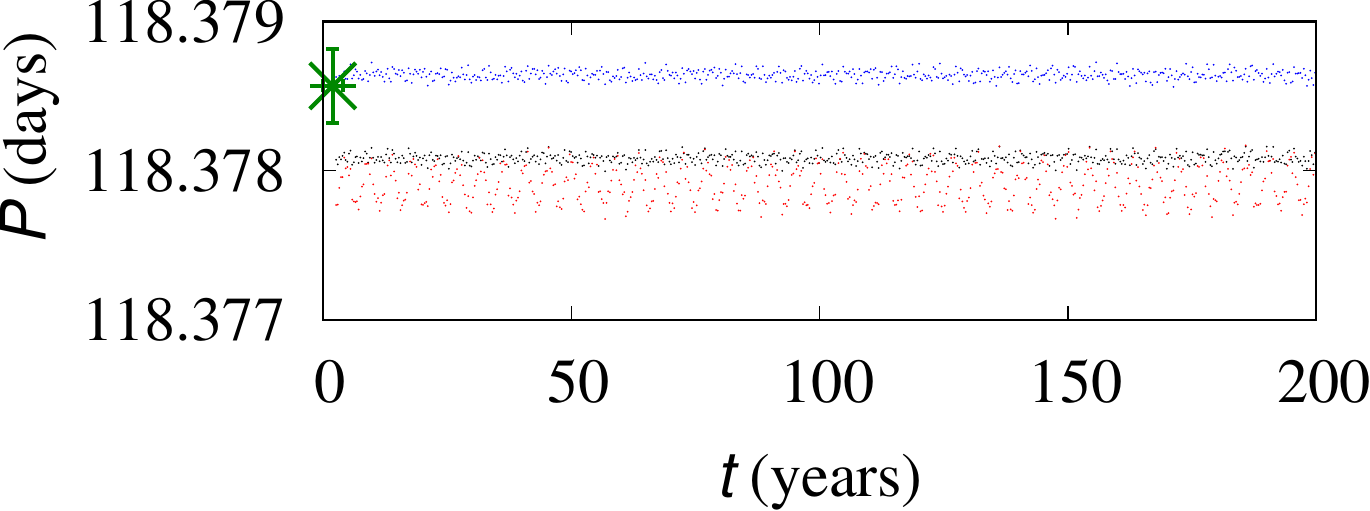}  
      \includegraphics [width = 2.2in, angle =0]{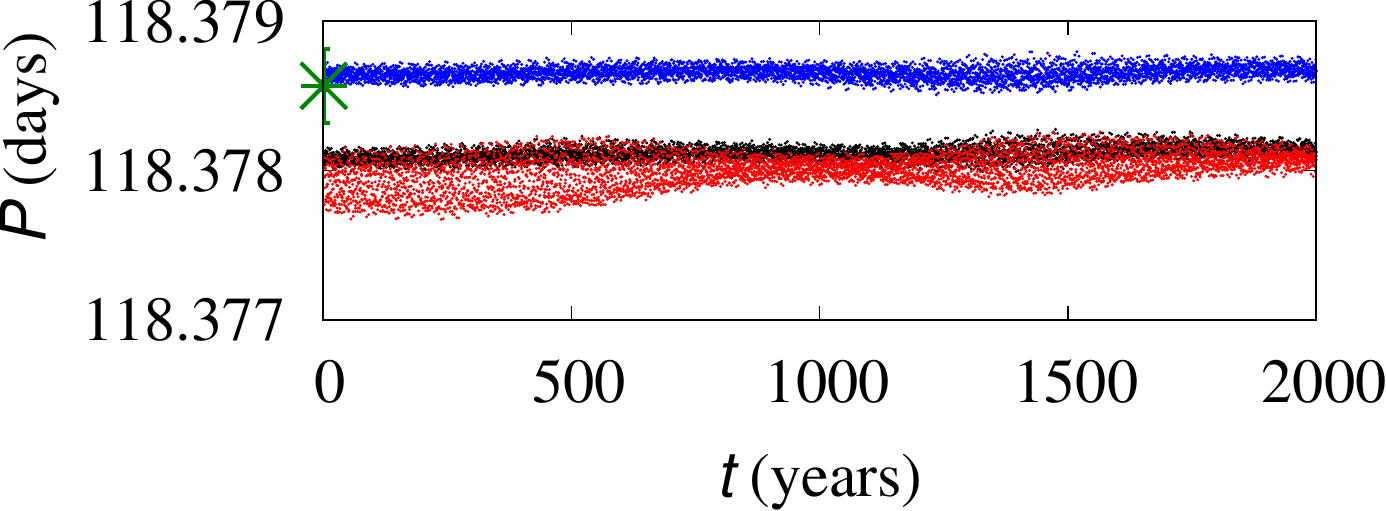}    
 \caption{Transit-to-transit and 4-year average periods for each of the planets known to orbit Kepler-11, with panels ordered from top to bottom by increasing orbital period. The small dots in the panels on the left show transit-to-transit orbital periods (the length of the time interval between the midpoint of one transit and the midpoint of the subsequent transit) from three samples of the 101 solutions used to compute the period estimates and uncertainties listed in Table \ref{tab:periods}. Specifically, the 101 solutions are ordered by increasing mean long-term ($10^6$~days) orbital period of KOI-157.01 = Kepler-11~c, and blue represents the 17$^{\rm th}$ sample (one standard deviation below the median), black the 51$^{\rm st}$ sample (the median) and red the solution that is 85$^{\rm th}$ (one standard deviation above the median) on this list. The more brightly colored points with a larger symbol size in the left panels and all points in the middle and right panels represent the running average of 4-year segments centered on the given time.  Time is measured from the beginning of \ik science operations. The green $\times$'s near the left of each panel represent our fits to the \ik transit times assuming constant period (Table \ref{tab:planetcatalog}). Blue points are plotted first, then black and finally red points on top. {The upper five panels in the right column have been thinned to show only every ninth, seventh, fourth, third and second transit-to-transit interval to limit the size of the manuscript file. }\label{fig:Kep11}  } 
\end{figure}

Figure \ref{fig:Kep11} shows orbital period evolution of all six planets as calculated for three of the 101 simulations, selected by rank according to the mean period of Kepler-11 c (KOI-157.01) over the $10^6$ day interval as discussed above.   Note that four-year running averages of the periods of the dynamically-active five inner planets vary substantially more than the uncertainties in the measured mean periods during the \ik epoch, especially for the two shortest-period planets.

\subsubsection{Kepler-36 = KOI-277}

Kepler-36 has two planets, each more than four times as massive as Earth, on 
orbits moderately close to the 7:6 mean motion resonance.  The planets are not in a low-order mean-motion resonance, but their proximal orbits lead to strong dynamical interactions.  The outer planet, which is a bit less than twice as massive as the inner one, nonetheless has 15 times the volume.  
Some planetary parameters allowed by short-term fits to the \ik data in the system discovery paper by  \cite{Carter:2012} were subsequently eliminated by imposing the requirement for long-term stability  \citep{Deck:2012}. 

We performed a photodynamical analysis of the data using the PhotoDynamical Multiplanet Model (PhoDyMM, Ragozzine et al., in prep.) similar to other analyses in the past \citep[e.g.,][]{MacDonald:2021}. Each of the 101 samples from the posterior distribution was then integrated using REBOUND's WHFast integrator \citep{Rein:2015} with a timestep of 2\% of the inner planet's orbit for $10^6$ inner planet orbits. Inspection of the final orbital states for all 101 systems showed that they were still on stable orbits very similar to those of the current system. While our integrations are substantially shorter than those of \cite{Carter:2012}, they are still much longer than the typical Lyapunov time of 10 years \citep{Deck:2012}.  Therefore, it seems likely that the change in stability properties results from having a more accurate set of initial conditions that were derived from the full \ik lightcurve, which included an additional year of Short Cadence data. 

\begin{figure}[!hbt]
\includegraphics [width = 2.4 in, angle =0]{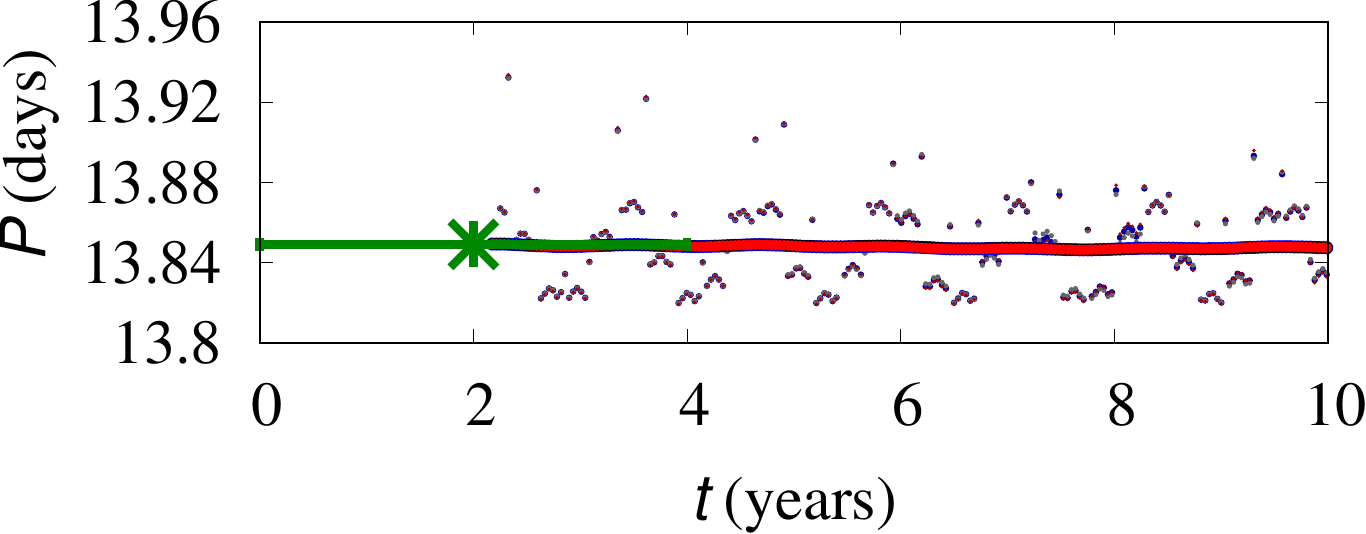}
  \includegraphics [width = 2.4 in, angle =0]{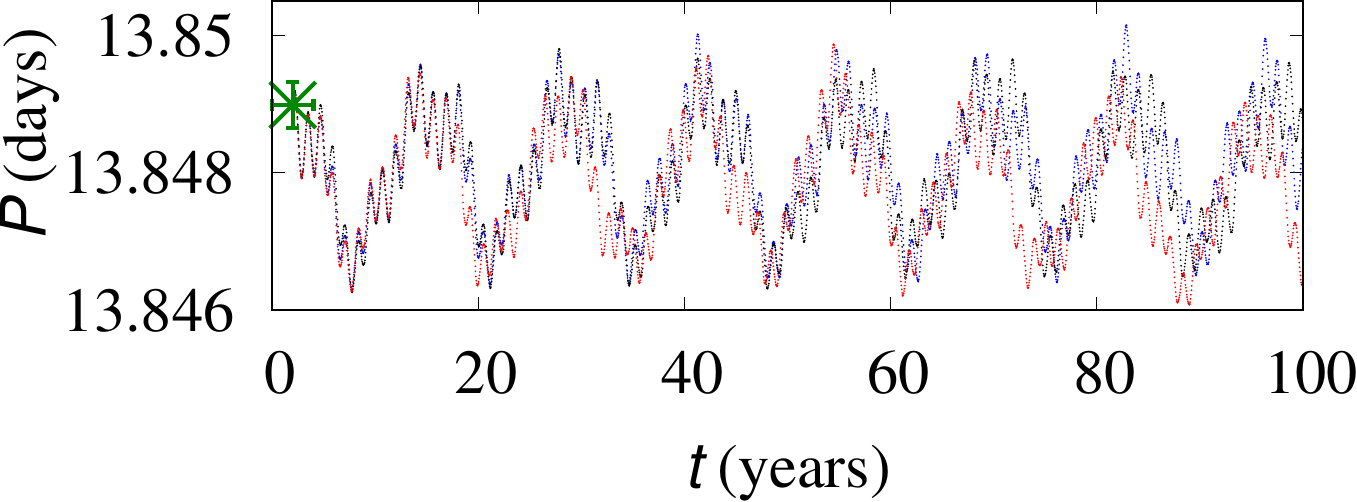}
  \newline
 \includegraphics [width = 2.4 in, angle =0]{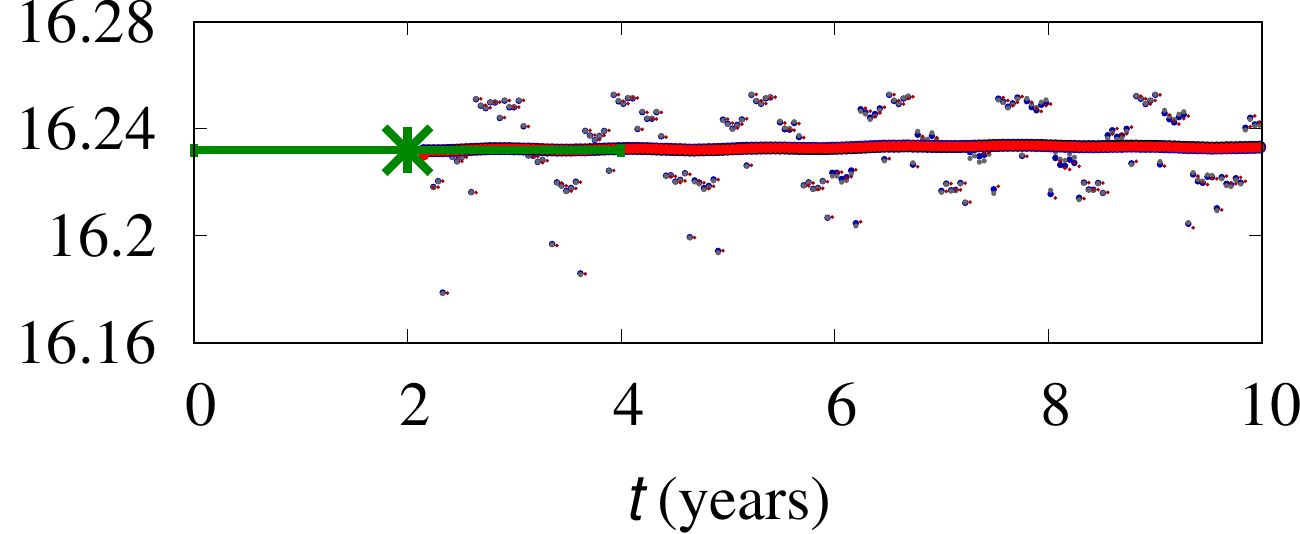}
   \includegraphics [width = 2.4 in, angle =0]{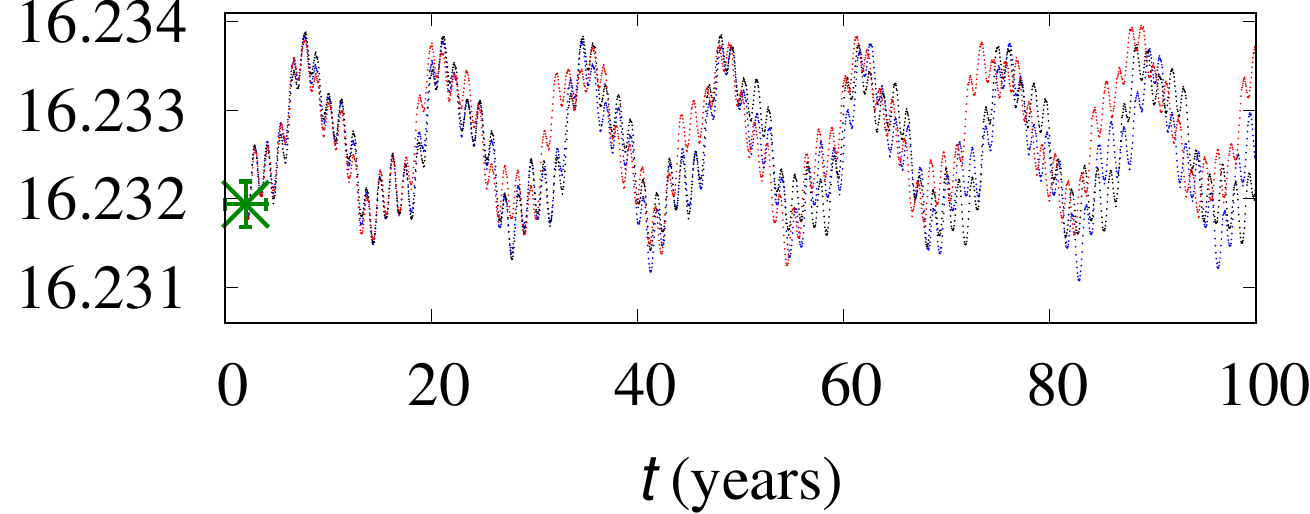}
 \caption{Transit-to-transit and 4-year average periods for Kepler-36 b and Kepler-36 c. The dots in panels on the left show transit-to-transit orbital periods from three samples following photodynamical fits to the lightcurve. The solid curves in the panels on the left and the curves on the right show the average of 4-year segments centered on the given time. Black represents the sample with the median long-term (1000 years) average period of KOI-277.01 (= Kepler-36~c), blue the sample with the 17$^{\rm th}$ lowest average period and red the one that is   85$^{\rm th}$ on this list. Time is measured from the beginning of \ik science operations. The green $\times$ near the left side of all panels represent our fits to the measured transit times assuming constant period (i.e., the orbital period and uncertainty of that planet listed in Table \ref{tab:planetcatalog}).
 \label{fig:Kep36} } 
\end{figure}

In analogy with Fig.~\ref{fig:Kep11}, Figure \ref{fig:Kep36} shows orbital period evolution for three of the 101 simulations, selected as described in the caption. Note that the orbital periods of both planets averaged over 4 year intervals vary by almost 10 times as much as the uncertainties in their orbital periods measured during the \ik epoch.
 
\subsubsection{Kepler-80 = KOI-500}
Kepler-80 harbors six known transiting planets. 
The innermost is a USP planet that is (at least on the time scales relevant here) dynamically detached from its siblings. The middle four planets, with periods of 3.1, 4.6, 7.1, and 9.5 days, were studied extensively by \cite{MacDonald:2016}. Two years later, a sixth transiting planet, with $P = 14.6$~days,
 was discovered by \cite{Shallue:2018}.
 
 All six planets were included in a full photodynamical analysis by \citet{MacDonald:2021}. This photodynamical analysis was performed using the PhotoDynamical Multiplanet Model (PhoDyMM, Ragozzine et al., in prep.), with all six masses allowed to float, but all the longitudes of ascending node were fixed at 0$^{\circ}$. The middle four planets form a resonance chain, with each neighboring pair having  period ratio $\sim 1 - 3$\% larger than either 3/2 or 4/3 and each neighboring threesome librating within a three-body resonance. The orbital period of outer planet suggests that it, too, is a member of the resonance chain. 
However, transits of the outer planet are much shallower, and only 14\% of
the fits to the \ik data by \citet{MacDonald:2021} find it to be
 librating within a resonance.
 
\begin{figure}[!hbt]
 \includegraphics [width = 2.1 in, angle =0]{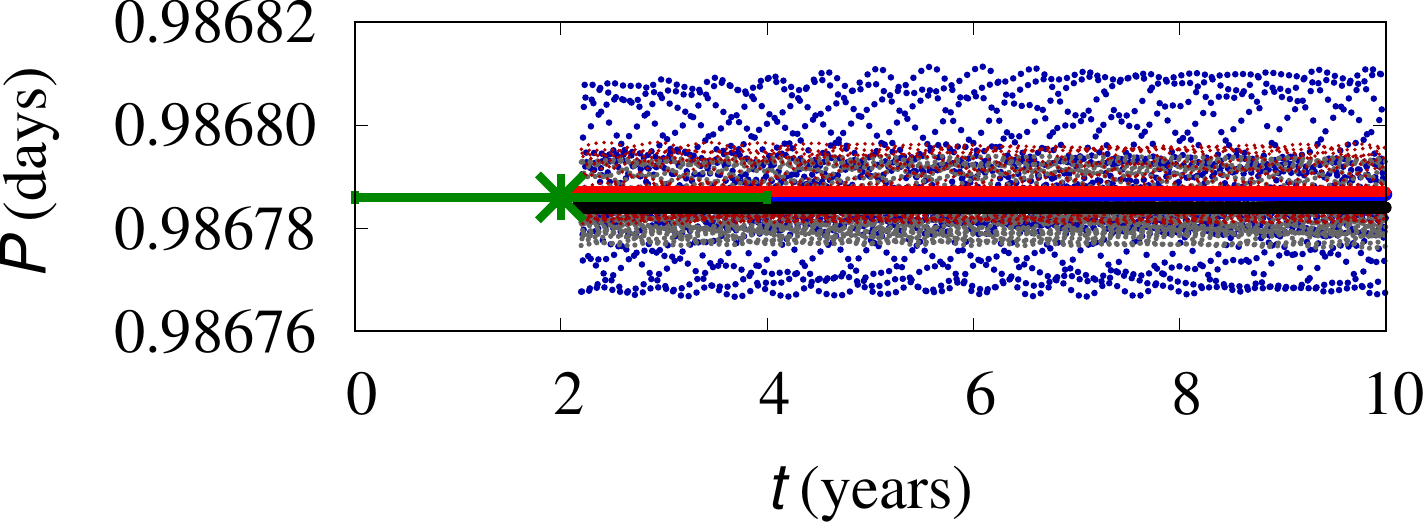}
  \includegraphics [width = 2.1 in, angle =0]{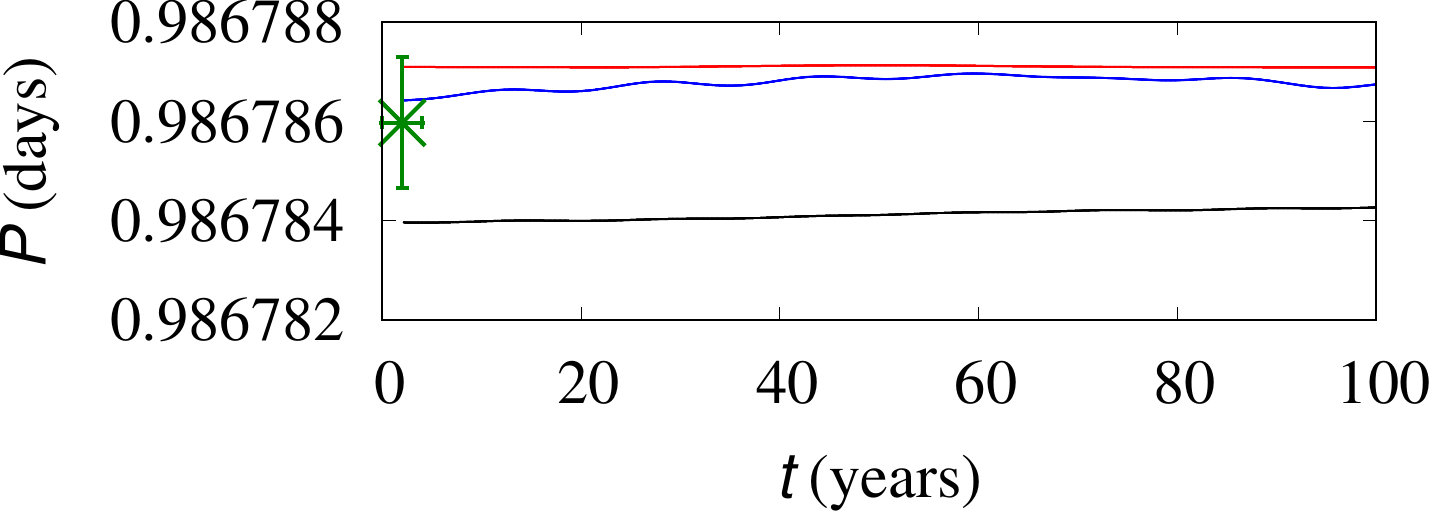}
  
   \includegraphics [width = 2.1 in, angle =0]{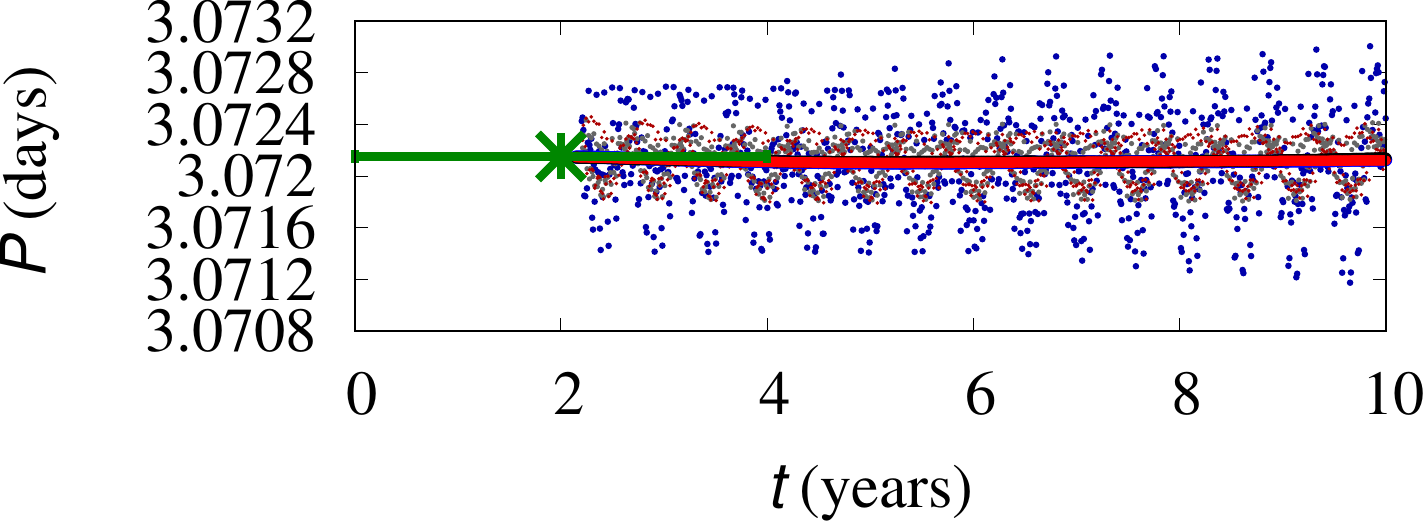}
  \includegraphics [width = 2.1 in, angle =0]{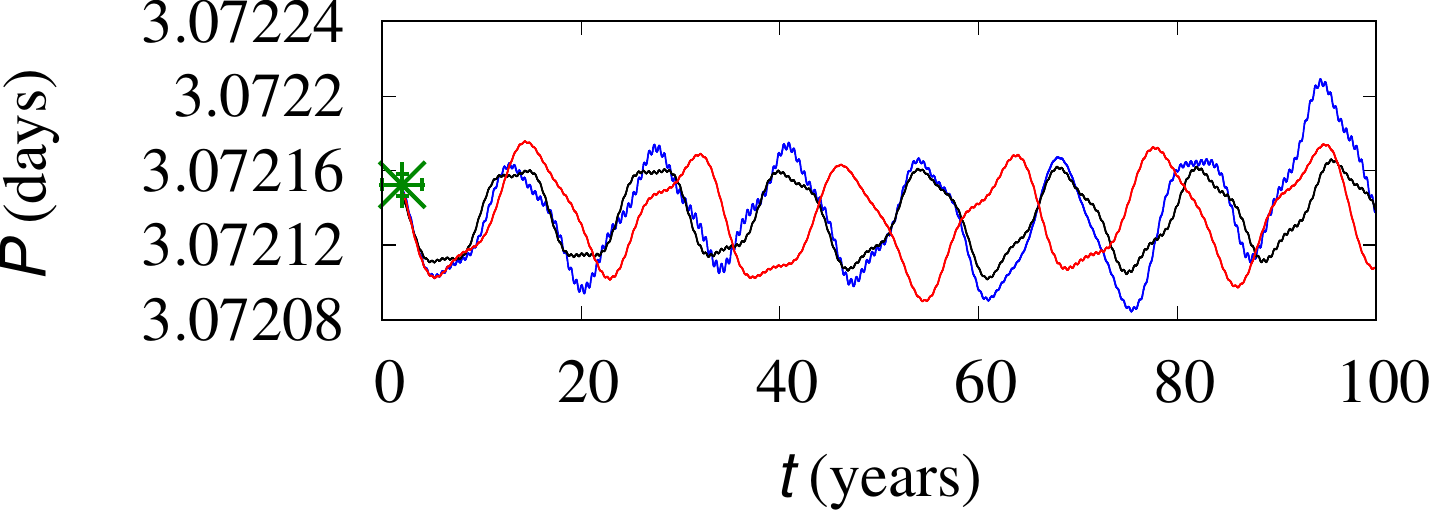}
 
   \includegraphics [width = 2.1 in, angle =0]{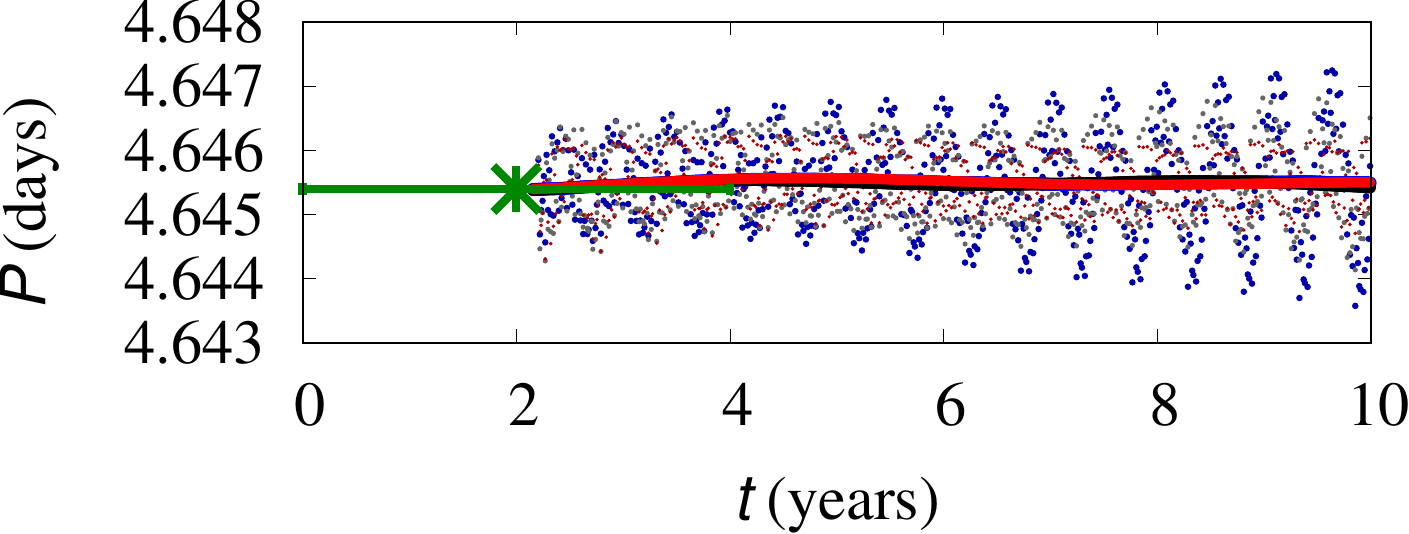}
  \includegraphics [width = 2.1 in, angle =0]{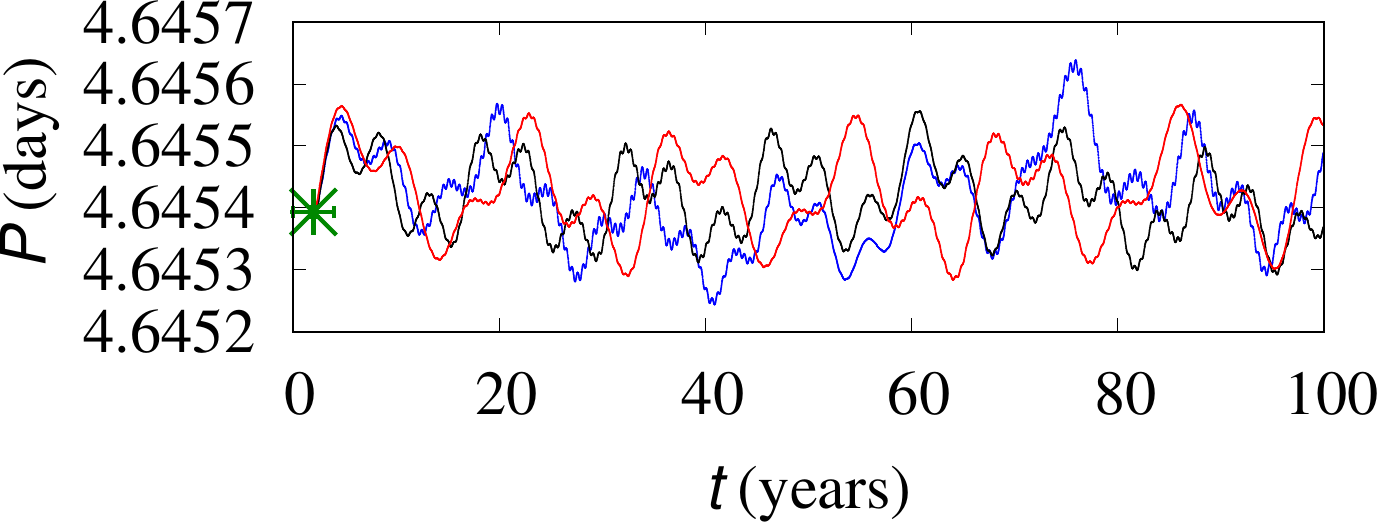}
 
   \includegraphics [width = 2.1 in, angle =0]{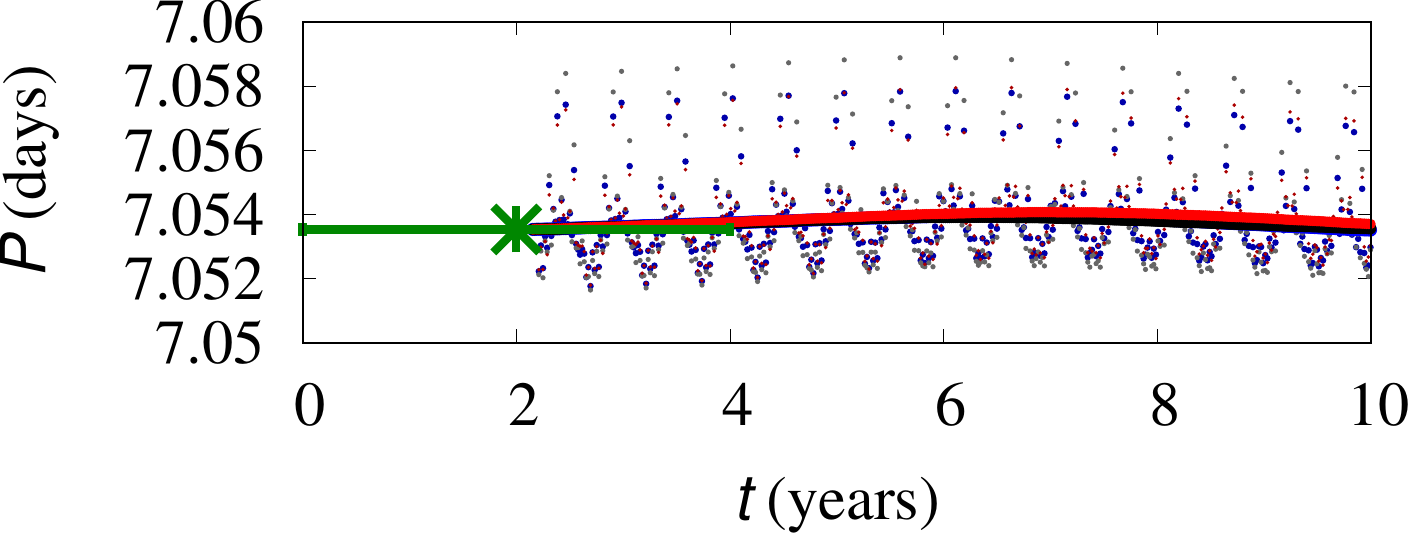}
  \includegraphics [width = 2.1 in, angle =0]{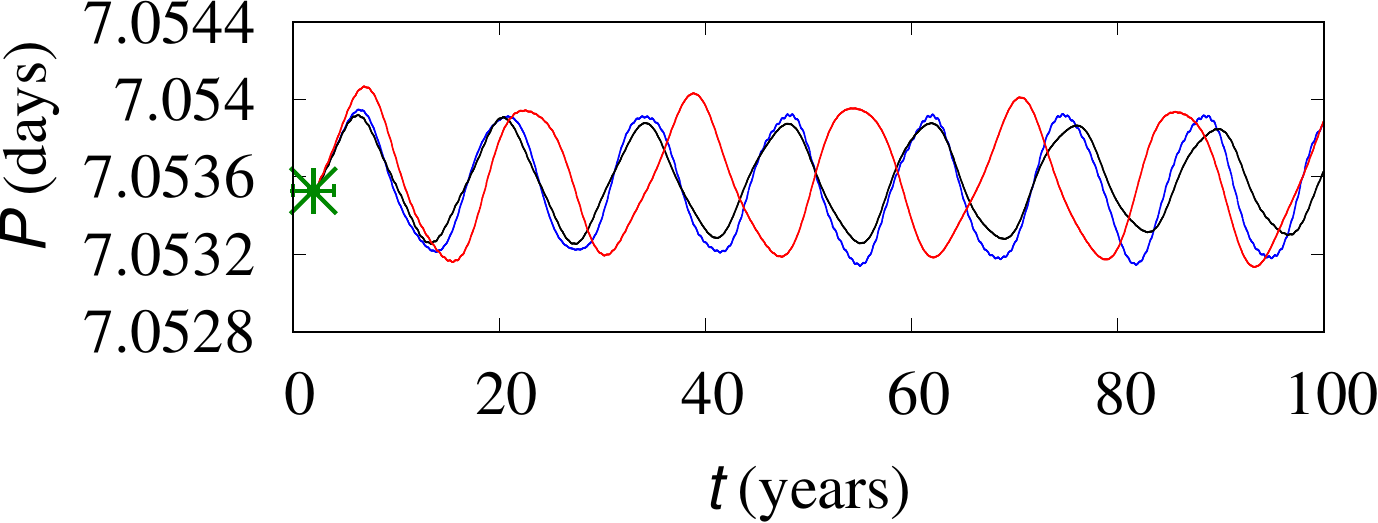}
 
  \includegraphics [width = 2.1 in, angle =0]{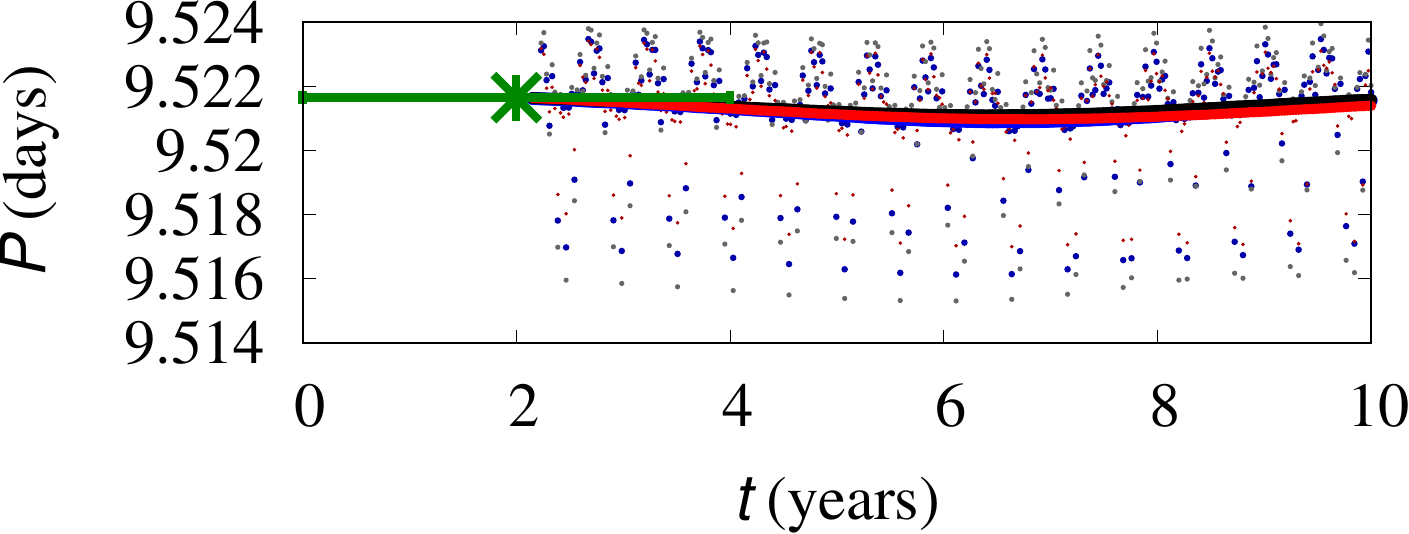}
   \includegraphics [width = 2.1 in, angle =0]{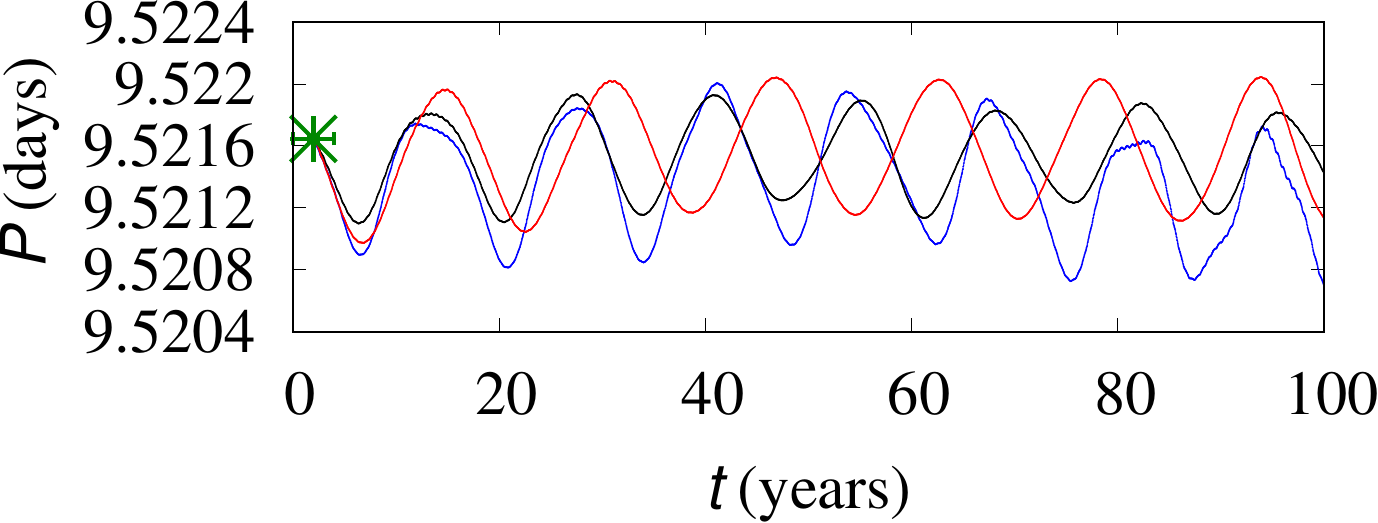}

   \includegraphics [width = 2.1 in, angle =0]{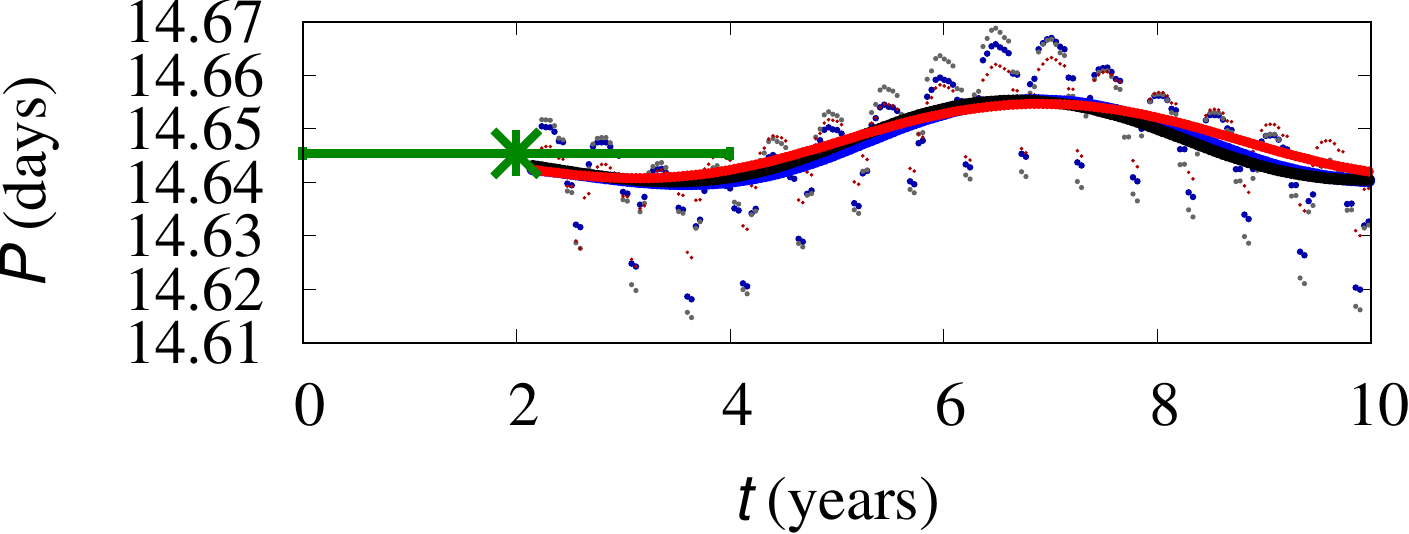}
  \includegraphics [width = 2.1 in, angle =0]{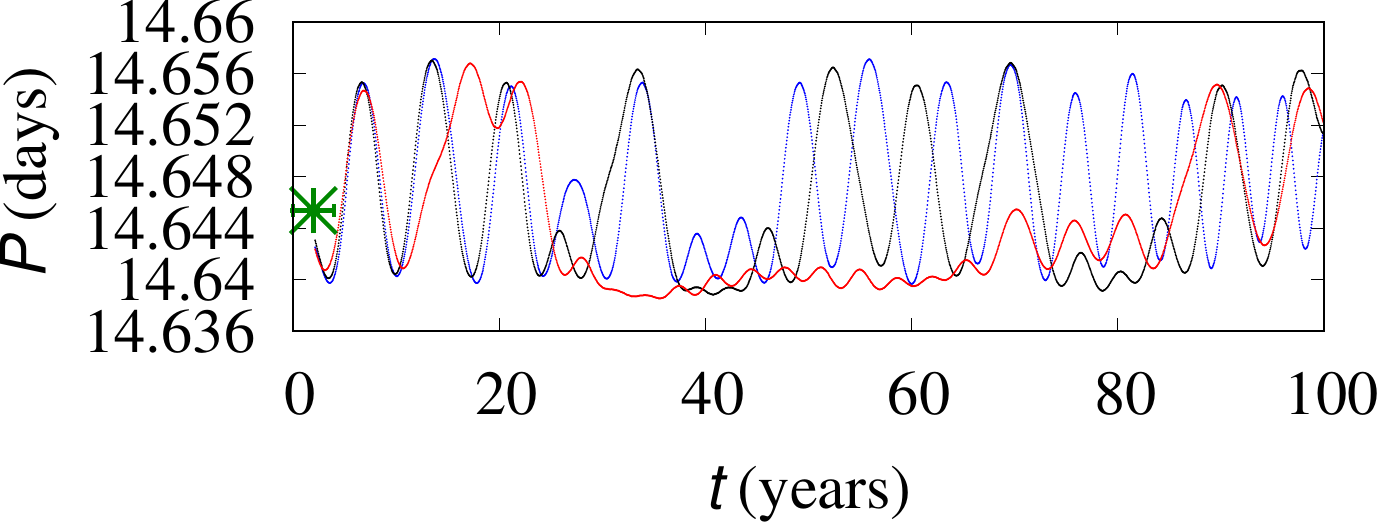}
 \caption{Transit-to-transit and 4-year average periods for each of the planets known to orbit Kepler-80, ordered by increasing orbital period. The dots in panels on the left show transit-to-transit orbital periods from three samples of the solution posteriors, as in Fig.~\ref{fig:Kep36}. Time is measured from the beginning of \ik science operations. The green $\times$ at the midpoint of \ik observations represent our fits to the TTs assuming constant period (Table \ref{tab:planetcatalog}).  {The upper right panel has been thinned to show every fourth mean transit-to-transit interval.}  \label{fig:Kep80}  } 
\end{figure}

As for Kepler-36,
101 samples were taken from the posterior distribution from PhoDyMM. Each of these mock systems was integrated for 100 years (nearly $4\times10^4$ orbits of the innermost planet). The $9^{th}$ through $14^{th}$ numerical rows of
Table \ref{tab:periods} list the mean and dispersion of the periods of each planet in this sample at epoch $T_{\rm epoch}$ = 800.0 days and averaged over 100 years. Figure \ref{fig:Kep80} shows period variations of each of the planets derived from our integrations that used three samples from the posteriors calculated by Ragozzine et al.~(in prep.). The four-year averaged periods of the five planets in the resonance chain have fractional variations of $\sim 10^{-5} - 10^{-4}$.


\begin{figure}[!hbt]

\end{figure}

\subsubsection{Kepler-223 = KOI-730}

Kepler-223 is a system of four planets in mean-motion resonance, where the three-body angles are consistent with libration, and the TTVs have a few-hour amplitude with a timescale of a few years \citep{Mills:2016}.  The planets have higher eccentricities than typical for closely-spaced planetary systems, and the period ratios of the planets are extremely  close to ratios of small integers, with departures of order 0.1\% or less \citep{Lissauer:2011b,Mills:2016}. 

To understand how the observed transit timing variations manifest on a timescale beyond the \ik time series, we draw samples from the posterior of the photodynamic fits to the data.  \cite{Mills:2016} produced a sample of solutions that assure the two Laplace angles,  one relating the inner three planets and the other relating the outer three planets, librate with small amplitude over a 100 year span, and called it the C3 posterior.  Values of their osculating orbital periods for Kepler-223's planets are given at $T_{\rm epoch}$ = 800.0 days.  

\cite{Mills:2016} ran $N$-body simulations of 300 systems from the C3 posterior for $10^7$~years, with outputs every $10^4$~years, and found all of them to be stable.  We inspected the osculating semimajor axes over that time.  In many cases, the oscillations in semimajor axis on 100 year timescale are small, but then grow by a factor of several or even an order of magnitude, and simultaneously lose the libration of the three-body angles, indicating that these solutions were not in secular steady-state.

We found that only 12 systems out of the 300 in the C3 posterior seemed to keep the same semimajor axis behavior from the first 100 years for all $10^7$ years. 
If the other trajectories were taken as models of the observed system, then the system must have been observed at a special time, which violates the Copernican Principle and thus is highly unlikely.  
Therefore, we consider the 12 continually-librating solutions likely represent better models of the system, and we restrict our attention to them for our analysis of long-term orbital period variations of this system.  

\begin{figure}[!hbt]

 \includegraphics [width = 2.38 in, angle =0]{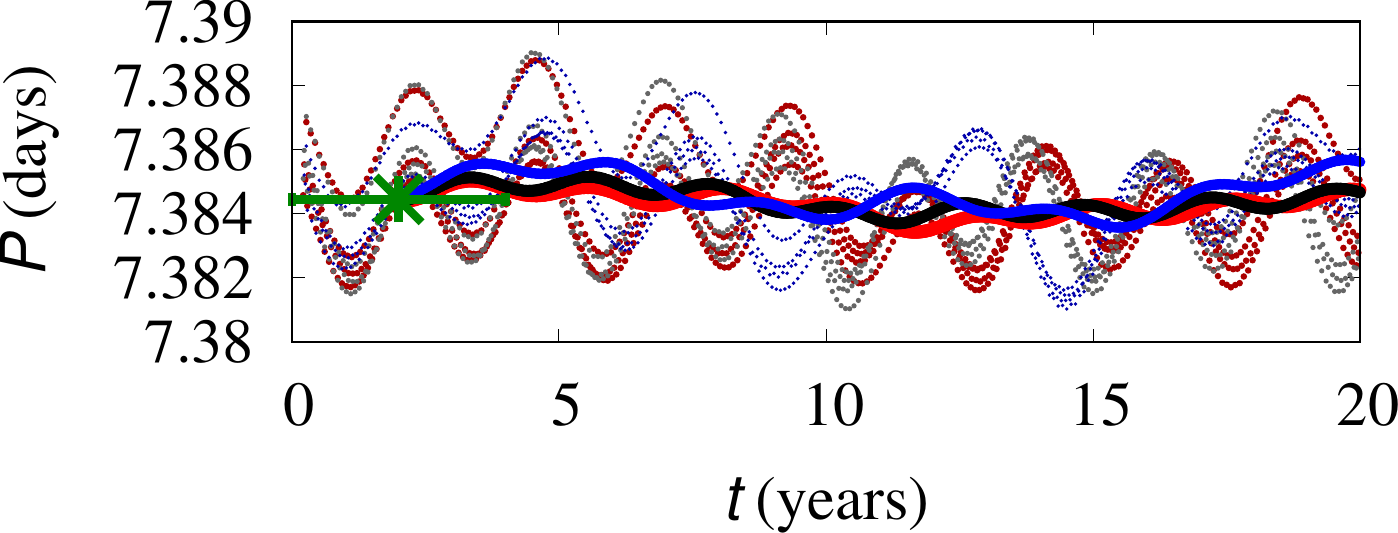}
  \includegraphics [width = 2.38 in, angle =0]{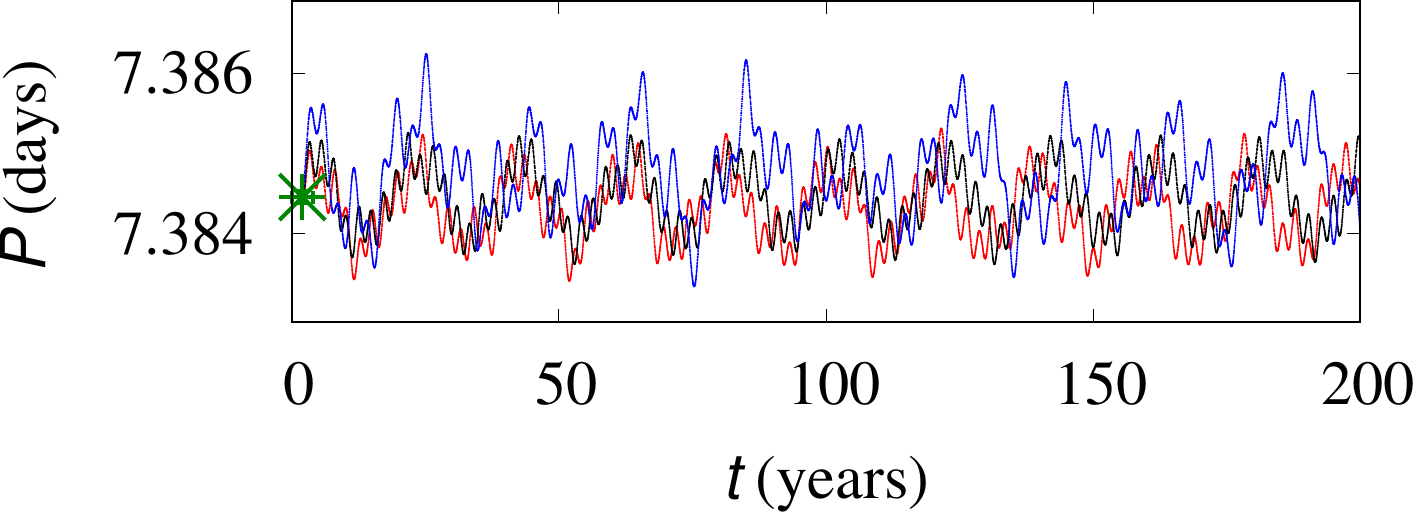}
\newline

 \includegraphics [width = 2.38 in, angle =0]{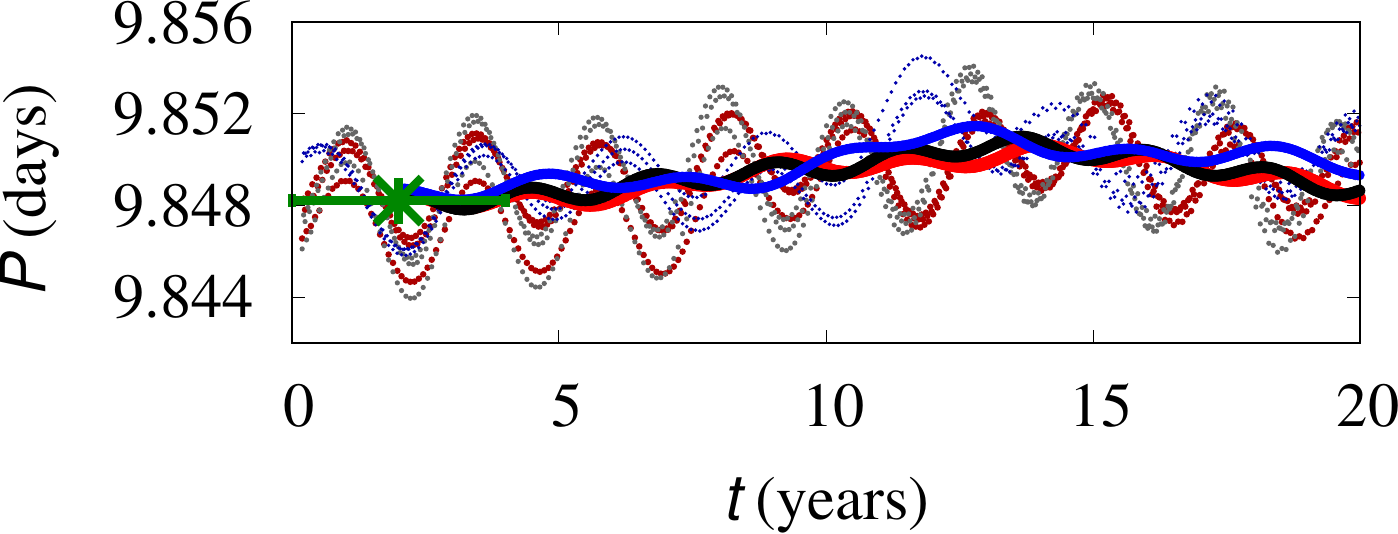}
   \includegraphics [width = 2.38 in, angle =0]{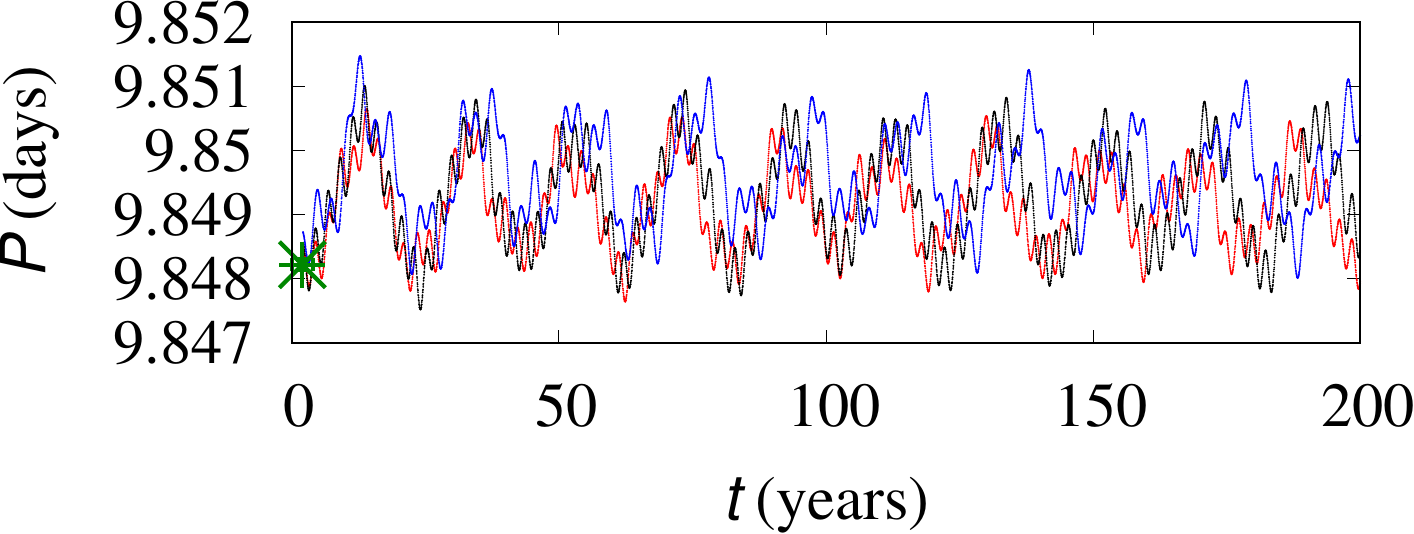}
 \newline

  \includegraphics [width = 2.38 in, angle =0]{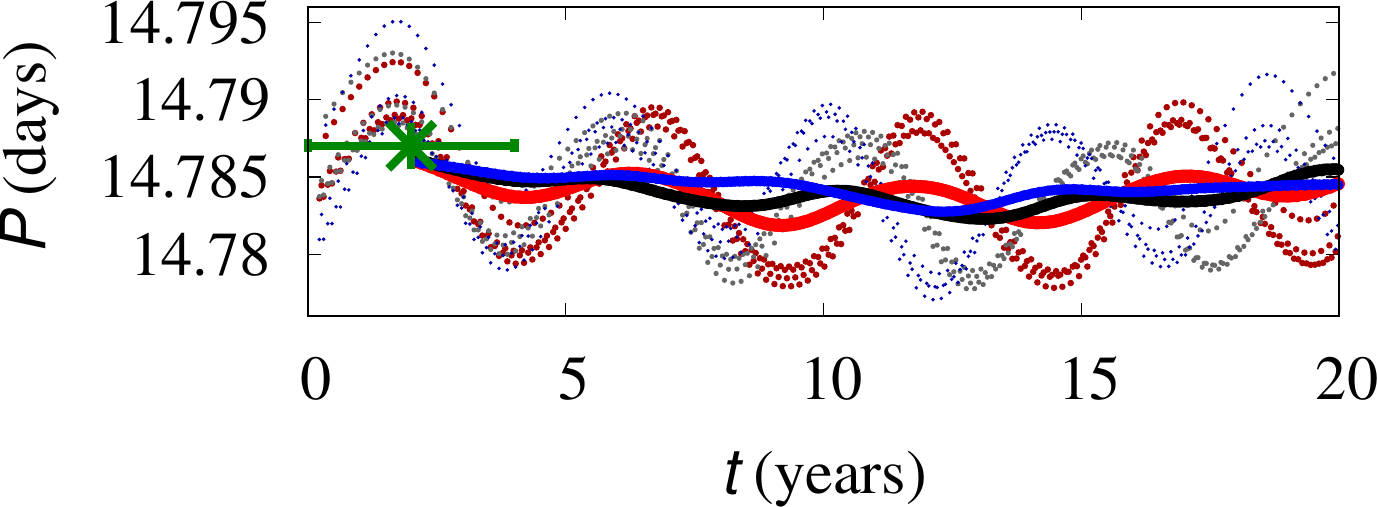}
    \includegraphics [width = 2.38 in, angle =0]{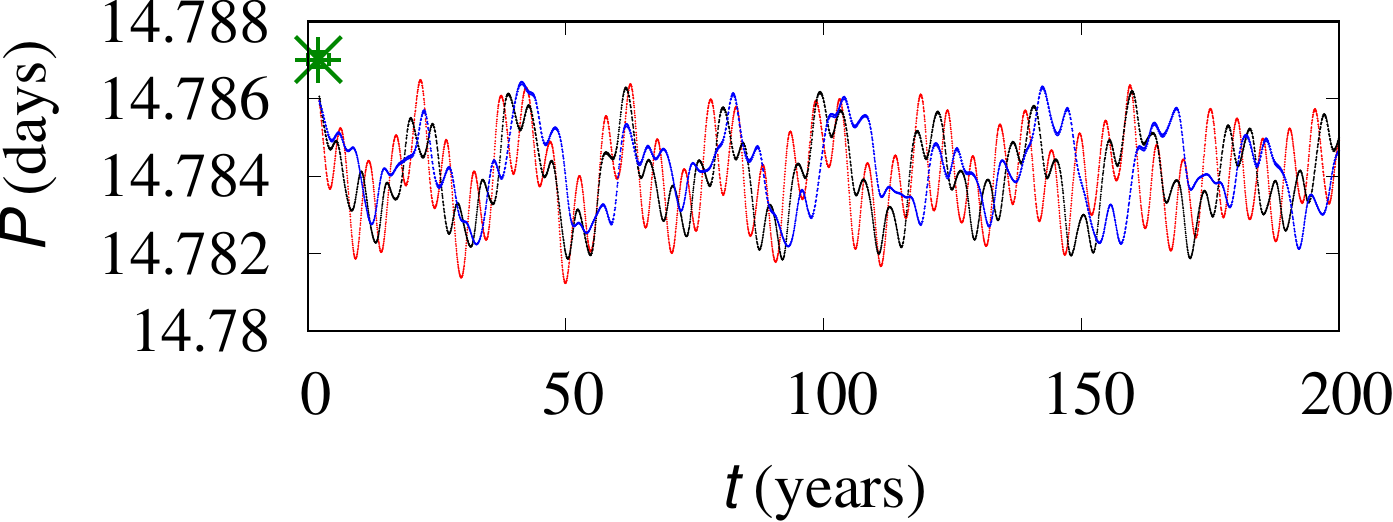}
   \newline

\includegraphics [width = 2.38 in, angle =0]{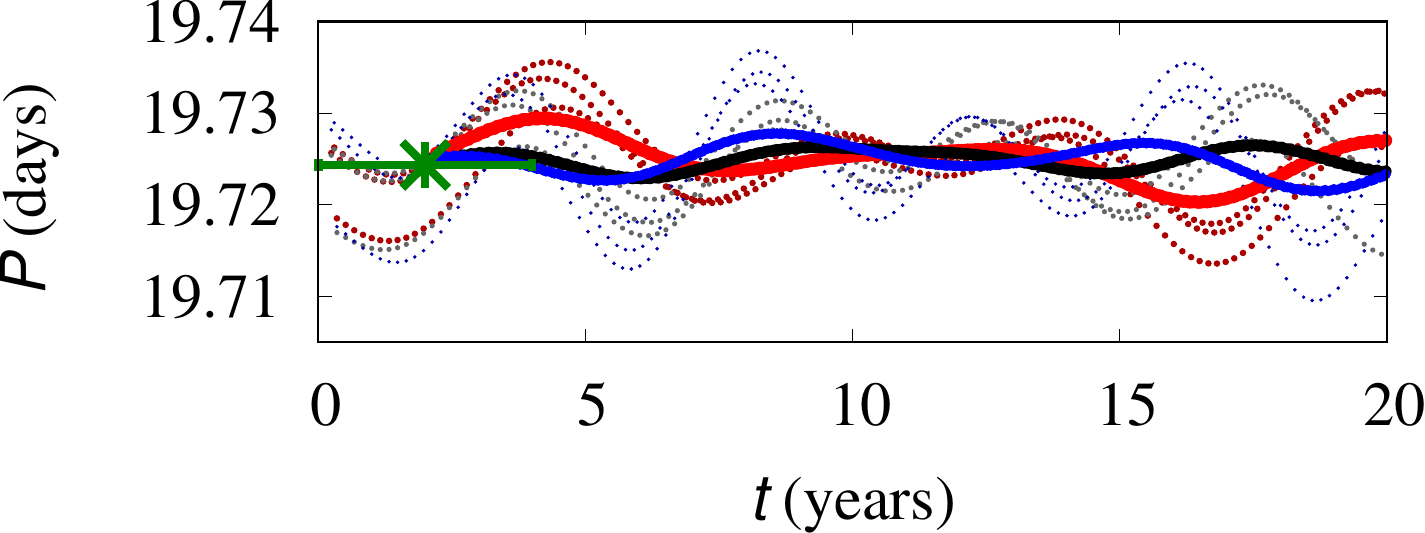}
   \includegraphics [width = 2.38 in, angle =0]{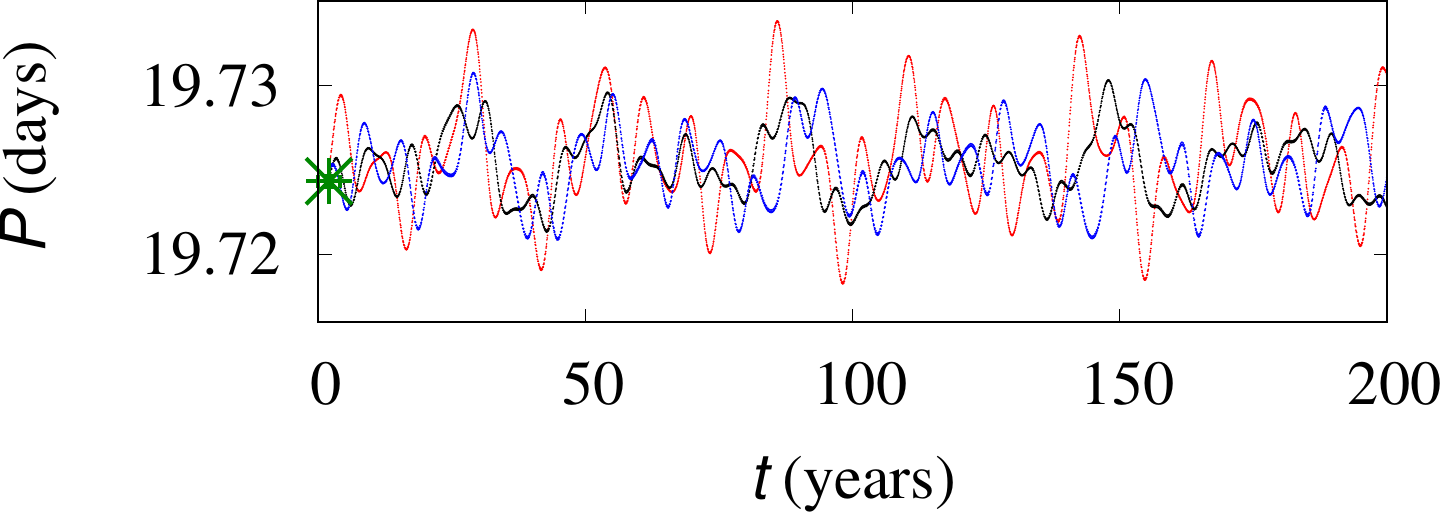}
\caption{Transit-to-transit and 4-year average periods for each of the planets known to orbit Kepler-223, ordered by increasing orbital period. The dots in panels on the left show transit-to-transit orbital periods from three samples of the C3 posterior from the photodynamical models of \cite{Mills:2016} for which the three-body resonant arguments of the inner and outer threesomes of planets remain in libration for $10^7$ years. The solid curves show the average of 4-year segments centered on the given time. Black represents the sample with the median long-term (1000 years) average period of KOI-730.01 among the 12 samples that remained in libration (since this number is even, we broke the degeneracy by choosing the one whose period was closer to the average of the ensemble); the sample with the second shortest average period of this planet is shown in blue, and red represents the one with the second longest average period.  Time is measured from the beginning of \ik science operations. The green $\times$ at the beginning represent our fits to the measured TTs assuming constant period (Table \ref{tab:planetcatalog}). \label{fig:Kep223}  } 
\end{figure}

The long-term (1000 year) mean periods of each of Kepler-223's four planets and the standard deviations thereof are listed in Table \ref{tab:periods}.  Figure \ref{fig:Kep223} shows the period variations of three of these possible realizations, with black representing the system with median mean period of KOI-730.01 (we broke the degeneracy caused by the even number of systems in the sample by selecting the one with period closer to the mean of the 12 samples), and blue and red showing the sample systems for which KOI-730.01 has the second shortest and second longest mean values, respectively.

On shorter timescales, due to orbital fluctuations, the period ratios cross back and forth across the nominal period ratios of 4:3 and 3:2 -- this is true both when computing the ratios with either osculating periods \citep[Extended Data Fig. 5]{Mills:2016} or modelled transit periods. The four-year average periods vary by a few parts in $10^{-4}$, and the period ratio for the inner pair of planets drops slightly below 4/3 for some portions of the simulated interval in each of the 12 samples. In contrast, the ratio of periods of the middle pair of planets stay slightly above these small integer ratios.

\subsubsection{Kepler-29 = KOI-738}
We took 101 samples from the TTV posteriors of Kepler-29 from \cite{Jontof:2021} and simulated transit times over 10$^6$ days. The system contains two sub-neptunes on proximate, dynamically-interacting orbits with orbital periods close to the 9:7 mean motion resonance \citep{Migaszewski:2017}. 


\begin{figure}[!hbt]
\includegraphics [width = 2.4 in, angle =0]{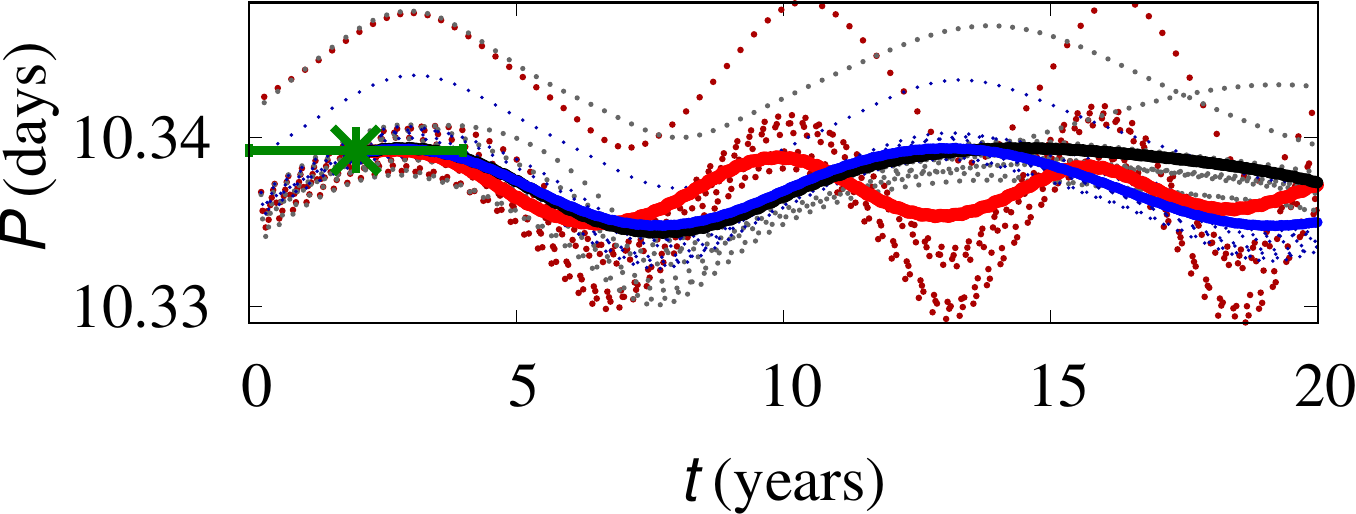}
  \includegraphics [width = 2.4 in, angle =0]{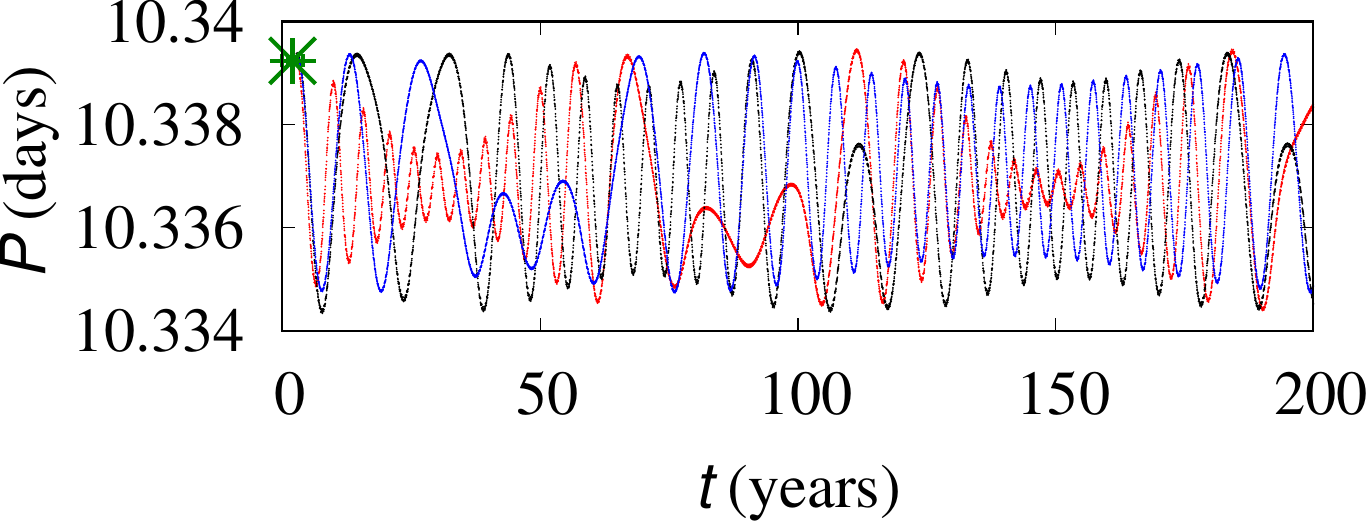}

  \includegraphics [width = 2.4 in, angle =0]{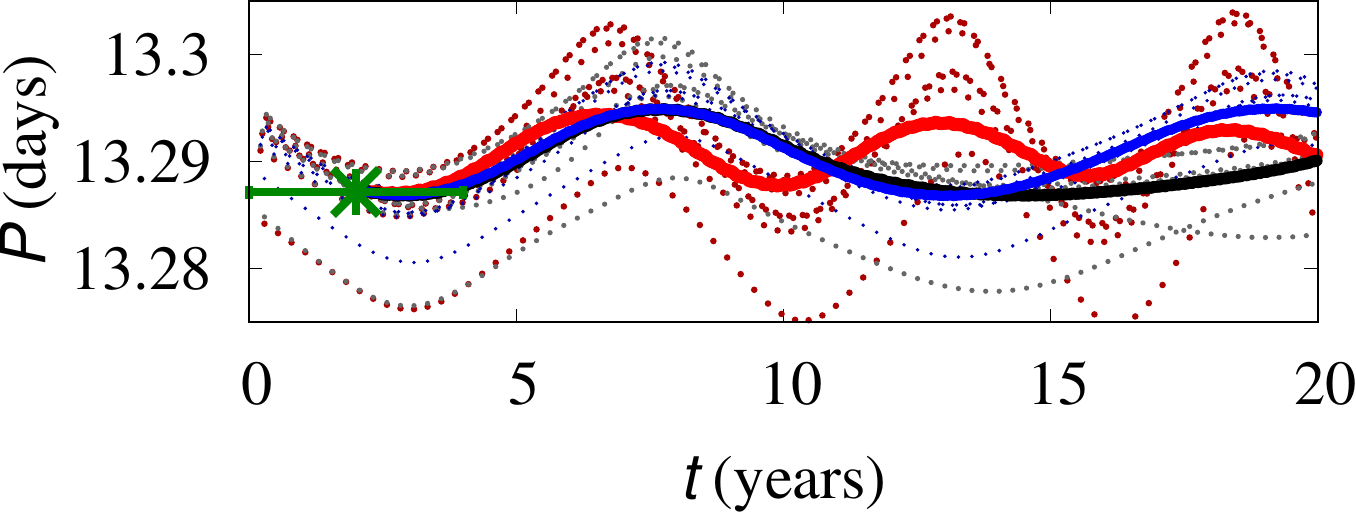}
   \includegraphics [width = 2.4 in, angle =0]{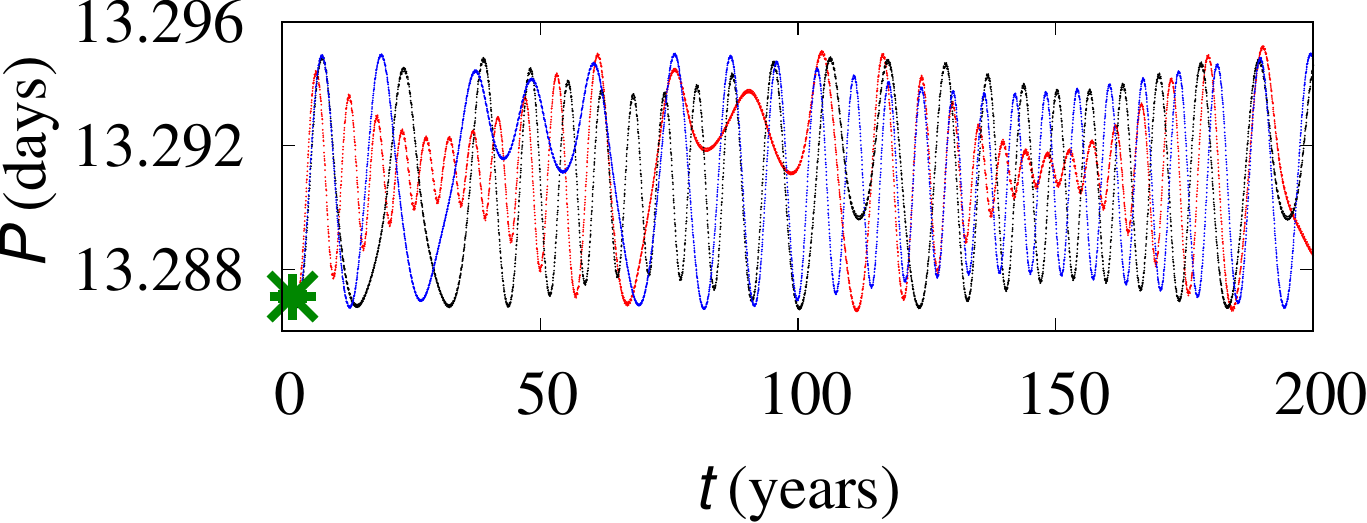}
 \caption{Transit-to-transit and 4-year average periods for both of the planets known to orbit Kepler-29, ordered by increasing orbital period. The dots in panels on the left show transit-to-transit orbital periods from three samples following \cite{Jontof:2021}'s dynamical fits to the long cadence transit times of \cite{Rowe:2015b}. The solid curves show the average of 4-year segments centered on the given time. Black represents the sample with the median long-term (1000 years) average period of KOI-738.01; blue the sample with the 17$^{\rm th}$ lowest average period and red the one that is   85$^{\rm th}$ on this list. Time is measured from the beginning of \ik science operations. The green $\times$ near the left side of all panels represent our fits to the TTs assuming constant period (Table \ref{tab:planetcatalog}).  
 \label{fig:Kep29} } 
\end{figure}

In Figure~\ref{fig:Kep29}, we see that \ik observed Kepler-29 when the transit-to-transit periods of Kepler-29 b and c were near their extrema. Four-year mean periods vary by a few parts in $10^{-4}$, with the periods of the two planets being highly anti-correlated.  The ratio of orbital periods over the \ik baseline is 1.2851, slightly (about 1 part in 2000) \emph{less than} that of the 9:7 commensurability (1.285714).  The average period ratio over 1000 years of the samples shown in Figure~\ref{fig:Kep29} are 1.285812 (17$^{\rm th}$), 1.285754 (51$^{\rm st}$), 1.285729 (85$^{\rm th}$) respectively, all just above commensurability. Indeed, much of the variation averages out on decadal time scales, but four year average period ratios oscillate about 9/7 and the long-term average period ratio in this system is several standard deviations away from the value  implied by the orbital periods for the planets listed in our catalog.  

\cite{Holczer:2016} found strong TTVs for both planets in KOI-738, but each planet's TTs were fit by a polynomial rather than a sine wave, so mean orbital periods were not estimated.  This fit was likely selected because the orbital periods of the planets oscillate with a periodicity of about ten years, and their TTVs during the four years of the \ik mission resemble parabolas; see the panels on the left in Fig.~\ref{fig:Kep29}.

\subsubsection{Kepler-419 = KOI-1474}

Two planets are known to orbit Kepler-419: a transiting inner planet (Table \ref{tab:planetcatalog}) on a quite eccentric orbit ($e \approx 0.8$) and a  non-transiting super-jupiter that has a period almost ten times as long but nonetheless induces large TTVs because of its high mass and the large eccentricity of the transiting planet's orbit.  The system is well characterized by both RV and TTV analysis (see Table \ref{tab:nontransiting} for the properties of the non-transiting planet). The transiting planet, Kepler-419~b, displays a quintessential ``photoeccentric effect'' \citep{Dawson:2012}, wherein a lower bound on the eccentricity of a planet can be estimated from the shape and duration of the transit lightcurve. The period ratio of these two planets is $\sim$~9.657. 

\begin{figure}[!hbt]
 \includegraphics [width = 2.2 in]{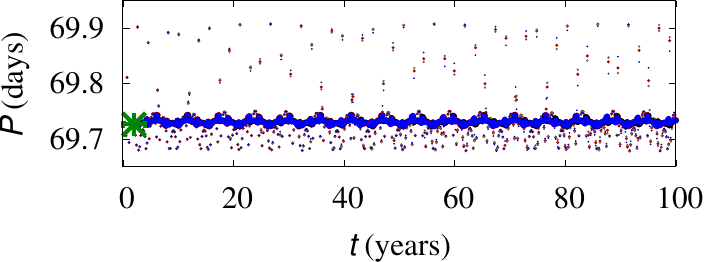}
  \includegraphics [width = 2.2 in]{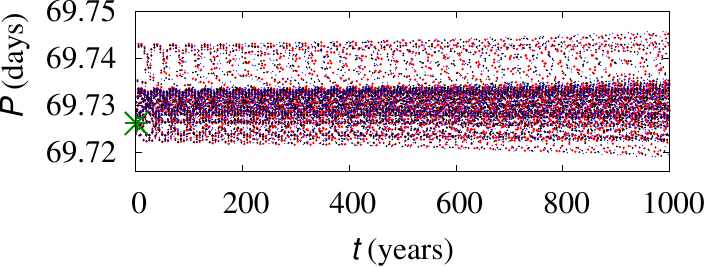}
  \includegraphics [width = 2.2 in]{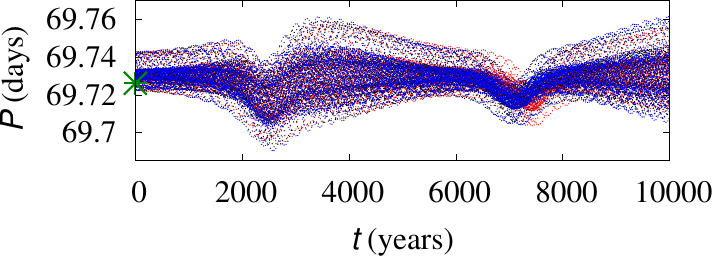}
 \caption{Transit-to-transit and 4-year average periods for Kepler-419~b, the only known \emph{transiting} planet in this system. We integrated 4044 samples of system parameters from \citet{Almenara:2018} for  $10^4$ years, then sorted them by averaged orbital period in Kepler-419~b, and display results from the 642$^{\rm nd}$ (blue), 2023$^{\rm rd}$ (black) and 3403$^{\rm rd}$ (red) members of the resultant list.  Left panel: the small points mark transit-to-transit period, while the larger points mark a 4-year running average period over 100 years. The right two panels include only the 4-year running averages over intervals of 1000 yr and $10^4$ yr, respectively. {The panel on the right has been thinned to show every fourth mean transit-to-transit interval.} 
 \label{fig:Kep419Periods}} 
\end{figure}

Table \ref{tab:periods} lists the value of the osculating orbital period for Kepler-419~b at $T_{\rm BJD} = 2454958$ from \cite{Almenara:2018}.  We took the 4044 posterior samples of the system parameters from the combined RV/photodynamical fits of \cite{Almenara:2018} and ran simulations for 10,000 years. For each of these samples, we computed a long-term orbital period by taking a linear fit to simulated transit times. We tabulate the average and standard deviation of these 4044 long-term periods. The transit-to-transit interval of the inner planet and its four year running average for three of the samples are plotted in Figure~\ref{fig:Kep419Periods}.  Note that the plots in this figure cover longer timescales than those for other planets considered in this subsection because the longer orbital periods of the planets lead to larger characteristic time intervals over which major variations are observed.  



At the present epoch, Kepler-419~b is close to periapse when its transits. The dips in orbital period near 2700 and 7300 years occur when the planet passes through apoapse near the time when it transits. This illustrates the variations predicted from Equation (\ref{eqn:periodsII}) with $\varpi \ll 1$.

We examined the intervals between times when the center of the non-transiting planet Kepler-419~c was closest to the sky-projected location of its host star Kepler-419 and closer to the Solar System than is Kepler-419. These intervals varied with the same periodicity and opposite phase as the corresponding variations of the averaged transit-to-transit period of Kepler-419~b. However, unlike its transiting companion (right panel of Fig.~\ref{fig:Kep419Periods}), this planet's variations are nearly sinusoidal. These behaviors are consistent with \cite{Almenara:2018}'s findings that the  periapse locations of the two planets oscillate about anti-alignment with a small amplitude and the orbital eccentricity of Kepler-419~c is small ($e < 0.2$) whereas Kepler-419~b has $e \approx 0.8$. 

\subsubsection{Kepler-60 = KOI-2086}

\begin{figure}[!hbt]
 \includegraphics [height = 0.9 in]{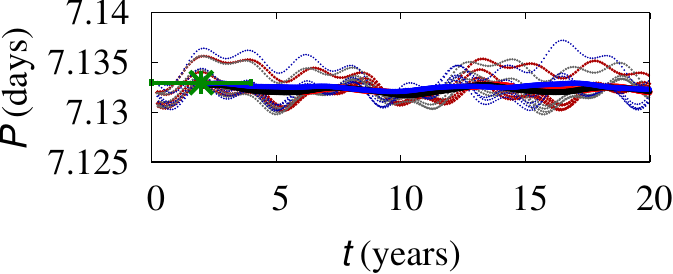}
  \includegraphics [height = 0.9 in]{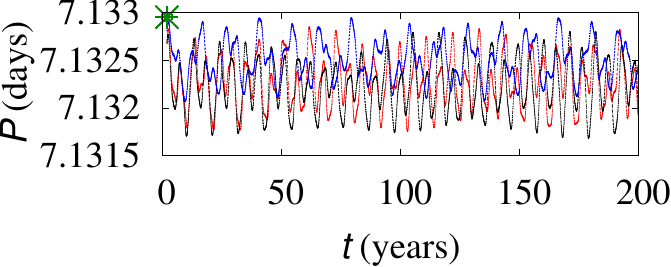}
  \includegraphics [height = 0.9 in]{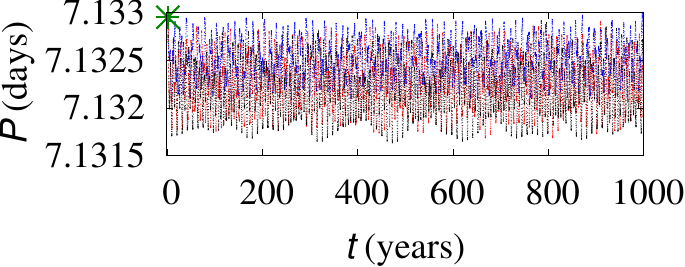}
  \newline
   \includegraphics [height = 0.9 in]{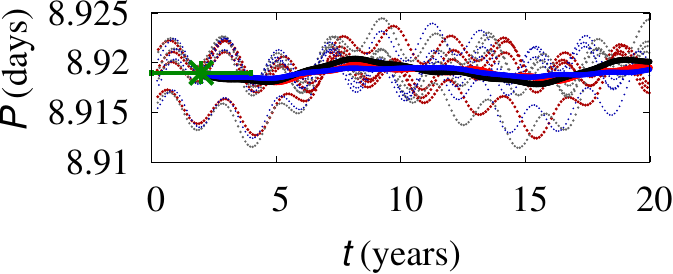}
  \includegraphics [height = 0.9 in]{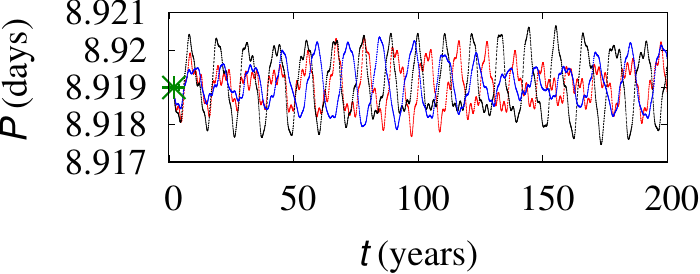}
  \includegraphics [height = 0.9 in]{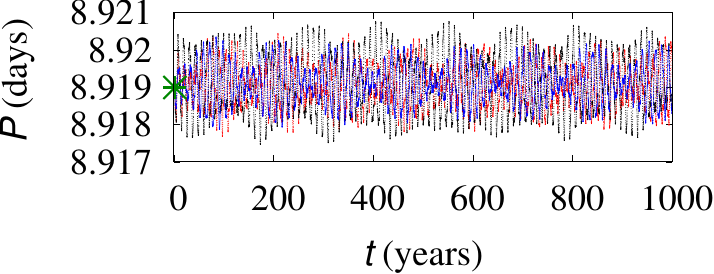}
  \newline
   \includegraphics [height = 0.9 in]{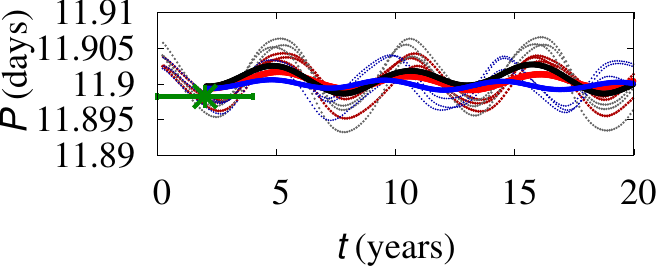}
  \includegraphics [height = 0.9 in]{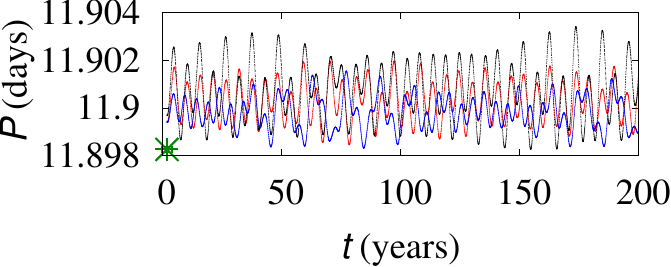}
  \includegraphics [height = 0.9 in]{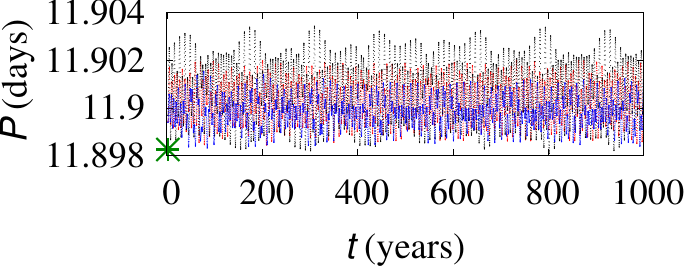} 
 \caption{Transit-to-transit and 4-year average periods for each of the planets known to orbit Kepler-60, ordered by increasing orbital period. The dots in panels on the left show transit-to-transit orbital periods from three samples (of the 101 considered) following dynamical fits to the long cadence transit times of \cite{Rowe:2015b}. The solid curves represent the average of 4-year segments centered on the given time. Black represents the sample with the median long-term (1000 years) average period of KOI-2086.01; blue the 17$^{\rm th}$ sample in the  list ranked by  average period and red the  85$^{\rm th}$ member of the list. Time is measured from the beginning of \ik science operations. The green $\times$ at the beginning represent our fits to the lightcurve assuming constant period (Table \ref{tab:planetcatalog}).   {The top panel on the right  has been thinned to show every fourth mean transit-to-transit interval, whereas the middle and bottom right panels show every third mean transit-to-transit interval.} 
 \label{fig:Kep60Periods}} 
\end{figure}

The Kepler-60 system contains three confirmed planets with periods between 7 and 12 days, with both neighboring pairs orbiting near first-order MMRs. The lightcurve also reveals an unverified planet candidate with an orbital period of 336 days that we do not consider in our analysis of the inner threesome.

We took 101 samples from the TTV posteriors of Kepler-60 from \cite{Jontof:2021}. From simulations of these 101 samples, we estimated uncertainties on the period ratios of Kepler-60 c/b (5:4) and Kepler d/c (4:3) given the $17^{\rm th}$, $51^{\rm st}$ and $85^{\rm th}$ period ratio after averaging period ratios over 1000-year simulations and sorting.  We found the inner pair have a period ratio of 1.250473 $\pm~0.000015 $ and the outer pair have period ratio 1.334176 $\pm~0.000027$. Averaged over four years, none of the neighboring planet pairs dropped below the ratio of small integers signifying their resonance in any of the three samples plotted in Figure \ref{fig:Kep60Periods}.

\pagebreak

\section{Conclusions} \label{sec:conclusions}

We have assembled in Table \ref{tab:planetcatalog} a catalog of \ik planet candidates that prioritizes completeness 
and makes use of additional information to improve accuracy whenever practical rather than providing a sample that has been defined and analyzed homogeneously, as done for the final PC catalog produced by the \ik Project \citep{Thompson:2018}. We have also listed an alternative set of planetary properties  (\S\ref{sec:unified}) for most planet candidates that inputs the more uniformly-derived set of stellar properties from \cite{Berger:2020a}.  \cite{Berger:2020a}'s results are available for $\sim 95$\% of \ik target stars that host one or more PCs {(and also for the vast majority of other \ik targets)}, and selecting the values listed in the appropriate columns in Table \ref{tab:planetcatalog} therefore yields measurements of planetary properties that are well-suited for studies of occurrence rates (see Appendix \ref{sec:App_ORtables} for details). {Figure \ref{fig:perrad} displays the planet candidates on the orbital period-planetary radius plane, showing the multiplicity of the system in which each PC resides. Figure \ref{fig:gallery} shows the periods of the planets in each of the multi-planet systems included in our catalog.}

Table \ref{tab:planetcatalog} presents an extensive set of stellar and planetary properties for each of almost 9700 KOIs (\ik Objects of Interest), almost half of which are considered viable planet candidates. Section \ref{sec:unified} provides a column-by-column list of types of data presented in Table \ref{tab:planetcatalog}, and more details on the derivation of many key planetary and stellar properties are provided elsewhere in \S\ref{sec:catalog}. 
A less comprehensive listing of the properties of 
non-transiting planetary companions to transiting \ik planets is provided in Table 
\ref{tab:nontransiting}; see Appendix \ref{sec:App_nontransiting}.

\begin{figure}
    \centering
    \includegraphics[scale=.175]{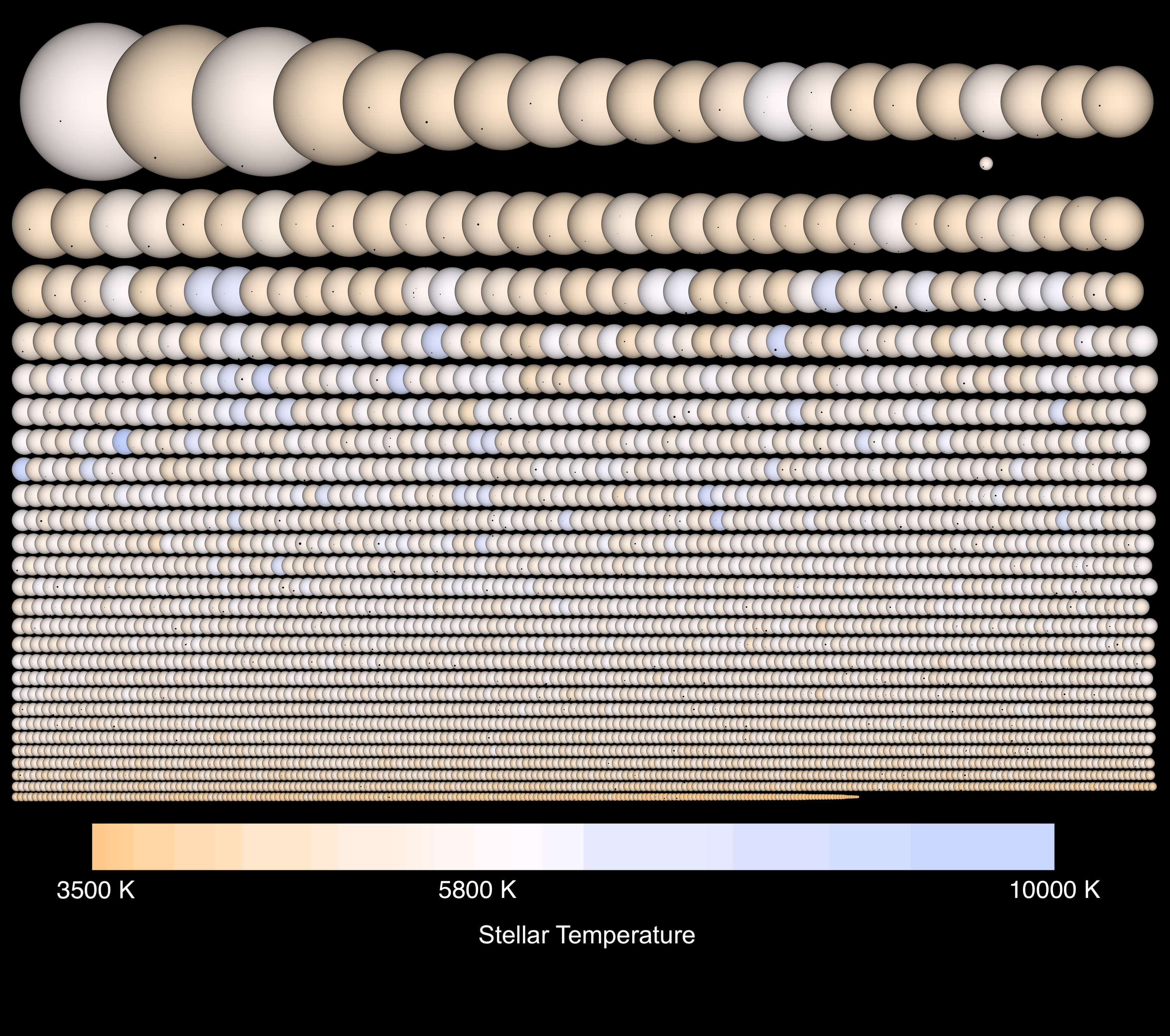}
    \caption{{Visualization of all planet candidates listed in Table \ref{tab:planetcatalog} transiting in front of their host stars.  The radii of all of the planets and stars are shown to the same scale, and the vertical distance of each planet from the center of its star shows the impact parameter of the transit. Systems are ordered by stellar size, and the color of each stellar disk represents the star's $T_{\rm eff}$, with the color scale shown below. The single star located between the top two rows on the right shows Jupiter and the Earth transiting the solar disk (for scale). The 12 host stars that have $T_{\rm} < 3500$~K and the 2 with $T_{\rm} > 10000$~K are represented by colors shown at the extrema of the scale bar. The version of this image included within the pdf manuscript is compressed to reduce the size of the file; a full-resolution version that clearly displays each of the transiting planets upon the disk of its host star is provided within the electronic version of the manuscript.}}
    \label{fig:transitingplanets}
\end{figure}

{Table \ref{tab:planetcatalog} is superior to previous cumulative catalogs of \ik planet candidates in that it provides a more complete listing of KOIs, more accurate and diverse dispositions of KOIs (for details see item \#64 in the list of tabulated properties provided in \S\ref{sec:unified}), and more accurate stellar and derived planetary properties. {Because we utilize information from previous \ik planet candidate catalogs, community studies and our own analyses, our assessments of dispositions should be at least as reliable as those of any previous \ik PC catalog for the portion of the sample listed in both catalogs.} The most substantial improvements in planetary properties are for orbital periods of planets exhibiting TTVs (\S\S\ref{sec:periods}, \ref{sec:Pconclusions}), as well as transit models and calculated radii of planets with grazing transits (\S\ref{sec:transitmodel}) and/or substantially-revised estimates of host star size.

Figure \ref{fig:transitingplanets} illustrates all planet candidates in our catalog as they transit their stars. This image is the successor to diagrams released as part of press packages for some of the official \ik project catalogs of planet candidates; we show it here to emphasize the fact that with substantial improvements in estimates of stellar radii and planetary impact parameters, the current version now has substantial scientific content.}

\subsection{Estimating Orbital Periods}  \label{sec:Pconclusions}

We have made special efforts, both in data analysis (\S\ref{sec:periods}) and theoretically (\S\ref{sec:periodtheory}), to improve the accuracy of planetary orbital periods to aid in ephemeris predictions and dynamical studies. The values of $P$ listed in Table \ref{tab:planetcatalog} are generally of equal or higher accuracy than those in previous tabulations, with estimates for many of the planets exhibiting TTVs significantly improved.
Fractional uncertainties quoted for orbital periods of the majority of planet candidates listed in recent \ik catalogs, including those presented herein, are $<10^{-5}$, with values $\sim 10^{-6}$ (corresponding to 2 minutes per 4 years) being typical.  These small uncertainties suggest {(in some cases misleadingly)} that ephemeris predictions for most \ik planets are robust for decades to come.  However, the tabulated periods and quoted uncertainties are for the mean times between midpoints of successive transits during the time interval in which transits were observed, and do not reflect possible long-timescale TTVs (Section \ref{sec:periodtheory}). 

Transit timing variations (TTVs) produce errors in estimates of some planets' orbital periods that need to be accounted for in certain dynamical investigations and ephemeris predictions.
Periodic sinusoidal TTVs with timescales short compared to the interval of \ik observations largely average out and do not produce significant errors in estimates of orbital periods.  TTVs with timescales comparable to the four year interval of \ik observations have been fit for dozens of \ik planet candidates to estimate long-term average orbital periods by \cite{Holczer:2016}, and more detailed dynamical models have been used to estimate long-term average periods of a small number of well-studied planets, including the seven systems presented herein (\S\ref{sec:individual}, Figures \ref{fig:Kep11}~--~\ref{fig:Kep60Periods}). 
{These figures show that in many cases the four-year average transit-to-transit orbital period deviates from the long-term average orbital period by a factor many times as large as the formal uncertainty of the four-year average.} Most \ik planets that show large TTVs are near mean-motion orbital resonances with other planets. The largest effect for planets moderately close to two-body resonances is due to rotation of the forced eccentricity vector by resonant perturbations; the timescale of this precession for most \ik planets is short compared to  the four years of \ik observations, so the variations tend to average out.

Libration of planets locked in resonances typically occurs on timescales longer than the \ik baseline, but most \ik planets do not appear to be resonantly-locked. The more general but smaller (during the era of \ik observations) effect is caused by secular precession of the planets' free eccentricities, which usually takes much longer than the four-year baseline of the \ik observations to complete a revolution, so it is not accounted for in estimates of mean periods or uncertainties.  This precession causes a discrepancy between \ik era mean orbital period and long-term mean orbital period to exist even for planets having TTVs that are too small to be observable during the epoch of \ik observations, and the magnitude of this discrepancy increases with the eccentricity of the planet's orbit. Thus, some systems/planets that do not show TTVs, for which an observer might assume that the linear ephemeris is reliable, could have deviations on longer timescales, although for most planets without clear TTVs the \ik predictions are likely to be very good. The main concern is systems where the variations on 4-year timescales are too small to be detected, but where the amplitude of the decadal timescale TTVs is substantial. {Because of the very complicated and poorly-quantified selection biases of the \ik sample, as well as small number statistics applying to some classes of dynamical configurations, we do not attempt to quantify the numbers of systems that exhibit the types of behavior seen within these systems.}
 Nonetheless, it should thus be kept in mind that the actual uncertainties in future TTs have broad (albeit low) tails that are not captured by tabulated uncertainties and are growing in length and height with time until additional transits are observed.

{We identify multiplanet candidates that have periods that are too close to each other to remain stable, which we use 
to estimate the percentage of apparent multiplanet systems wherein the planets are distributed  between two blended stars as $\sim~2.6\%$. Similarly, we use an error of a factor of two in the estimated period of a planet candidate that we identified by stability considerations to estimate the number of orbital periods that are aliases of the true periods to be $\sim~0.36\%$ (Section \ref{sec:falsemultis}). Other evidence suggests that a somewhat larger fraction of planets with estimated orbital periods of $\lesssim 1$~day actually complete two or more complete orbits within the period listed in catalogs (Section \ref{sec:selection}).} 

\subsection{Correlations Between Planetary Properties and Systems Multiplicity}  \label{sec:Mconclusions}

We find that the vast majority of PCs with low S/N are candidate single planets rather than being in multiplanet systems (multis).  In contrast, for those with moderate S/N, there are similar numbers of PCs in multis and in singles (Fig.~\ref{fig:snr_comparison}). Since the fraction of actual planetary detections that are in multis probably is similar for PCs with low S/N and those with moderate S/N, we suspect that a substantial fraction of these low S/N single planet candidates are false positives. 

An early \ik result was that the fraction of large planets among systems with multiple transiting planets is smaller than among lone transiting planets \citep{Latham:2011}. A few years later it was pointed out that  multis are more concentrated to the period range $1.6-100$ days than are singles \citep{Lissauer:2014,Rowe:2014}.   

The size distributions of singles and multis are quite similar over the range $\sim$~0.5~--~3~R$_\oplus$ (Fig.~\ref{fig:radius_multis_singles}), although there is a hint of a larger fraction of planets in multis below the radius valley at $
\sim 1.7$~R$_\oplus$. The size distributions of singles and multis are also indistinguishable over the range $\sim$~5~--~10~R$_\oplus$  (Fig.~\ref{fig:large_radius_multiplicities_split}), although the fraction of planets with sizes 5~--~10 R$_\oplus$ that are found in multis is only about two-thirds that among small planets; planets in the 5~--~10 R$_\oplus$ size interval range from low-mass super-puffs with tens of percent H/He by mass to cool giant planets hundreds or even thousands of times as massive as the Earth{; mature brown dwarfs and the smallest main sequence stars also fall within this size range}.  The transition between this multis/singles abundance ratio is gradual in the range 3~--~5 R$_\oplus$, which may imply a fuzzy boundary or simply be the result of errors in estimated planetary sizes (primarily small planets having sizes overestimated, since there are far more small planets than large ones). Placing the boundary between ``small'' and ``large'' planets near the size of Neptune is consistent with the results of the contemporaneous study of \cite{Ghezzi:2021}, who investigated correlations between stellar metallicity and maximum radius of observed transiting planets. Above 10 R$_\oplus$, the number of multis drops off steeply relative to singles, with few planets of radius $R_p > 12$~R$_\oplus$ found in multi-transiting system. The concentration of the largest planets in singles is partly due to inflated hot jupiters rarely having close companions, but the singles also appear to have inflated jupiters at longer periods~--~perhaps on eccentric orbits that bring them close.  Alternatively, it may be that most PCs significantly larger than Jupiter with $P>10$~days are FPs caused by eclipsing binary stars. Both the size distribution and the period distribution of planets in two-planet systems are intermediate between the distributions of single planets and those in systems with more than two transiting planets. 

\subsection{Planetary Eccentricities}
We analyze the distributions of normalized transit durations (Eq.~\ref{eqn:tdur}) to confirm the previous result that single transiting planets are more likely to have high eccentricity than are planets in multiply transiting systems. We extend this result by demonstrating that planets in systems with two transiting planets are typically more eccentric than those in systems with three transiting planets, and the orbits of PCs in systems with four or more transiting planets tend to be even less elongated (Fig.~\ref{fig:EccMulti}). 

Planets with orbital periods $P > 6$~days are typically more eccentric than  short-period transiting planets.  In contrast, we find no other clear trends in the eccentricity distribution with orbital period (Fig.~\ref{fig:EccPeriod}). 

Transiting planets in the rocky size range ($R_p <$~1.6~R$_\oplus$) have lower average $e$ than do sub-neptunes and neptunes, which in turn are typically less eccentric than planets with $R_p >$~5~R$_\oplus$ (Fig.~\ref{fig:EccRadius}). However, no such trend exists within the population of planets in systems with both large and small transiting planets.

\subsection{Epilogue}
It has been more than a decade since \ik ceased its collection of data from its prime field of view. Nevertheless, the list of \ik planet candidate remains the largest and most homogeneous collection of exoplanets known. Our new catalog contributes to the understanding of \ik planet candidates, especially with our focus on the information-rich systems with multiple transiting planets. Improvements in estimates of orbital periods and a better understanding of processes that alter apparent orbital periods on a variety of time scales advance our understanding of planetary dynamics and improve ephemerides for prediction of future transits. More accurate impact parameters, identification of correlations with multiplicity, and identifying trends with eccentricities also provide new avenues for research into the formation and evolution of planetary systems.

\begin{acknowledgements}

\section*{Acknowledgements}
We thank Travis Berger for informative discussions regarding the merits and shortcomings of various catalogs of \ik stellar properties and for providing tabulations of calculated stellar parameters in advance of publication.
Steve Bryson provided us with valuable information regarding the properties of the DR25Supp catalog.
Jose Manuel Almenara and Rosemary Mardling kindly provided data from their analysis of the Kepler-419 system that we used as input for our calculations of the long-term average period of Kepler-419~b.
Tony Dobrovolskis, Kevin Zahnle and two anonymous referees made constructive  comments on earlier versions of this manuscript. 

Primary support for authors in the USA was provided from NASA's Astrophysics Data Analysis Program 16-ADAP16-0034. 
J.F.R. acknowledges Canada Research Chair program and NSERC Discovery, which also covered funding for the publication of this article.
E.B.F. acknowledges NASA \ik Participating Scientist Program Cycle II grant NNX14AN76G, and NASA Origins of Solar Systems grant \#NNX14AI76G.  
The Center for Exoplanets and Habitable Worlds is supported by the Pennsylvania State University and the Eberly College of Science.  Compute Canada and Calcul Quebec computing resources supported the analysis of \ik observations.

\end{acknowledgements}

\vspace{5mm}
\facilities{\ik space telescope}

\smallskip

\appendix

\section{Tabulated Properties Useful for Planetary Occurrence Studies}\label{sec:App_ORtables}

As noted in Section \ref{sec:catalog}, our primary stellar and planetary properties catalog (first 63 columns of Table \ref{tab:planetcatalog}) prioritized accuracy over uniformity. In contrast, replacing the values given in columns 39~--~44 and 46~--~61 of Table \ref{tab:planetcatalog} by those given in columns 64~--~85 provides an analogous listing using the stellar parameters from \cite{Berger:2020a} for all KOIs whose stellar parameters are listed in \cite{Berger:2020a}'s tabulation. \cite{Berger:2020a} provides stellar properties of $> 90$\% of \ikt's targets, unbiased by whether or not they host planet candidates. Stellar properties from the \cite{Fulton:2018} catalog incorporated information from spectra taken by the Keck 1 telescope, which are only available for a tiny fraction of \ik target stars, most of which are KOIs and either host multiple PCs or are brighter than $K_p=14.2$. Most of the planetary properties listed in the Table \ref{tab:planetcatalog} that were derived using the stellar parameters in \cite{Berger:2020a} or are independent of stellar parameters were derived in a uniform manner that makes them suitable for use in planetary occurrence rate studies.  We provide specific recommendations for such studies in this Appendix.

When performing occurrence rate studies, we recommend that
researchers only include planetary systems associated with target stars that pass a uniform set of selection criteria that do not contain an
implicit dependence on presence of KOIs. We also recommend using those stellar and planetary properties that are listed in columns 64~--~85 of Table \ref{tab:planetcatalog}, which are based upon tabulations of \cite{Berger:2020a}, rather than the heterogeneous listing presented in columns  39~--~44 and 46~--~61. Additionally, our best-available dispositions of KOIs, given by the first letter of the 63$^{\rm rd}$ column in this table, were derived by heterogeneous methods, thus are not appropriate {to adopt without modification. Specifically, no KOIs other than those found and classified as planet candidates by a homogeneous and well-characterized process such as that used for DR25 should be counted. Nonetheless, it may be of use to include our dispositions of DR25 PCs in assessing the reliability of the sample, i.e., in rejecting some KOIs that were classified as PCs in DR25}. Analogously,
dispositions from the DR25 supplemental catalog should not be
used to add candidates that were not listed as such in DR25, as their selections, like our own choices, were not identified by a fully automated and reproducible process. 
The \ik DR25 planet catalog  \citep{Thompson:2018} is the premier catalog derived from a uniform and
systematic analysis of \ik lightcurves, and the dispositions from DR25 are given by the third letter of our the four-letter code given in the 63$^{\rm rd}$ column in Table \ref{tab:planetcatalog}.  

\cite{Hsu:2019}
performed an analysis of planet occurrence rates that makes
use of stellar properties from \gaia DR2.  The \cite{Hsu:2019}
target star criteria is just one example of a set of selection criteria
that do not have an implicit dependence on whether KOIs were
identified for a given target.  Future studies may wish to make use of
other large surveys (e.g., LAMOST, \gaia DR3 and beyond) that provide stellar information
for most of the \ik planet search targets.  When updating stellar
parameters, care must be exercised to update derived quantiles
self-consistently.  For example, the measured transit epoch, depth,
and duration do not depend on the stellar properties, but the inferred
planet size, semimajor axis, incident flux, and orbital inclination
would need to be updated to be consistent with the alternative set of
stellar properties.

While we recommend that the selection of planet candidates be based on
DR25 data products due to their automated detection and vetting process,
the planetary parameters from DR25 can be improved upon while
still maintaining a nearly homogeneous analysis, e.g., by using those values listed in the abovementioned columns of our Table \ref{tab:planetcatalog}.  In particular, the
DR25 planet properties table was based on a ``best fit" and did not
make use of MCMC simulations to characterize the uncertainties in
planetary properties.  While MCMC posterior samples were provided for
all DR25 planet candidates, these are not the basis for the catalog
values.  Therefore, statistical analyses can likely be improved upon
by updating the planet parameters with information from the MCMC
chains.   The MCMC posterior samples provided herein feature
important improvements that are advantageous for occurrence rates
studies.  First, our MCMC posterior samples correct a
bug that caused a biased distribution of impact parameters in the
previously released MCMC posteriors.  Additionally, our
results are derived from simultaneously fitting all the identified
planets in system, rather than by iteratively fitting one planet at a
time and masking out observations near transits of previous planets.
Therefore, we expect that the precision and accuracy of our MCMC
posterior samples represent an improvement on those originally
provided with \ik DR25.  Occurrence rate studies may thus
choose to update measured parameters (e.g., transit depths, durations,
impact parameters) with the results from this study.

\section{Supplemental Catalog of Non-transiting Planets}\label{sec:App_nontransiting}

The planet catalog presented in Table \ref{tab:planetcatalog} does not include circumbinary planets (CBPs) found by \ikt, nor does it list photometrically-identified non-transiting planets, nor non-transiting planets found around stars known to also host transiting planets found by \ikt.  For completeness, we provide references to lists of the first two classes of PCs and a tabulation of non-transiting companions to transiting \ik planets in this Appendix.  

Circumbinary transiting planets are searched for and analyzed quite differently from \ik planets that orbit around just one star (whether or not said star has more distant stellar companions).  Table 1 of \cite{Martin:2022} summarizes the properties of all 12 confirmed \ik CBPs. \cite{Welsh:2019}  identifies one additional candidate transiting \ik CBP.  Only one \ik multi-planet CBP system is known.

Phase variation photometry has been used to identify non-transiting hot jupiter candidates around stars that do not have transiting planet candidates.  A few of these candidates have been confirmed via radial velocity observations. See \cite{Lillo-Box:2021} and references therein for lists of these objects. 

Table \ref{tab:nontransiting} lists non-transiting \ik planets found from TTVs and RVs. Only planets with at least moderately well-constrained orbital periods are included;  planet candidates with poorly-constrained periods (from multiple possible TTV solutions or just a lower bound from RV data) are omitted.  All of the listed planets are in multiple planet (although not necessarily multi-transiting) systems, since the detection of one or more transiting planet(s) motivated further study.
Note that KOI 1442.10 is quite massive, and may be above the giant planet/brown dwarf boundary.

\begin{table}[!hbt]
    \centering
    \begin{tabular}{|c|c||c|c||c|c|}
        \multicolumn{6}{c}{\textbf{LIST OF NON-TRANSITING \ik PLANETS}}\\
        \hline
        Source & Method & KOI & Kepler- &  $P$ (days) &$M_p$ or $M_p \sin{i}$ (M$_\oplus$)  \\
        \hline
Y18 & RV  & 3 & 3~c & $3407^{+360}_{-190}$ & $507^{+30}_{-27}$ \\
B16  & RV  & 70 & 20~g &  $34.940^{+0.038}_{-0.035}$ &  $19.96^{+3.08}_{-3.61}$ \\ 
M17 & TTV+RV & 84 & 19~c & $28.731^{+0.012}_{-0.005}$ & 13.1 $\pm$ 2.7 \\
M17 & TTV+RV & 84 & 19~d & $62.95^{+0.04}_{-0.30}$ & $22.5^{+1.2}_{-5.6}$\\
M19 & RV & 85 & 65~e & $258.8^{+1.5}_{-1.3}$ & $260^{+200}_{-50}$\\
M14  & RV  & 104 & 94~c  &  820.3 $\pm$ 3  &  3126 $\pm$ 200   \\ 
W20 & TTV+RV & 142 & 88~c & $22.2649 \pm 0.0007$ & $214 \pm 5$\\
W20 & RV & 142 & 88~d & $1403 \pm 14$ & $965 \pm 44$\\
M14  & RV  & 148 & 48~e &  982 $\pm$ 8 &  657 $\pm$ 25   \\ 
E14 & RV & 214 & 424~c & 223 & 2215 \\
M19  & RV  & 244 & 25~d &  $122.4^{+0.8}_{-0.7}$ &  72 $\pm$ 10   \\ 
B23  & RV  & 246 & 68~d &  $632.62 \pm 1.03$ &  238 $\pm$ 5  \\ 
B23  & RV  & 246 & 68~e &  $3455^{+348}_{-169}$ &  86 $\pm$ 10  \\ 
B23  & RV  & 273 & 454~c &  $524.19 \pm 0.20$ &  1433 $\pm$ 38  \\ 
B23  & RV  & 273 & 454~d &  $4073^{+399}_{-186}$ &  $734 ^{+86}_{-51}$  \\ 
F19 & TTV & 448 & 159~d & $88.73^{+0.60}_{-0.05}$ & $121^{+5}_{-4}$\\
S17 & TTV & 872 & 46~c & $57.325^{+0.116}_{-0.098}$ &$115 \pm 5$\\
Fr19 & TTV & 880  & 82~f & 75.732 $\pm$ 0.012 & 20.9 $\pm$ 1.0  \\
N14 & TTV & 884 & 247~e & 60.05 $\pm$ 0.1 & $850 \pm 200$\\
O16  & RV  & 1241 & 56~d & $1002 \pm 5$ & $1784 \pm 120$\\
Q15 & RV & 1299 & 432~c & $406.2^{+3.9}_{-2.5}$ & $772^{+70}_{-76}$\\
M14  & RV  & 1442 & 407~c &  3000 $\pm$ 500 &  4000 $\pm$ 2000 \\ 
A18 & RV+TTV & 1474 & 419~c &  $673.35 \pm 0.84$ & 2432 $\pm$ 86\\
S19 & TTV & 1781 & 411~e & $31.509728 \pm 0.000085$ & $10.8 \pm 1.1$\\
 \hline
    \end{tabular}
    \caption{References (sources of the parameters reported herein, not necessarily the discovery paper): Y18 = \cite{Yee:2018}, B16 = \cite{Buchhave:2016}, M17 = \cite{Malavolta:2017}, M19 = \cite{Mills:2019}, M14 = \cite{Marcy:2014}, W20 = \citet{Weiss:2020}, 
    E14 = \cite{Endl:2014}, B23 = \cite{Bonomo:2023}, 
    F19 = \cite{Fox:2019}; substantially longer period and larger mass solutions for this planet are consistent with the data, but all of them both provide poorer fits to the TTV data and are less likely a priori based on planetary demographics, S17 = \cite{Saad:2017}, Fr19 = \cite{Freudenthal:2019}, N14 = \cite{Nesvorny:2014}
    , O16 = \cite{Otor:2016}, Q15 = \cite{Quinn:2015},  A18 = \cite{Almenara:2018}, S19 = \cite{Sun:2019}. Only the integer portions of KOI numbers are given; the \ik project's protocol was to use decimals beginning with .20 for non-transiting planet candidates, but the \cite{Marcy:2014} study used .10 and most other sources do not use any decimal numerical designators appended to KOI numbers for non-transiting planets.  Letters in most of the \ik names are those assigned by the authors, even if they didn't use \ik numbers; the 60 day period planet orbiting Kepler-247 is designated `e' despite it being referred to as KOI-884~c in the discovery paper because three transiting planets were announced prior to the publication of said paper. Planet masses are given for discoveries using TTVs; $M_p \sin{i}$ is listed for  RV  detections. 
}
    \label{tab:nontransiting}
\end{table}

The orbital periods of most non-transiting planets found by radial velocity measurements are not known to high precision, and the ability to detect non-transiting planets from TTVs strongly depends on period ratios, leading to a biased sample. Also, radii of non-transiting \ik planets have not been measured. Thus, we don't  use any of the planets listed in Table \ref{tab:nontransiting} 
for our statistical studies, even when computing the multiplicity of the systems hosting their sibling transiting planets, nor are they represented in any of the figures within this article.

\end{document}